\documentclass[aps,prb,longbibliography, twocolumn,tbtags,superscriptaddress,floatfix]{revtex4-2}
\usepackage{amsmath,amssymb,amsthm}
\usepackage{mathrsfs} 
\usepackage{stmaryrd}
\usepackage{graphicx}
\usepackage{bbm}
\usepackage{color,soul}
\usepackage[T1]{fontenc}
\usepackage{ae,aecompl}
\usepackage{float}
\usepackage{verbatim}
\usepackage[colorlinks=true,urlcolor=blue,linkcolor=blue,citecolor=blue]{hyperref}
\usepackage{xspace}
\usepackage[dvipsnames]{xcolor}
\usepackage{mathtools,xparse}
\usepackage{multirow}

\DeclareMathOperator{\Tr}{Tr\,}
\DeclarePairedDelimiterX{\bra}[1]{\langle}{\rvert}{#1\,}
\DeclarePairedDelimiterX{\ket}[1]{\lvert}{\rangle}{\,#1}
\DeclarePairedDelimiterX{\makebraket}[1]{\langle}{\rangle}{#1}
\newcommand{\lket}[1]{\left. \left| #1 \right> \! \right>} 
\newcommand{\lbra}[1]{\left< \! \left<  #1 \right|\right.} 

\NewDocumentCommand{\braket}{som}{
	\begingroup\activatebraketbar
	\IfBooleanTF{#1}
	{\makebraket*{#3}}
	{\IfNoValueTF{#2}{\makebraket{#3}}{\makebraket[#2]{#3}}}
	\endgroup
}
\makeatletter
\newcommand{\braketbar}{
	\,\delimsize\vert\@ifnextchar|{\!}{\,}}
\makeatother
\newcommand{\activatebraketbar}{
	\begingroup\lccode`~=`|\lowercase{\endgroup\let~}\braketbar
	\mathcode`|="8000
}

\begin{document}
	
	\title{Reflections on Quantum Reflectometry: \\Quantum and Tunneling capacitances as well as Sisyphus and Hermes resistances} 
	\date{\today }
	\author{O.~Yu.~Kitsenko}
	\email{kitsenko.sasha1212@gmail.com}
	\affiliation{B. Verkin Institute for Low Temperature Physics and Engineering of the National Academy of Sciences of Ukraine, Kharkiv
		61103, Ukraine} 
	\affiliation{V.~N.~Karazin Kharkiv National University,
		Kharkiv 61022, Ukraine}
	\author{S.~N.~Shevchenko}
	\email{sshevchenko@ilt.kharkov.ua}
	\affiliation{B. Verkin Institute for Low Temperature Physics and Engineering of the National Academy of Sciences of Ukraine, Kharkiv
		61103, Ukraine}
	\affiliation{Department of Mathematics, Kyiv School of Economics, 03113 Kyiv, Ukraine}
	\author{L.~Peri}
	\affiliation{Quantum Motion, London, UK}
	\affiliation{Cavendish Laboratory, University of Cambridge, Cambridge, UK}
	\author{Franco~Nori}
	\affiliation{Quantum Computing Center, RIKEN, Wako-shi, Saitama, 351-0198,
		Japan} 
	\affiliation{Physics Department, The University of Michigan, Ann
		Arbor, MI 48109-1040, USA}
	
	\begin{abstract}
		When a quantum electronic device is coupled to an electrical resonator, admittance changes of the quantum subsystem may be detected. The effective reactance may include capacitive and inductive terms that incorporate geometric, quantum, and tunneling components; while the effective resistance may be composed of Sisyphus and Hermes terms linked to relaxation and decoherence, respectively. Such reflectometry is usually studied when all characteristic times of the quantum system are much
		shorter than the resonator's period, in which case only stationary quantum
		states are probed. We present a
		rigorous description of a driven-dissipative qu$d$it-resonator system. Our
		approach demonstrates how to strictly introduce quantum and tunneling
		capacitances as well as Hermes and Sisyphus resistances, and how these values are modified when the dynamics of the
		subsystems becomes mutually dependent. We present the cases of a
		Cooper-pair box, a single-Cooper-pair transistor, a double quantum dot, and a single-electron box. Our
		approach can be applied to describe any quantum system coupled to any
		classical resonator.
	\end{abstract}
	\maketitle

	\section{Introduction}
	
	\label{Sec:Introduction}
	
	\subsection{Effective reactance and resistance}

	The issue of coupling quantum systems to classical ones appeared from the very first days of quantum physics. One notable example is the photon box thought experiment~\cite{Haroche2006}. This thought experiment, originally proposed by Einstein, was part of his effort to demonstrate that quantum mechanics is incomplete. He aimed to challenge the uncertainty principle by suggesting that, in principle, one could measure both the energy of the photon and the exact time of its emission with arbitrary precision. The idea was that by weighing the box before and after the photon escapes, using the deflection of the box suspended on a \textit{classical } spring, one  could determine the photon's energy. Simultaneously, by controlling the timing of the shutter, it would be possible to  specify the time of emission, apparently violating the time-energy uncertainty relation.  Bohr countered using general relativity: weighing the box causes gravitational time dilation, so localizing its center of mass introduces time uncertainty, preserving the principle.  While such thought experiments could not be realized a century ago, when quantum mechanics was still being formulated, experimentalists now routinely work with coupled classical and quantum systems.

	In contrast to natural quantum systems (like photons), artificial ones
	are often operated dynamically and are being coupled to driving and readout 
	electronics~\cite{Zagoskin2011, Huard2014, Shevchenko2019, Boutin2025}. In this work we are
	interested in dissipative semiclassical systems, where a classical electromagnetic field is used to probe a quantum electronic system.  A driven quantum system is the object of
	which the state is monitored by a resonant circuit  by means of measuring its
	response to a weak probe signal~\cite{Johansson2006, Sillanpaeae2006, Vigneau2023}.
	
	The quantum electronic systems considered in this work are built on a substantial body of prior research. The theoretical foundations of single-electron tunneling and Coulomb blockade in mesoscopic circuits were established in the late 1980s and 1990s~\cite{Averin1986, Likharev1996, Averin1991, Likharev1999, Schoen1990, Schoeller1994}, and were subsequently extended to Josephson-junction-based devices and quantum-state engineering with superconducting circuits~\cite{Makhlin2001, Devoret2000}. 
	
	The development of circuit quantum electrodynamics demonstrated that superconducting artificial atoms can be coherently coupled to microwave resonators, establishing a powerful framework for quantum information processing~\cite{You2005, You2011, Nigg2012, Gu2017, Krantz2019, Kockum2019a, Kjaergaard2020, Blais2021}. In parallel, semiconductor quantum dots confining single electrons or holes have emerged as a complementary platform~\cite{Zwanenburg2013, Burkard2023}. 
	
	Across 	all these platforms, radio-frequency reflectometry has established itself as the primary technique for fast and sensitive readout of quantum devices, from its early realization with the RF single-electron transistor~\cite{Schoelkopf1998} to its broad application in gate-based sensing of quantum dots, spin qubits, and topological devices~\cite{House2016, Zirkle2020, Filmer2022, Frake2015, Zhang2024, Loo2026, Driel2024}.
	
	While the resonant circuit serves as a classical probe, the effective reactance and resistance extracted from its response are properties of the quantum subsystem itself. These effective parameters emerge as a linear-response characterization of the driven quantum system to a weak classical probe and can be accessed experimentally through the measurable response of the coupled resonator. 
	
	For example, for capacitive coupling between the quantum and classical subsystems, the resonator's response is defined by the induced charge in the quantum system. A quantum two-level system with ground $\left\vert E_{-}\right\rangle $ and excited state $\left\vert E_+\right\rangle $ will exhibit two corresponding capacitance responses $C_\text{Q}^{(-)}$ and $C_\text{Q}^{(+)}$, respectively. It is further intuitively clear that, within the linear response of the quantum system, the impact on the circuit of a superposition state is defined by the probabilities of the respective cases, $p_{-}$ and $p_{+}=1-p_{-}$, so that $C_\text{Q} =  p_{-} C_\text{Q}^{(-)}+p_{+} C_\text{Q}^{(+)}$. Similarly, for a $d$-level system, instead of $p_{\pm}$ the expressions will contain occupation probabilities of each energy level $p_i$.
	
	In such a quantum-mechanical system, the effective capacitance may have three contributions: 
	
	(i) a term proportional to the state probabilities, $p_i$, the so called \textit{quantum capacitance} $C_\text{Q}(p_i)$, 
	
	(ii) a term proportional to the derivative of the probabilities $p'_i$, the so-called \textit{tunneling capacitance} $C_\text{T}(p'_i)$, and 
	
	(iii) the \textit{geometric capacitance}, $C_\text{geom}$ (defined as if there were no quantum effects). 
	
	 We note that, if the coupling is inductive, one would introduce corresponding additions to the inductance~\cite{Shnyrkov2006, Shevchenko2008, Frey2012, Shnyrkov2009a, Shnyrkov2012}, while otherwise the consideration is similar.
	
	While capacitive components have been extensively studied, the resistive contributions are less known. When the quantum system is coupled to a dissipative environment, it acquires a non-trivial \textit{effective resistance}~\cite{Persson2010, GZ2015, Shevchenko2015, Peri2023b}. Historically, a term related to quantum state relaxation has been dubbed the \textit{Sisyphus resistance}~\cite{Grajcar2008, Persson2010, GZ2015, Malinowski2022}. The Sisyphus mechanism of dissipation occurs when periodic driving partly excites the system out of its instantaneous ground state, after which environmental relaxation returns it back, resulting in repeated excitation-relaxation cycles and net energy loss per period, analogous to the myth of Sisyphus~\cite{CohenTannoudji1990}. 
	
	In addition, a term related to decoherence has been dubbed the \textit{Hermes resistance}~\cite{Peri2023b}. As we show in this work, the Sisyphus term $R_{\mathrm{Sis}}(p_\pm', T_1)$ depends on the derivatives of the state populations and on the relaxation times of the system, whereas the Hermes resistance $R_{\mathrm{Hrm}}(p_\pm, T_2)$ depends directly on the populations themselves and on the decoherence times of the system. Similarly to the Sisyphus mechanism, the Hermes process results in energy dissipation, arising, however, from the cost of maintaining quantum coherence --- an ever-turning ($\pi$o$\lambda$$\acute{\upsilon}$$\tau$$\rho$o$\pi$o$\varsigma$) cycle of coherence formation and dephasing.

\begin{figure}[t]
	\includegraphics[width=8.7 cm]{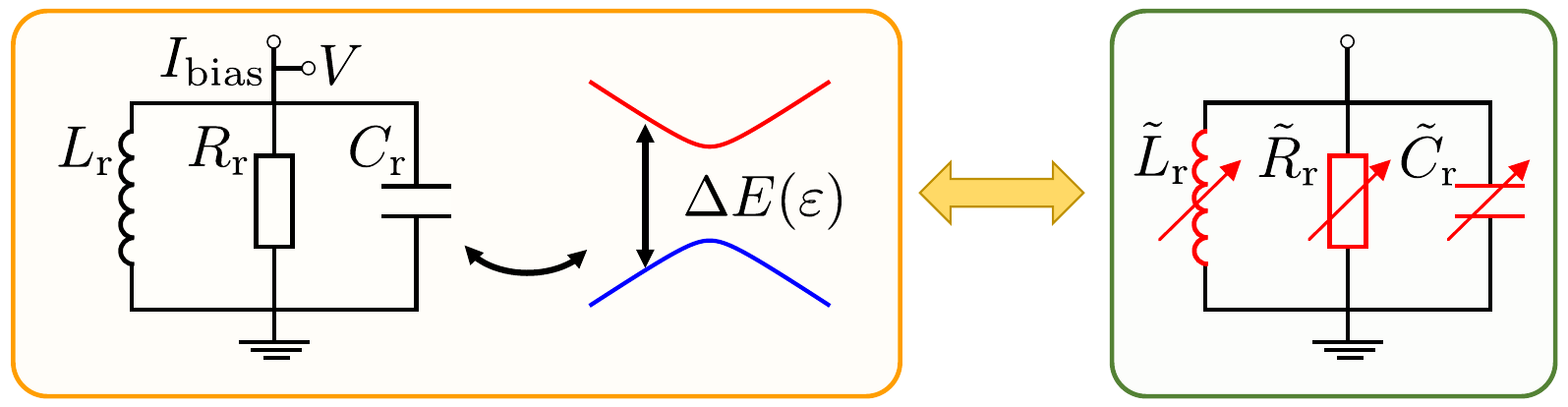}
	\caption{\textbf{Quantum reflectometry}. A classical resonator (with capacitance $C_{\text{r}}$,
		inductance $L_{\text{r}}$, and resistance $R_{\text{r}}$) is coupled to a generic quantum
		system. While the latter can be a multi-level system (see Sec.~\ref{Sec:Qudit}), here we show only a two-level system, with the energy levels
		separated by $\Delta E$, controlled by the energy bias $\protect\varepsilon $%
		. The resonator is biased by the current $I_{\mathrm{bias}}$ and the
		reflected signal is probed via the voltage $V$. The aggregate system can be
		described as an effective circuit to the right with the modified elements defined as $\tilde{C}_\text{r} = C_\text{r}+C_\text{eff}$, $\tilde{R}^{-1}_\text{r} = R^{-1}_\text{r}+R^{-1}_\text{eff}$, $\tilde{L}^{-1}_\text{r} = L^{-1}_\text{r}+L^{-1}_\text{eff}$, where the effective capacitance $C_\text{eff}$, inductance $L_\text{eff}$, and resistance $R_\text{eff}$ are associated with the quantum system.}
	\label{Fig:intro}
\end{figure}

	%
	Importantly, such reduced description of the quantum system impact on a
	classical resonator can be introduced only  if all the characteristic quantum
	times are much smaller than the resonator's period. Arguably, the simplest way to see this mathematically is the Krylov-Bogolyubov asymptotic expansion technique~\cite{Bogolyubov-Mitropolskii, Shevchenko2012}.
	This is schematically demonstrated in Fig.~\ref{Fig:intro}, which shows the
	essence of the asymptotic expansion technique: the nonlinear system (orange
	box to the left in Fig.~\ref{Fig:intro}) can be reduced to oscillations in
	an equivalent linear system (green box to the right in Fig.~\ref{Fig:intro}).
%

\subsection{Reasoning behind the definition of effective circuit elements}
Consider now the literature which will bring us to the formulation of the problems to solve. The notion of a quantum capacitance can be introduced in different ways, as e.g. in Refs.~\cite{Zagoskin2011} and~\cite{Vigneau2023}. It seems reasonable to call a quantum capacitance what appears as an addition to a classical geometric $C_{\mathrm{geom}}$ component. However, this quantum correction can be split into two components, the (pure) quantum capacitance $C_{\mathrm{Q}}$ and the tunneling capacitance $C_{\mathrm{T}}$, for a review see Ref.~\cite{Vigneau2023}.

To  avoid confusions here, following Refs.~\cite{Gonzalez-Zalba2016, Mizuta2017, Esterli2019, Vigneau2023}, we introduce both corrections, $C_{\mathrm{Q}}$ and $C_{\mathrm{T}}$, for the total effective capacitance $C_\text{eff}$, and define them as follows: the \textit{quantum capacitance} $C_{\mathrm{Q}}$ is the contribution defined by the adiabatic charge transitions and non-zero curvature of the energy bands; the \textit{tunneling capacitance} $C_{\mathrm{T}}$ is defined by the population redistribution processes. As a result of such definitions~\cite{Gonzalez-Zalba2016}, the former contribution, $C_\text{Q}$, is proportional to the quantum state occupation probability and the latter, $C_\text{T}$, is proportional to its derivative in respect to the energy bias, so the full effective capacitance can be written as
\begin{equation}
	C_{\mathrm{eff}}=C_{\mathrm{geom}}+C_{\mathrm{Q}}(p_\pm)+C_{\mathrm{T}
	}(p_\pm').  \label{Ceff}
\end{equation}
Equation~(\ref{Ceff}) will be described in more detail further down. 

The quantum capacitance $C_\text{Q}$ was studied in superconducting Cooper-pair boxes~\cite{Sillanpaeae2005, Duty2005, Ivakhnenko2023}, quantum-dot based transistors~\cite{Gonzalez-Zalba2016}, two-dimensional materials such as graphene~\cite{Xia2009}, carbon nanotubes~\cite{Ilani2006}, Majorana-based topological qubits~\cite{Peri2023b, Boutin2025, Microsoft2025, Loo2026}, surface electrons on liquid helium~\cite{Jennings2025}, double quantum dots~\cite{Prosko2024}, and quantum-dot arrays~\cite{Secchi2023,Secchi2023_1}. The quantum and tunneling components, both capacitive and inductive ones, were argued to be useful for non-destructive readout of quantum states~\cite{Sillanpaeae2005}. As a special point, we note that both the quantum and tunneling capacitances, as well as the quantum inductance, can be significantly \textit{larger} than the corresponding geometric components~\cite{Johansson2006, Shevchenko2008b}.

The dissipative contribution associated with relaxation processes in a quantum system was denoted as the Sisyphus resistance~\cite{Grajcar2008, Nori2008, Persson2010, GZ2015, Malinowski2022}. A more detailed analysis also reveals the emergence of the Hermes resistance~\cite{Peri2023b}. In this work, we demonstrate that, similarly to the decomposition of the effective capacitance into quantum and tunneling components, the effective conductance can be written as
\begin{equation}
	R_\text{eff}^{-1}= R_\text{Hrm}^{-1}(p_\pm,T_2)+R_\text{Sis}^{-1}(p_\pm',T_1), \label{Qubit_R_eff}
\end{equation}
where the Hermes resistance arises as a consequence of decoherence processes in the system and is proportional to the occupation probabilities of the energy levels; while the Sisyphus resistance originates from relaxation processes and depends on the derivatives of these probabilities~[see Secs.~\ref{Sec:Theory} and \ref{Sec:Qudit}].

What we described above mostly relates to semiclassical studies, when a quantum system is much faster than a slow resonator, treated as a classical circuit. Recently, similar studies~\cite{Peri2023a, Peri2023b, Peri2025} were done to describe both the quantumness of the resonator and strong coupling to the environment.

To summarize, in most studies of quantum reflectometry the authors
naively or phenomenologically introduce the quantum capacitance, inductance,
and/or resistance, ignoring the \textit{dynamics} of the quantum subsystem. More
elaborated studies give either numerical or bulky results. This brings us to the questions: how to correctly obtain the effective
capacitance, inductance, and resistance? How to extend these considerations to multi-level systems? How are the effective circuit elements influenced by relaxation and dephasing,
by temperature, and by driving? It is thus
important to develop a natural, universal, and systematic theory
of quantum reflectometry, which is the subject of our research here.

\begin{table}[t]
	\centering
	\renewcommand*{\arraystretch}{3}
	\begin{tabular}{|c|c|c|}
		\hline \textbf{Regime} &  \textbf{validity} &  $C_\text{eff}$ \\ \hline

		good qubit &  $T_{1}^{-1} \ll \omega_{\text{rf}} \ll \omega_\text{q}$ &  
		$-\overline{\dfrac{\partial^2 E}{\partial V^2 }}_\text{th}$
		\\  \hline
		
		\parbox[m]{5em}{bad qubit} &  $\omega_\text{rf} \ll \omega_{\text{q}} \ll T_{1}^{-1} $ & $ -\dfrac{\partial }{\partial V }\overline{ \dfrac{\partial E}{\partial V }}_\text{th} $ \\ \hline
		
		ideal qubit & $T_{1,2} \rightarrow \infty,\, k_\text{B}T \rightarrow 0$   &$ -\overline{\dfrac{\partial^2 E}{\partial V^2 }}_\text{S} $ \\ \hline	
		
		good qu$d$it &  $T_{\mathrm{rel},i}^{-1} \ll \omega_\text{rf}\ll \omega_{\text{qd}}$ & $ -\overline{ \dfrac{\partial^2 E}{\partial V^2 }}_\text{th}  $ \\ \hline
		
		bad qu$d$it &  $\omega_\text{rf} \ll \omega_{\text{qd}} \ll T_{\mathrm{rel},i}^{-1}$  & $-\dfrac{\partial }{\partial V }\overline{ \dfrac{\partial E}{\partial V }}_\text{th} $ \\ \hline
		
		ideal qu$d$it &  $T_{\mathrm{rel},i},  T_{\mathrm{dec},j} \rightarrow \infty,\,\, k_\text{B}T \rightarrow 0$ & $ -\overline{\dfrac{\partial^2 E}{\partial V^2 }}_\text{S} $  \\ \hline
	\end{tabular}
	\caption{\textbf{Brief summary for the effective capacitance} $C_\text{eff}$ of qubits and multi-level systems~(=qu$d$its). The table shows the formulas for the effective capacitance in various limiting cases.  Here $V$ is a classical probing voltage that perturbs the quantum system. This voltage is generated by a driven $RLC$ resonator with a period $T_\text{rf} = 2\pi\omega_\text{rf}^{-1}$.  Here, $E$ represents the energy levels $E_i$ of a quantum system, the definitions of averages $\overline{A}_{\text{S},\text{th}}$ are given in the main text. The times $T_{1,2}$ denote the relaxation and decoherence times of the qubit, $T_{\mathrm{rel},i}$ and $T_{\mathrm{dec},j}$ denote the corresponding times of a qu$d$it, and $\omega_{\text{q/qd}}$ is the characteristic qubit/qu$d$it frequency.  Note that in many papers the authors phenomenologically use what appears to be the expression for the non-coherent quantum system, which is valid only for  $\omega_\text{q/qd} \ll T_{1,2}^{-1}$, and thus requires a correction in the general case.}
	\label{Table1}
\end{table}

\section{Summary of the main results and structure of the paper}
\label{Intro:results}
Our approach to describe hybrid semiclassical systems is done in
four steps:

\begin{table*}[tp]
	\begin{center}
		\normalsize
		\centering
		\everymath{\displaystyle}
		\begin{tabular}{|p{9cm}|p{9cm}|}
			\hline
			\parbox[c]{9cm}{\centering\vspace{2ex}
				\fontsize{11}{13}\selectfont\textbf{Capacitance}
				\vspace{2ex}}
			&
			\parbox[c]{9cm}{\centering\vspace{1ex}
				\fontsize{11}{13}\selectfont\textbf{Conductance ($=1/\text{Resistance}$)}
				\vspace{1ex}}
			\\ \hline
			\rule{0pt}{4ex}\parbox[c]{9cm}{\centering\normalsize{$C_\text{eff}=C_\text{geom}+C_\text{Q}(\chi_\text{th})+C_\text{T}(\chi'_\text{th},T_1) $}} & \parbox[c]{9cm}{\centering\normalsize{$ R_\text{eff}^{-1}=R_\text{Hrm}^{-1}(\chi_\text{th},T_2)+R_\text{Sis}^{-1}(\chi'_\text{th},T_1)$}}\\[2ex] \hline

			\rule{0pt}{5ex}\parbox[c]{9cm}{\centering\normalsize{$C_\text{geom} = -\dfrac{\partial^2 E_\text{geom}}{\partial V^2}$, $\,\,\, C_{\text{Q}0} = \dfrac{(\alpha e)^2}{2\Delta}$, $\,\,\, \alpha =\dfrac{1}{e} \dfrac{\partial \varepsilon}{\partial V}$}} & \parbox[c]{9cm}{\centering\normalsize{$R_{\text{Q}0}^{-1} = \dfrac{\pi}{2} R_\text{K}^{-1}
				\alpha^2, \,\,\,\,\,\, R_\text{K}^{-1}=\dfrac{e^2}{h}$}}\\[3ex] \hline

			\rule{0pt}{5ex}\parbox[c]{9cm}{\centering\normalsize{$C_\text{Q}=
				C_{\text{Q}0}\dfrac{\Delta ^3}{\Delta E_0^3} \chi_\text{th}$, $\,\,\, \chi_\text{th} = \tanh \left(\dfrac{\Delta E_0}{2k_\text{B}T}\right)$}} & \parbox[c]{9cm}{\centering\normalsize{$R_\text{Hrm}^{-1}= 2 R_{\text{Q}0}^{-1} \dfrac{\hbar \omega_\text{rf}}{\Delta}
				\dfrac{\Delta^3}{\Delta E_0^3}  \chi_\text{th} \dfrac{T_2 \omega_{\text{rf}}}
				{1+T_2^2 \omega_{\text{q}0}^2}$}} \\[3ex] \hline
			\rule{0pt}{5ex}\parbox[c]{9cm}{\centering\normalsize{$C_{\text{T}}=C_{\text{Q}0}\dfrac{\Delta \, \varepsilon_0}{\Delta E_0}\chi'_\text{th} \dfrac{1}{1+T_1^{2}\omega_\text{rf}^2}$, $\,\,\, \chi_\text{th}' = \dfrac{\partial \chi_\text{th}}{\partial \varepsilon_0}$}} & \parbox[c]{9cm}{\centering\normalsize{$  R_\text{Sis}^{-1}= R_{\text{Q}0}^{-1} \dfrac{\hbar \omega_\text{rf}}{\Delta}\dfrac{\Delta \, \varepsilon_0}{\Delta E_0}\chi'_\text{th}
				\dfrac{ T_1 \omega_\text{rf}}
				{1+T_1^2 \omega_\text{rf}^2}$}} \\[3ex] \hline
		\end{tabular}
	\end{center}
	\caption{\textbf{Summary of the results for two-level systems}. The left column presents expressions related to the effective capacitance $C_\text{eff}$, and the right column corresponds to the effective conductance $G_\text{eff} = R_\text{eff}^{-1}$. In the first row, the general formulas for $C_\text{eff}$ and $R_\text{eff}^{-1}$ are given. The second row shows the geometrical capacitance $C_\text{geom}$ and the maximum quantum capacitance $C_{\text{Q}0}$, as well as the characteristic value of the effective conductance $R_{\text{Q}0}^{-1}$; $V$ is a classical probing voltage, applied to the quantum system. The last two rows list the expressions for the quantum and tunneling capacitances, and for the Hermes and Sisyphus conductances. Here $\Delta$ is the minimal energy splitting of the two-level system, $\varepsilon_0$ is the energy bias controlled by the dc component of the probing voltage, $\alpha$ is the lever arm, $R_\text{K}=h/e^2$ is the von Klitzing constant, and $\Delta E_0 = \sqrt{\Delta^2 + \varepsilon_0^2}$ is the difference of ground and excited states energies. The geometric energy $E_{\text{geom}}$, which gives rise to the geometric capacitance $C_{\text{geom}}$, is defined as the term proportional to the identity operator in the system's Hamiltonian. For further details see Sec.~\ref{Sec:Theory}.}

	\label{Table:TL_results}
\end{table*}

(1) start from the Lagrangian $\mathcal{L}$ of the full system and turn the formalism to Hamiltonian for quantum degrees of freedom with the help of the Routh function $\mathcal{R}$;

(2) quantize the Routh function $\mathcal{R} \rightarrow \hat{\mathcal{H}}$ to obtain the quantum Hamiltonian, and write down both classical and quantum 
equations of motion for the corresponding degrees of freedom;

(3) by considering the classical probing signal perturbatively, obtain analytical
results for quantum reflectometry introducing the effective capacitance,
inductance, and resistance;

(4) by comparing the numerical solution of the motion equations (in step 2) and
analytics (from step 3), clarify the validity of the phenomenological and
analytical results.

To illustrate the developed formalism, we apply it to a specific example --- a Josephson-junction-based charge qubit~\cite{Johansson2006}. Still, our approach is quite general: the main results were obtained \textit{for both isolated and open $d$-level systems} with an arbitrary dependence on the energy bias~$\varepsilon$.

In this work, we analytically solve the Schr\"{o}dinger equation for isolated systems and the Gorini-Kossakowski-Sudarshan-Lindblad (GKSL) equation~\cite{Breuer2010, Lidar2019, Manzano2020, Minganti2024a} for open systems, considering both two- and multi-level quantum systems. In what follows, we refer to the quantum system described by the Schr\"{o}dinger equation as an isolated or \textquotedblleft \textit{ideal}" system. 
The effective capacitance obtained from the Schrödinger equation is naturally referred to as the capacitance of the  isolated system. Unlike the ideal case, where the solution  is valid for any time $t$, the analytical solution obtained using the GKSL equation is valid only in the asymptotic limit $t \gg T_{\mathrm{rel},i}$, $T_{\mathrm{dec},j}$, where $T_{\mathrm{rel},i}$ and $T_{\mathrm{dec},j}$ denote the relaxation and decoherence times of a qu$d$it respectively. In the particular case of a qubit, these correspond to the standard relaxation and decoherence times $T_{1}$ and $T_{2}$. In this regime all coherent effects have vanished and the system has relaxed 
to a thermal equilibrium state which is weakly probed by a classical resonant circuit.

There are two possible asymptotic scenarios for open systems. 

(i) First, when both the characteristic qubit/qu$d$it frequency $\omega_{\text{q}/\text{qd}}$ and the probing frequency $\omega_\text{rf}$  are much smaller than the relaxation and decoherence rates \text{$\omega_\text{rf} \ll \omega_{\text{q}/\text{qd}}  \ll T_{\mathrm{rel},i}^{-1}, T_{\mathrm{dec},j}^{-1}$}, the system undergoes fast dissipation  and we refer to it  as a \textquotedblleft \textit{bad}" open system. As we show later, many previously known phenomenological results 
are indeed correct but only in this limiting case~\cite{ Duty2005,  Shevchenko2012, Gonzalez-Zalba2016, Prosko2024, Shnyrkov2006,  Shevchenko2008b,  Shevchenko2012a, Lundberg2024}; however, such phenomenology gives no dependence on the 
relaxation rates. In contrast, in this work we obtain results valid for arbitrary relaxation rates.

(ii) Second, the opposite limiting case \text{$T_{\mathrm{rel},i}^{-1}, T_{\mathrm{dec},j}^{-1} \ll \omega_\text{rf} \ll \omega_{\text{q}/\text{qd}}$}, which we refer to as \textquotedblleft \textit{good}", is of significant interest because of its applications to quantum computing and engineering.

The results of the different regimes studied in this paper are briefly summarized in Table~\ref{Table1}.  The notation $\overline{A}$ denotes the statistical average of a physical quantity $A$, defined as $\overline{A} = \sum_i p_i A_i$, where $p_i$ is the probability of the system occupying the state $\ket{E_i}$ with energy $E_i$. In particular, for a system in thermal equilibrium, $\overline{A}_{\text{th}} = \sum_i p_i^\text{th} A_i$, where the probabilities $p_i^{\text{th}}$ correspond to the thermal equilibrium distribution, while for an isolated system described by the Schrödinger equation, $\overline{A}_\text{S} = \sum_i p_i A_i$, where $p_i$ corresponds to the quantum-mechanical occupation probability of the eigenstate $\ket{E_i}$.

The ready-to-use analytical expressions for the effective capacitance and resistance of an arbitrary two-level system coupled capacitively to a classical  resonator, obtained from the solution of the GKSL equation, are summarized in Table~\ref{Table:TL_results}.

\begin{figure}[!t]
	\includegraphics[width=8.3 cm]{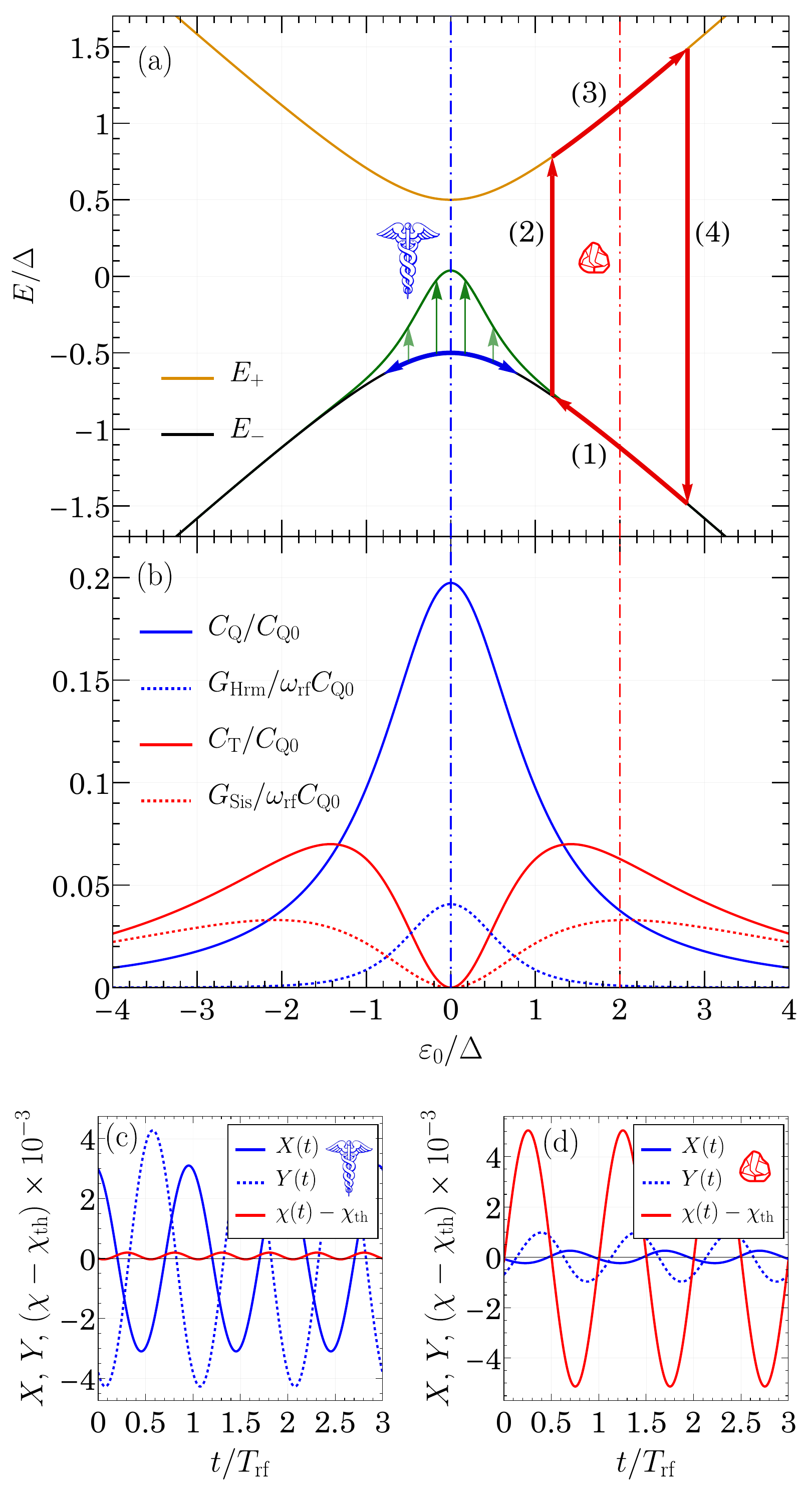}
	\caption{\textbf{\textquotedblleft Sisyphus" and \textquotedblleft Hermes" processes in a two-level system}. (a)~Energy-level diagram of a two-level system. Hermes- and Sisyphus-type processes are indicated in blue and red throughout. The Hermes process at around $\varepsilon_0/\Delta = 0$ is denoted by the blue caduceus symbol, and the Sisyphus process by the red stone symbol at, say, $\varepsilon_0/\Delta = 2$. The blue double-headed arrow and red arrows~(1)--(4) illustrate the Hermes and Sisyphus processes, respectively. The green curve and arrows schematically indicate thermal excitation leading to a partial population of the upper energy level. (b)~Corresponding Hermes ($C_\text{Q}$, $G_\text{Hrm}$) and Sisyphus ($C_\text{T}$, $G_\text{Sis}$) admittance components, calculated from Table~\ref{Table:TL_results}. (c),~(d)~Time dependence of the Bloch vector components $\vec{R}=(X,\,Y,\,\chi)$ for the Hermes ($\varepsilon_0/\Delta=0$) and Sisyphus ($\varepsilon_0/\Delta=2$) processes, respectively. The Bloch vector $\vec{R}$ is obtained from the numerical solution of the GKSL equation for the density matrix $\hat{\rho}=(\hat{\mathbf{1}}+\vec{R}\cdot\hat{\vec{\sigma}})/2$ in the instantaneous energy representation.  Parameters: $\hbar\Gamma^\text{c}/\Delta = 0.2$, $\hbar\gamma_\phi/\Delta = 0.01$, $k_\text{B}T/\Delta = 2.5$, $\delta\varepsilon_\text{rf}/\Delta = 0.03$, $\hbar \omega_\text{rf}/\Delta = 0.5$. For further details see Sec.~\ref{Sec:Theory}.}
	\label{Fig:response}
\end{figure}

As can be seen from Eqs.~(\ref{Ceff}, \ref{Qubit_R_eff}) and Table~\ref{Table:TL_results}, the quantum $C_\text{Q}$ and tunneling $C_\text{T}$ capacitances, and the Hermes $R_\text{Hrm}$ and Sisyphus $R_\text{Sis}$ resistances all arise from only two physically distinct processes, illustrated in Fig.~\ref{Fig:response}: the ``Hermes''- and ``Sisyphus''-type processes, dominant near the avoided crossing ($\varepsilon_0/\Delta \approx 0$) and away from this region ($|\varepsilon_0|/\Delta \gtrsim 1$), respectively. The physical origin of this distinction becomes transparent from the dynamics of the Bloch vector \text{$\vec{R} = (X, Y, \chi)$} shown in Figs.~\ref{Fig:response}(c, d). 

In the Hermes-type regime, the probe signal primarily drives the coherences $X$ and $Y$, while the population $\chi$ stays close to $\chi_\text{th}$ [see Fig.~\ref{Fig:response}(c)]. As a result, the system response is governed by the equilibrium populations $\chi_\text{th}$ and the decoherence time $T_2$ [see the third row of Table~\ref{Table:TL_results}]. 

In the Sisyphus-type regime the probe signal primarily changes $\chi$, driving it away from equilibrium $\chi_\text{th}$~[see Fig.~\ref{Fig:response}(d)]. The resulting dynamics traces the cyclic excitation--relaxation Sisyphus cycle~(1)--(4) in Fig.~\ref{Fig:response}(a). In this case, the system response is governed by the derivative of the equilibrium population $\chi'_\text{th}$ and the relaxation time $T_1$ [see the fourth row of Table~\ref{Table:TL_results}].

The rest of the paper is organized as follows. In Sec.~\ref{Sec:Naive} we review the phenomenological approach to introduce the effective capacitance as a  response to the dynamics of the quantum subsystem. In Sec.~\ref{Sec:Theory} we develop a rigorous theoretical approach to study the coupled dynamics of a quantum subsystem with a classical $RLC$ resonator. Using this formalism, we study the responses of the stationary states of the qubits as well as the superposition states either in thermal equilibrium or created by microwave resonant driving. Then, in Sec.~\ref{Sec:Qudit}, we present the effective capacitance and resistance results for isolated and open multi-level systems. Finally, Sec.~\ref{Sec:Other} is devoted to examples of other quantum systems, such as a single-Cooper-pair transistor, a double quantum dot and a single-electron box. The main text ends with the Conclusions. Appendices~\ref{AppendixA}--\ref{Appendix:DQD} present details of the calculations.

\section{Conventional approach to effective capacitance}
\label{Sec:Naive}
\subsection{Phenomenological introduction of the  effective capacitance}

\begin{figure}[t]
	\includegraphics[width=8 cm]{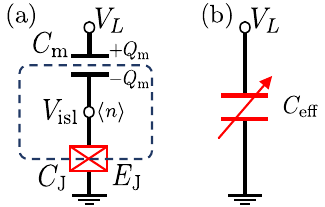}
	\caption{As an example, (a) shows a  Cooper-pair box built from a measurement capacitor $C_{\text{m}}$ and a Josephson junction with capacitance $C_{\text{J}}$ and Josephson energy $ E_{\text{J}}$. The state of the system can be probed via the voltage $V_L$, which couples the charge qubit with a $RLC$ resonator. (b) In the two-level approximation, the qubit can be represented as an effective capacitance $C_{\text{eff}}$. }
	\label{Fig2}
\end{figure}

For concreteness, we now consider a Cooper-pair box (CPB)~\cite{Johansson2006},
shown in Fig.~\ref{Fig2}, as a simple and suitable example of a
quantum subsystem. The superconducting island, with voltage $V_\text{isl}$, is formed by a Josephson junction (JJ) characterized by Josephson energy $E_{\text{J}}$ and capacitance $C_{\mathrm{J}}$, which is capacitively coupled to a measurement electrode via the capacitor $C_{\text{m}}$. In a similar manner, as will be shown in Secs.~\ref{Sec:Theory} - \ref{Sec:Other}, the effective capacitance and resistance for a
wide variety of other physical systems can also be easily derived.

We start by noting that the probing signal induces a charge on the top capacitor, $C_\text{m}$. 
The probing voltage, 
\begin{equation}
	V_L=V_{L0}+\delta V_{\text{rf}} \cos (\omega _{\text{rf}}t),
\end{equation}%
is assumed to contain a constant dc component $V_{L0}$ that tunes the
state of the qubit and an ac component $\delta V_{\text{rf}}
\cos( \omega _{\text{rf}}t)$, allowing the rf phase shift and amplitude change of the probing signal to be extracted. Following Refs.~\cite{Duty2005,Mizuta2017}, we define the effective capacitance as the first derivative of the charge $Q_\text{m}$  on the capacitor $C_\text{m}$  with respect to the voltage (i.e., the differential capacitance) 
\begin{equation}
	C_{\text{eff}}\ =\dfrac{\partial Q_\text{m}}{\partial V_{L}}.
\end{equation}

The total charge on the island is
\begin{equation}
	2en=-Q_\text{m}+Q_{\text{J}}=C_{\text{m}}(V_\text{isl}-V_{L})+C_{\text{J}}V_\text{isl},
\end{equation}
where $Q_\text{J}$ is the charge on the capacitance $C_\text{J}$ of the Josephson junction, and $V_{\text{isl}}$ is the voltage of the superconducting island. Thus, the average charge seen from the top electrode $\langle Q_\text{m}\rangle $ can be expressed in terms of the average number of Cooper pairs on the island $\langle n\rangle $, so that  the effective capacitance becomes
\begin{equation}
	C_{\text{eff}}
	=
	\underbrace{\dfrac{C_{\text{m}} C_{\text{J}}}{C_\Sigma}}_{C_{\text{geom}}}
	-
	\underbrace{2 e \dfrac{C_{\text{m}}}{C_\Sigma} \dfrac{\partial \langle n \rangle}{\partial V_{L} }}_{C_{\text{Q}} + C_{\text{T}}} \label{Cap_eff_naive}
\end{equation}
where $C_{\Sigma }=C_{\text{m}}+C_{\text{J}}$ is the total capacitance of the superconducting island  and $C_\text{geom} = C_\text{m} C_\text{J}/C_\Sigma$ is the geometric capacitance, representing the series combination of $C_\text{m}$ and $C_\text{J}$. 

To obtain expressions  for the quantum and tunneling capacitances one should know how  $\langle n\rangle $ depends on $V_{L}$. For this, in the next subsections we consider three important situations:  (i) non-driven isolated quantum system, (ii) an open quantum system at finite temperature, and (iii) a driven quantum isolated system.

\subsection{Effective capacitance of a non-driven isolated system}
\label{Sec:Phenomenology_isolated}
An instructive example of the emergence of the quantum capacitance can be provided by a simple two-level model.  When the Josephson energy $E_{
	\text{J}}$ is much smaller than the charging energy $E_{\text{C}
}=e^{2}/2C_{\Sigma }$, the superconducting island behaves as a well-defined
two-level system with energy levels given by~\cite{Shevchenko2019}  \begin{equation}
	E_{\pm }=\pm \frac{1}{2}\sqrt{\Delta ^{2}+\varepsilon_0 ^{2}},
\end{equation}
where  
\begin{equation}
	\Delta =E_{\text{J}}, \,\,\,\,\,\, \varepsilon_0 =4E_\text{C}(1-2n_{\text{g}0}) \label{Delta_and_varepsilon_0}
\end{equation}
and $n_{\text{g}0}=-C_{\text{m}}V_{L0}/2e$ being a dimensionless charge. The bias $\varepsilon_0$ is tuned by the dc component of the probing voltage $V_{L}$.

The expectation value of the charge operator $\langle \hat{n}\rangle $ for such qubit can be written as  
\begin{equation}
	\langle \hat{n}\rangle =\frac{1}{2}\left[ 1-\frac{\varepsilon_0}{\Delta E_0}\chi%
	\right] ,  \label{naive_n}
\end{equation}
where $\Delta E_0=E_{+}-E_{-}$ is the difference of energies and \text{$\chi=p_{-}-p_{+}$}, assumed constant here,
is the difference between occupation probabilities of the ground and excited
states. Substituting Eq.~(\ref{naive_n}) into Eq.~(\ref{Cap_eff_naive}) we obtain 
\begin{equation}
	C_{\text{eff}}=\dfrac{C_{\text{m}}C_{\text{J}}}{C_{\Sigma }}-\overline{\dfrac{\partial ^{2}E}{\partial V_{L}^{2}}}_S=C_{\text{geom}}+C_{\text{Q}}(\varepsilon_0),   \label{C_ideal_phenomenological}
\end{equation}
where the quantum capacitance is given by
\begin{equation}
	C_{\text{Q}}(\varepsilon_0)= C_{\text{Q}0}\dfrac{\Delta ^{3}}{\Delta
		E^3_0}\chi,\label{Phenomenology:quantum_capacitance}
\end{equation}
with
\begin{equation}
	C_{\text{Q}0}= 4\dfrac{C_{\text{m}}^{2}}{ C_{\Sigma } }\dfrac{E_\text{C}}{\Delta}=\dfrac{1}{2\Delta}\left(2e\dfrac{C_\text{m}}{C_\Sigma}\right)^2. \label{CQO_phenomenology}
\end{equation}
In particular, for the ground $\chi=1$ and excited $\chi=-1$ states, we obtain the capacitances of these states: 
\begin{equation}
	C_{\text{Q}}^{(\pm)}(\varepsilon_0)=\mp C_{\text{Q}0}\dfrac{\Delta ^{3}}{\Delta
		E^3_0}.
\end{equation}

The fact that for an isolated system the quantum capacitance of the $n$-th
level is proportional to the second derivative of the energy band, $C_{\text{%
		Q}}^{(n)}\propto \partial ^{2}E_{n}/\partial \varepsilon_0^{2}$, is a general
result that will be proven in Sec. \ref{Sec:Qudit} for arbitrary bias  dependence
in the Hamiltonian $\hat{\mathcal{H}}=\hat{\mathcal{H}}(\varepsilon )$. We note that the quantum capacitance $C_\text{Q}$ can be of the same order of magnitude or larger than the geometric capacitance $C_\text{geom}$ for typical parameter values, especially at the avoided crossing point, where it attains its maximum value $C_{\text{Q}0}$~\cite%
{Duty2005,Johansson2006}.

To summarize, in the phenomenological approach, one assumes that a classical $RLC$ circuit does not influence the dynamics of the quantum subsystem at all. This is why we can find the unperturbed evolution of the quantum subsystem and evaluate the quantum capacitance by taking partial derivatives with respect to the dc component of the probing voltage $V_{L0}$.
\subsection{Impact of temperature}
\label{Sec:Naive3}
When we take into account the interaction with the dissipative environment by solving  the two-level Bloch equations, the expectation value $\langle \hat{n} \rangle$ is still given by Eq.~(\ref{naive_n}) \cite{Shevchenko2015}, where the difference between occupation probabilities is given by thermal probabilities  $p_{\pm}^\text{th}$ with
\begin{equation}
	\chi=\chi_\text{th}=p_{-}^\text{th}-p_{+}^\text{th}=\tanh\left(\dfrac{\Delta E_0}{2k_\text{B} T}\right). \label{chi_thermal}	
\end{equation}
Then the phenomenological expression will coincide with the one in Ref.~\cite{Duty2005}
\begin{equation}
	C_{\text{eff}}^{\text{th}}=C_{\text{geom}}-\dfrac{\partial }{\partial V_{
			L}}\overline{\dfrac{\partial E}{\partial V_{L}}}_\text{th}  ,  \label{CQ_naive}
\end{equation}%
where $\overline{\partial E/\partial V_{L}}_\text{th} $ stands
for the Boltzmann-weighted average of the charge $\langle \hat{n}\rangle_\text{th}$~\cite{Shevchenko2015}. Such effective capacitance gives rise not only to a quantum capacitance
\begin{equation}
	C_{\text{Q}}(\varepsilon_0,T)= C_{\text{Q}0}\dfrac{\Delta ^{3}}{\Delta
		E^{3}_0} \tanh \left(\dfrac{\Delta E_0}{2k_\text{B}T}\right), \label{CQ_phenomenology}
\end{equation}
which is directly related to the previous expression~\eqref{Phenomenology:quantum_capacitance} by replacing $\chi$ with the thermal occupation $\chi_\text{th}$, but also to the tunneling one, which is proportional to $\partial \chi_\text{th}/ \partial V_{L}$,
\begin{equation}
	C_{\text{T}}(\varepsilon_0,T)=C_{\text{Q}0}\dfrac{\Delta}{2 k_\text{B}T}\dfrac{ \varepsilon_0^2}{\Delta E_0^2}\cosh^{-2}\left(\dfrac{\Delta E_0}{2 k_\text{B}T}\right). \label{CT_phenomenology}
\end{equation}

From Eqs.~(\ref{CQ_phenomenology}) and (\ref{CT_phenomenology})  it is clear that for low temperatures we obtain the quantum capacitance of the ground state and zero tunneling capacitance
\begin{equation}
	\dfrac{k_\text{B}T}{\Delta} \rightarrow 0 \,\,\, \Rightarrow \,\,\,
	\begin{cases}
		C_\text{Q}(\varepsilon_0,T) \rightarrow C_\text{Q}^{(-)}, \\ 
		C_\text{T}(\varepsilon_0,T)\rightarrow 0.
	\end{cases} \label{Ceff_limit_at_0_temperature}
\end{equation}%

\subsection{Microwave driving}
Furthermore, in addition to the weak classical probe signal $\delta \varepsilon_\text{rf}(t) = \delta \varepsilon_\text{rf} \cos( \omega_\text{rf} t)$, we consider the dynamics induced by a strong resonant drive $\varepsilon _{\text{d}}(t)=A\cos (\omega _{\text{d}} t)$, with the driving frequency $\omega _{\text{d}}$ close to the qubit frequency $\omega _{\text{q}}=\Delta E/\hbar $, when the system experiences Rabi oscillations. The effective capacitance is given by Eq.~(\ref{Cap_eff_naive}), where the difference $\chi$ of the energy level occupation probabilities will be given by the Rabi solution 
\begin{equation}
	\chi_\text{R}=\dfrac{\delta \omega^2}{\Omega _\text{R}^2},
\end{equation}
where $\Omega _\text{R}=\sqrt{\Omega^{2}+\delta \omega^2}$ is the generalized Rabi frequency, $\Omega =A\Delta /(2\hbar \Delta E_0)$ is the Rabi frequency at the resonance point, and $\delta \omega = \omega _{\text{d}}-\omega _{\text{q0}}$ is the frequency detuning. 

The effective capacitance now also consists of the quantum and tunneling parts,
\begin{equation}
	C_{\text{Q}}(\varepsilon_0,A,\omega_\text{d})= C_{\text{Q}0} \dfrac{\Delta^3 }{\Delta E_0^3}\dfrac{\delta \omega^2}{\Omega _\text{R}^2}, \label{CQ_drive}
\end{equation}
\begin{equation}
	C_{\text{T}}(\varepsilon_0,A,\omega_\text{d})= 2C_{\text{Q}0}  \dfrac{\varepsilon_0^2 \Delta}{\Delta E_0^3 } \dfrac{\Omega^{2} (\omega_\text{q}-\delta\omega)\delta\omega}{\Omega _\text{R}^4}. \label{CT_drive}
\end{equation}
If the detuning frequency $\delta \omega$ is much larger $\delta \omega \gg \Omega$ than the Rabi frequency $\Omega$, the above expressions [Eqs.~(\ref{CQ_drive}, \ref{CT_drive})] give zero tunneling capacitance, like in the thermal equilibrium case, while the quantum capacitance will tend to the quantum capacitance of the ground state, as in Eq.~(\ref{Ceff_limit_at_0_temperature}).

This phenomenological approach of defining an effective response of the quantum subsystem can give some useful  results for geometric, quantum, and tunneling capacitances. However, it is also important to study conditions of validity for the derived expressions: which restrictions on the amplitude $\delta \varepsilon _{\text{rf}}$ and frequency $\omega _{\text{rf}}$ of the probing signal and other parameters should be there in order for these results to be valid? Conventional phenomenological approaches cannot indicate whether these results can be applied or not for systems with large and small relaxation $T_{1}$ and decoherence $T_{2}$ times. Therefore, in Sections \ref{Sec:Theory} and \ref{Sec:Qudit} we will develop a rigorous theoretical description of the interaction between classical and quantum subsystems in order to establish the regions of validity, study the influence of the temperature $T$, relaxation and decoherence times $T_{1,2}$ for two- and multi-level systems when there is stationary probe and microwave driving.

\section{Coupled dynamics of classical and quantum subsystems}

\label{Sec:Theory} In this section, we develop the theory of interaction between classical and quantum subsystems and will use it to obtain the effective capacitance and resistance of qubits with and without resonant driving.

\subsection{Classical equations of motion}
\label{Sec:Theory:Hamilton}

To derive the coupled classical and quantum equations for an arbitrary system, we start from the classical Lagrangian $\mathcal{L}(\varphi_{\text{cl}}, \dot{\varphi}_{\text{cl}}, \varphi_{\text{q}}, \dot{\varphi}_{\text{q}})$, which depends on both classical $\varphi_{\text{cl},i}$ and quantum $\varphi_{\text{q},j}$ degrees of freedom, indexed by $i$ and $j$ sub indices, respectively. It is convenient to decompose the Lagrangian into two parts: a purely classical component $\mathcal{L}_\text{cl}(\varphi_{\text{cl}}, \dot{\varphi}_{\text{cl}})$, which does not depend on the quantum degrees of freedom, and a quantum part $\mathcal{L}_\text{q}(\varphi_{\text{cl}}, \dot{\varphi}_{\text{cl}}, \varphi_{\text{q}}, \dot{\varphi}_{\text{q}})$, which depends on both classical and quantum variables, such that $\mathcal{L} = \mathcal{L}_\text{cl} + \mathcal{L}_\text{q}$.

In the presence of both classical and quantum variables, the Lagrangian formulation must be partially transformed into the Hamiltonian formalism to enable the canonical quantization of the quantum degrees of freedom. This is achieved by introducing the Routhian function~\cite{Landau1976}
\begin{equation}
	\mathcal{R}(\varphi_{\text{q}}, p_{\text{q}}, \varphi_{\text{cl}}, \dot{\varphi}_{\text{cl}})
	= \sum_j p_{\text{q},j}\dot{\varphi}_{\text{q},j} - \mathcal{L}
	= \mathcal{H} - \mathcal{L}_\text{cl}, \label{Routhian}
\end{equation}
where $p_{\text{q},j} = \partial \mathcal{L}_\text{q} / \partial \dot{\varphi}_{\text{q},j}$. 
Upon decomposing the Routhian as $\mathcal{R} = \mathcal{H} - \mathcal{L}_\text{cl}$, the first term becomes the quantum Hamiltonian after quantization, $\mathcal{H} \rightarrow \hat{\mathcal{H}}$. In contrast, the classical contribution $\mathcal{L}_\text{cl}$ gives rise only to a term proportional to the identity operator and therefore does not affect the quantum dynamics.

After quantization, one obtains the following classical equations of motion from the Routhian for capacitively coupled quantum and classical systems [see Appendix~\ref{AppendixA}]:
\begin{equation}
	\frac{d}{dt}\frac{\partial \mathcal{L}_\text{cl}}{\partial \dot{\varphi}_{\text{cl},i}}
	- \frac{\partial \mathcal{L}_\text{cl}}{\partial \varphi_{\text{cl},i}}
	+ \frac{\partial \mathcal{F}}{\partial \dot{\varphi}_{\text{cl},i}}
	- \frac{d}{dt}\Big\langle \frac{\partial \hat{\mathcal{H}}}{\partial \dot{\varphi}_{\text{cl},i}} \Big\rangle
	= 0,
	\label{ClEOM}
\end{equation}
where $\mathcal{F}$ is a classical dissipation function and $\langle \hat{A} \rangle$ denotes the quantum-mechanical expectation value of an observable $\hat{A}$. The first three terms of Eq.~(\ref{ClEOM}) form the Euler-Lagrange equation, while the last term corresponds to the coupling with the quantum subsystem. We emphasize that the developed formalism is equally applicable to systems with inductive coupling [see Appendix~\ref{AppendixA}].

Having developed a general formalism applicable to arbitrary coupled classical-quantum systems, in the next subsection we illustrate the approach using a concrete example: a Josephson-junction-based charge qubit.

\subsection{Quantization of the Cooper-pair box}
\label{Sec:Theory:Quantization}
\begin{figure}[t]
	\includegraphics[width=8 cm]{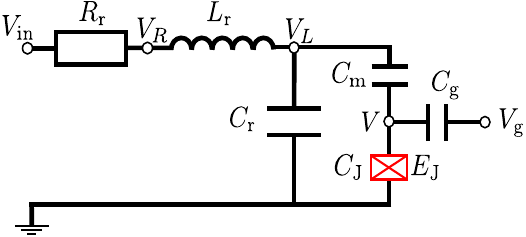}
	\caption{Classical resonant $RLC$ circuit capacitively coupled to the CPB through the capacitor $C_{\text{m}}$. The JJ-based charge qubit has
		Josephson energy $E_{\text{J}}$ and capacitance $C_{\text{J}}$. The system
		can be driven through the gate capacitor $C_{\text{g}}$ via the applied
		voltage $V_{\text{g}}$. Classical voltages in the nodes
		between the resistor, inductor, and capacitor  are $V_{R}$ and $%
		V_{L}$, respectively. A probe signal $V_{\text{in}}(t)=V_{\text{in}%
		}\cos (\omega _{\text{rf}}t)$ is applied to the resistance, and the
		gate  voltage $V_{\text{g}}=V_{\text{g}0}+V_{\text{d}}\cos (\omega _{\text{d}}t)$  has a dc component $V_{\text{g}0}$  which tunes the qubit and
		an ac component exciting the system. }
	\label{Fig3}
\end{figure}

In order to study coupled dynamics, we start by adding a classical $RLC$ circuit to a CPB in Fig.~\ref{Fig2}. Additionally, in order to include a resonant microwave drive, we tune the qubit via an additional gate capacitor $C_{\text{g}}$, as shown in Fig.~\ref{Fig3}, as in Refs.~\cite{Johansson2006, Persson2010_2}. The corresponding gate voltage 
$V_{\text{g}} = V_{\text{g}0} + V_{\text{d}} \cos (\omega_{\text{d}} t)$ 
contains a dc component $V_{\text{g}0}$ that sets the qubit operating point and an ac component that resonantly excites the system. Note that alternatively to the $RLC$ circuit, a quantum system can be probed via a nanomechanical resonator~\cite{LaHaye2009, Shevchenko2012a}.

We start by writing the classical Lagrange function $\mathcal{L}(\varphi _{i}, \dot{\varphi}_{i})$ for the electrical circuit in Fig.~\ref{Fig3}. To do this, we choose generalized velocities to be proportional to voltages at the nodes $V_{i}$, so that the nodal phases $\varphi_i =\frac{2e}{\hbar }\int V_i dt $  are the generalized coordinates~\cite{Zagoskin2011}.  Equivalently, one may work with the nodal fluxes \text{$\Phi_i = (\Phi_0  / 2\pi)\varphi_i $}~\cite{Blais2021}, where $\Phi_0 = h/2e$ is the magnetic flux quantum.

To be specific, $V_{R,L}$ are the classical voltages between the resistance $R_{\text{r}}$, inductance $L_{\text{r}}$, and capacitor $C_{\text{r}}$, $V_{\text{in}}$ is a weak probe rf signal at the input, $V_{\text{g}}$  is a parameter that controls the qubit state. These voltages correspond to the phases $\varphi_{L,R}$, $\varphi_{\text{in}}$, and $\varphi_{\text{g}}$, respectively. So, the Lagrangian can be written as 
\begin{gather}
	\left( \frac{2e}{\hbar }\right) ^{2}\mathcal{L}=\frac{C_{\text{r}}}{2}\dot{\varphi}
	_{L}^{2}-\frac{1}{2L_{\text{r}}}\left( \varphi _{L}-\varphi _{R
	}\right) ^{2}+\frac{C_{\text{g}}}{2}\left( \dot{\varphi}-\dot{\varphi}_{
		\text{g}}\right) ^{2}+  \notag  \label{Theory:lagrangian} \\
	+\frac{C_{\text{J}}}{2}\dot{\varphi}^{2}+\frac{C_{\text{m}}}{2}(\dot{\varphi}
	-\dot{\varphi}_{L})^{2}+\left( \frac{2e}{\hbar }\right) ^{2}E_{\text{J
	}}\cos \varphi , \label{Lagrangian}
\end{gather}%
which contains \textquotedblleft kinetic"  terms corresponding to the electrostatic energy of the capacitors and the \textquotedblleft potential"  terms due to a magnetic energy of the inductance and the energy of the JJ.

As discussed in the previous subsection, the superconducting phase of the island $\varphi$ must be quantized. Therefore, we introduce the Routhian function $\mathcal{R} = p \dot{\varphi} - \mathcal{L}$, where $p = \partial \mathcal{L} / \partial \dot{\varphi}$ [see Eq.~(\ref{Routhian}), Appendix~\ref{Appendix:Quantization}, and Ref.~\cite{Sillanpaeae2005}].  

Quantization is done replacing variables related to the quantum subsystem with corresponding 
operators, so that the Hamiltonian $\hat{\mathcal{H}}$ is 
\begin{equation}
	\hat{\mathcal{H}}=4E_{\text{C}}(\hat{n}-n_{\text{g}})^{2}-E_{\text{J}}\cos 
	\hat{\varphi}
	-\dfrac{C_\text{g} V_\text{g}^2}{2}-\dfrac{C_\text{m} V_L^2}{2},  \label{Theory:Qhamiltonian}
\end{equation}
where $E_{\text{C}}=e^{2}/2C_{\Sigma }$ is the charging energy with the total
capacitance of the island $C_{\Sigma }=C_{\text{J}}+C_{\text{g}}+C_{\text{m}%
} $, $\hat{n}=\hat{p}/\hbar $ is the dimensionless momentum with the
commutation relation $[\hat{\varphi},\hat{n}]=i$, and $n_{\text{g}}$ is an
effective dimensionless charge 
\begin{equation}
	n_{\text{g}}=-\frac{\hbar }{4e^{2}}\left( C_{\text{g}}\dot{\varphi}_{\text{g}%
	}+C_{\text{m}}\dot{\varphi}_{L}\right) .  \label{Theory:ng}
\end{equation}%

The dimensionless charge consists of three terms \text{$n_{\text{g}}=n_{\text{g}0}+\delta n_{\text{rf}}+\delta n_{\text{d}}$}, where $n_{\text{g}0}$ is the stationary bias formed by $V_\text{g0}$ [see Eq.~(\ref{Delta_and_varepsilon_0})], $\delta n_{\text{rf}}$ is associated with the weak coupling to the classical resonator arising from $\dot \varphi_L$, and $\delta n_{\text{d}}\propto \cos (\omega_\text{d}t)$ represents the strong resonant drive applied to the system. 

In the two-level approximation, valid for $E_{\text{J}}/4E_{\text{C}}\ll 1$, the Hamiltonian can be written in the pseudo-spin form 
\begin{equation}
	\hat{\mathcal{H}}=E_\text{geom}[\varepsilon(t)]\hat{\mathbf{1}}-\frac{\Delta }{2}\hat{\sigma}_{x}-\frac{\varepsilon(t) }{2}\hat{\sigma}_{z}  \label{Theory:2levelHam}
\end{equation}
with $\varepsilon(t)=4E_\text{C}(1-2 n_\text{g})$ being the voltage-controlled energy bias and $\Delta$ describing the tunneling between the nearest charge states. The term $E_\text{geom}$ denotes the geometric energy, and is defined in Eq.~(\ref{B:Egeom}). Although it does not influence the quantum dynamics, it gives rise to a geometric capacitance, thereby providing a more complete description of the interaction between the subsystems.

The resulting classical equations for $\varphi_{L,R}$, are essentially the Kirchhoff equations for the electrical currents $I_{L,R}$ flowing through the corresponding circuit nodes. These are:
\begin{equation}
	\begin{cases}
		\dfrac{\dot{\varphi}_{\text{in}}-\dot{\varphi}_{R}}{R_{\text{r}}}=\dfrac{
			\varphi _{R}-\varphi _{L}}{L_{\text{r}}}, \\ 
		C_\text{r}\ddot{\varphi}_{L}-\left(\dfrac{2e}{\hbar}\right)^2 \dfrac{d}{dt}\Big\langle \dfrac{\partial \hat{\mathcal{H}}}{\partial \dot\varphi_L}\Big \rangle=\dfrac{\varphi_R-\varphi_L
		}{L_{\text{r}}}
	\end{cases}
	\label{Theory:CPB_clas_eqs}
\end{equation}
and must be supplemented by a quantum master equation that determines the quantum current
\begin{equation}
	\langle \hat{\mathcal{I}}\rangle = - \dfrac{d}{dt}\Big\langle \dfrac{\partial \hat{\mathcal{H}}}{\partial V_L}\Big \rangle = - \dfrac{\partial \varepsilon}{\partial V_L}\dfrac{d}{dt}\Big\langle \dfrac{\partial \hat{\mathcal{H}}}{\partial \varepsilon}\Big \rangle. \label{quantum_current_exp}
\end{equation}

As will be shown in Secs.~\ref{Theory:GKSL} and \ref{Sec:Qudit:GKSL}, \textit{the quantum current gives rise to both the effective capacitance and the effective resistance}
\begin{equation}
	\dfrac{2e}{\hbar}\langle \hat{\mathcal{I}}\rangle= C_\text{eff} \ddot \varphi_L+\dfrac{\dot \varphi_L}{R_\text{eff}}. \label{quantum_current}
\end{equation}
This equality should be understood as time-averaged on the timescale of the quantum subsystem, whose frequency $\omega_{\text{q}/\text{qd}}$ is much greater $ \omega_{\text{q/qd}} \gg \omega_\text{rf}$ than the frequency of the classical subsystem $\omega_\text{rf}$. This results in an effective response only on the timescale of the classical probing device~\cite{Barajas2026, Barajas2025a}.

So far, we have developed a framework that consistently couples the classical electrical resonator to the quantum subsystem, for the specific example of a charge qubit.The coupled classical equations (\ref{Theory:CPB_clas_eqs}) need to be supplemented by a quantum master equation. In the following subsections we will see how the effective response of the quantum system arises both for the isolated (Sec.~\ref{Theory:isolated}) and open (Sec.~\ref{Theory:GKSL}) two-level systems.

\subsection{Quantum capacitance of an ideal qubit}
\label{Theory:isolated}
In contrast to the previous subsection, where we focused on the specific system shown in Fig.~\ref{Fig3}, here we consider a general two-level quantum system with a pseudo-spin Hamiltonian given by Eq.~(\ref{Theory:2levelHam}). We assume that the system is being probed by a classical voltage $V$ that can be treated perturbatively. We consider a non-driven isolated system, so
that there are no effects of temperature, relaxation, and decoherence. This
is described by the Schr\"{o}dinger equation for the evolution operator $i\hbar \hat{\dot{U}}=\hat{\mathcal{H}}\hat{U}$ with Hamiltonian (\ref{Theory:2levelHam}) and bias
\begin{equation}
	\varepsilon (t)=\varepsilon
	_{0}+\delta \varepsilon _{\text{rf}}(t),
\end{equation}
where the rf component of the bias is related to the probing voltage via $\delta \varepsilon_\text{rf} = \alpha e \delta V$ with the lever arm
\begin{equation}
	\alpha =\dfrac{1}{e} \dfrac{\partial \varepsilon}{\partial V}. \label{lever_arm}
\end{equation}
The evolution operator $\hat{U}$ evolves the initial state $|\psi_0\rangle$ to the state $|\psi\rangle = \hat{U} |\psi_0\rangle$.

We assume that the coupling between the subsystems is small, so that the probe signal $\delta \varepsilon _{\text{rf}}(t)$ can be treated as a perturbation in the Schr\"{o}dinger equation at 
\begin{equation}
	\delta \varepsilon _{\text{rf}}, \, \hbar \omega_\text{rf}\ll \Delta. \label{2-level_perturbation_conditions}
\end{equation}%

We seek a solution for the classical
phase $\varphi_{L}$ in the form $\varphi _{L}=\varphi _{L%
	0}\sin (\omega _{\text{rf}}t)$. Then, from definition~(\ref{Theory:ng}) and that of the time-dependent energy bias $\varepsilon(t)$, we obtain
\begin{equation}
	\delta \varepsilon_{\text{rf}}(t)=\hbar \dfrac{C_{\text{m}}}{C_{\Sigma }}\dot \varphi_L \eqqcolon \delta
	\varepsilon _{\text{rf}} \cos (\omega _{\text{rf}}t). \label{Theory:delta_varepsilon_rf}
\end{equation}

The solution of the Schr\"{o}dinger equation for the evolution operator $\hat{U} $ with the Hamiltonian (\ref{Theory:2levelHam}) can be obtained using time-dependent perturbation theory [see Ref.~\cite%
{LandauLishitz_QM} and Appendix~\ref{Appendix:IdQb}, Eqs.~(\ref{AppendixB_U0_solution}-\ref{AppendixB:WFsolution})]. After this, we
can calculate the time derivative of the average number of Cooper pairs 
\begin{equation}
	\dfrac{d\langle \hat{n}\rangle }{dt}=\dfrac{i}{\hbar }\big\langle [\hat{%
		\mathcal{H}},\hat{n}]\big\rangle =\dfrac{i}{\hbar }\langle \psi_0 |\hat{U}^\dagger[\hat{%
		\mathcal{H}},\hat{n}]\hat{U}|\psi_0 \rangle  \label{dndt}
\end{equation}%
up to the first order in perturbation. 

For an arbitrary initial superposition state 
\begin{equation}
	\ket{\psi_0}=\sqrt{\dfrac{1+\chi}{2}} \ket{E_{-}}+\sqrt{\dfrac{1-\chi}{2}} e^{i\vartheta} \ket{E_{+}},
\end{equation}
the resulting expression for the quantum capacitance is
\begin{equation}
	C_\text{Q}
	=
	C_\text{Q0}
	\frac{\Delta^3}{\Delta E_0^3}
	\chi
	\frac{1}{1 - \left( \omega_\text{rf}/\omega_\text{q0} \right)^2 }.
\end{equation}
This expression coincides with the phenomenological result, Eq.~(\ref{Phenomenology:quantum_capacitance}), up to terms of the order $O\!\left( (\omega_{\text{rf}}/\omega_{\text{q0}})^2 \right)$. Here, the states $\ket{E_{\pm}}$ are defined as eigenstates corresponding to the eigenenergies 
\begin{equation}
	E_\pm =E_\text{geom}^{(0)} \pm \dfrac{\hbar \omega_{\text{q}0}}{2}
\end{equation} 
of the Hamiltonian (\ref{Theory:2levelHam})  with the time-independent energy bias $\varepsilon(t) = \varepsilon_0$, while $\vartheta$ is an arbitrary phase difference. 

So far, our formalism yields correct results for the effective capacitance that coincides with previously known phenomenological results. In the next subsection we investigate how the situation changes when the quantum system can no longer be considered isolated. In particular, we show how the expression for the quantum capacitance must be modified, and how both the tunneling capacitance and the effective resistance arise.

In Sec.~\ref{Sec:Qudit}, we will consider a more general formulation and show that, for an arbitrary multi-level system, the quantum capacitance is exactly proportional to the second derivative of the energy with respect to the energy bias, and we will also show how to obtain general expressions for the tunneling capacitance and the effective resistance.

\subsection{Impact of temperature, relaxation and decoherence}

\label{Theory:GKSL}
Now our task is to describe the interaction of quantum open systems with classical subsystems. 
\subsubsection{GKSL equation}
To account for temperature, relaxation, and decoherence effects, we must use the quantum master equation, because the Schr\"{o}dinger equation can describe
only unitary dynamics. Instead, we now consider the GKSL equation for the density matrix $\hat{\rho}$. This equation must be written in the instantaneous representation, e.g.~\cite{Malinowski2022, Yamaguchi2017}, obtained by the unitary transformation $\hat{S}$ [see Eq.~(\ref{2lvl_unitary_transformation_S})], which diagonalizes the original Hamiltonian
\begin{gather}
	\hat{\mathcal{H}}_\text{inst}=\hat{S}^{\dag }\hat{\mathcal{H}}\hat{S}-i\hbar 
	\hat{S}^{\dag }\dot{\hat{S}}=\label{Hinst_qb}\\
	\nonumber=E_\text{geom}\hat{\mathbf{1}}-\frac{\Delta E(t)}{2}\hat{\sigma}_{z}+\frac{\hbar
		\Delta \dot{\varepsilon}}{2\Delta E^{2}(t)}\hat{\sigma}_{y}, 
\end{gather}
where $\Delta E(t)=\sqrt{\Delta^2+\varepsilon^2(t)}$. The GKSL equation reads
\begin{equation}
	\hat{\dot\rho} = -\dfrac{i}{\hbar}[\hat{\mathcal{H}}_\text{inst},\hat{\rho}]+\sum_\alpha \gamma_\alpha \left[ \hat{L}_\alpha \hat{\rho} \hat{L}_\alpha ^\dagger - \dfrac{1}{2}\left\{\hat{L}_\alpha^\dagger  \hat{L}_\alpha ,\hat{\rho}\right\}\right]
	.  \label{Lindblad_eq}
\end{equation}
We choose the relaxation rates $\gamma_\alpha = \{\gamma_{\pm},\gamma_\phi\}$ and the Lindblad jump operators $\hat{L}_\alpha = \{\hat{\sigma}_{\pm},\hat{\sigma}_z\}$, imposing the instantaneous detailed balance condition 
\begin{equation}
	\dfrac{\gamma_{+}}{\gamma_{-}}=\exp \left[\dfrac{\Delta E(t)}{k_\text{B}T}\right] , \label{two_level_detailed_balance}
\end{equation}
which leads to the Bloch equation. The relaxation rates are chosen in such a way so that they reproduce the relaxation and decoherence times,
\begin{equation}
	T_1^{-1}=\gamma=\gamma_++\gamma_- \quad\text{and}\quad T_2^{-1}=2\gamma_\varphi+T_1^{-1}/2,
\end{equation}
in the Bloch equation [see Appendix~\ref{Appendix_bloch_qubit}]. The expressions for the effective capacitance and resistance will remain the same in such case.

For small probing amplitude $\delta \varepsilon_\text{rf}$, both the Hamiltonian $\hat{\mathcal{H}}_{\text{inst}}$ and the relaxation rates $\gamma_\alpha$ can be expanded up to  first order as 
\begin{equation}
	\hat{\mathcal{H}}_{\text{inst}}=\hat{\mathcal{H}}_{\text{inst}}^{(0)}+\hat{V}_{\delta \varepsilon}\delta \varepsilon_\text{rf}(t)+\hat{V}_{\delta \dot\varepsilon}\delta \dot\varepsilon_\text{rf}(t) \label{Hinst_pert_decomposition_2lvl}
\end{equation}
and 
\begin{equation}
	\gamma_\alpha(\varepsilon) = \gamma_\alpha^{(0)}+\delta \varepsilon_\text{rf}(t) \dfrac{\partial \gamma_\alpha^{(0)}}{\partial \varepsilon}. \label{gamma_series}
\end{equation}
For the detailed definitions of $\hat{\mathcal{H}}_{\text{inst}}^{(0)}$, $\hat{V}_{\delta \varepsilon,\delta \dot \varepsilon}$, $\gamma_\alpha^{(0)}$, and $\partial_\varepsilon \gamma_\alpha^{(0)}$, see Eqs.~(\ref{2lvl_H_pert}, \ref{2lvl_gamma_decomposition}).

The solution for the density matrix is obtained by finding the first-order corrections to the stationary thermal equilibrium state at times much longer than the relaxation and decoherence times $t \gg T_1, T_2$:
\begin{equation}
	\hat{\rho} =\hat{\rho}^{(0)}+\hat{\rho}^{(1)}=\hat{\rho}^{(0)} +\hat{A} \delta \varepsilon_\text{rf}(t)+\hat{B} \delta \dot\varepsilon_\text{rf}(t), \label{Lindblad_anzatz}
\end{equation} 
where $\hat{\rho}^{(0)}$ is the thermal equilibrium density matrix at the unperturbed energy bias $\varepsilon_0$. Expanding the GKSL equation (\ref{Lindblad_eq}) yields two coupled linear matrix equations for the unknown matrices $\hat{A}$ and $\hat{B}$ [see Eq.~(\ref{system6ab})], which can be found analytically. Using these matrices, we compute the first-order correction to the quantum current~(\ref{quantum_current}). As a result, we obtain expressions for the effective capacitance~(\ref{Appendix_bloch_qubit_Ceff}) and the effective  resistance~(\ref{Appendix_bloch_qubit_Reff}), which are analyzed in Secs.~\ref{Sec:Theory:cap} and \ref{Sec:Theory:res}, respectively. 

\subsubsection{Effective capacitance}
\label{Sec:Theory:cap}
The effective capacitance~(\ref{Appendix_bloch_qubit_Ceff}) consists of geometric $C_\text{geom}$, quantum $C_\text{Q}$, and tunneling $C_\text{T}$ components. The geometric capacitance can be written as~[see Appendix \ref{AppendixA} and Sec.~\ref{Sec:Other}]
\begin{equation}
	C_\text{geom} = -\dfrac{\partial^2 E_\text{geom}}{\partial V^2}, \label{C_geom}
\end{equation}
where $V$ is a classical probe voltage. In this and the next sections, we consider $E_i = E_\text{geom} + E_i'$ to be the total eigenvalues  of the Hamiltonian, including the geometric energy.

From Eq.~(\ref{CQ_GKSL_2}) at $\omega_\text{rf} \ll \omega_{\text{q}0}$ the quantum capacitance is
\begin{equation}
	C_\text{Q}= 
	C_{\text{Q}0}\dfrac{\Delta ^3}{\Delta
		E_0^3} \tanh \left(\dfrac{\Delta E_0}{2k_\text{B}T}\right),  \label{qubit_CQ}
\end{equation}
where 
\begin{equation}
	C_{\text{Q}0} = \dfrac{(\alpha e)^2}{2\Delta} \label{CQ0} 
\end{equation}
is the maximum value of the quantum capacitance that can be obtained from the known dependence of the energy bias on the classical probe voltage, with $\alpha$ being the lever arm defined in Eq.~(\ref{lever_arm}).  The result in Eq.~(\ref{CQ0}), unlike the expression in Eq.~(\ref{CQO_phenomenology}), is valid for an arbitrary two-level system. In particular, by substituting $\alpha = 2 C_\text{m} / C_\Sigma$ into Eq.~(\ref{CQ0}), one recovers the maximum quantum capacitance of the CPB, Eq.~(\ref{CQO_phenomenology}). In the quantum capacitance~(\ref{qubit_CQ}) the dependence on the decoherence time $T_2$ appears in the next order of $\omega_\text{rf}/ \omega_\text{q0}$ [see Eq.~(\ref{Appendix_bloch_qubit_Ceff})].

The tunneling capacitance reads
\begin{equation}
	C_{\text{T}}=C_{\text{Q}0}\dfrac{\Delta \, \varepsilon_0^2}{2 k_\text{B} T \Delta E_0^2}\cosh^{-2}\left(\dfrac{\Delta E_0}{2 k_\text{B}T}\right) \dfrac{1}{1+T_1^{2}\omega_\text{rf}^2}. \label{qubit_CT} 
\end{equation}

From Eq.~(\ref{qubit_CT}) we see that $C_{\text{T}}$ originates from the dynamical redistribution of the level occupations, which occurs on the relaxation timescale $T_1$. For the bad qubits with $\omega_\text{rf} \ll T_1^{-1}$, the populations efficiently follow the classical probe, the redistribution of the populations of the energy levels remains quasi-equilibrated, and the tunneling capacitance reaches its maximal value. We note that only in this regime the phenomenological expression~(\ref{CT_phenomenology}) is applicable. 

In contrast, when the probe signal becomes too fast compared to the timescales of relaxation, $T_1^{-1} \ll \omega_\text{rf} $, the populations are effectively frozen out of equilibrium, the redistribution of the populations is dynamically blocked, and $C_{\text{T}}$ is strongly suppressed. As it was observed in Eq.~(\ref{Ceff_limit_at_0_temperature}), as the temperature tends to absolute zero, the tunneling capacitance still vanishes, while the quantum capacitance tends to the capacitance of the ground state.

The general result for the effective capacitance can be simplified for the limiting cases of a bad qubit ($\omega_\text{rf} \ll \omega_\text{q} \ll T_1^{-1}$)
\begin{equation}
	C^{(\text{bad})}_\text{eff}=-
	\dfrac{\partial }{\partial V }\overline{ \dfrac{\partial E}{\partial V }}_\text{th} \label{qubit_C_BAD}
\end{equation}%
and a good qubit (\text{$T_1^{-1} \ll \omega_\text{rf} \ll \omega_\text{q}$})
\begin{equation}
	C^{(\text{good})}_\text{eff}=-\overline{ \dfrac{\partial^2 E}{\partial V^2 }}_\text{th}, \label{qubit_C_GOOD}
\end{equation}
where $E$ denotes the eigenenergies $E_\pm$ of the Hamiltonian~(\ref{Theory:2levelHam})~[see Sec.~\ref{Intro:results} for the definition of average $\overline{A}_\text{th}$]. 

These two limits reflect the same physics discussed above: for a bad qubit, the occupations follow the drive and the response involves the derivative of the thermally averaged slope of the energy. Whereas for a good qubit the populations are effectively frozen and the capacitance is determined solely by the curvature of the eigenenergies. 

The fact that such results can be written in terms of qubit eigenenergies will be proven in Sec.~{\ref{Sec:Qudit}}. We will also show that expressions (\ref{qubit_C_BAD}, \ref{qubit_C_GOOD}) remarkably are unchanged for multi-level systems, as it is sometimes implied~\cite{Lundberg2024, Microsoft2025}.

For good qubits at $T_1^{-1} \ll \omega_\text{rf}$  the tunneling capacitance vanishes from Eq.~(\ref{qubit_C_GOOD}) and we can then deduce that the quantum capacitance can be defined as
\begin{equation}
	C_\text{Q}=-\overline{ \dfrac{\partial^2 E'}{\partial V^2 }}_\text{th}. \label{CQoverline}
\end{equation}
For bad qubits  at $\omega_\text{rf} \ll T_1^{-1}$, the tunneling capacitance reaches its maximum value, as can be seen by combining Eqs.~(\ref{qubit_C_BAD}) and (\ref{qubit_C_GOOD})
\begin{equation}
	C_\text{T,max}=\overline{ \dfrac{\partial^2 E}{\partial V^2 }}_\text{th} - \dfrac{\partial }{\partial V }\overline{ \dfrac{\partial E}{\partial V }}_\text{th}. \label{CToverline}
\end{equation}

\subsubsection{Effective resistance}
\label{Sec:Theory:res}
So far, the system was never allowed to dissipate any energy. However, open systems may exchange energy with their environments, which leads to dynamical dissipation. Alternatively, from a mathematical perspective, the GKSL master equation introduced a finite response time in the system. At the thermal equilibrium state, this creates a phase lag between the probe voltage at the input and the electrical current on the output. Thus, the system is dissipating power, which must be accounted for with a resistive term $R_\text{eff}$.

The effective conductance, Eq.~(\ref{Appendix_bloch_qubit_Reff}), can be split in two terms, as in Eq.~(\ref{Qubit_R_eff}): the Sisyphus conductance \cite{Grajcar2008, Persson2010, Esterli2019},
\begin{equation}
	G_\text{Sis} = R_\text{Sis}^{-1}= R_{\text{Q}0}^{-1} \frac{\hbar \omega_\text{rf} }{k_\text{B}T}\dfrac{\varepsilon_0^2}{\Delta E_0^2}
	\dfrac{ T_1 \omega_\text{rf}}
	{1+T_1^2 \omega_\text{rf}^2}\cosh^{-2}\left(\dfrac{\Delta E_0}{2 k_\text{B}T}\right), \label{R_Sisyphus}
\end{equation}
and the Hermes conductance \cite{Peri2023b},
\begin{equation}
	G_\text{Hrm} = R_\text{Hrm}^{-1}= 2 R_{\text{Q}0}^{-1}  \dfrac{\omega_\text{rf}^2 }{\omega_{\text{q}0}^2}
	\dfrac{\Delta^2}{\Delta E_0^2} \dfrac{T_2 \omega_{\text{q}0}}
	{1+T_2^2 \omega_{\text{q}0}^2} \tanh \left(\dfrac{\Delta E_0}{2k_\text{B}T}\right).
	\label{R_Hermes}
\end{equation} We have introduced
\begin{equation}
	R_{\text{Q}0}^{-1} = \dfrac{\pi}{2} R_\text{K}^{-1}
	\alpha^2 \label{RQ0} 
\end{equation} 
as the characteristic value of the effective conductance and $R_\text{K}=h/e^2$ is the von Klitzing constant. 

When the temperature tends to absolute zero, the effective conductance simplifies to:
\begin{equation}
	\begin{cases}
		G_\text{Sis} = R_\text{Sis}^{-1} \,\,\rightarrow\,\, 0, \\ 
		G_\text{Hrm} = R_\text{Hrm}^{-1} \,\,\rightarrow\,\, 2 R_{\text{Q}0}^{-1}  \dfrac{\omega_\text{rf}^2 }{\omega_{\text{q}0}^2} 
		\left(\dfrac{\Delta}{\Delta E_0}\right)^2 
		\dfrac{T_2 \omega_{\text{q}0}}
		{1+T_2^2 \omega_{\text{q}0}^2}.
	\end{cases} \label{Reff_limit_at_0_temperature}
\end{equation}%
Moreover, if the qubit satisfies either $T_2\rightarrow 0 $ or $T_2 \rightarrow \infty$, then the Hermes conductance also vanishes.

We note that the effective capacitance $C_\text{eff}$ and the effective resistance $R_\text{eff}$ defined in Eqs.~(\ref{Appendix_bloch_qubit_Ceff}, \ref{Appendix_bloch_qubit_Reff}) depend explicitly on the probing frequency $\omega_\text{rf}$, and therefore $C_\text{eff}$ and $R_\text{eff}$ should not be interpreted as true circuit elements in the strict sense. A rigorous identification of the underlying circuit representation would require decomposing the admittance $Y = R_\text{eff}^{-1} + i\omega_\text{rf} C_\text{eff}$ into elementary rational functions of $\omega_\text{rf}$~\cite{Solgun2014}.

\subsubsection{Graphical results}
\begin{figure}[t]
	\includegraphics[width=8.6 cm]{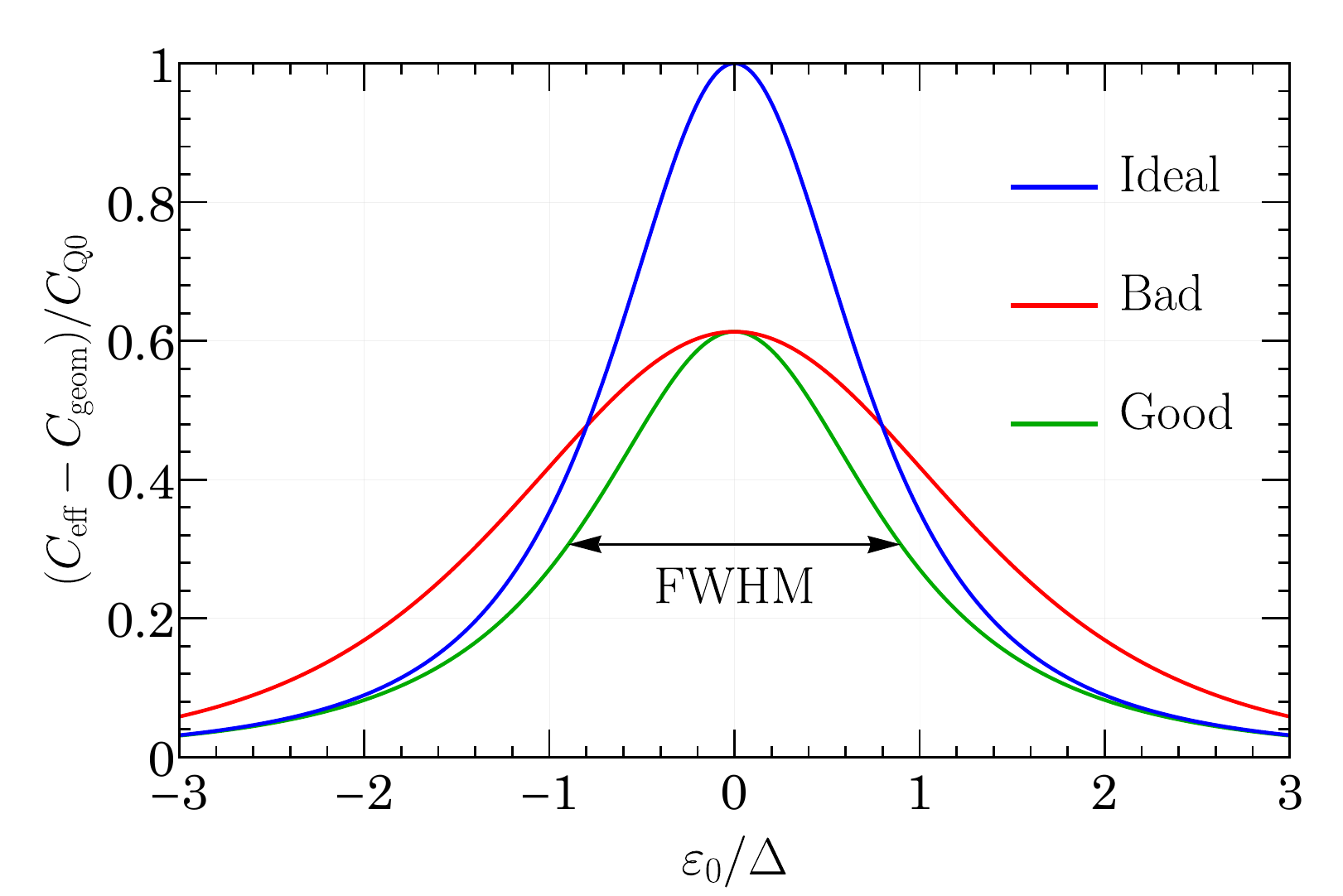}
	\caption{Effective capacitances (without the geometric contribution, $C_\text{eff}-C_\text{geom}$) of an ideal qubit in the ground state (blue curve), of a high-coherence (good) qubit (green curve), and of a low-coherence (bad) qubit (red curve). Expressions for the effective capacitance of the bad and good qubits are given in Eqs.~(\ref{qubit_C_BAD}) and~(\ref{qubit_C_GOOD}). Deviations from the isolated-system case illustrate the effects of decoherence and energy-level broadening on the effective capacitance. The temperature for good and bad qubits is set to be $k_{\text{B}}T/\Delta = 0.7$. }
	\label{Fig:ideal}
\end{figure}

\begin{figure}[!t]
	\includegraphics[width=8.6 cm]{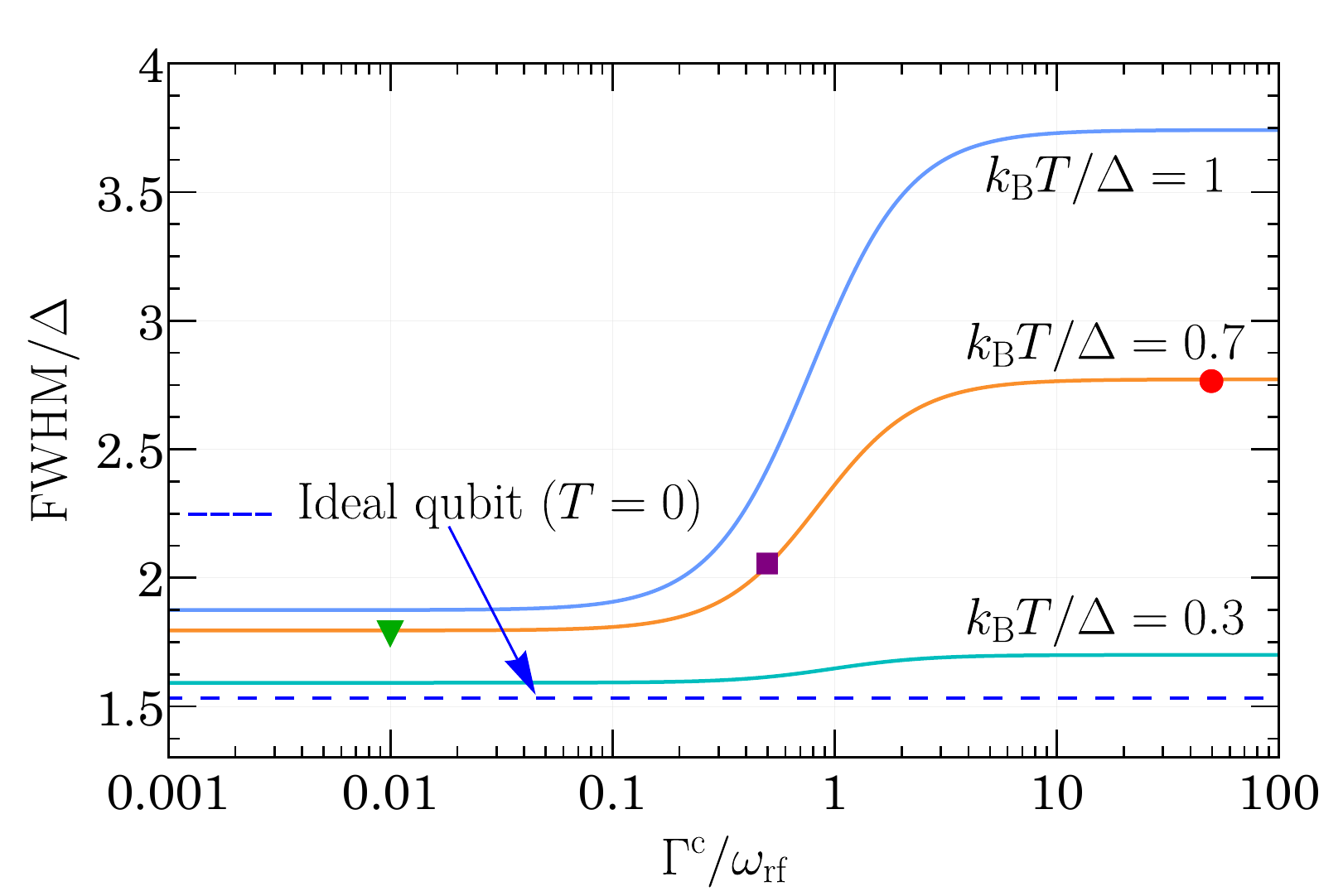}
	\caption{Dependence of the effective capacitance full width at half-maximum
		(FWHM) on the normalized relaxation rate $\Gamma^\text{c}/\omega_\text{rf}$  for different temperatures. 
		In the region $\Gamma^\text{c}/\omega_\text{rf} \ll 1$, on which the marker \color[rgb]{0,0.666,0}$\blacktriangledown \text{ }$\color{black} is placed, the qubit is in the good (coherent) limit and the curve saturates to a constant value. In contrast, for $\Gamma^\text{c}/\omega_\text{rf} \gg 1$, on which the marker \color{red}$\bullet \text{ }$\color{black} is placed, the qubit is bad (strongly dephased), and the width reflects the broadening of the effective capacitance at small relaxation times. The marker \color[rgb]{0.5,0,0.5}$\blacksquare \text{ }$\color{black} is for an intermediate situation, when the qubit is neither bad nor good. For the cyan, orange, and blue curves the temperatures are $k_\text{B}T/\Delta=0.3, \,0.7 \text{, and }  1,$ respectively. For the green, purple, and red markers, the relaxation rates are $\Gamma^\text{c}/\omega_\text{rf} = 0.01$, $0.5$, and $50$, respectively.  The blue dashed curve corresponds to the case of an ideal qubit, where there is no dependence of the FWHM on the relaxation time. }
	\label{Fig:thermal}
\end{figure}

\begin{figure}[t]
	\includegraphics[width=8.6 cm]{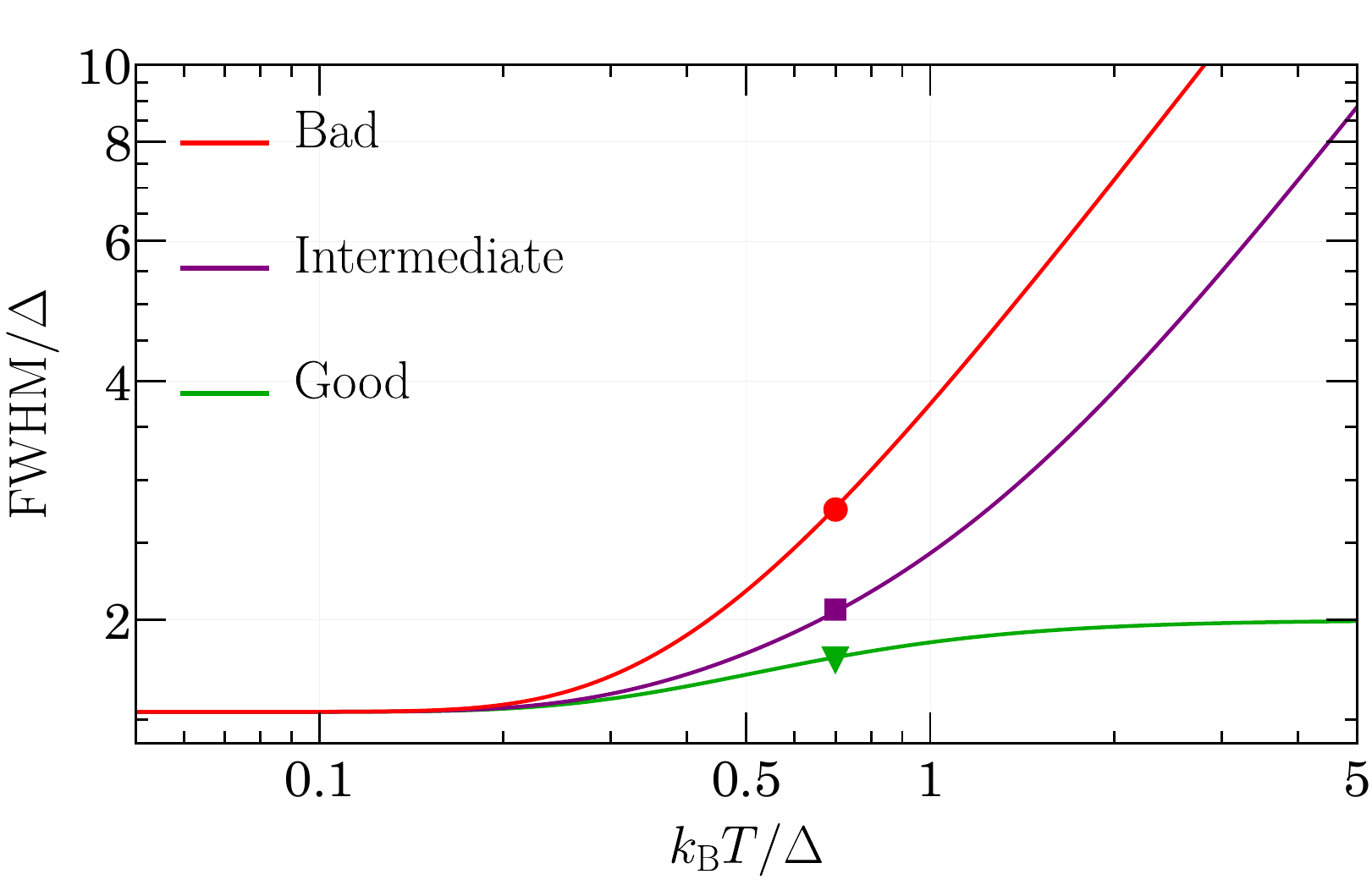}
	\caption{Dependence of the effective capacitance full width at half-maximum on
		the normalized temperature $k_{\text{B}}T/\Delta $ for good, 
		intermediate, and bad cases (corresponding to the respective markers in Fig.~\ref{Fig:thermal}). All parameters are the same as in Fig.~\protect\ref{Fig:ideal}.} 
	\label{Fig:FWHM}
\end{figure}

\begin{figure}[t]
	\includegraphics[width=8.9 cm]{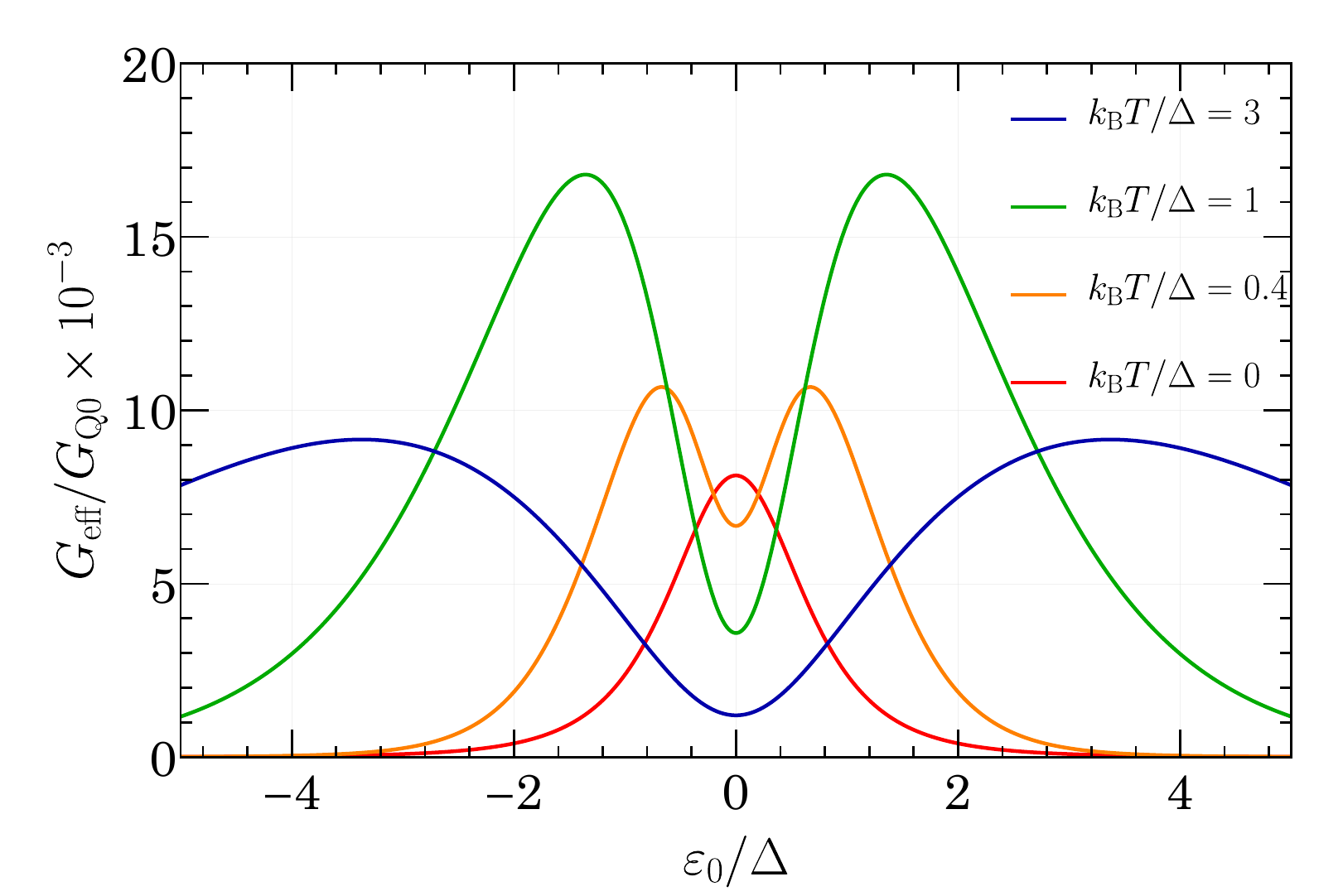}
	\caption{Dependence of the effective conductance $G_\text{eff}=R_\text{eff}^{-1}$, in units of $G_\text{Q0}=R_\text{Q0}^{-1}$, on the energy bias $\varepsilon_0$ for different temperatures $T$.  The parameters used in this figure are $\hbar \Gamma^{\text{c}} / \Delta = \hbar \omega_\text{rf} / \Delta = 0.1$ and $T_\phi \Delta / \hbar = 0.5$.} 
	\label{Fig:Resistance1}
\end{figure}

\begin{figure}[t]
	\includegraphics[width=8.9 cm]{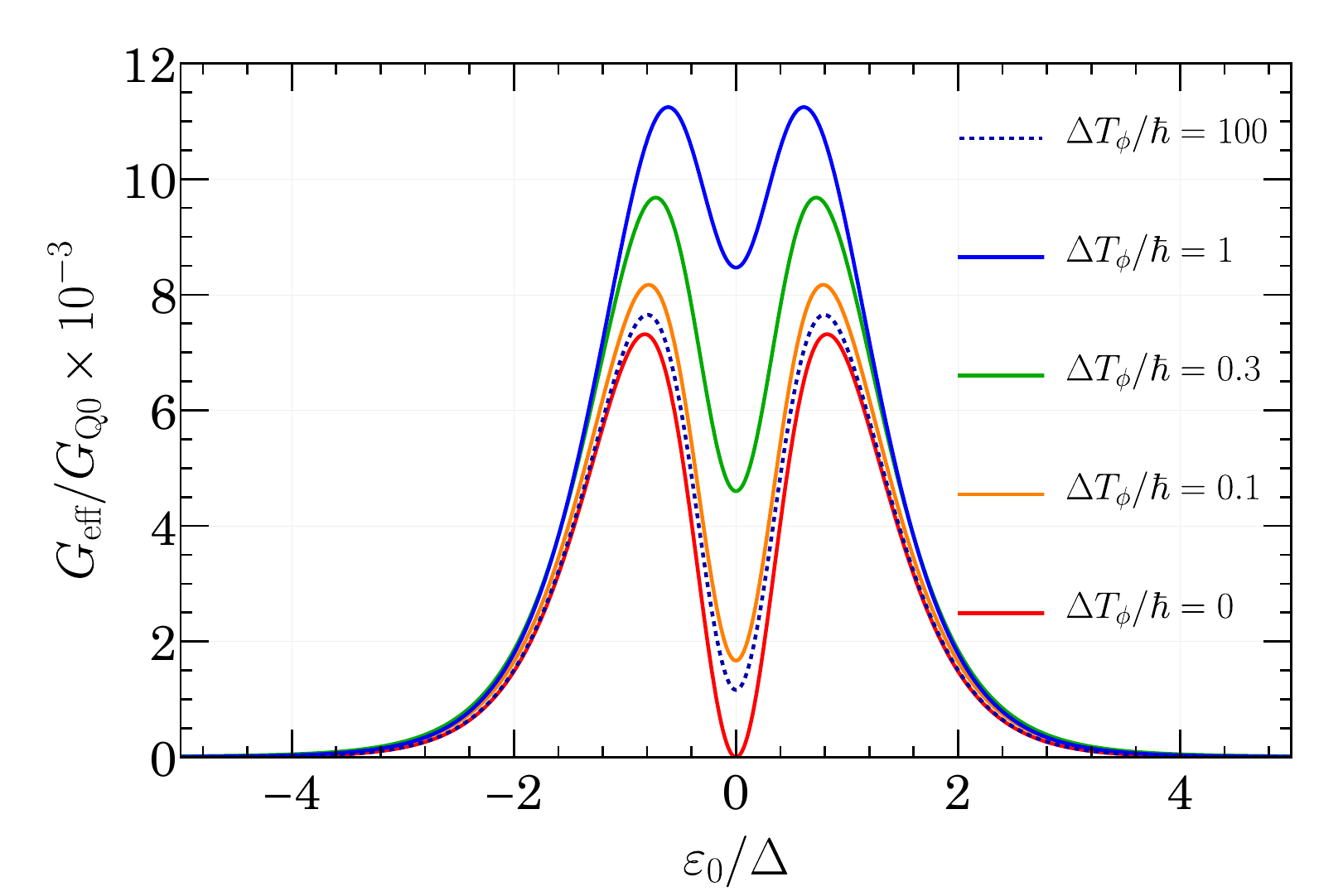}
	\caption{Dependence of the effective conductance $G_\text{eff}=R_\text{eff}^{-1}$ on the energy bias $\varepsilon_0$ at different times of the pure dephasing $T_\phi \Delta/\hbar$. The temperature is set to be $k_\text{B} T/\Delta = 0.4$. Other parameters are the same as in Fig.~\ref{Fig:Resistance1}.} 
	\label{Fig:Resistance2}
\end{figure}

\begin{figure}[t]
	\includegraphics[width=8.6 cm]{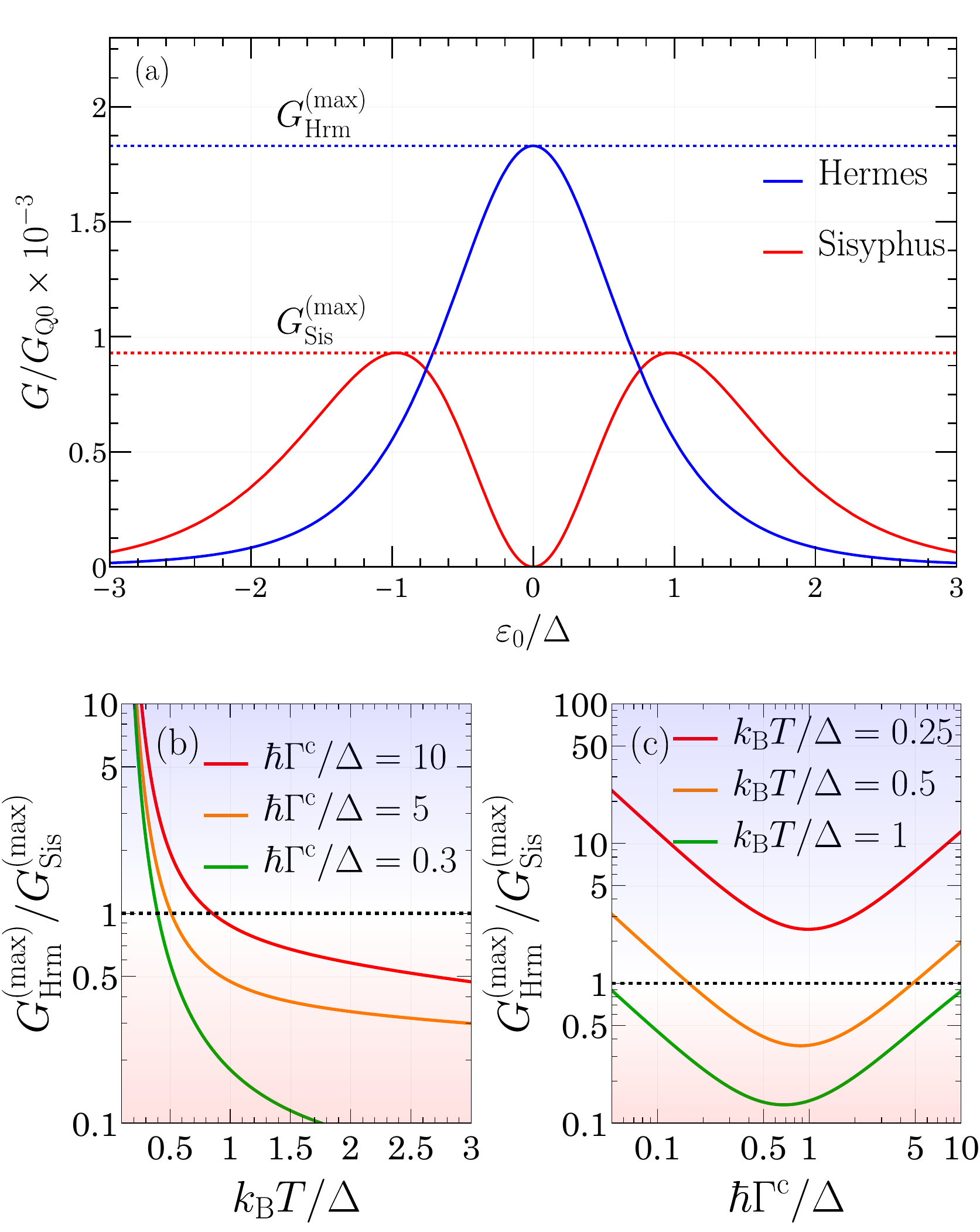}
	\caption{(a) The Sisyphus (red curve) and Hermes (blue curve) conductances 
		as functions of the energy bias $\varepsilon_0$ for 
		$k_\text{B}T/\Delta = 0.5$, $\hbar\Gamma^\text{c}/\Delta = 10$,  $\hbar/T_\phi\Delta = 0.5$, and $\hbar \omega_\text{rf}/\Delta=0.05$. 
		The dotted lines indicate the maximum values of the Sisyphus and 
		Hermes conductances, $G_\text{Sis}^{(\text{max})}$ and 
		$G_\text{Hrm}^{(\text{max})}$, respectively.
		(b) The ratio $G_\text{Hrm}^{(\text{max})}/G_\text{Sis}^{(\text{max})}$ 
		as a function of the normalized temperature $k_\text{B}T/\Delta$ 
		for several values of the relaxation rate 
		$\hbar\Gamma^\text{c}/\Delta$.
		(c) The ratio $G_\text{Hrm}^{(\text{max})}/G_\text{Sis}^{(\text{max})}$ 
		as a function of the normalized relaxation rate 
		$\hbar\Gamma^\text{c}/\Delta$ for several values of 
		$k_\text{B}T/\Delta$.
		The horizontal dotted black lines in panels (b) and (c) refer to $G_\text{Hrm}^{(\text{max})}/G_\text{Sis}^{(\text{max})}=1$, 
		indicating equal maximum conductances.
		The red (blue) shading indicates the Sisyphus-dominated (Hermes-dominated) regime, where $G_\text{Hrm}^{(\text{max})}/G_\text{Sis}^{(\text{max})} < 1$ ($G_\text{Hrm}^{(\text{max})}/G_\text{Sis}^{(\text{max})} > 1$).
	} 
	\label{Fig:Resistance3}
\end{figure}

As a next step, we present and analyze the obtained analytical expressions for the effective capacitance in Figs.~\ref{Fig:ideal}, \ref{Fig:thermal}, \ref{Fig:FWHM}, and for the effective conductance in Figs.~\ref{Fig:Resistance1}, \ref{Fig:Resistance2}, and \ref{Fig:Resistance3} for different relevant parameters. 

Figure~\ref{Fig:ideal} shows the quantum capacitance of an isolated quantum two-level system (blue curve)  and the effective capacitance with the geometric contribution subtracted, $C_\text{eff}-C_\text{geom}$, shown for both limiting cases of good ($T_1^{-1} \ll \omega_\text{rf}$, green curve) and bad ($\omega_\text{rf} \ll T_1^{-1}$, red curve) qubits. 

Since the green curve corresponds to the good qubit, it represents the pure quantum capacitance $C_\text{Q}$ in accordance with Eq.~(\ref{CQoverline}), and the difference between the green and red curves illustrates the maximum value of the tunneling capacitance $C_\text{T,max}$,  Eq.~(\ref{CToverline}). As the temperature approaches absolute zero $k_\text{B}T/\Delta \rightarrow 0$ (not shown in Fig.~\ref{Fig:ideal}), both the good- and bad-qubit curves converge to the quantum capacitance of the ground state. The maximum value of the effective capacitance at the quasi-crossing point is the same for both limiting cases and is reduced by a factor of $\tanh(\Delta/2k_\text{B}T)$ compared to $C_{\text{Q}0}$ in the ideal case (at $k_\text{B}T/\Delta=0$).

We can better understand the obtained results by also examining the full width at half maximum (FWHM) of the effective capacitance $C_\text{eff}$. Figure~\ref{Fig:thermal} shows its dependence on the relaxation rate $\Gamma^\text{c}$ 
for different temperatures, where the relaxation time is modeled as~\cite{Mizuta2017, Esterli2019}
\begin{equation}
	T_1^{-1} = \Gamma^\text{c} \coth \left(\dfrac{\Delta E}{2k_\text{B}T}  \right). \label{T1model}
\end{equation}

To obtain the FWHM  curves, we solved the equation $C_{\text{eff}}(\varepsilon_0)-C_\text{geom}=(C_{\text{eff}}|_{\varepsilon_0=0}-C_\text{geom})/2$ for $\varepsilon_0$. The difference between the two numerical solutions,  $|\varepsilon_0^{(2)}-\varepsilon_0^{(1)}|$, gives the FWHM [see Fig.~\ref{Fig:ideal}]. Interaction with a dissipative environment broadens the effective capacitance curve. In the limit $k_\text{B}T/\Delta \rightarrow 0$, the FWHM curve approaches a constant value corresponding to the FWHM of an ideal qubit (blue dashed line).

In Fig. \ref{Fig:FWHM} we analyze how the FWHM depends on temperature for different relaxation times, corresponding to the good, bad, and intermediate regimes. All curves tend to the same asymptotic constant value as $k_\text{B}T \rightarrow 0$, equal to the FWHM of the ideal qubit. Furthermore, for the considered model of the relaxation time $T_1$~(\ref{T1model}), 
the FWHM increases asymptotically linearly with temperature, except in the case 
$\Gamma^\text{c} = 0$, corresponding to a good qubit. 

At sufficiently high temperatures, the two-level approximation breaks down, as the presence of additional quasi-crossings bounds the maximum value of the FWHM. Despite this limitation, by using the results from Sec.~\ref{Sec:Qudit}, we can still obtain the correct values for the FWHM, as it cannot exceed unity due to the periodic structure of both the energy levels and the effective capacitance with respect to the dimensionless gate charge~$n_\text{g}$.

We also analyze the obtained expressions for the effective conductance $G_\text{eff} = R_\text{eff}^{-1}$ 
in units of $G_\text{Q0} = R_\text{Q0}^{-1}$. Here, the relaxation time is modeled as in Eq.~(\ref{T1model}), and the decoherence time is given by $T_2^{-1} = T_\phi^{-1} + (2T_1)^{-1}$. First, we study its temperature dependence in Fig.~\ref{Fig:Resistance1}. From Eq.~(\ref{Reff_limit_at_0_temperature}), in the limit $k_\text{B}T/\Delta \rightarrow 0$, the effective conductance is determined by the Hermes term. As the temperature increases, the Sisyphus conductance reaches its maximum value at $k_\text{B}T \sim \Delta$. With a further increase in temperature, $k_\text{B}T \gg \Delta$, the effective conductance tends to zero, as one can see from Eqs.~(\ref{R_Sisyphus}, \ref{R_Hermes}).

As follows from Eq.~(\ref{R_Sisyphus}), the Sisyphus conductance reaches its maximum exactly at $T_1 \omega_\text{rf}=1$ for constant $T_1$, in agreement with Ref.~\cite{Grajcar2008}, where it was assumed that the maximum occurs around $T_1 \omega_\text{rf} \sim 1$. This indicates that losses in the quantum system are maximal when the classical probing signal excites the system on the same timescale as strong relaxation occurs. 

By varying the pure dephasing time in Fig.~\ref{Fig:Resistance2}, one can observe how the Hermes conductance is affected. In particular, for $T_\phi = 0$, Eq.~(\ref{R_Hermes}) yields $R_\text{Hrm}^{-1} = 0$; the maximum value of $R_\text{Hrm}^{-1}$ is achieved at $T_2 \omega_{\text{q}0} = 1$. As the dephasing time $T_\phi$ tends to infinity, $T_2 \rightarrow 2T_1$, the Hermes conductance saturates at a finite value, as one can see from Eq.~(\ref{R_Hermes}).

Furthermore, not only does the Hermes resistance represent a novel contribution 
to the effective conductance, but there also exists a parameter range in which 
the Hermes conductance significantly exceeds the Sisyphus one. 
In Fig.~\ref{Fig:Resistance3}(a), we plot both the Sisyphus and Hermes 
conductances for a representative set of parameters and identify the maximum 
values of the respective contributions, $G_\text{Sis}^{(\text{max})}$ and 
$G_\text{Hrm}^{(\text{max})}$. 

To determine which contribution dominates, 
we plot the ratio $G_\text{Hrm}^{(\text{max})}/G_\text{Sis}^{(\text{max})}$ 
as a function of the temperature $k_\text{B}T/\Delta$ and the relaxation rate $\hbar\Gamma^\text{c}/\Delta$ 
in Figs.~\ref{Fig:Resistance3}(b) 
and~\ref{Fig:Resistance3}(c), respectively. 
The horizontal dotted line marking unity separates the regime where the 
Sisyphus contribution dominates (below) from the novel regime where the 
Hermes contribution dominates (above), the latter occurring for either large or small relaxation rates or low temperatures $k_\text{B}T/\Delta \lesssim 1$.

\subsection{Microwave driving} 

Now, consider a microwave resonant signal to be applied to the gate capacitor $C_\text{g}$ (see Fig.~\ref{Fig3}). The corresponding quantum current in Eq.~(\ref{quantum_current_with_drive}) should be substituted into the system Eq.~(\ref{Theory:CPB_clas_eqs}) with 

\begin{equation}
	V_\text{g} = V_{\text{g}0} + \delta V_\text{mw} \cos (\omega_\text{d} t).
\end{equation}

It appears that this problem cannot be treated using conventional perturbation theory due to the three-timescale structure of the system, $\omega_\text{rf} \ll \Omega_\text{R} \ll \omega_\text{q}$, and the applicability conditions of perturbation theory near resonance, which give rise to Rabi oscillations and the emergent Rabi timescale $T_\text{R}=2\pi/\Omega_\text{R}$. However, this problem can be addressed using multi-timescale quantum averaging theory (QAT) \cite{Barajas2026, Barajas2025a}. The essence of QAT lies in systematically separating the dynamics of a quantum system in accordance with its different timescales. The procedure begins by representing the evolution operator as a product of a fast, rapidly oscillating unitary part $\hat{U}_\text{fast} = \exp (i \hat{\Phi})$ expressed as a Magnus expansion with $\hat{\Phi} = \sum_n \hat{\Phi}^{(n)}$ and an effective evolution operator $\hat{U}_\text{eff}$ that evolves only on slower timescales. The effective Hamiltonian $\hat{H}_\text{eff}$ can be found by re-normalizing the fast dynamical phase $\hat{\Phi}^{(n)}$ in every order.  By iteratively applying this separation, one can progressively isolate the fastest, intermediate, and slowest dynamics in the system. This approach effectively filters out the fast oscillations and allows one to construct an averaged description of the system that captures the slower, physically relevant dynamics (in our case the timescales of the classical resonant circuit) without losing information about the influence of the faster processes.

Within the framework of QAT, it becomes possible to write explicit expressions for the quantum and tunneling capacitances while also rigorously identifying the limits where phenomenological approaches remain valid \cite{my_upcoming_work_QAT}. QAT thus provides a controlled and systematic way to handle multi-timescale quantum dynamics that cannot be treated accurately with standard perturbative methods. As a result, we obtained formulas like Eqs.~(\ref{CQ_drive}, \ref{CT_drive}) for the quantum and tunneling capacitances~\cite{my_upcoming_work_QAT}. These results will be presented elsewhere.

\section{Effective capacitance and resistance of a multi-level system}

\label{Sec:Qudit}

In this section we prove several general results related to qu$d$its. Section~\ref{Qudit:isolated} is devoted to the general treatment of the isolated system dynamics. There we prove that an arbitrary $d$-level Hamiltonian, which is dependent on the energy bias $\varepsilon(t)$, has the quantum capacitance proportional to the second derivative of the energy levels. Section~\ref{Sec:Qudit:GKSL} is devoted to the study of quantum open systems. There we show how to generalize the approach of Sec.~\ref{Theory:GKSL} onto $d$-level systems described by the GKSL master equation with arbitrary dependence on relaxation rates and Lindblad jump operators. We show how the effective capacitance and resistance could be obtained and we analyze interesting limits of good and bad qu$d$its. In Sec.~\ref{Qudit:d=2}, we demonstrate that the general formalism developed for $d$-level systems reduces to the qubit case for $d=2$, reproducing the previously obtained results for the effective capacitance and resistance.

In our approach, we start from the classical Lagrangian, which is dependent on several classical $\varphi_{\text{cl},i}$ and quantum $\varphi_{\text{q},j}$ degrees of freedom. Using the formalism developed in Sec.~\ref{Sec:Theory:Hamilton}, we obtain the Hamiltonian after quantization, $\hat{\mathcal{H}}$. The total effective capacitance (including the geometric, quantum, and tunneling contributions) as well as the total effective resistance (including the Sisyphus and Hermes components) can be derived from the quantum current $\langle \hat{\mathcal{I}}\rangle \propto d/dt\langle \partial \hat{\mathcal{H}}/\partial \varepsilon \rangle$ [see Eq.~(\ref{quantum_current_exp})].

\subsection{Effective capacitance of an isolated qu\textit{d}it}
\label{Qudit:isolated}
Let us now consider the Schrödinger equation for the evolution operator $i\hbar \hat{\dot{U}}= \hat{\mathcal{H}} \hat{U}$, with the Hamiltonian $\hat{\mathcal{H}} = \hat{\mathcal{H}}(\varepsilon(t))$, where the energy bias includes a small probing signal, $\varepsilon(t) = \varepsilon_0 + \delta\varepsilon_\text{rf}(t)$. The system can be described correctly using time-dependent perturbation theory, for which the formal conditions are that the amplitude and frequency of the probing signal are much smaller than the minimal distance between the energy levels:
\begin{equation}
	\delta \varepsilon_\text{rf},  \hbar \omega_\text{rf} \ll \hbar \omega_\text{qd}\coloneqq \min_{\varepsilon,i,j} |E_i(\varepsilon) - E_j(\varepsilon)|, \label{QuditA_applicability_for_perturbation_theory}
\end{equation}
where $E_i(\varepsilon)$ are the eigenenergies of the Hamiltonian, defined by $\hat{\mathcal{H}}(\varepsilon)\ket{E_i(\varepsilon)} = E_i(\varepsilon)\ket{E_i(\varepsilon)}$. Since the Hamiltonian includes the geometric energy term proportional to $E_\text{geom} \hat{\mathbf{1}}$, as in Sec.~\ref{Sec:Theory} we write $E_i = E_\text{geom}+E_i'$.

Following the procedure outlined in Sec.~\ref{Theory:GKSL}, we begin by applying the unitary transformation $\hat{S}$ [see Eq.~(\ref{S_inst})] to express the Hamiltonian in the instantaneous eigenbasis. Using the Hellmann-Feynman theorem, we then obtain the corresponding instantaneous Hamiltonian $\hat{\mathcal{H}}_\text{inst}$, whose explicit form is presented in Appendix~\ref{Appendix:IdQd}, Eq.~(\ref{H_inst}).

Since the instantaneous Hamiltonian depends on the time-dependent energy bias, we expand it perturbatively in the small probe $\delta \varepsilon_\text{rf}$ and its time derivative  $\delta \dot \varepsilon_\text{rf}$, as in Eq.~(\ref{Hinst_pert_decomposition_2lvl}). Explicit expressions for $\hat{\mathcal{H}}_{\text{inst}}^{(0)}$, $\hat{V}_{\delta \varepsilon}$ and $\hat{V}_{\delta \dot \varepsilon}$ can be found in Eq.~(\ref{perturbative_operators}).

As shown in Sec.~\ref{Sec:Theory:Hamilton}, the interaction term between the classical and quantum subsystems in the classical equations of motion (the quantum current) can be written as $\langle \hat{\mathcal{I}}\rangle \propto d/dt\langle \partial \hat{\mathcal{H}}/\partial \varepsilon \rangle$~[see Eq.~(\ref{quantum_current_exp})], since there is a linear dependence of the energy bias $\varepsilon$ on the classical probing voltage $V$. This means that we have to evaluate the time derivative operator in the instantaneous representation, $(\partial \hat{\dot{\mathcal{H}}}/ \partial \varepsilon)_\text{inst}$, to determine the effective capacitance of the system. The procedure is straightforward and is discussed in Appendix~\ref{Appendix:IdQd} [see Eq.~(\ref{Qudit_isolated_dhdeps_inst}) for the $(\partial_\varepsilon \hat{\dot{\mathcal{H}}})_\text{inst}$].

After all the preparation steps, we are ready to construct time-dependent perturbation theory. First, using Eq.~(\ref{Hinst_pert_decomposition_2lvl}) for the Hamiltonian, we find the first-order corrections to the evolution operator $\hat{U}$. Next, we calculate the quantum current, defined as the quantum-mechanical average $\langle \psi_0|\hat{U}^\dagger (\partial_{\varepsilon}\hat{\dot{\mathcal{H}}})_\text{inst}\hat{U}|\psi_0 \rangle$. Up to first order, we obtain
\begin{equation}
	\dfrac{d}{dt}\Big\langle \dfrac{\partial \hat{\mathcal{H}}}{\partial V} \Big\rangle = \dfrac{dV}{dt}  \overline{ \dfrac{\partial^2 E}{\partial V^2}}_\text{S}, \label{qudit_ideal_response}
\end{equation}
where $V$ is the classical voltage that probes the quantum system. The straightforward procedure for converting from the energy bias $\varepsilon$ to the probing voltage $V$ is described in Appendix~\ref{Appendix:IdQd:CQ}.  We note that the response of the quantum system, given by Eq.~(\ref{qudit_ideal_response}), includes a statistical average over the Hamiltonian eigenstates, where the probabilities $p_i$ correspond to the occupations of the energy levels $\ket{E_i}$.

The effective capacitance consists of geometric and quantum components. The tunneling capacitance is zero, because there are no redistribution processes yet. While the expression for the geometric capacitance (\ref{C_geom}) remains valid for qu$d$its, the quantum capacitance contribution associated with the $n$-th energy level is equal to the second derivative of its energy with respect to the probing voltage (when the geometric energy $E_\text{geom}$ coming from the diagonal part of the original Hamiltonian is not taken into account):
\begin{equation}
	C_{\text{Q}}^{(n)} =-\dfrac{\partial^2 E'_n}{\partial V^2}.
\end{equation}
The total quantum capacitance is given as the sum of partial contributions from each energy level
\begin{equation}
	C_{\text{Q}}=\sum\limits_{n=1}^d p_n C_\text{Q}^{(n)}=-  \overline{ \dfrac{\partial^2 E'}{\partial V^2}}_\text{S}. \label{CQ_isolated_qd}
\end{equation}

\subsection{Effective capacitance and resistance of a dissipative qu\textit{d}it}
\label{Sec:Qudit:GKSL}
We start with the GKSL master equation (\ref{Lindblad_eq}) written in the instantaneous representation of the initial Hamiltonian $\hat{\mathcal{H}} = \hat{\mathcal{H}}(\varepsilon(t))$. Analogously to Sec.~\ref{Theory:GKSL}, we first find perturbative correction  for the density matrix, and then obtain the effective capacitance and resistance, assuming that both the Lindblad jump operators and the relaxation rates in the GKSL equation may be bias-dependent: $\hat{L}_\alpha=\hat{L}_\alpha(\varepsilon)$ and $\gamma_\alpha = \gamma_\alpha(\varepsilon)$.

To solve this problem, we use the Liouville space \cite{Gyamfi2020}, where the GKSL equation takes the form of a linear first-order differential equation 
\begin{equation}
	\lket{\dot \rho} = \mathscr{L}\!\! \lket{\rho} \label{GKSL_Liouville}
\end{equation}
with respect to the superket $|\rho\rangle\!\rangle$, that is obtained by applying the vectorization map to the density operator, which consists in reshaping the $d \times d$ matrix $\hat{\rho}$ into $d^2$-dimensional column vector by stacking its columns sequentially. The definition of the Liouvillian superoperator is given in Eq.~(\ref{exact_liouvillian}).

Similarly to the previous subsections, the Liouvillian $\mathscr{L}$ can be perturbatively expanded within the small probe and its derivative 
\begin{equation}
	\mathscr{L} = \mathscr{L}_0+\mathscr{L}_{\delta \varepsilon} \delta \varepsilon_\text{rf}(t)+\mathscr{L}_{\delta \dot \varepsilon} \delta \dot \varepsilon_\text{rf}(t), \label{Liov_upto1}
\end{equation}  
provided that the corresponding assumptions are satisfied, as discussed below. A short introduction to the Liouville space, discussion of the unperturbed Liouvillian $\mathcal{L}_0$ properties and perturbation decomposition $\mathscr{L}_{\delta \varepsilon,\delta \dot \varepsilon}$ could be found in the Appendices~\ref{Appendix:GKSL:Liouville}, \ref{Appendix:GKSL:properties} and Eqs.~(\ref{def1}-\ref{def4}) respectively.

The system can be correctly described using perturbation theory, provided that the conditions (\ref{QuditA_applicability_for_perturbation_theory}) are satisfied. Additionally, to perform a series expansion of the relaxation bias-dependent terms $\gamma_\alpha(\varepsilon)$ in Eq.~(\ref{gamma_series}), the probing amplitude needs to be small enough 
\begin{equation}
	\delta \varepsilon_\text{rf} \ll k_\text{B} T \dfrac{\gamma_\alpha^{(0)} }{|\partial \gamma_\alpha/\partial x_i| |\partial E_i/\partial \varepsilon|}, \label{pert_gamma}
\end{equation} 
where we assume that the relaxation rates may depend on dimensionless combinations of the form $x_i=E_i/k_\text{B}T$. Since $|\partial E_i/\partial \varepsilon|$ is a dimensionless quantity typically of the order of unity (not in the vicinity of quasi-crossing points, where the next non-zero series-expansion term needs to be written), the  problem may happen primarily at very low temperatures $T \lesssim \delta \varepsilon_\text{rf}/k_\text{B}$ for a fixed amplitude of the classical probing signal. However, because the relaxation rates tend to constant values at low temperatures $k_\text{B} T \ll \hbar \omega_\text{qd}$, the corresponding derivatives $|\partial \gamma_\alpha/\partial x_i|$ vanish, and the perturbation theory remains valid under the conditions (\ref{QuditA_applicability_for_perturbation_theory}).

The solution for the density matrix can be found in the same form as in Eq.~(\ref{Lindblad_anzatz})  by searching for the first-order corrections to the thermal equilibrium state $| \rho^{(0)} \rangle\!\rangle$ at \text{$t \rightarrow\infty$} for times that are much larger than all relaxation and decoherence times of the system $t \gg T_{\text{rel}, i}, T_{\text{dec}, j}$
\begin{equation}
	\lket{\rho} = \lvert  \rho^{(0)} \rangle\!\rangle + \delta \varepsilon_\text{rf}(t)\lket{A}+ \delta \dot\varepsilon_\text{rf}(t) \lket{B}. \label{new_anzatz}
\end{equation}
All relaxation and decoherence times can be defined as $T_\alpha = -\text{Re}\,(l_\alpha)$, such that $T_\alpha > 0$, where $\ell_\alpha$ are eigenvalues of the unperturbed Liouvillian $\mathscr{L}_0$, $\mathscr{L}_0 \lket{\ell_{\text{R},\alpha}} = \ell_\alpha \lket{\ell_{\text{R},\alpha}}$, corresponding to the right eigenvectors $\lket{\ell_{\text{R},\alpha}}$ [see Appendix~\ref{Appendix:GKSL:properties}]. It is also possible to find corrections to the dynamics of the quantum system analytically for times $t$ which are not restricted by such strong inequality $t \gg T_{\text{rel}, i}, T_{\text{dec}, j}$ \cite{my_upcoming_work_relaxation}.

By substituting the ansatz (\ref{new_anzatz}) into the GKSL equation (\ref{GKSL_Liouville}) with the Liouvillian being expanded within the small classical probe up to first order (\ref{Liov_upto1}), we obtain two coupled equations on unknown superkets $|A\rangle\!\rangle$ and $|B \rangle\!\rangle$ [see Eq.~(\ref{system_AB})]. Thus both $|A,B\rangle\!\rangle$ can be found [see Eq.~(\ref{sol_AB})].

The quantum current $\langle \hat{\mathcal{I}}\rangle$ is still proportional to $d/dt\langle \partial \hat{\mathcal{H}}/\partial  \varepsilon \rangle$ [see Eq.~(\ref{quantum_current_exp})]. The first-order corrections with respect to the small classical probing signal results in both the effective capacitance
\begin{widetext}
	\begin{gather}
		C_\text{eff} = -(\alpha e)^2
		\Big[  \langle\!\langle \partial_\varepsilon (\partial_\varepsilon \mathcal{H})_\text{inst}^{(0)}
		\,|\, \rho^{(0)} \rangle\!\rangle +\label{QD_GKSL_Ceff}  \langle\!\langle (\partial_\varepsilon \mathcal{H})_\text{inst}^{(0)}
		| \rho^{(0)}_{\delta \varepsilon} \rangle\!\rangle +
		\langle\!\langle (\partial_\varepsilon \mathcal{H})_\text{inst}^{(0)}
		| \left[1+(\mathscr{L}_0/\omega_\text{rf})^2 \right]^{-1}  [\mathscr{L}_{\delta \dot \varepsilon}| \rho^{(0)} \rangle\!\rangle-| \rho^{(0)}_{\delta \varepsilon} \rangle\!\rangle]
		\Big],
	\end{gather} 
	and the effective conductance
	\begin{gather}
		R_\text{eff}^{-1} = -(\alpha e)^2  \langle\!\langle (\partial_\varepsilon \mathcal{H})_\text{inst}^{(0)}  | \mathscr{L}_0 \left[1+(\mathscr{L}_0/\omega_\text{rf})^2 \right]^{-1}  [\mathscr{L}_{\delta \dot \varepsilon}| \rho^{(0)} \rangle\!\rangle-| \rho^{(0)}_{\delta \varepsilon} \rangle\!\rangle]\label{QD_GKSL_Reff}.
	\end{gather}
	The effective admittance $Y_\text{eff}=  i \omega_\text{rf} C_\text{eff}+R_\text{eff}^{-1}$ can be found as
	\begin{gather}
		Y_\text{eff} = -i \omega_\text{rf}(\alpha e)^2 \Big[ \langle\!\langle \partial_\varepsilon (\partial_\varepsilon \mathcal{H})_\text{inst}^{(0)}
		\,|\, \rho^{(0)} \rangle\!\rangle+ \langle\!\langle (\partial_\varepsilon \mathcal{H})_\text{inst}^{(0)}
		| \dfrac{1}{1+i\mathscr{L}_0/\omega_\text{rf}}  \mathscr{L}_{\delta \dot \varepsilon}| \rho^{(0)} \rangle\!\rangle +i\langle\!\langle (\partial_\varepsilon \mathcal{H})_\text{inst}^{(0)}
		| \dfrac{\mathscr{L}_0/\omega_\text{rf}}{1+i\mathscr{L}_0/\omega_\text{rf}}  | \rho^{(0)}_{\delta \varepsilon} \rangle\!\rangle\Big] \label{QD_GKSL_Yeff}.
	\end{gather}
\end{widetext}
Equations~(\ref{QD_GKSL_Ceff}), (\ref{QD_GKSL_Reff}), and (\ref{QD_GKSL_Yeff}) constitute the central result of this work: they provide fully general analytical expressions for the effective capacitance, resistance, and admittance of an open multi-level quantum system coupled to a classical resonator within the GKSL framework. These expressions are valid for arbitrary system Hamiltonians $\hat{\mathcal{H}}(\varepsilon)$, relaxation rates $\gamma_\alpha (\varepsilon)$, and jump operators $\hat{L}_\alpha (\varepsilon)$, when the conditions~(\ref{pert_gamma}) and (\ref{QuditA_applicability_for_perturbation_theory}) for the applicability of perturbation theory are satisfied.

Consider now how different terms of Eqs.~(\ref{QD_GKSL_Ceff}) and (\ref{QD_GKSL_Reff}) contributes to the geometric, quantum, and tunneling capacitances, as well as to the Sisyphus and Hermes conductances.

The first term of Eq.~(\ref{QD_GKSL_Ceff}) can be computed exactly [see Eq.~(\ref{first_term_ceff_evaluated})], and contributes to both the geometric and quantum capacitances,
\begin{gather}
	C_\text{geom}+C_\text{Q} = -\overline{ \dfrac{\partial^2 E}{\partial V^2 }}_\text{th}.\label{Cgeom+CQ}
\end{gather}
This results in the same geometric capacitance as in Eq.~(\ref{C_geom}), while the quantum capacitance reads
\begin{gather}
	C_\text{Q} = -\overline{ \dfrac{\partial^2 E'}{\partial V^2 }}_\text{th},
\end{gather}
and can be seen as a generalization of Eq.~(\ref{CQ_isolated_qd}) [for the derivation see Appendix~\ref{Appendix:GKSL:Ceff}]. The tunneling capacitance consists of the last two terms of Eq.~(\ref{QD_GKSL_Ceff}),
\begin{gather}
	C_\text{T} = -(\alpha e)^2
	\Big[   \langle\!\langle (\partial_\varepsilon \mathcal{H})_\text{inst}^{(0)}
	| \rho^{(0)}_{\delta \varepsilon} \rangle\!\rangle + \label{CT_qd_main}\\
	\nonumber+
	\langle\!\langle (\partial_\varepsilon \mathcal{H})_\text{inst}^{(0)}
	| \left[1+(\mathscr{L}_0/\omega_\text{rf})^2 \right]^{-1}  [\mathscr{L}_{\delta \dot \varepsilon}| \rho^{(0)} \rangle\!\rangle-| \rho^{(0)}_{\delta \varepsilon} \rangle\!\rangle]
	\Big].
\end{gather}
The first term of Eq.~(\ref{CT_qd_main}) can be computed using Eq.~(\ref{second_term_ceff_evaluated}), and thus from its qubit analogy, Eq.~(\ref{CToverline}), can be associated with the maximum possible value of the tunneling capacitance
\begin{equation}
	C_\text{T,max} = -\sum\limits_n  \dfrac{\partial E_n}{\partial V} \dfrac{\partial p_n^\text{th}}{\partial V}.
\end{equation}
The effective conductance consists of two terms. The term proportional to the occupation probabilities $|\rho^{(0)}\rangle\!\rangle$, by analogy with the qubit case~(\ref{R_Hermes}), generalizes the Hermes conductance,
\begin{gather}
	R_\text{Hrm}^{-1} = \label{Rhrm_main} 	 -(\alpha e)^2
	\langle\!\langle (\partial_\varepsilon \mathcal{H})_\text{inst}^{(0)}  | \mathscr{L}_0 \left[1+(\mathscr{L}_0/\omega_\text{rf})^2 \right]^{-1}  \!\!\mathscr{L}_{\delta \dot \varepsilon}| \rho^{(0)} \rangle\!\rangle,
\end{gather}
while the term proportional to the derivatives of the occupation probabilities $|\rho^{(0)}_{\delta \varepsilon}\rangle\!\rangle$, with its qubit analogue given in Eq.~(\ref{R_Sisyphus}), yields the generalized Sisyphus conductance,
\begin{equation}
	R_\text{Sis}^{-1}= (\alpha e)^2
	\langle\!\langle (\partial_\varepsilon \mathcal{H})_\text{inst}^{(0)}  | \mathscr{L}_0 \left[1+(\mathscr{L}_0/\omega_\text{rf})^2 \right]^{-1}  | \rho^{(0)}_{\delta \varepsilon} \rangle\!\rangle. \label{Rsis_main}
\end{equation}

The structure of Eq.~(\ref{QD_GKSL_Yeff}) provides an intuitive picture of the origin of each contribution to the effective admittance. The first term gives rise to the geometric and quantum capacitances~(\ref{Cgeom+CQ}). The second term, in the good qu$d$it limit $T_{\text{rel},i}^{-1}, T_{\text{dec},j}^{-1} \ll \omega_\text{rf}$, ensures the correct definition of the admittance [see Eq.~(\ref{AppendixB:CQ})], and simultaneously determines the Hermes conductance~(\ref{Rhrm_main}). The third term gives rise to both the tunneling capacitance~(\ref{CT_qd_main}) and the Sisyphus conductance~(\ref{Rsis_main}). This reveals a common origin of the quantum capacitance and the Hermes resistance, as well as of the tunneling capacitance and the Sisyphus resistance [see Fig.~\ref{Fig:response}].

The results for the effective capacitance and resistance can be significantly simplified in the limiting cases of good and bad qu$d$its. To obtain such expressions one needs to estimate the magnitudes of each term in Eqs.~(\ref{QD_GKSL_Ceff}) and (\ref{QD_GKSL_Reff}) and drop terms $\propto \omega_\text{rf}, \, \omega_\text{qd}$ for the case of a bad qu$d$it, or drop terms $\propto \omega_\text{rf}, \, T_{\text{rel}, i}^{-1}, T_{\text{dec}, j}^{-1}$ for the case of a good qu$d$it [see Appendix~\ref{Appendix:GKSL_GOOD_BAD} for detailed calculations]. The resulting effective capacitance of a bad qu$d$it ($\omega_\text{rf} \ll \omega_\text{qd} \ll T_{\text{rel}, i}^{-1}, T_{\text{dec}, j}^{-1}$) becomes
\begin{equation}		
	C_\text{eff}^{(\text{bad})} =-\dfrac{\partial}{\partial V}\overline{ \dfrac{\partial E}{\partial V}}_\text{th}, 
	\label{C_qd_BAD}
\end{equation}
while for a good qu$d$it ($T_{\text{rel}, i}^{-1}, T_{\text{dec}, j}^{-1} \ll \omega_\text{rf} \ll \omega_\text{qd}$)
\begin{equation}		
	C_\text{eff}^{(\text{good})} =-\overline{ \dfrac{\partial^2 E}{\partial V^2}}_\text{th}. 
	\label{C_qd_GOOD}
\end{equation}
The effective conductance $R_\text{eff}^{-1}$ vanishes in both limiting cases.

\begin{figure}[t]
	\includegraphics[width=8.6 cm]{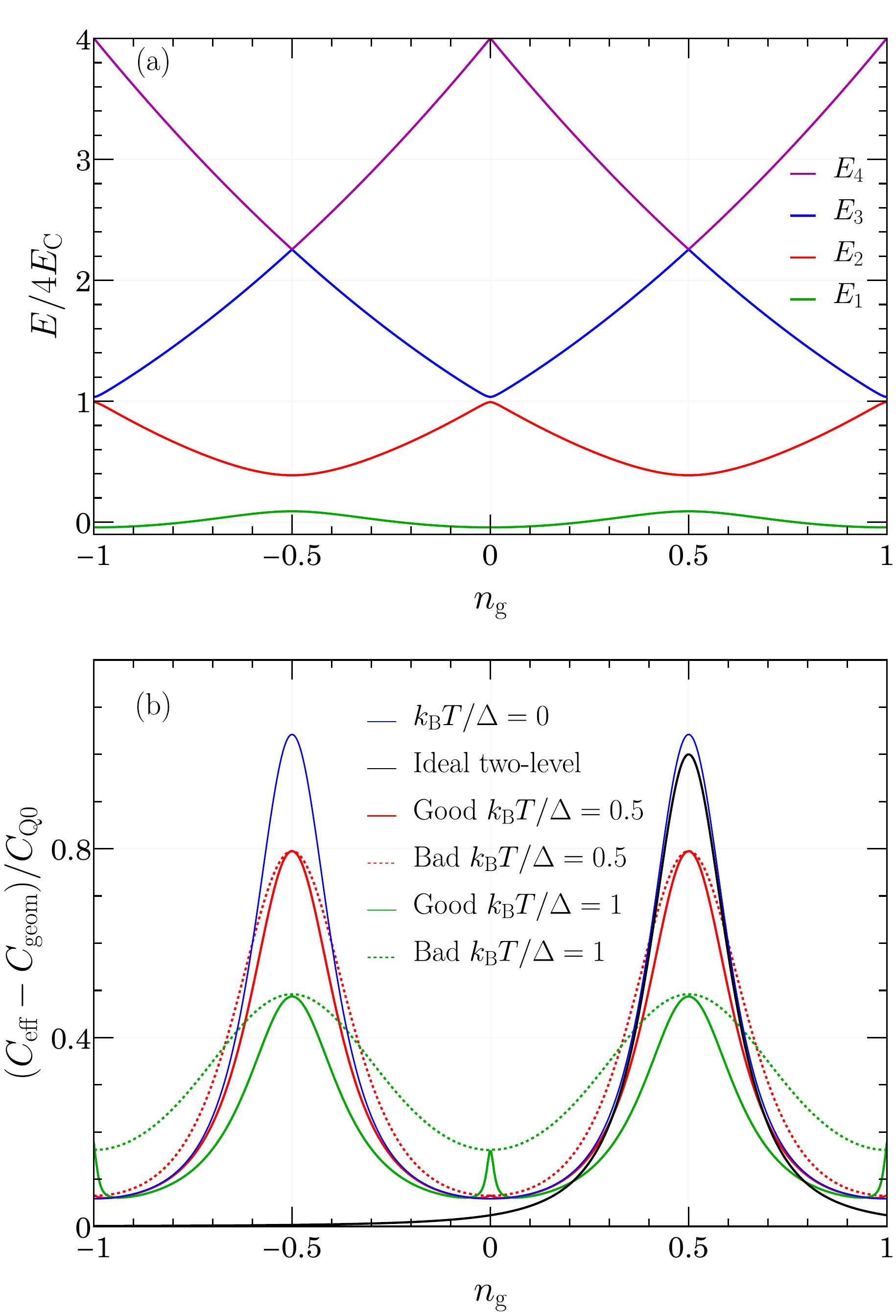}
	\caption{(a) Energy levels of a charge qu$d$it at $E_\text{J}/4E_\text{C}=0.3$. (b) The effective capacitance minus the geometric capacitance, $C_\text{eff} - C_\text{geom}$,  of the four-level qu$d$it. Considering that the
		upper levels effective capacitances of both good and bad qu$d$its saturates at
		non-zero value because capacitance is proportional to the second derivative of
		the energy levels $C_{\text{Q}}\propto \partial ^{2}E/\partial \protect%
		\varepsilon ^{2}$ and the energy levels are essentially parabolic.
		The dependencies are shown for three different temperatures, namely $k_\text{B}T/\Delta=0, \,0.5 \text{ and }  1$. Capacitances of good and bad qu$d$its coincide for the
		limit $k_\text{B}T/\Delta = 0$. The black curve depicts the quantum capacitance in the ground state within the two-level approximation.}
	\label{Fig:Plotq2}
\end{figure}

For the graphical analysis, we consider a multi-level superconducting island in the regime where the condition $E_\text{J}/4E_\text{C} \ll 1$ is no longer satisfied, such that the two-level approximation breaks down. Figure~\ref{Fig:Plotq2}(a) shows the first four energy levels for the ratio $E_\text{J}/4E_\text{C}=0.3$. Figure~\ref{Fig:Plotq2}(b) presents the effective capacitance in the limiting cases of a good and a bad qu$d$it at different temperatures. The black curve corresponds to the quantum capacitance of the ground state within the two-level approximation centered around $n_\text{g}=0.5$. The blue curve shows the exact quantum capacitance of the ground state, while the red and green curves correspond to the effective capacitances at finite temperatures $k_\text{B}T/\Delta = 0.5$ and $k_\text{B}T/\Delta = 1$, respectively, with solid and dashed lines representing the good and bad qu$d$it limits. For a similar realization of a multi-level system --- a multi-electron double quantum dot --- see Ref.~\cite{Lundberg2024}.

Pronounced additional peaks at $n_\text{g}=0, \, \pm 1$ emerge due to the contribution of higher excited states at high-enough temperature, in particular, the transition between the second and the third energy levels, which is absent in the two-level model. Our theory for qu$d$its explains the origin of the experimental observations reported in Ref.~\cite{Persson2010_2}. For the parameters of that experiment, $T_1^{-1} \ll \omega_{\text{rf}}$, so that the qu$d$it is good, relaxation-induced broadening is weak, and additional peaks at integer values of $n_\text{g}$ can be resolved. This clearly demonstrates that multi-level effects can strongly modify the effective capacitance, even at relatively small $E_\text{J}/4E_\text{C}$. In contrast to the two-level prediction, the effective capacitance does not vanish far from the quasi-crossing points. This behavior originates from the parabolic structure and finite curvature of the energy levels.

Just like for qubits, at finite temperature, the thermal population of the excited states reduces the maximum value of the effective capacitance and increases the FWHM, while the difference between the good and bad qu$d$it limits becomes more pronounced, reflecting the contribution of relaxation effects into the tunneling capacitance.

\subsection{From qu$d$its to qubits: $d=2$}
\label{Qudit:d=2}
It is instructive to point out that the general results of the analysis above for $d$-level systems can be directly mapped onto the previously obtained results in Sec.~\ref{Theory:GKSL} for qubits.

To see this, one takes the expressions for the effective capacitance (\ref{QD_GKSL_Ceff}) and resistance (\ref{QD_GKSL_Reff}) of a $d$-level system with a known Hamiltonian $\hat{\mathcal{H}}$, relaxation rates $\gamma_\alpha(\varepsilon)$, and the Lindblad jump operators $\hat{L}_\alpha(\varepsilon)$ for $d=2$. The Liouville space then consists of four-dimensional superkets, super-bras, and $4 \times 4$ superoperators~[see Eqs.~(\ref{deHinst_superket}-\ref{inverse_superoperator})]. One can obtain the expressions for the effective capacitance and resistance simply by calculating the dot products of the relevant objects in Eqs.~(\ref{QD_GKSL_Ceff}) and (\ref{QD_GKSL_Reff}). The final results, Eqs.~(\ref{cgeom+cq}-\ref{Rsis_app}), with the use of Eqs.~(\ref{CQ0}) and (\ref{RQ0}), totally agree with the effective capacitance~(\ref{Appendix_bloch_qubit_Ceff}) and effective conductance~(\ref{Appendix_bloch_qubit_Reff}) obtained in Sec.~\ref{Theory:GKSL} and Appendix~\ref{Appendix_bloch_qubit}.

\section{Other quantum systems}
\label{Sec:Other}

\begin{figure}[htbp]
	\includegraphics[width=8.5cm]{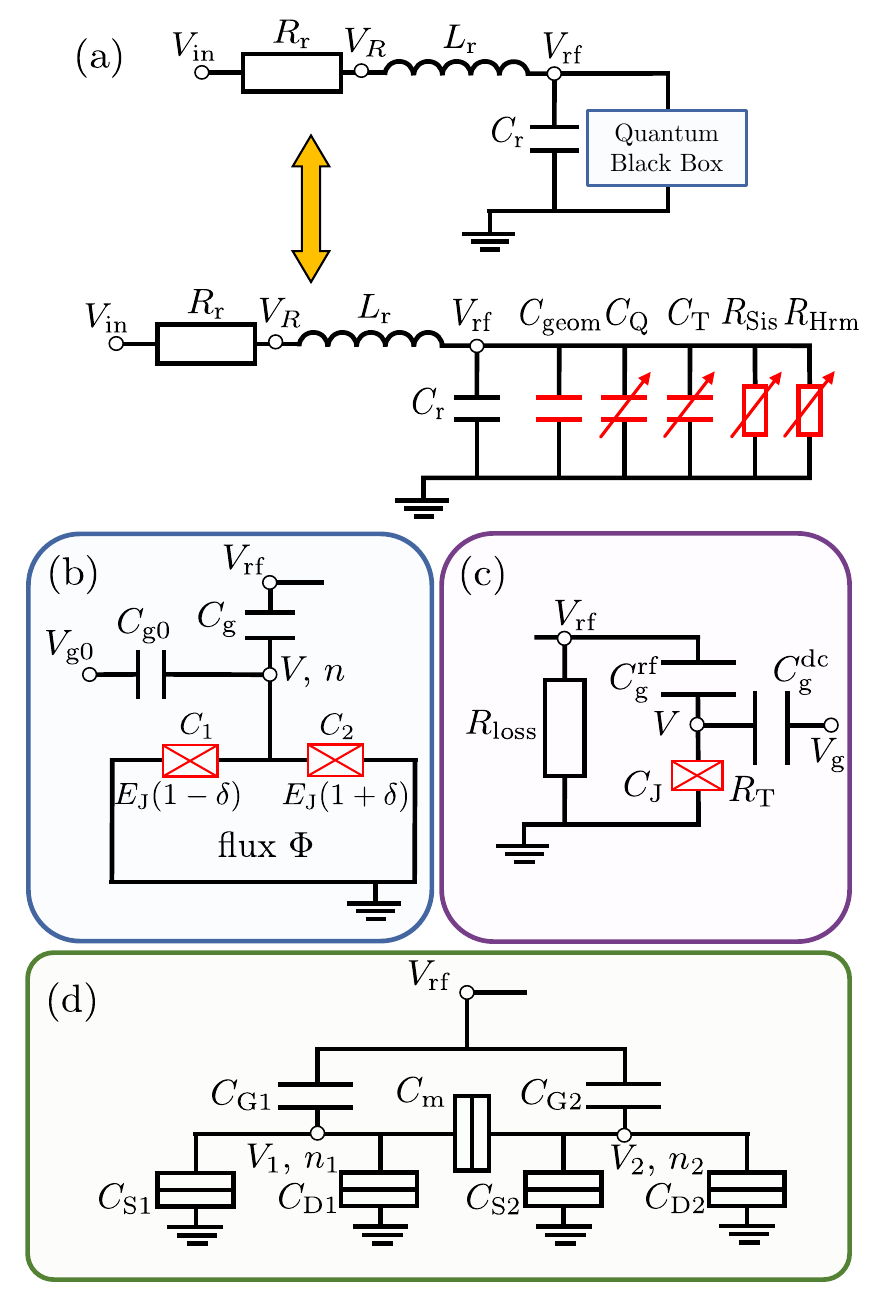}
	\caption{ 	
		(a) Any quantum system (\textquotedblleft black box\textquotedblright) coupled capacitively to an $RLC$ circuit can be represented as a collection of geometric $C_\text{geom}$, quantum $C_\text{Q}$, and tunneling $C_\text{T}$ capacitances, as well as Sisyphus $R_\text{Sis}$ and Hermes $R_\text{Hrm}$ resistances, connected in parallel. In the circuit diagrams (b), (c), and (d), the classical resonator is not shown explicitly; it is coupled to the quantum systems as depicted in (a) via the rf voltage $V_\text{rf}$. 
		(b) A single-Cooper-pair transistor (SCPT) coupled to a classical $RLC$ circuit \cite{Sillanpaeae2005}. The SCPT consists of a superconducting island formed by two Josephson junctions (JJs)  with energies $E_\text{J}(1\pm \delta)$ and capacitances $C_{1,2}$, forming a loop threaded by a magnetic flux $\Phi$. We introduce the superconducting phase difference $\varphi$ across the two junctions. The island is coupled to a gate voltage $V_\text{g0}$ via the capacitor  $C_\text{g0}$, from which the energy bias $\varepsilon$ could be controlled. The SCPT is coupled to the classical circuit via  the capacitor  $C_\text{g}$.
		(c) A single-electron box (SEB) coupled to a classical $RLC$ circuit. The capacitor  $C_\text{g}^\text{dc}$ tunes the working point, while the capacitor  $C_\text{g}^\text{rf}$ is coupled to the classical probing device. The SEB is terminated to ground via a JJ with capacitance $C_\text{J}$ and losses  $R_\text{T}$ and $R_\text{loss}$, which arise due to dissipation in resonators normal top layer \cite{Persson2010}.
		(d) Schematic representation of a double quantum dot (DQD) connected to a classical resonant $RLC$ circuit \cite{Mizuta2017, Esterli2019}. The islands with phases $\varphi_{1,2}$ correspond to quantum dots, coupled to each other via a tunnel barrier with mutual capacitance $C_\text{m}$. Each island is connected to source and drain tunnel barriers with capacitances $C_{\text{S}1,2}$ and $C_{\text{D}1,2}$, respectively. The dots are coupled to the classical circuit via gate capacitors  $C_{\text{G}1,2}$.
	}
	\label{Fig_other}
\end{figure}

In this section, we demonstrate that an arbitrary quantum system coupled capacitively to a classical resonator can be modeled as a collection of the effective capacitance and resistance, connected in parallel, as illustrated in Fig.~\ref{Fig_other}(a). This \textquotedblleft quantum black box\textquotedblright \, representation provides a universal framework for describing the back-action of quantum systems on classical probing circuits.

Using our results for qubits [see Eqs.~(\ref{C_geom}--\ref{qubit_CT}) for the effective capacitance, and Eqs.~(\ref{Qubit_R_eff}--\ref{RQ0}) for the effective resistance] and qu$d$its [see Eqs.~(\ref{QD_GKSL_Ceff}--\ref{Rsis_main})], we show how to derive analytical expressions for the effective impedance for other quantum systems. For convenience, all qubit results are summarized in Table~\ref{Table:TL_results}.

Once the effective capacitance and resistance for a given system have been determined, one can readily calculate the phase shift $\Theta$ acquired by the reflected signal~\cite{Paila2009},
\begin{equation}
	\Gamma = |\Gamma|e^{i\Theta} = \dfrac{Z - Z_0}{Z + Z_0},
\end{equation}
where $Z_0 = 50\,\Omega$ is the characteristic impedance of the transmission line. The load impedance $Z$ is given by
\begin{equation}
	Z = \left[R_\text{eff}^{-1} + \left(\dfrac{-i}{\omega_\text{rf} C_\text{eff}}\right)^{-1} + \left(\dfrac{-i}{\omega_\text{rf} C_\text{r}}\right)^{-1}\right]^{-1} + i\omega_\text{rf} L_\text{r} + R_\text{r},
	\label{total_impedance}
\end{equation}
since the geometric, quantum, and tunneling capacitances together with the Sisyphus and Hermes resistances all enter the effective circuit connected in parallel.

In what follows, we consider four particular systems [see Fig.~\ref{Fig_other}]. First, we start with the Cooper-pair box, which was described in Sec.~\ref{Sec:Theory},  Fig.~\ref{Fig3} and Ref.~\cite{Johansson2006}, by demonstrating that all the results from Table~\ref{Table:TL_results} are self-consistent. After that, we explore the single-Cooper-pair transistor \cite{Sillanpaeae2005} in Sec.~\ref{Sec:Other:SCPT}, the double quantum dot \cite{Mizuta2017,Esterli2019} in Sec.~\ref{Sec:Other:DQD}, and the single-electron box \cite{Persson2010, Cochrane2024} in Sec.~\ref{Sec:Other:SEB}. 

\begin{table*}[t]
	\begin{center}
		\normalsize
		\begin{tabular}{|p{3cm}|p{6cm}|p{8cm}|}
			\hline
			
			\parbox[c]{3cm}{\centering\vspace{1.5ex}
				\fontsize{11}{13}\selectfont\textbf{Quantum system}
				\vspace{1.5ex}}
			& \parbox[c]{6cm}{\centering\vspace{1.5ex}
				\fontsize{11}{13}\selectfont\textbf{System parameters} $\varepsilon$, $\Delta$, $T_1$, $T_2$
				\vspace{1.5ex}}
			& \parbox[c]{8cm}{\centering\vspace{1.5ex}
				\fontsize{11}{13}\selectfont\textbf{Expressions for} $C_{\text{geom}}$, $C_{\text{Q}0}$, $R_{\text{Q}0}$
				\vspace{1.5ex}}
			\\ \hline
			
			\parbox[c]{3cm}{\centering\vspace{1.5ex} Cooper-pair box from Sec.~\ref{Sec:Theory} \cite{Johansson2006}, Fig.~\ref{Fig3}\vspace{1.5ex}}
			&
			\parbox[c]{6cm}{\centering\vspace{1.5ex}
				$\varepsilon(t)=4E_\text{C}(1-2n_\text{g})$, $\,\,\,\alpha_\text{CPB} = 2 \dfrac{C_\text{m}}{C_\Sigma}$,\\[1ex]
				$\Delta = E_\text{J}$.
				\vspace{1.5ex}}
			&
			\parbox[c]{8cm}{\centering\vspace{1.5ex}
				$C_\text{geom} = C_\text{m} \left(1-\dfrac{C_\text{m}}{C_\Sigma}\right)$,\\[1ex] $C_{\text{Q0}}=\dfrac{2e^2}{\Delta}\left(\dfrac{C_\text{m}}{C_\Sigma}\right)^{2}$,  $\,\,\, R_{\text{Q}0}^{-1}=\dfrac{e^2}{\hbar} \left(\dfrac{C_\text{m}}{C_\Sigma}\right)^{2}$
				\vspace{1.5ex}}
			\\ \hline
			
			\parbox[c]{3cm}{\centering\vspace{1.5ex} Single-Cooper-pair transistor \cite{Sillanpaeae2005}, Fig.~\ref{Fig_other}(b)\vspace{1.5ex}}
			&
			\parbox[c]{6cm}{\centering\vspace{1.5ex}
				$\varepsilon(t)=4E_\text{C}(1-2n_\text{g})$, $\,\,\,\alpha_\text{SCPT} =2 \dfrac{C_\text{g}}{C_\Sigma}$,\\[1ex]
				$\Delta =2E_\text{J}\Big|\cos \dfrac{\phi}{2}\Big|\sqrt{1+\delta^2 \tan^2 \dfrac{\phi}{2}}$.
				\vspace{1.5ex}}
			&
			\parbox[c]{8cm}{\centering\vspace{1.5ex}
				$C_\text{geom}=C_\text{g}\left(1-\dfrac{C_\text{g}}{C_\Sigma} \right)$,\\[1ex] $C_{\text{Q}0}=\dfrac{e^2}{\Delta}\left( \dfrac{C_\text{g}}{C_\Sigma}\right)^{2}$, $\,\,\, R_{\text{Q}0}^{-1}=\dfrac{e^2}{\hbar} \left(\dfrac{C_\text{g}}{C_\Sigma}\right)^{2}$
				\vspace{1.5ex}}
			\\ \hline
			
			\parbox[c]{3cm}{\centering\vspace{1.5ex} Double quantum dot \cite{Esterli2019}, Fig.~\ref{Fig_other}(d)\vspace{1.5ex}}
			&
			\parbox[c]{6cm}{\centering\vspace{1.5ex}
				$\varepsilon=E_{\text{C}1}(1-2n_{\text{g}1})-E_{\text{C}2}(1-2n_{\text{g}2})$,\\[1ex]
				$\alpha_\text{DQD} = \alpha'$, $\,\,\,\Delta = \Delta_\text{c}$,\\[1ex]
				$T_1^{-1} = \Gamma^\text{c} \coth \dfrac{\Delta E}{2 k_\text{B}T}$, $\,\,\,T_2=0$.
				\vspace{1.5ex}}
			&
			\parbox[c]{8cm}{\centering\vspace{1.5ex}
				$C_\text{geom}= C_{\text{SD}1} \dfrac{C_{\text{G}1}}{C_{\Sigma 1}}+ C_{\text{SD}2} \dfrac{C_{\text{G}2}}{C_{\Sigma 2}}$,\\[1ex]
				$C_{\text{Q}0}=\dfrac{(e \alpha')^2}{2\Delta}$, $\,\,\, R_{\text{Q}0}^{-1}=\dfrac{(e \alpha')^2}{4\hbar}$
				\vspace{1.5ex}}
			\\ \hline
			
			\parbox[c]{3cm}{\centering\vspace{1.5ex} Single-electron box \cite{Persson2010}, Fig.~\ref{Fig_other}(c)\vspace{1.5ex}}
			&
			\parbox[c]{6cm}{\centering\vspace{1.5ex}
				$\varepsilon(t)=E_\text{C}(1-2n_\text{g})$, $\,\,\, \alpha_\text{SEB} = \dfrac{C_\text{g}^\text{rf}}{C_\Sigma}$,\\[1ex]
				$T_1^{-1} = \dfrac{\Delta E_0}{h} \dfrac{R_\text{K}}{R_\text{T}}\coth \dfrac{\Delta E_0}{2 k_\text{B}T}$,\\[1ex]
				$\Delta = 0$, $\,\,\, T_2 = 0$.
				\vspace{1.5ex}}
			&
			\parbox[c]{8cm}{\centering\vspace{1.5ex}
				$C_\text{geom} = C_\text{g}^\text{rf} \left(1-\dfrac{C_\text{g}^\text{rf}}{C_\Sigma}\right)$,\\[1ex]
				$C_\text{Q0}=\infty$, $\,\,\, R_{\text{Q}0}^{-1} = \dfrac{e^2}{4\hbar} \left( \dfrac{C_\text{g}^\text{rf}}{C_\Sigma}\right)^{2}$
				\vspace{1.5ex}}
			\\ \hline
			
		\end{tabular}
	\end{center}
	\caption{\textbf{Summary of parameters for different quantum systems}. For each system listed in the first column, the second column  contains the corresponding expressions for the energy bias $\varepsilon(t)$, minimal energy splitting $\Delta$, relation between the energy bias and the classical probing voltage $\alpha e=\partial \varepsilon/ \partial V$, as well as relaxation and decoherence times $T_{1,2}$. The third column contains expressions for the resulting geometric capacitance $C_\text{geom}$, maximum quantum capacitance $C_{\text{Q}0}$, and characteristic conductance value $R_{\text{Q}0}^{-1}$. By substituting these parameters from the second and third columns into the general two-level expressions summarized in Table~\ref{Table:TL_results} yields the effective capacitance and conductance for each system. For the double quantum dot the expression for the geometric capacitance and lever arm $\alpha_\text{DQD}$ are valid only for $C_\text{m} \ll C_{\Sigma i}$ [see Eq.~(\ref{C_geom_DQD}) for exact geometric capacitance].}
	\label{Table:Conclusions}
\end{table*}

One can obtain expressions with the use of Table~\ref{Table:TL_results} when the parameters of the system of interest are completely mapped onto the energy bias $\varepsilon(t)$, its dependence on the probing voltage $\alpha e=\partial \varepsilon/ \partial V$, minimal energy splitting $\Delta$, relaxation time $T_1$, and decoherence time $T_{2}$. To this parameter mapping procedure we mainly devote Table~\ref{Table:Conclusions}: the resulting expressions for the effective capacitance and conductance can be obtained by substituting the corresponding parameters from Table~\ref{Table:Conclusions} into expressions in Table~\ref{Table:TL_results}. 

In the literature, e.g. Refs.~\cite{Esterli2019, Persson2010}, it is common to see theoretical treatments developed by solving the two-level quantum master rate equation, which is applicable only for bad quantum systems [see Appendix~\ref{Appendix:GKSL_to_rate}]. We show how these results can be readily obtained from the GKSL formalism and highlight the changes that arise with a more precise theoretical treatment.

We show that all our results perfectly coincide, within the given limiting cases, with the expressions for the effective capacitance and resistance reported in~\cite{Johansson2006, Sillanpaeae2005, Esterli2019, Persson2010}. Moreover, our results, obtained from the solution of the GKSL equation, are more general. In particular, we can show how the predictions change when the rate equation approach is no longer applicable.

\subsection{Cooper-pair box}
\label{Sec:Other:Charge_qubit}

For the CPB in Fig.~\ref{Fig3} from Sec.~\ref{Sec:Theory:Quantization}, the main parameters of the Hamiltonian are $\varepsilon(t)=4E_\text{C}(1-2n_\text{g})$ and $\Delta = E_\text{J}$. The dependence of the energy bias on the probing voltage $V_\text{rf}$ can be obtained from Eq.~(\ref{Theory:ng}) as
\begin{equation}
	\alpha_\text{CPB} = \dfrac{1}{e}\dfrac{\partial \varepsilon}{\partial V_\text{rf}} = 2 \dfrac{C_\text{m}}{C_\Sigma}, \label{depsdv_charge_qubit}
\end{equation}
while the relaxation time should be taken as in Eq.~(\ref{T1model}) and the decoherence time as $T_2^{-1} = T_\phi^{-1} + (2T_1)^{-1}$ to reproduce the results in Figs.~\ref{Fig:thermal}--\ref{Fig:Resistance3}.

\subsection{Single-Cooper-pair transistor}
\label{Sec:Other:SCPT}
Now we consider a single-Cooper-pair transistor (SCPT) \cite{Sillanpaeae2005}, Fig.~\ref{Fig_other}(b). We start from the classical Lagrangian, quantize the system and consider the two-level approximation. After the unitary transformation, the pseudo-spin Hamiltonian can be written in the form~(\ref{Theory:2levelHam})  [detailed discussion can be found in the Appendix~\ref{Appendix:SCPT}]. 

The geometric energy $E_\text{geom}$, given in Eq.~(\ref{Egeom_SCPT}), results in a geometric capacitance $C_\text{geom} = C_\text{g}(1-C_\text{g}/C_\Sigma)$ [see Eq.~(\ref{Cgeom_SCPT})]. The energy bias has the standard form for a superconducting system $\varepsilon(t) = 4E_\text{C}(1-2n_\text{g})$ and the minimal energy splitting 
\begin{equation}
	\Delta =2E_\text{J}\Big|\cos \dfrac{\phi}{2}\Big|\sqrt{1+\delta^2 \tan^2 \dfrac{\phi}{2}} \label{delta_SCPT}
\end{equation}
that can be tuned by external magnetic flux $\phi = 2\pi \Phi/\Phi_0$.

The energy bias $\varepsilon$ and the probing voltage $V_\text{rf}$ are related as
\begin{equation}
	\alpha_\text{SCPT} =\dfrac{1}{e}\dfrac{\partial \varepsilon}{\partial V_\text{rf}}=2 \dfrac{C_\text{g}}{C_\Sigma},
\end{equation}
 so that the maximum quantum capacitance $C_{\text{Q0}}$ is
\begin{gather}
	C_{\text{Q0}}=\dfrac{e^2}{\Delta}\left( \dfrac{C_\text{g}}{C_\Sigma}\right)^2.
\end{gather}

To compare with Ref.~\cite{Sillanpaeae2005} we set $T_{1(2)} = \infty$ and $k_\text{B}T=0$, which   results only in a quantum capacitance of the ground state $C_\text{Q}=C_{\text{Q}0} (\Delta/\Delta E_0)^3$. Thus the effective capacitance $C_\text{eff}=C_\text{geom}+C_\text{Q}$ totally coincides with Eqs.~(1) and (2) of Ref.~\cite{Sillanpaeae2005}, while the tunneling capacitance, the Sisyphus and Hermes conductances are zero for such theoretical treatment.

\subsection{Double quantum dot}
\label{Sec:Other:DQD}

As a next example, we consider a double quantum dot (DQD) \cite{Mizuta2017, Esterli2019}. Its capacitive model is shown in Fig.~\ref{Fig_other}(d). The DQD is coupled to a classical $RLC$ circuit in parallel with the classical capacitor  $C_\text{r}$ just like in the previous subsection. 

In the same manner we start from the classical Lagrangian and find the Routhian function for quantum phases of the two quantum dots $\varphi_{1,2}$. The detailed procedure can be found in the Appendix~\ref{Appendix:DQD}. After quantization in the two-level approximation, the Hamiltonian reduces to Eq.~(\ref{Theory:2levelHam}), where the geometric energy $E_\text{geom}$ is defined in Eq.~(\ref{Egeom_DQD}).

The energy bias $\varepsilon$ is tuned by the applied voltage on the input \text{$V_\text{in}=V_0+\delta V_\text{rf} \cos (\omega_\text{rf}t)$}. The dc component of the voltage tunes the working point $\varepsilon_0$ and the ac component $\delta \varepsilon_\text{rf}$ probes the system. The relation between the energy bias and the probing voltage $V_\text{G}$ in the typical experimental situation with $C_\text{m} \ll C_{\Sigma i}$ is given by
\begin{equation}
	\alpha_\text{DQD} =\dfrac{1}{e}\dfrac{\partial \varepsilon}{\partial V_\text{rf}}=  \alpha'=\alpha_1-\alpha_2, \label{alpha_DQD}
\end{equation}
where $\alpha_i=C_{\text{G}i}/C_{\Sigma i}$ is the lever arm of the $i$-th QD and $C_{\Sigma i}=C_{\text{G}i}+C_{\text{S}i}+C_{\text{D}i}+C_{\text{m}}$ [results with no restrictions on $C_\text{m}$ can be found in Appendix~\ref{Appendix:DQD}].

In order to compare the results with the quantum master balance equation approach of Refs.~\cite{Mizuta2017, Esterli2019}, we set 
\begin{equation}
	\gamma_{\varphi}=\infty, \,\,\,\,\,\,  \gamma_{+}=\Gamma^\text{c} n_\text{ph}, \,\,\,\,\,\, \gamma_{-}=\Gamma^\text{c}(1+n_\text{ph}),
\end{equation}
where $n_\text{ph}=[\exp (\Delta E/k_\text{B}T)-1]^{-1}$. This results in relaxation time 
\begin{equation}
	T_1^{-1} = \gamma_{+}+\gamma_{-}=\Gamma^\text{c} \coth \left(\dfrac{\Delta E}{2 k_\text{B}T}\right),
\end{equation}
as in Eq.~(\ref{T1model}), and decoherence time $T_2=0$.  After this we can immediately calculate the effective capacitance and resistance with the use of Table~\ref{Table:Conclusions}.


The geometric capacitance is \text{$C_\text{geom}=\sum_i \alpha_i C_{\text{SD}i}$}, the quantum capacitance is the same as presented in Table~\ref{Table:TL_results} with $C_{\text{Q}0}=(e \alpha')^2/2\Delta$ being the maximum quantum capacitance of the ground state, while the expressions for the tunneling capacitance can also be readily obtained.

Since $T_2=0$, only the Sisyphus component of Eq.~(\ref{Qubit_R_eff}) contributes to the effective resistance. Results for the effective capacitance and resistance are in total agreement with Eqs.~(5) and (18) of Ref.~\cite{Mizuta2017}, and Eqs.~(11--13) of Ref.~\cite{Esterli2019}. 

We note that in the limit $\gamma \rightarrow \infty$, we recover the effective capacitance of a bad qubit [see Eq.~(\ref{qubit_C_BAD}) and Eq.~(20) of Ref.~\cite{Mizuta2017}], while for $\gamma \rightarrow 0$ we obtain capacitance of a good qubit [see Eq.~(19) of Ref.~\cite{Mizuta2017} and our Eq.~(\ref{qubit_C_GOOD})]. However, our results are obtained within the GKSL formalism thus could be used also in the case, $\Gamma^\text{c} \gg \omega_\text{qb}$, where the rate equation does not hold~[see Appendix~\ref{Appendix:GKSL_to_rate}]. Moreover, since expressions (\ref{C_qd_BAD}) and (\ref{C_qd_GOOD}) for the effective capacitance of good and bad systems depends only on the energy levels structure and  are valid for any multi-level system we can see how the expressions (19) and (20) of Ref.~\cite{Mizuta2017} algebraically coincide with our more general expressions for given energy levels $E_i$.

In these two limiting cases (of a good and bad qubits) the conductance still tends to zero $R_\text{eff}^{-1} \rightarrow 0$, as we have shown in Sec.~\ref{Sec:Qudit:GKSL}. Since in the equivalent scheme [shown in Fig.~\ref{Fig_other}(a)] all of the capacitances and the resistances are coupled in parallel, this simply means that no current flows through the resistance and the response of the quantum subsystem is purely capacitive. The Sisyphus conductance is zero in the quasi-crossing point $\varepsilon_0=0$. The effective conductance becomes non-zero in the point $\varepsilon_0=0$ if we account for finite rate of dephasing, since the Hermes contribution in Eq.~(\ref{Qubit_R_eff}) can no longer be omitted.

\subsection{Single-electron box}
\label{Sec:Other:SEB}

As a final example we consider the single-electron box (SEB) \cite{Persson2010}, shown in Fig.~\ref{Fig_other}(c). It turns out that the system description is very similar to that in Sec.~\ref{Sec:Other:Charge_qubit}. The only difference is that for the SEB, we consider a normal (non-superconducting) system, for which the minimal energy splitting is $\Delta = 0$. This results in the energy bias $\varepsilon = E_\text{C}(1-2n_\text{g})$, which differs by a factor of four compared to $\varepsilon$ in the superconducting case. The coupling of the bias with respect to the probing voltage is also reduced by a factor of two compared to Eq.~(\ref{depsdv_charge_qubit}):
\begin{equation}
	\alpha_\text{SEB} =\dfrac{1}{e}\dfrac{\partial \varepsilon}{\partial V_\text{rf}} = \dfrac{C_\text{g}^\text{rf}}{C_\Sigma}.
\end{equation}

In order to compare with Ref.~\cite{Persson2010}, we set 
\begin{equation}
	T_1^{-1}=\gamma = \dfrac{\Delta E_0 R_\text{K}}{h R_\text{T}} \coth\left(\dfrac{\Delta E_0}{ 2 k_\text{B}T}\right),
\end{equation}
and $T_2=0$, since the theoretical treatment is based on the rate-equation approach. This results in the effective resistance
\begin{equation}
	R_\text{eff} =R_\text{Sis}= \dfrac{2R_\text{T}}{\alpha_\text{SEB}^2}  \dfrac{k_\text{B}T}{\Delta E_0} \sinh \left(\dfrac{\Delta E_0}{k_\text{B}T}\right) \dfrac{\gamma^2+\omega_\text{rf}^2}{\gamma^2},
\end{equation}
which coincides with Eq.~(4) of Ref.~\cite{Persson2010}. 

The capacitance $C_{\text{Q}0}$ is ill-defined,  as $\Delta \rightarrow 0$ [see Table~\ref{Table:Conclusions}], but still the quantum capacitance and the Hermes conductance are zero due to $\Delta=0$ and $T_2=0$, respectively.  Remarkably, the tunneling capacitance does not vanish,
\begin{equation}
	C_\text{T}=\dfrac{(e \alpha_\text{SEB})^2}{4 k_\text{B}T} 
	\cosh^{-2}\left(\dfrac{\Delta E_0}{2 k_\text{B}T}\right)
	\dfrac{\gamma^2}{\gamma^2+\omega_\text{rf}^2},
\end{equation}
but for the given values of the experimental parameters one finds
$R_\text{Sis}^{-1} \gg \omega_\text{rf} C_\text{T}$, such that the Sisyphus conductance dominates the response of the quantum system.

The finite value of $C_\text{T}$ can be explained as follows. The quantum capacitance is proportional to the second derivative of the energy levels with respect to the probing voltage,
$C_\text{Q} \propto \partial^2 E_i/\partial V_\text{rf}^2$,
whereas the tunneling capacitance is proportional only to the first derivative,
$C_\text{T} \propto \partial E_i/\partial V_\text{rf}$ (for a bad qubit), and therefore does not vanish until the qubit becomes sufficiently good, $ T_1^{-1} \ll \omega_\text{rf} $. Our theory remains applicable for systems with energy-level crossings $\Delta = 0$, provided that the energy bias satisfies
$\left| \varepsilon_0 \right| \eqcolon \Delta E_0 \gg \hbar \omega_\text{rf}$
[see Eqs.~(\ref{2-level_perturbation_conditions}) and (\ref{QuditA_applicability_for_perturbation_theory})].

\section{Conclusions}
We developed a rigorous four-step framework to describe hybrid semiclassical qubit/qu$d$it-resonator systems. First, we derive the Routh function from the full Lagrangian to separate classical and quantum variables. Second, we quantize the Routh function and obtain coupled classical-quantum equations of motion. Third, we treat the probe signal perturbatively and derive analytical expressions for the effective capacitance, inductance, and resistance. Finally, we compare the analytical theory with numerical solutions of the full dynamics, establishing the range of validity of the phenomenological description.

Although the formalism is general, we illustrated it using a Josephson-junction-based charge qubit in Sec.~\ref{Sec:Theory}. We showed that the influence of a quantum system on a classical resonator can be captured by a combination of geometric, quantum, and tunneling capacitances, as well as Sisyphus and Hermes resistances arising from coupling to a dissipative environment.

In Secs.~\ref{Sec:Theory} and \ref{Sec:Qudit}, we derived general expressions for the effective capacitance and resistance of arbitrary two- and multi-level isolated and open quantum systems with arbitrary relaxation rates $\gamma_\alpha(\varepsilon)$ and Lindblad jump operators $\hat{L}_\alpha(\varepsilon)$. In Sec.~\ref{Sec:Other}, we demonstrated how these results can be readily applied to different quantum systems, including the Cooper-pair box, single-Cooper-pair transistor, double quantum dot, and single-electron box. 

\begin{acknowledgments}
	We gratefully acknowledge fruitful discussions with M.F. Gonzalez-Zalba. O.Yu.K. and S.N.S. acknowledge financial support of the National Research Foundation of Ukraine (Grant No. 2025.07/0044). This work was supported in part by: the Japan Science and Technology Agency (JST) [via the CREST Quantum Frontiers program Grant No. JPMJCR24I2, the Quantum Leap Flagship Program (Q-LEAP), the ASPIRE program, the Moonshot RD Grant No. JPMJMS2061], the Office of Naval Research (ONR) Global (via Grant No. N62909-23-1-2074), Army Research Office (Grant No. W911NF-20-1-0261), the U.S. National Academy of Sciences (NAS) and the Office of Naval Research (ONR) in the framework of the IMPRESS-U project.
\end{acknowledgments}

\appendix

\section{Lagrangian and Hamiltonian formalisms}

\label{AppendixA}

In this Appendix we develop a general formalism to obtain equations of motion for classical degrees of freedom when a classical subsystem is coupled to a quantum one. Consider Lagrangian $\mathcal{L}(\varphi_{\text{cl}},\dot \varphi_{\text{cl}},\varphi_{\text{q}},\dot \varphi_{\text{q}})=\mathcal{L}_\text{cl}(\varphi_{\text{cl}},\dot \varphi_{\text{cl}})+\mathcal{L}_\text{q}(\varphi_{\text{cl}},\dot \varphi_{\text{cl}},\varphi_{\text{q}},\dot \varphi_{\text{q}})$ with both classical $\varphi_{\text{cl},i}$ and quantum $\varphi_{\text{q},j}$ degrees of freedom [see Sec.~\ref{Sec:Theory:Quantization}]. 

Now we determine the Routhian function $\mathcal{R}$ with use of Eq.~(\ref{Routhian}) in order to write down classical equations of motion with conjugated quantum variables. We note that in the definition of the canonical momentum $p_{\text{q},j}=\partial \mathcal{L}/\partial \dot \varphi_{\text{q},j}$ we can exchange $\mathcal{L}$ with $\mathcal{L}_\text{q}$ since only the quantum part of the Lagrangian depends on the quantum degrees of freedom. We further can write down the Routhian as $\mathcal{R} = \mathcal{H} - \mathcal{L}_\text{cl}$ with 
\begin{equation}
	\mathcal{H} = \sum\limits_j p_{\text{q},j}\dot{\varphi}_{\text{q},j}-\mathcal{L}_\text{q}
\end{equation}
being essentially quantum part of the Routhian function. With such definition, the equation of motion for the classical degrees of freedom with the dissipation function $\mathcal{F}$ is
\begin{equation}
	\frac{d}{dt}\frac{\partial \mathcal{R}}{\partial \dot{\varphi}_{i}}-\frac{%
		\partial \mathcal{R}}{\partial \varphi _{i}}=\frac{\partial \mathcal{F}}{%
		\partial \dot{\varphi}_{i}} 
\end{equation}
and can be rewritten as
\begin{equation}
	\frac{d}{dt}\frac{\partial \mathcal{L}_\text{cl}}{\partial \dot \varphi_{\text{cl},i}}-\frac{
		\partial \mathcal{L}_\text{cl}}{\partial \varphi_{\text{cl},i}}+\frac{\partial \mathcal{F}}{
		\partial \dot\varphi_{\text{cl},i}}-\frac{d}{dt}\frac{\partial \mathcal{H}}{\partial \dot \varphi_{\text{cl},i}}+\frac{\partial \mathcal{H}}{\partial  \varphi_{\text{cl},i}}=0, \label{A:GCEOM}
\end{equation}
Here we suppose that the dissipation function $\mathcal{F}$ depends only on classical degrees of freedom, however the approach can be easily generalized for the explicit dependence on quantum degrees of freedom as well. After the quantization procedure the last two terms of Eq.~(\ref{A:GCEOM}) should be exchanged with corresponding quantum averages
\begin{equation}
	\frac{d}{dt}\frac{\partial \mathcal{H}}{\partial \dot \varphi_{\text{cl},i}} \rightarrow \frac{d}{dt}\Big\langle \frac{\partial \hat{\mathcal{H}}}{\partial \dot \varphi_{\text{cl},i}}\Big \rangle, \,\,\, \frac{\partial \mathcal{H}}{\partial  \varphi_{\text{cl},i}} \rightarrow \Big \langle \frac{\partial \hat{\mathcal{H}}}{\partial  \varphi_{\text{cl},i}} \Big \rangle, \label{A:quantum_exchange}
\end{equation}
and we obtain the final system of classical equations
\begin{equation}
	\frac{d}{dt}\frac{\partial \mathcal{L}_\text{cl}}{\partial \dot \varphi_{\text{cl},i}}-\frac{
		\partial \mathcal{L}_\text{cl}}{\partial \varphi_{\text{cl},i}}+\frac{\partial \mathcal{F}}{
		\partial \dot\varphi_{\text{cl},i}}-\frac{d}{dt}\Big\langle \frac{\partial \hat{\mathcal{H}}}{\partial \dot \varphi_{\text{cl},i}}\Big \rangle+\Big \langle \frac{\partial \hat{\mathcal{H}}}{\partial  \varphi_{\text{cl},i}} \Big \rangle=0. \label{classical_equations}
\end{equation}

For capacitively coupled systems, the Hamiltonian does not depend explicitly on the classical coordinate $\varphi_{\text{cl},i}$, so that the fifth term in Eq.~(\ref{classical_equations}) vanishes and the equation reduces to Eq.~(\ref{ClEOM}). In contrast, for inductively coupled systems the Hamiltonian does not depend on $\dot{\varphi}_{\text{cl},i}$, and therefore the fourth term is zero instead.

\subsection{Toy example: quantum $LC$ oscillator}

To better understand the meaning of the quantum contributions in 
Eq.~(\ref{classical_equations}), consider a simple $LC$ circuit as an illustrative 
example. We start from the Lagrangian approach to the problem 
in which the energy of a capacitor 
$E_{\dot\varphi} = \left(\frac{2e}{\hbar}\right)^2\frac{C\dot{\varphi}^2}{2}$ 
is associated with the ``kinetic'' energy and the energy of an inductor 
$E_\varphi = \left(\frac{2e}{\hbar}\right)^2\frac{\varphi^2}{2L}$ is associated 
with the ``potential'' energy. Following the convention introduced in 
Sec.~\ref{Sec:Theory:Quantization}, the Lagrangian reads
\begin{equation}
	\left(\frac{2e}{\hbar}\right)^2 \mathcal{L} = \frac{C}{2}\dot{\varphi}^2 
	- \frac{1}{2L}\varphi^2,
\end{equation}
where $\varphi = \frac{2e}{\hbar}\int V dt$ is the node phase. The conjugate momentum is
\begin{equation}
	p = \frac{\partial \mathcal{L}}{\partial \dot{\varphi}} = 
	\left(\frac{\hbar}{2e}\right)^2 C\dot{\varphi} = \frac{\hbar}{2e} Q,
\end{equation}
which is proportional to the charge $Q = CV$ on the capacitor. The Hamiltonian is then
\begin{equation}
	\mathcal{H} = p\dot{\varphi} - \mathcal{L} = \frac{Q^2}{2C} + \frac{\Phi^2}{2L},
\end{equation}
recovering the standard result~\cite{Blais2021}, where $\Phi = \frac{\hbar}{2e}\varphi$ 
is the flux threading the inductor. Upon quantization, $\varphi \to \hat{\varphi}$ and $p \to \hat{p}$ 
with the commutation relation $[\hat{\varphi}, \hat{p}] = i\hbar$, or equivalently 
$[\hat{\Phi}, \hat{Q}] = i\hbar$, and the system becomes a quantum harmonic oscillator 
with frequency $\omega = 1/\sqrt{LC}$. In accordance with Eq.~(\ref{A:quantum_exchange}), 
the two quantum contributions evaluated at the quantum variable $\varphi$ read
\begin{equation}
	\frac{d}{dt}\frac{\partial \mathcal{H}}{\partial \dot{\varphi}} \rightarrow  \frac{d \langle \hat{Q} \rangle}{dt},
	\qquad
	\frac{\partial \mathcal{H}}{\partial \varphi} \rightarrow 
	\frac{\hbar}{2e} \frac{\langle\hat{\Phi}\rangle}{L} = \frac{V_L}{L},
\end{equation}
which are simply the current through the capacitor and the inductor, 
respectively, recovering the standard Kirchhoff equations for an $LC$ circuit.

From the above observation, the two quantum contributions in Eq.~(\ref{classical_equations}) 
have a clear interpretation: the term $d\langle\partial \hat{\mathcal{H}}/\partial \dot \varphi_{\text{cl},i} \rangle/dt$ 
arises from the response of the charge degrees of freedom of the quantum subsystem, 
while $\langle \partial \hat{\mathcal{H}}/\partial \varphi_{\text{cl},i} \rangle$ 
arises from the flux degrees of freedom. In the low-frequency limit, for an isolated 
quantum system, these two contributions give rise to the effective capacitance and 
inductance, respectively, while for open systems both terms additionally contribute 
to the effective resistance [see Secs.~\ref{Theory:GKSL}, \ref{Sec:Qudit:GKSL}, 
and \ref{Sec:Other}].

\section{Quantization of a superconducting island}

\label{Appendix:Quantization}
Consider now the quantization of a superconducting island. The total Lagrangian of the system shown in Fig.~\ref{Fig3} is given by Eq.~(\ref{Lagrangian}). The classical part of the Lagrangian $\mathcal{L}_\text{cl}$ consists of the first two terms of Eq.~(\ref{Lagrangian}) and  $\mathcal{L}_\text{q}$ contains the remaining part associated with the quantum subsystem. Resistance $R_{\text{r}}$ can be included via the dissipation function 
\begin{equation}
	\mathcal{F}=\left( \frac{\hbar }{2e}\right) ^{2}\frac{\left( \dot{\varphi}_{%
			R}-\dot{\varphi}_{\text{in}}\right) ^{2}}{2R_{\text{r}}}.
	\label{A:dissipation}
\end{equation}

Since the phases $\varphi _{L}$ and $\varphi _{R}$ are purely
classical variables, we solve the Euler-Lagrange equations for them, but the phase 
$\varphi $ is a quantum degree of freedom so it needs to be quantized. For
that we must first turn to the Hamiltonian representation with respect to
the phase $\varphi $. We write down the Routhian function $\mathcal{R}$~(\ref{Routhian})
\begin{equation}
	\mathcal{R}(\varphi ,p,\varphi _{i},\dot{\varphi}_{i})=p\dot{\varphi}-%
	\mathcal{L}(\varphi _{i},\dot{\varphi}_{i}),
\end{equation}%
where $p=\partial \mathcal{L}/\partial \dot{\varphi}=n\hbar$. The Routhian is given
by 
\begin{gather}
	\mathcal{R}=\mathcal{H}-\mathcal{L}_\text{cl}=4E_{\text{C}}(n-n_{\text{g}})^{2}-E_{\text{J}}\cos \varphi -
	\label{A:routhian} \\
	-\left( \frac{\hbar }{2e}\right) ^{2}\left[ \frac{C_{\text{r}}+C_{\text{m}}}{2}\dot{%
		\varphi}_{L}^{2}+\frac{C_{\text{g}}}{2}\dot{\varphi}_{\text{g}}^{2}-%
	\frac{1}{2L_{\text{r}}}\left( \varphi _{L}-\varphi _{R}\right) ^{2}%
	\right]   \notag
\end{gather}%
with $n_\text{g}$ from Eq.~(\ref{Theory:ng}). The classical equations for the generalized coordinates $\varphi _{L,%
	R}$ are given by Eq.~(\ref{ClEOM}) with the classical Hamiltonian
\begin{equation}
	\mathcal{H} = 4E_\text{C} (n-n_\text{g})^2-E_\text{J} \cos \varphi  -\dfrac{C_\text{g} V_\text{g}^2}{2}-\dfrac{C_\text{m} V_L^2}{2}, \label{A:Hq}
\end{equation}
in which we have no dependence on classical phases $\varphi_{L,R}$, so that $ \partial \mathcal{H}/\partial  \varphi_{L,R}=0$.

The first two terms in both $\mathcal{R}$ and $\mathcal{H}$ relate to the Hamiltonian of the charge qu$d$it. However, up to this point all equations have been purely classical. To quantize the system, we replace the classical variables $\varphi \rightarrow \hat{\varphi}$ and $p \rightarrow \hat{p}$, imposing the commutation relation $[\hat{\varphi},\hat{p}] = i \hbar$. The resulting quantum Hamiltonian is then given by Eq.~(\ref{Theory:Qhamiltonian}).

We emphasize that upon quantization $\mathcal{H} \rightarrow \hat{\mathcal{H}}$, the terms proportional to the identity operator are deliberately retained in the Hamiltonian. Although such terms do not influence the quantum dynamics, they give rise to the geometric capacitance $C_\text{geom}$ and are therefore essential for a unified and systematic derivation of all three contributions to the effective capacitance.

In the two-level approximation the Hamiltonian reduces to Eq.~(\ref{Theory:2levelHam}) with the geometric energy
\begin{equation}
	E_\text{geom}=2E_\text{C}(1-2n_\text{g}+2n_\text{g}^2)-\dfrac{C_\text{g}V_\text{g}^2}{2}-\dfrac{C_\text{m}V_L^2}{2}. \label{B:Egeom}
\end{equation}

Next, in the second equation of the system~(\ref{Theory:CPB_clas_eqs}), with the use of Eqs.~(\ref{Theory:2levelHam}) and (\ref{B:Egeom}), the quantum current can be computed as
\begin{gather}
	\nonumber -\left( \dfrac{2e}{\hbar }\right) ^{2}	\dfrac{d}{dt}\Big \langle \dfrac{\partial \hat{\mathcal{H}}}{\partial \dot \varphi_L} \Big \rangle  = C_\text{geom} \ddot \varphi_L -\\
	- \dfrac{8 E_\text{C} C_\text{m}}{C_\Sigma}\dfrac{d \langle \hat{n} \rangle}{dt}+\dfrac{C_\text{m} C_\text{g}}{C_\Sigma} \ddot \varphi_\text{g}, \label{quantum_current_with_drive}
\end{gather}
where
\begin{equation}
	C_{\text{geom}}=-\dfrac{\partial^2 E_\text{geom}}{\partial V_L^2}= \frac{(C_{\text{g}}+C_{\text{J}})C_{\text{m}}}{%
		C_{\Sigma }}
\end{equation}
is the geometric capacitance of the quantum subsystem, while the second term of Eq.~(\ref{quantum_current_with_drive}) $\propto d \langle \hat{n} \rangle / dt$ gives both quantum and tunneling capacitances, as well as Sisyphus and Hermes resistances, and the last term describes the capacitive coupling of the time-dependent gate drive to the dynamics of the classical subsystem. We note that the geometric capacitance is obtained from the diagonal part of the Hamiltonian $\hat{\mathcal{H}}$   proportional to the identity matrix $\hat{\mathbf{1}}$.

\section{Quantum capacitance for ideal two-level system}
\label{Appendix:IdQb}
\subsection{Perturbation theory for the evolution operator}
Consider the Schr\"{o}dinger equation with the Hamiltonian (\ref{Theory:2levelHam}) and the bias
$\varepsilon(t)=\varepsilon_{0}+\delta\varepsilon_{\text{rf}}\cos(\omega_{\text{rf}} t)$.
This detuning leads to the Hamiltonian $\hat{\mathcal{H}}=\hat{\mathcal{H}}_{0}+\hat{V}$, with
\begin{equation}
	\hat{\mathcal{H}}_{0}=-\frac{\Delta}{2}\sigma_{x}
	-\frac{\varepsilon_{0}}{2}\sigma_{z}
\end{equation}
and
\begin{equation}
	\hat{V}=-\dfrac{\delta\varepsilon_{\text{rf}}\cos(\omega_{\text{rf}} t)}{2}\sigma_{z}
	\label{AppendixB_perturbation}
\end{equation}
being the unperturbed Hamiltonian and the perturbation, respectively.

It is beneficial and pedagogical to build a perturbation theory for the evolution operator instead of using the standard state-based treatment. Of course, the conventional perturbation theory formulated directly for the state vectors yields the same corrections to the expectation values of all observable operators, and therefore leads to the same quantum capacitance.. However, the approach presented below turns out to be more compact and elegant, and allows for a more transparent derivation of the final expressions.

The evolution operator $\hat{U}$ evolves the initial state $|\psi(t_0)\rangle$ to the state $|\psi\rangle = \hat{U}(t,t_0) |\psi(t_0)\rangle$. For this work, we always take $t_0=0$. The evolution operator up to the first order  in the classical probe can be written as $\hat{U}=\hat{U}_0+\hat{U}_1$. The solution for the unperturbed problem,
\text{$i \hbar  \hat{\dot U}_0=\hat{\mathcal{H}}_0 \hat{U}_0$},
with the initial condition $\hat{U}_0(0,0)=\hat{\mathbf{1}}$, is given by
\begin{equation}
	\hat{U}_0=\hat{\mathbf{1}} \cos\dfrac{\omega_{\text{q0}} t}{2} 
	+i \left[ \dfrac{\Delta}{\Delta E_0}\hat{\sigma}_x+ \dfrac{\varepsilon_0}{\Delta E_0}\hat{\sigma}_z \right]
	\sin \dfrac{\omega_{\text{q0}} t}{2},
	\label{AppendixB_U0_solution}
\end{equation}
where $\omega_{\text{q0}}=\Delta E_0/\hbar = \sqrt{\Delta^2+\varepsilon_0^2}/\hbar$.

To solve the Schr\"{o}dinger equation with a perturbation, it is convenient to seek the first-order correction in the form 
\begin{equation}
	\hat{U}_1 = \hat{U}_0 \hat{\tilde{U}}. \label{U1ansatz}
\end{equation}
This substitution is equivalent to transforming to the interaction picture with respect to the unperturbed propagator $\hat{U}_0$. 
It leads to 
\begin{equation}
	i \hbar \dfrac{d \hat{\tilde{U}}}{dt} =\hat{U}_0^\dagger \hat{V} \hat{U}_0.
\end{equation} 
The solution of this equation is
\begin{equation}
	\hat{\tilde{U}}=-\dfrac{i}{\hbar} \int\limits_0^t \hat{U}_0^\dagger \hat{V} \hat{U}_0\, dt,
\end{equation}
which explicitly reads
\begin{gather}
	\nonumber\hat{\tilde{U}}=\dfrac{i \delta \varepsilon_\text{rf}}{2 \hbar}
	\int\limits_0^t \cos (\omega_\text{rf} t) \bigg[
	- \dfrac{\Delta \varepsilon_0}{\Delta E_0^2}\left(1-\cos (\omega_\text{q0} t)\right)\hat{\sigma}_x \\
	-\dfrac{\Delta}{\Delta E_0} \sin (\omega_\text{q0} t) \hat{\sigma}_y
	+\dfrac{\varepsilon_0^2+\Delta^2 \cos (\omega_\text{q0} t)}{\Delta E_0^2}
	\bigg] dt . \label{Utilde_sol}
\end{gather}
The state vector can then be written as
\begin{equation}
	\ket{\psi} = \hat{U}_0 \left[\hat{\mathbf{1}}-\dfrac{i}{\hbar} \int\limits_0^t \hat{U}_0^\dagger \hat{V} \hat{U}_0\, dt \right]\ket{\psi_0},
	\label{AppendixB:WFsolution}
\end{equation}
where $\ket{\psi_0} = \ket{\psi(t=0)}$ is the initial state.

\subsection{First-order corrections to the quantum current}
As a next step we need to find the first-order corrections to the expectation value of the rate of change of the Cooper-pair numbers on the superconducting island  $d \langle{\hat{n}}\rangle/dt$, and from Eqs.~(\ref{Theory:CPB_clas_eqs}) and (\ref{quantum_current}) determine the quantum capacitance. We note that only terms $\propto \exp(\pm i \omega_\text{rf} t)$ in $d \langle{\hat{n}}\rangle/dt$ will contribute to the quantum capacitance, since terms $\propto \exp[i( \pm\omega_\text{rf} +k \omega_\text{q0})  t)]$ will average out to zero for integers $k \neq 0$ on the much faster timescale $ \omega_\text{rf}\ll \omega_\text{q0} $. 

First, the time derivative of the excess number of Cooper pairs can be written as
\begin{equation}
	\dfrac{d \langle \hat{n} \rangle}{dt}
	= \dfrac{i}{\hbar}\langle \psi | [\hat{\mathcal{H}},\hat{n}] |\psi \rangle
	= \dfrac{i}{\hbar}\langle \psi_0 | \hat{U}^\dagger [\hat{\mathcal{H}}_0,\hat{n}] \hat{U} |\psi_0 \rangle,
\end{equation}
where we have used the fact that $[\hat{V},\hat{n}]=0$.  

We now introduce the decomposition $\hat{\dot n} = \hat{\dot n}^{(0)} + \hat{\dot n}^{(1)}$, such that
$\langle \psi_0| \hat{\dot n} |\psi_0\rangle \equiv d\langle \hat{n} \rangle / dt$, with
\begin{gather}
	\nonumber \hat{\dot n}^{(0)} 
	= \dfrac{i}{\hbar} \hat{U}_0^\dagger [\hat{\mathcal{H}}_0,\hat{n}] \hat{U}_0
	= \dfrac{\Delta}{2\hbar} \hat{\sigma}_y \cos (\omega_{\text{q0}} t) +\\
	+ \dfrac{\Delta}{2 \Delta E_0 \hbar}
	\big(\Delta \hat{\sigma}_z-\varepsilon_0 \hat{\sigma}_x  \big)
	\sin (\omega_{\text{q0}} t) ,
\end{gather}
and
\begin{gather}
	\nonumber \hat{\dot n}^{(1)} 
	= \dfrac{i}{\hbar}
	\left[
	\hat{U}_1^\dagger [\hat{\mathcal{H}}_0,\hat{n}] \hat{U}_0
	+ \hat{U}_0^\dagger [\hat{\mathcal{H}}_0,\hat{n}] \hat{U}_1
	\right] =\\
	= \hat{\tilde{U}}^\dagger \hat{\dot n}^{(0)}
	+ \hat{\dot n}^{(0)} \hat{\tilde{U}},
\end{gather}
where we have used $\hat{U}_1 = \hat{U}_0 \hat{\tilde{U}}$.

We are interested only in those terms of $\hat{\dot n}^{(1)}$ that have the time dependence of the form $\exp(\pm i \omega_{\text{rf}} t)$. 
To isolate these contributions, we expand $\hat{\dot n}^{(1)}$ as
\begin{gather}
	\hat{\dot n}^{(1)} = \sum\limits_{k} \hat{\dot n}^{(1)}_k \exp(i k \omega_{\text{q0}} t),
\end{gather}
such that only the component $\hat{\dot n}^{(1)}_0$ contains terms proportional to $\exp(\pm i \omega_{\text{rf}} t)$. 
We find
\begin{gather}
	\hat{\dot n}^{(1)}_0 
	= - \dfrac{\Delta^2 \, \delta \varepsilon_{\text{rf}} \, \omega_{\text{rf}}}
	{\Delta E_0^2 \left(\Delta E_0^2 - \hbar^2 \omega_{\text{rf}}^2 \right)}
	\hat{\mathcal{H}}_0 \sin (\omega_{\text{rf}} t) . \label{AppendixB:n1}
\end{gather}

\subsection{Quantum capacitance}

Since $\hat{\dot n}^{(1)}_0 \propto \hat{\mathcal{H}}_0$, it is convenient to work in the eigenbasis of the unperturbed Hamiltonian $\hat{\mathcal{H}}_0$. Therefore, we expand the initial state $\ket{\psi_0} $ in the energy eigenstates $\ket{E_\pm}$ of $\hat{\mathcal{H}}_0$. At the initial moment of time the state $|\psi_0\rangle$ can be written as
\begin{equation}
	\ket{\psi_0} =\sqrt{\frac{1+\chi}{2}}\ket{E_{-}}
	+\sqrt{\frac{1-\chi}{2}} e^{i \vartheta}\ket{E_{+}}.
\end{equation}

Finally with the use of Eqs.~(\ref{Theory:CPB_clas_eqs}), (\ref{Theory:delta_varepsilon_rf}), and (\ref{AppendixB:n1}) we find 
\begin{equation}
	-\dfrac{8E_{\text{C}} C_{\text{m}}}{\hbar }\dfrac{%
		d \langle \hat{n}\rangle}{dt}  = C_\text{Q} \ddot \varphi_L,
\end{equation}
where the quantum admittance is
\begin{equation}
	C_\text{Q} =  C_\text{Q0} \dfrac{\Delta^3}{\Delta E_0^3 }\chi\dfrac{1}{1- \left(\dfrac{\omega_\text{rf}}{\omega_\text{q0}} \right)^2 }. \label{AppendixB:CQ}
\end{equation}
The capacitance $C_\text{Q0}$ is given by Eq.~(\ref{CQO_phenomenology}). For the limiting case $\omega_\text{rf} \ll \omega_\text{q0}$ we obtain Eq.~(\ref{Phenomenology:quantum_capacitance}).

\section{Effective capacitance and resistance of an open two-level system}

\label{Appendix_bloch_qubit}

\subsection{GKSL equation}

We will start from solving the GKSL equation which has to be written in the
instantaneous energy representation. By introducing the unitary transformation $\hat{S}$
such that $|\psi \rangle =\hat{S}|\psi ^{\prime }\rangle $: 
\begin{equation}
	\hat{S}=\lambda_{+}\hat{\mathbf{1}}+i\lambda_{-}\hat{\sigma}_{y} \label{2lvl_unitary_transformation_S}
\end{equation}
with 
\begin{equation}
	\lambda_{\pm}=\pm \frac{1}{\sqrt{2}}\sqrt{1\pm \frac{\varepsilon (t)}{\sqrt{%
				\Delta ^{2}+\varepsilon ^{2}(t)}}},
\end{equation}
which results in the instantaneous Hamiltonian~(\ref{Hinst_qb}). The GKSL equation for the density matrix is given by Eq.~(\ref{Lindblad_eq}). We introduce the Bloch vector to parameterize the density matrix

\begin{equation}
	\rho =\dfrac{1}{2}\left[ 1+\vec{R}\vec{\sigma}\right] ,
\end{equation}%
where $\vec{R}=(X,Y,\chi)$. Similarly, we rewrite the Hamiltonian in the
form $\hat{\mathcal{H}}_\text{inst}=-\frac{\hbar }{2}\vec{\mathcal{H}}_\text{inst}\hat{\vec{\sigma}}$, with
\begin{equation}
	\vec{\mathcal{H}}_\text{inst}=\left( 0,-\frac{\Delta \dot{\varepsilon}}{\Delta
		E^{2}(t)},\frac{\Delta E(t)}{\hbar }\right).
\end{equation}%
The GKSL equation reads
\begin{equation}
	\dot{\vec{R}}=\vec{R}\times \vec{\mathcal{H}}_\text{inst}-\left( \dfrac{X}{T_{2}},\dfrac{Y}{T_{2}},\dfrac{\chi-\chi_{0}(t)}{T_{1}}\right) , \label{GKSL_2level_vector_form}
\end{equation}%
where from Eq.~(\ref{two_level_detailed_balance})
\begin{equation}
	\chi_{0}(t)=\dfrac{\gamma_+-\gamma_-}{\gamma_+ +\gamma_-}=\tanh \left(\dfrac{\Delta E(t)}{2k_{\text{B}}\ T}\right), \label{Z0t}
\end{equation}
where 
\begin{gather}
	\gamma_++\gamma_- = \gamma = T_1^{-1}, \,\,\,\,\,\, T_2^{-1}=2\gamma_\varphi+\dfrac{1}{2T_1}, 
\end{gather}
and $\chi_0(t)$ is the instantaneous difference in the level occupations. From Eq.~(\ref{GKSL_2level_vector_form}), we emphasize that both the relaxation and decoherence times, $T_{1,2} = T_{1,2}[\varepsilon(t)]$, can depend on the energy bias.

\subsection{Quantum current operator}
The impact of the quantum subsystem on the classical one in Eq.~(\ref%
{Theory:CPB_clas_eqs}) is realized through the quantum current $\propto d\langle \partial \hat{\mathcal{H}}/\partial \dot \varphi_\text{cl}\rangle/dt = d\langle \partial \hat{\mathcal{H}}/\partial \varepsilon\rangle/dt \cdot \partial \varepsilon / \partial \dot \varphi_\text{cl}$ [see Eq.~(\ref{quantum_current})], where $\dot \varphi_\text{cl} = 2e V/\hbar$. From Eq.~(\ref{Theory:2levelHam})
\begin{equation}
	\dfrac{d}{dt}\Big\langle \dfrac{\partial \hat{\mathcal{H}}}{\partial \varepsilon}\Big\rangle = \dfrac{\partial^2 E_\text{geom}}{\partial \varepsilon^2}\dot \varepsilon-\dfrac{1}{2}\dfrac{d\langle \hat{\sigma}_z\rangle}{dt},
\end{equation}
so that the first term results in a geometric capacitance~(\ref{C_geom}).

The Pauli matrix $\hat{\sigma}_z$ has to be transformed in the same way as any operator with the use of the unitary transformation $\hat{S}$, the current can be rewritten with the use of the GKSL equation 
\begin{gather}
	\nonumber-\dfrac{1}{2}\dfrac{d\langle \hat{\sigma}_z\rangle}{dt}  =-\dfrac{1}{2}\dfrac{d}{dt}\text{Tr}\left( S^{\dag}\hat{\sigma}_z S\rho \right)=\\
	\nonumber=\frac{d}{dt}\frac{1}{2}\left[ 1+\frac{\Delta X-\varepsilon \chi}{\Delta E(t)}\right] = \\=\frac{1}{2}\left[ \frac{\Delta Y}{\hbar }-\frac{\Delta X}{T_{2}\Delta E(t)}+\frac{\varepsilon \left( \chi-\chi_{0}(t)\right) }{T_{1}\Delta E(t)}\right] .	\label{Appendix_bloch_qubit_dndt}
\end{gather}

\subsection{Perturbation theory}

In this subsection we will develop a perturbation theory for a small classical
probe $\delta \varepsilon _{\text{rf}}$ at long times $t\rightarrow
\infty $. 

The instantaneous Hamiltonian can be decomposed as in Eq.~(\ref{Hinst_pert_decomposition_2lvl}), with 
\begin{gather}
	\hat{\mathcal{H}}_\text{inst}^{(0)}=E_\text{geom}^{(0)}\hat{\mathbf{1}}-\dfrac{\Delta E_0}{2}\hat{\sigma}_{z}, \label{2lvl_H_pert}\\
	\nonumber 
	\hat{V}_{\delta \varepsilon} =\dfrac{\partial E_\text{geom}^{(0)}}{\partial\varepsilon_0}\hat{\mathbf{1}}-\dfrac{\varepsilon_0}{2\Delta E_0}\hat{\sigma}_{z},\,\,\,\,\,\, \hat{V}_{\delta \dot \varepsilon}=\dfrac{\hbar
		\Delta}{2\Delta E_0^2}\hat{\sigma}_{y},
\end{gather}
while the relaxation rates can be written as
\begin{gather}
	\gamma_\pm =\dfrac{1 \pm \chi_0(t)}{2 T_1(\varepsilon)},  \,\,\,\,\,\, \gamma_\varphi = \dfrac{1}{2T_2(\varepsilon)}-\dfrac{1}{4T_1(\varepsilon)} \label{2lvl_gamma_decomposition}.
\end{gather}
From Eqs.~(\ref{Z0t}) and (\ref{2lvl_gamma_decomposition}) the derivatives of the relaxation rates $\partial_\varepsilon \gamma_\alpha^{(0)}$ can be readily obtained. 
The stationary solution for the Bloch vector is $\vec{R}=(0,0,\chi_\text{th})$ with $\chi_\text{th}=\tanh \left( \Delta
E_0/2k_\text{B} T\right) $. The first-order correction to the density matrix can be found in the form~(\ref{Lindblad_anzatz}). Since $\Tr \hat{A} = \Tr \hat{B}=0$ to preserve unit trace of the density matrix, $\Tr \hat{\rho} =1$, we can parameterize both $\hat{A}$ and $\hat{B}$ as
\begin{equation}
	\hat{A}=\sum\limits_{i=1}^3 A_i \hat{\sigma}_i, \,\,\, \hat{B}=\sum\limits_{i=1}^3 B_i \hat{\sigma}_i. \label{AB_qudit_sub}
\end{equation}

By substituting Eq.~(\ref{AB_qudit_sub}) into the ansatz~(\ref{Lindblad_anzatz}), and after this, by substituting into the GKSL equation~(\ref{Lindblad_eq}), and by expanding up to the first order with the use of $\delta \ddot \varepsilon  = -\omega_\text{rf}^2 \delta \varepsilon$ this gives a system of six linear equations for unknown coefficients $A_i$ and $B_i$ 
\begin{equation}
	\begin{pmatrix}
		T_{2,0}^{-1} & -\omega_{\text{q}0} & 0 & -\omega_\text{rf}^2 & 0 & 0 \\
		\omega_{\text{q}0} & T_{2,0}^{-1} & 0 & 0 & -\omega_\text{rf}^2 & 0 \\
		0 & 0 & T_{1,0}^{-1} & 0 & 0 & -\omega_\text{rf}^2 \\
		1 & 0 & 0 & T_{2,0}^{-1} & -\omega_{\text{q}0} & 0 \\
		0 & 1 & 0 & \omega_{\text{q}0} & T_{2,0}^{-1} & 0 \\
		0 & 0 & 1 & 0 & 0 & T_{1,0}^{-1}
	\end{pmatrix}
	\begin{pmatrix}
		A_1 \\
		A_2 \\
		A_3 \\
		B_1 \\
		B_2 \\
		B_3
	\end{pmatrix} = 
	\begin{pmatrix}
		0 \\
		0 \\
		v_0 \\
		v_1 \\
		0 \\
		0
	\end{pmatrix}, \label{system6ab}
\end{equation}
where $T_{i,0}^{-1} = 1/T_i(\varepsilon_0)$, and
\begin{equation}
	v_0=  \dfrac{1}{T_1(\varepsilon_0)}\dfrac{\varepsilon_0}{4 k_\text{B} T \Delta E_0}\left(1-\chi_\text{th}^2\right), \,\,\, v_1 = \dfrac{\Delta}{2 \Delta E_0^2}\chi_\text{th}.
\end{equation}
The system~(\ref{system6ab}) can be solved analytically; however, the explicit solution is omitted for brevity.

\subsection{Effective capacitance and resistance}

By substituting the density matrix up to the first order into Eq.~(\ref{Appendix_bloch_qubit_dndt}), and also by expanding this equation up to first order we obtain
\begin{equation}
	-	\dfrac{d}{dt}\Big \langle \dfrac{\partial \hat{\mathcal{H}}}{\partial V} \Big \rangle =C_{\text{eff}}\dot V+\dfrac{V}{R_{\text{eff}}}
\end{equation}
with the effective capacitance
\begin{widetext}
	\begin{gather}
		\label{Appendix_bloch_qubit_Ceff}
		C_\text{eff}=C_\text{geom}+
		C_\text{Q0} \Bigg\{\dfrac{\Delta^3}{\Delta E_0^3}\tanh \left( \dfrac{\Delta E_0}{2 k_\text{B}T}\right)
		+\dfrac{\varepsilon_0^2 }{2k_B T \Delta E_0^2(1+T_1^2 \omega_\text{rf}^2)} \cosh^{-2} \left(\dfrac{\Delta E_0}{2 k_\text{B}T}\right)-\\
		\nonumber-\dfrac{\Delta^3}{\Delta E_0^3}\dfrac{\omega_\text{rf}^2}{\omega_\text{q0}^2}\dfrac{ T_2^2 \omega_\text{q0}^2 \left[1-T_2^2(\omega_{\text{q}0}^2-\omega_\text{rf}^2)\right]}{1+T_2^4(\omega_\text{rf}^2-\omega_{\text{q}0}^2)^2+2T_2^2(\omega_\text{rf}^2+\omega_{\text{q}0}^2)}\tanh \left(\dfrac{\Delta E_0}{2 k_\text{B}T}\right)\Bigg\}
	\end{gather}
	
	and the effective resistance
	
	\begin{gather}
		\label{Appendix_bloch_qubit_Reff}
		R_\text{eff}^{-1}= R_\text{Q0}^{-1}\Bigg\{\dfrac{\hbar \omega_\text{rf}}{k_\text{B} T}\dfrac{\varepsilon_0^2}{\Delta E_0^2}\dfrac{T_1 \omega_\text{rf}}{   1+T_1^2 \omega_\text{rf}^2}\cosh^{-2} \left(\dfrac{\Delta E_0}{2 k_\text{B}T}\right)
		+2 \dfrac{\Delta^2}{\Delta E_0^2} \dfrac{\omega_\text{rf}^2}{\omega_\text{q0}^2} \dfrac{T_2\omega_{\text{q}0} \left[1+T_2^2(\omega_\text{rf}^2+\omega_{\text{q}0}^2)\right] }{1+T_2^4(\omega_\text{rf}^2-\omega_{\text{q}0}^2)^2+2T_2^2(\omega_\text{rf}^2+\omega_{\text{q}0}^2)}\tanh \left(\dfrac{\Delta E_0}{2 k_\text{B}T}\right) \Bigg\},
	\end{gather}
\end{widetext}
where $C_\text{Q0}$ and $R_\text{Q0}^{-1}$ are defined in Eqs.~(\ref{CQ0}) and (\ref{RQ0}) respectively.

From Eq.~(\ref{Appendix_bloch_qubit_Ceff}) we see that at $T_{1,2} \rightarrow \infty$ without any restrictions on $\omega_\text{rf}$ and $\omega_\text{q0}$ one obtains
\begin{equation}
	C_\text{Q} =  C_\text{Q0} \dfrac{\Delta^3}{\Delta E_0^3 }\dfrac{1}{1- \left(\dfrac{\omega_\text{rf}}{\omega_\text{q0}} \right)^2 } \tanh \left(\dfrac{\Delta E_0}{2 k_\text{B}T}\right), \label{CQ_GKSL_2}
\end{equation}
which totally agrees with Eq.~(\ref{AppendixB:CQ}), when one considers populations given in the thermal equilibrium, $\chi=\chi_\text{th}$. 

Expression for the effective capacitance and conductance can be further simplified in the limit of
small probing frequency $\omega _{\text{rf}}\ll \omega _{\text{q}0}$. The effective capacitance then reads
\begin{gather}
	C_\text{eff}=C_\text{geom}+
	C_\text{Q0} \Bigg[\dfrac{\Delta^3}{\Delta E_0^3}\tanh \left(\dfrac{\Delta E_0}{2 k_\text{B}T}\right)
	+\label{Appendix_bloch_qubit_Ceff_simp}\\
	\nonumber+\dfrac{\varepsilon_0^2 }{2k_B T \Delta E_0^2(1+T_1^2 \omega_\text{rf}^2)} \cosh^{-2}\left( \dfrac{\Delta E_0}{2 k_\text{B}T}\right)\Bigg], 
\end{gather}
since the third term in Eq.~(\ref{Appendix_bloch_qubit_Ceff}) $\propto (\omega_\text{rf}/\omega_\text{q0})^2 f(T_2 \omega_\text{q0})$ with $f(x)=x^2(1-x^2)/(1+x^2)^2$, which is bounded $|f(x)|\leq 1$ for all $x$. The first term in Eq.~(\ref{Appendix_bloch_qubit_Ceff_simp}) is a quantum capacitance $C_\text{Q}$, while the second term is a tunneling capacitance $C_\text{T}$.

The effective conductance in the leading order of $\omega_\text{rf}/\omega_{\text{q}0}$  is
\begin{gather}
	R_\text{eff}^{-1}= R_\text{Q0}^{-1}\Bigg[\dfrac{\hbar \omega_\text{rf}}{k_\text{B} T}\dfrac{\varepsilon_0^2}{\Delta E_0^2}\dfrac{T_1 \omega_\text{rf}}{   1+T_1^2 \omega_\text{rf}^2}\cosh^{-2} \left(\dfrac{\Delta E_0}{2 k_\text{B}T}\right) \label{Appendix_bloch_qubit_Reff_simp}
	+\\
	\nonumber+2 \dfrac{\Delta^2}{\Delta E_0^2} \dfrac{\omega_\text{rf}^2}{\omega_\text{q0}^2} \dfrac{T_2\omega_{\text{q}0} }{1+T_2^2\omega_{\text{q}0}^2}\tanh \left(\dfrac{\Delta E_0}{2 k_\text{B}T}\right) \Bigg].
\end{gather}
Here both terms are $\propto \omega_\text{rf}^2$ and therefore can be of the same order [see Fig.~\ref{Fig:Resistance3}]. The first term is the relaxation-defined  Sisyphus conductance $R_\text{Sis}^{-1}$, while the second term is the decoherence-defined Hermes conductance $R_\text{Hrm}^{-1}$.

\subsection{Good and bad qubits}
Further, in the limit $\omega_\text{rf} \ll \omega _{\text{q}0} \ll T_{1}^{-1}, T_{2}^{-1}$ we obtain the effective capacitance of a bad qubit 
\begin{gather}
	C_\text{eff}^{(\text{bad})}=C_\text{geom}+
	C_\text{Q0} \Bigg[\dfrac{\Delta^3}{\Delta E_0^3}\tanh \left(\dfrac{\Delta E_0}{2 k_\text{B}T}\right)
	\label{Appendix_bloch_qubit_Ceff_bad}+\\
	\nonumber+\dfrac{\varepsilon_0^2 }{2k_B T \Delta E_0^2} \cosh^{-2} \left(\dfrac{\Delta E_0}{2 k_\text{B}T}\right)\Bigg], 
\end{gather}
As we show in Sec.~\ref{Sec:Qudit}, this result can be written in a
general form as
\begin{equation}
	C_{\text{eff}}^{(\text{bad})}=-\left(\dfrac{\partial \varepsilon}{\partial V}\right)^2 \dfrac{\partial }{\partial \varepsilon }\overline{ \dfrac{\partial E}{\partial \varepsilon }}_\text{th} = -\dfrac{\partial }{\partial V }\overline{ \dfrac{\partial E}{\partial V }}_\text{th} ,  \label{CQbad}
\end{equation}%

For a good qubit in the limit $T_{1}^{-1}, T_2^{-1}\ll\omega_\text{rf} \ll \omega _{\text{q}0}$ we
obtain 
\begin{equation}
	C_{\text{eff}}^{(\text{good})}=C_\text{geom}+C_\text{Q0} \dfrac{\Delta^3}{\Delta E_0^3}\tanh \left(\dfrac{\Delta E_0}{2 k_\text{B}T}\right),  \label{ACQ_good}
\end{equation}%
which can be rewritten in a general form as
\begin{equation}
	C_{\text{eff}}^{(\text{good})}=-\left(\dfrac{\partial \varepsilon}{\partial V}\right)^2\overline{ \dfrac{\partial^2 E}{\partial \varepsilon^2 }}_\text{th}=-\overline{ \dfrac{\partial^2 E}{\partial V^2 }}_\text{th} .
\end{equation}

From Eq.~(\ref{Appendix_bloch_qubit_Reff_simp}) we see that both for the cases of a good ($T_{1}^{-1},T_{2}^{-1}\ll\omega_\text{rf} \ll \omega _{\text{q}0}$) and bad (\text{$\omega_\text{rf} \ll \omega _{\text{q}0}  \ll T_{1}^{-1}, T_{2}^{-1} $}) qubits the effective conductance is zero:
\begin{equation}
	G_\text{eff}^{(\text{good})} = G_\text{eff}^{(\text{bad})}=0.
\end{equation}

In the bad qubit limit ($T_1, T_2 \rightarrow 0 $), the relaxation and decoherence times are much shorter 
than the probing period, so the energy levels populations follow the classical probe 
instantaneously and no dissipation occurs. Conversely, in the good qubit limit, 
$T_{1}, T_2 \rightarrow \infty$, the system cannot relax on the timescale of the 
probe signal, $T_\text{rf} = 2\pi/\omega_\text{rf}$, and the response is again purely reactive.
\section{Effective capacitance of an isolated $d$-level systems}

\label{Appendix:IdQd}
To solve the system of coupled quantum and classical equations of motion, we need
to develop a formalism that can be seen as a natural extension of Appendix~\ref{Appendix:IdQb}. 
A nontrivial generalization arises from the fact that the calculations 
need to be performed in a general form, without any particular knowledge of the Hamiltonian or the structure of its energy levels.

To this end, first, in Appendix~\ref{Appendix:IdQd:HF}, we obtain the necessary properties of the energy levels $E_i$ and eigenfunctions $|E_i \rangle$ when the Hamiltonian depends on a parameter, $\hat{\mathcal{H}}=\hat{\mathcal{H}}(\varepsilon)$, which will be physically interpreted as the energy bias. After that, we choose the instantaneous energy representation as the most convenient one for finding a solution. In this representation, we derive the quantum current operator responsible for the effective capacitance in Appendix~\ref{Appendix:IdQd:inst}, construct the perturbation theory for the evolution operator in Appendix~\ref{Appendix:IdQd:pert}, calculate the first-order corrections to the quantum current operator in Appendix~\ref{Appendix:IdQd:pert_op}, and present the final result for the effective capacitance of an \textit{arbitrary isolated qu$d$it} in Appendix~\ref{Appendix:IdQd:CQ}.

\subsection{Hellmann-Feynman theorem}
\label{Appendix:IdQd:HF}
We start from the Hellmann-Feynman theorem, which states that 
\begin{equation}
	\partial _{\varepsilon
	}E_{k} \coloneq \frac{\partial E_{k}}{\partial \varepsilon }=\langle E_{k}|\partial _{\varepsilon }\hat{\mathcal{H}}|E_{k}\rangle .
	\label{Hellmann-Feynman}
\end{equation}%
This relation can be obtained by differentiating the eigenenergy $E_k$, written in the form $E_k(\varepsilon) = \langle E_{k}(\varepsilon)|\hat{\mathcal{H}}(\varepsilon)|E_{k}(\varepsilon)\rangle$, with
respect to the bias $\varepsilon $, using the fact that
\begin{equation}
	\dfrac{\partial}{\partial \varepsilon}\langle E_{k}(\varepsilon)|E_{k}(\varepsilon)\rangle =0. \label{trivial_derivative}
\end{equation}

We are now interested in obtaining expressions for the second derivative of the energy levels, $ \partial ^{2}E_{k}/\partial \varepsilon ^{2}$, as well as the first derivatives of the energy eigenfunctions, $\partial |E_{k}\rangle / \partial \varepsilon$. 
Let us differentiate the eigenvalue equation $(\hat{\mathcal{H}} - E_{k})|E_{k}\rangle = 0$
and derive the expression for $\partial |E_{k}\rangle / \partial \varepsilon$ from it:
\begin{equation}
	(\hat{\mathcal{H}}-E_k)\dfrac{\partial |E_{k}\rangle}{\partial \varepsilon} = \left( \dfrac{\partial E_k}{\partial \varepsilon}-\dfrac{\partial \hat{\mathcal{H}}}{\partial \varepsilon} \right) |E_{k}\rangle.
\end{equation}
Multiplying by $\langle E_l|$ with $l \neq k$ yields
\begin{equation}
	\langle E_l|\frac{\partial  }{\partial \varepsilon } |E_{k}\rangle=
	\frac{\langle E_{l}|\partial _{\varepsilon }\hat{\mathcal{H}}|E_{k}\rangle }{
		E_{l}-E_{k}}.
\end{equation}
The coefficient for $l=k$ cannot be obtained in this way. 
However, we note that the phases of the energy eigenstates can always be redefined as
\begin{equation}
	|\tilde{E}_k(\varepsilon)\rangle \rightarrow \exp[i \chi_k(\varepsilon)]|E_k(\varepsilon)\rangle, \label{Appendix:IdQd:renorm}
\end{equation}
so that 
\begin{equation}
	\langle \tilde{E}_{k}|\frac{\partial }{\partial \varepsilon }|\tilde{E}_{k}\rangle =0,
	\label{Appendix:IdQd:constraint}
\end{equation}
while all other matrix elements with $l \neq k$ remain unchanged.  Indeed, substituting Eq.~(\ref{Appendix:IdQd:renorm}) into Eq.~(\ref{Appendix:IdQd:constraint}) gives
\begin{equation}
	\dfrac{\partial \chi_k}{\partial \varepsilon} = i\langle E_{k}|\frac{\partial }{\partial \varepsilon }|E_{k}\rangle \eqcolon  f_k(\varepsilon). 
\end{equation}
From Eq.~(\ref{trivial_derivative}) we see that $\mathrm{Im} f_k = 0$. 
By choosing $\chi_k(\varepsilon) = \int f_k(\varepsilon) \, d\varepsilon$, we satisfy Eq.~(\ref{Appendix:IdQd:constraint}). 
Hereafter, we assume that the phases $\chi_k(\varepsilon)$ have already been chosen such that~\cite{Duncan2025}
\begin{equation}
	\langle E_{k}|\frac{\partial }{\partial \varepsilon }|E_{k}\rangle =0.
\end{equation}%
Thus, we obtain 
\begin{equation}
	\label{dketEdeps}
	\frac{\partial |E_{k}\rangle }{\partial \varepsilon }=\sum\limits_{l\neq k}%
	\frac{\langle E_{l}|\partial _{\varepsilon }\hat{\mathcal{H}}|E_{k}\rangle }{%
		E_{k}-E_{l}}|E_{l}\rangle,
\end{equation}
with the analogous expression for the bra vectors:
\begin{equation}
	\label{dbraEdeps}
	\frac{\partial \langle E_{k}| }{\partial \varepsilon }=\sum\limits_{l\neq k}%
	\frac{\langle E_{k}|\partial _{\varepsilon }\hat{\mathcal{H}}|E_{l}\rangle }{%
		E_{k}-E_{l}}\langle E_{l}|.
\end{equation}

Differentiating Eq.~(\ref{Hellmann-Feynman}) with respect to $\varepsilon$ and using Eqs.~(\ref{dketEdeps}) and (\ref{dbraEdeps}), we obtain the expression for the second derivative of the eigenenergy:
\begin{equation}
	\frac{\partial^{2} E_{k}}{\partial \varepsilon^{2}}
	= \langle E_{k} | \partial_{\varepsilon}^{2} \hat{\mathcal{H}} | E_{k} \rangle
	+ 2 \sum\limits_{l \neq k}
	\frac{\left| \langle E_{k} | \partial_{\varepsilon} \hat{\mathcal{H}} | E_{l} \rangle \right|^{2}}
	{E_{k} - E_{l}} . \label{d2Edeps2}
\end{equation}

\subsection{Instantaneous energy representation}
\label{Appendix:IdQd:inst}
Now we consider the Hamiltonian $\hat{\mathcal{H}}(\varepsilon)$ with a time-dependent parameter $\varepsilon = \varepsilon(t) $. We transform the Hamiltonian into the instantaneous representation using the unitary matrix 
\begin{equation}
	\hat{S}=\sum\limits_{m}|E_{m}\rangle \langle m|,  \label{S_inst}
\end{equation}%
where $\ket{m}$ denotes the standard orthonormal basis vector, i.e., a column vector with a single nonzero entry 1 at the $m$-th position.

The Hamiltonian in the new representation is
\begin{equation}
	\hat{\mathcal{H}}_\text{inst}=
	\hat{S}^{\dagger }\hat{\mathcal{H}}\hat{S}-i\hbar \hat{S}^{\dagger }\hat{
		\dot{S}}, \label{Appendix:IdQd:Hinst}
\end{equation}
and can be written explicitly using Eqs.~(\ref{dketEdeps}) and (\ref{dbraEdeps}) as 
\begin{equation}
	\hat{\mathcal{H}}_\text{inst}=\sum\limits_{m}E_{m}\left\vert m\right\rangle
	\left\langle m\right\vert -i\hbar \dot{\varepsilon}\sum\limits_{m\neq
		k}\left\vert m\right\rangle \left\langle k\right\vert \frac{\left\langle
		E_{m}\right\vert \partial _{\varepsilon }\hat{\mathcal{H}}\left\vert
		E_{k}\right\rangle }{E_{k}-E_{m}},  \label{H_inst}
\end{equation}
where both the eigenenergies $E_m(\varepsilon)$ and eigenvectors $|E_m(\varepsilon)\rangle$ acquire time-dependence through the energy bias $\varepsilon = \varepsilon(t)$.

Since the expectation value of the quantum current operator is proportional to 
$\langle \hat{\mathcal{I}} \rangle \propto d\langle \partial_\varepsilon \hat{\mathcal H} \rangle / dt$, 
we need to determine the time derivative of the operator 
$\partial_\varepsilon \hat{\mathcal H}$ in the instantaneous energy representation.

For any operator $\hat{A}$, its time derivative in the instantaneous representation can be expressed as 
\begin{gather}
	\nonumber\hat{\dot{A}}_\text{inst}=\frac{\partial (\hat{S}^{\dag }\hat{A}\hat{S})}{%
		\partial t}+\frac{i}{\hbar }\left[ \hat{\mathcal{H}}_\text{inst},\hat{S}^{\dag
	}\hat{A}\hat{S}\right] =\\
	=\frac{\partial \hat{A}_\text{inst}}{\partial t}+\frac{i%
	}{\hbar }\left[ \hat{\mathcal{H}}_\text{inst},\hat{A}_\text{inst}\right] .
	\label{operator_inst}
\end{gather}%
For $\hat{A} = \partial_\varepsilon \hat{\mathcal{H}}$, its instantaneous representation is
\begin{equation}
	(\partial _{\varepsilon }\hat{\mathcal{H}})_\text{inst}=\sum\limits_{m,k}\left\vert m\right\rangle \left\langle k\right\vert
	\left\langle E_{m}\right\vert \partial _{\varepsilon }\hat{\mathcal{H}}%
	\left\vert E_{k}\right\rangle .  \label{He_inst}
\end{equation}%
Using Eqs.~(\ref{S_inst}), (\ref{Appendix:IdQd:Hinst}), (\ref{H_inst}), and (\ref{operator_inst}), we obtain
\begin{equation}
	(\partial _{\varepsilon }\hat{\dot{\mathcal{H}}})_\text{inst}=\sum\limits_{n,l}
	\left[ \dot{\varepsilon}(\partial_{\varepsilon }^{2}\hat{\mathcal{H}})_{nl}+
	i\omega_{nl}(\partial_{\varepsilon }\hat{\mathcal{H}})_{nl}
	\right] \left\vert n\right\rangle \left\langle l\right\vert, \label{Qudit_isolated_dhdeps_inst}
\end{equation}
where $\omega_{nl} = (E_n - E_l)/\hbar$, and, starting from this point, we will use the shorthand notation  $A_{nl} = \langle E_n | \hat{A} | E_l \rangle$ for matrix elements of $\hat{A}$.

\subsection{Perturbation theory for the evolution operator}
\label{Appendix:IdQd:pert}
In this subsection, we build a general perturbation theory for $d$-level systems. The energy bias in the Hamiltonian $\varepsilon(t) = \varepsilon_0 + \delta \varepsilon_\text{rf} \cos( \omega_\text{rf} t)$ contains a stationary point $\varepsilon_0$ and a small classical probing term. Since both the probing amplitude $\delta \varepsilon_\text{rf}$ and the probing frequency $\omega_\text{rf}$ are small [see Eq.~(\ref{QuditA_applicability_for_perturbation_theory})], we can find the first-order perturbative correction to the instantaneous Hamiltonian~(\ref{H_inst}) in the form~(\ref{Hinst_pert_decomposition_2lvl}), with
\begin{gather}
	\nonumber\hat{\mathcal{H}}_{\text{inst}}^{(0)}=  \sum\limits_{m}E_{m}^{(0)}\left\vert m\right\rangle
	\left\langle m\right\vert, \,\,\, \hat{V}_{\delta \varepsilon} =  \sum\limits_{m}\dfrac{\partial E_m}{\partial \varepsilon}\left\vert m\right\rangle
	\left\langle m\right\vert,\\
	\hat{V}_{\delta \dot\varepsilon} = i \sum\limits_{m\neq
		k} \dfrac{(\partial_\varepsilon \hat{\mathcal{H}})_{mk}}{\omega_{mk}^{(0)}} | m\rangle \langle k|. \label{perturbative_operators}
\end{gather}

Then the first order of Eq.~(\ref{Hinst_pert_decomposition_2lvl}) can be written as
\begin{equation}
	\hat{\mathcal{H}}_{\text{inst}}^{(1)} = \hat{V}_{\delta \varepsilon}\delta \varepsilon_\text{rf}(t)+\hat{V}_{\delta \dot\varepsilon}\delta \dot\varepsilon_\text{rf} = \hat{F}e^{-i \omega_\text{rf}t}+\hat{F}^\dagger e^{i \omega_\text{rf}t},
\end{equation}
where
\begin{equation}
	\hat{F}= \dfrac{\delta \varepsilon_\text{rf}}{2}\left[\hat{V}_{\delta \varepsilon}-i\omega_\text{rf} \hat{V}_{\delta \dot \varepsilon} \right]. \label{F_definition}
\end{equation}

As in Appendix~\ref{Appendix:IdQb}, we construct the perturbation theory for the evolution matrix $\hat{U}$. 
The zeroth-order term is
\begin{equation}
	\hat{U}_0 = \exp \left[ -\dfrac{i}{\hbar} \hat{\mathcal{H}}_{\text{inst}}^{(0)} t\right] = \sum\limits_m e^{-i E_{m}^{(0)} t/\hbar}|m\rangle \langle m |,  \label{qdU0}
\end{equation}
while the first-order term can be found in the same form as in Eq.~(\ref{U1ansatz}), $\hat{U}_1 =\hat{U}_0 \hat{\tilde{U}}$, yielding
\begin{gather}
	\nonumber \hat{\tilde{U}}=-\dfrac{i}{\hbar}\sum\limits_{m,k} |m\rangle \langle k| \bigg[F_{mk}\dfrac{e^{i(\omega_{mk}^{(0)}-\omega_\text{rf}) t}-1}{i(\omega_{mk}^{(0)}-\omega_\text{rf})} +\\
	+F_{km}^* \dfrac{e^{i(\omega_{mk}^{(0)}+\omega_\text{rf}) t}-1}{i(\omega_{mk}^{(0)}+\omega_\text{rf})}\bigg]. \label{qdUt}
\end{gather}

\subsection{First-order corrections to the quantum current}
\label{Appendix:IdQd:pert_op}
We note that $(\partial_\varepsilon \hat{\dot{\mathcal{H}}})_\text{inst}$ from Eq.~(\ref{Qudit_isolated_dhdeps_inst}) itself consists of both zeroth- and first-order terms:
\begin{equation}
	(\partial _{\varepsilon }\hat{\dot{\mathcal{H}}})_\text{inst} = (\partial _{\varepsilon }\hat{\dot{\mathcal{H}}})_\text{inst}^{(0)}+(\partial _{\varepsilon }\hat{\dot{\mathcal{H}}})_\text{inst}^{(1)},
\end{equation}
where the zeroth-order term arises from the second term of Eq.~(\ref{Qudit_isolated_dhdeps_inst}):
\begin{equation}
	(\partial _{\varepsilon }\hat{\dot{\mathcal{H}}})_\text{inst}^{(0)} = \sum\limits_{n,l} i\omega_{nl}^{(0)}(\partial_{\varepsilon }\hat{\mathcal{H}})_{nl}^{(0)}
	\left\vert n\right\rangle \left\langle l\right\vert, \label{deH0}
\end{equation}
while the first-order term arises from both the first and second terms of Eq.~(\ref{Qudit_isolated_dhdeps_inst}):
\begin{gather}
	\nonumber(\partial _{\varepsilon }\hat{\dot{\mathcal{H}}})_\text{inst}^{(1)} =\sum\limits_{n,l}
	\bigg[ \delta \dot\varepsilon_\text{rf}(\partial_{\varepsilon }^{2}\hat{\mathcal{H}})_{nl}^{(0)}+\\
	+
	i \delta \varepsilon_\text{rf}(t) \dfrac{\partial}{\partial \varepsilon}\big(\omega_{nl}(\partial_{\varepsilon }\hat{\mathcal{H}})_{nl} \big)^{(0)}
	\bigg] \left\vert n\right\rangle \left\langle l\right\vert. \label{deH1}
\end{gather}

Next, we search the first-order correction to the expectation value of the quantum current:
\begin{equation}
	\dfrac{d  \langle \partial_\varepsilon \hat{\mathcal{H}} \rangle}{dt} = \langle (\partial _{\varepsilon }\hat{\dot{\mathcal{H}}})_\text{inst}\rangle  = \langle \psi_0 | \hat{U}^\dagger (\partial _{\varepsilon }\hat{\dot{\mathcal{H}}})_\text{inst} \hat{U}|\psi_0 \rangle.
\end{equation}
This expectation value also contains both the zeroth- and first-order contributions: 
\begin{equation}
	\langle (\partial_\varepsilon \hat{\dot{\mathcal{H}}})_\text{inst} \rangle 
	= \langle (\partial_\varepsilon \hat{\dot{\mathcal{H}}})_\text{inst} \rangle^{(0)} 
	+ \langle (\partial_\varepsilon \hat{\dot{\mathcal{H}}})_\text{inst} \rangle^{(1)}.
\end{equation}
To distinguish terms of different orders, we introduce the operators $\hat{I}^{(i)}$ of the zeroth and first orders
\begin{equation}
	\langle \psi_0 | \hat{I}^{(i)} | \psi_0 \rangle\coloneqq \langle (\partial_\varepsilon \hat{\dot{\mathcal{H}}})_\text{inst} \rangle^{(i)}, \label{quantum_current_intro}
\end{equation}
from which
\begin{gather}
	\hat{I}^{(0)} = \hat{U}_0^\dagger (\partial _{\varepsilon }\hat{\dot{\mathcal{H}}})_\text{inst}^{(0)} \hat{U}_0,\\
	\hat{I}^{(1)} = \hat{\tilde{U}}^\dagger \hat{I}^{(0)}+ \hat{I}^{(0)} \hat{\tilde{U}}+\hat{U}_0^\dagger (\partial _{\varepsilon }\hat{\dot{\mathcal{H}}})_\text{inst}^{(1)} \hat{U}_0.
\end{gather} 

Operators $\hat{I}^{(0)}$ and  $\hat{I}^{(1)}$ can be obtained directly from Eqs.~(\ref{qdU0}), (\ref{qdUt}), (\ref{deH0}), and (\ref{deH1}). 
The final results read
\begin{equation}
	\hat{I}^{(0)} = \sum\limits_{n,l} i \omega_{nl}^{(0)}(\partial_\varepsilon \hat{\mathcal{H}})_{nl}^{(0)}e^{i \omega_{nl}^{(0)}t}   | n \rangle \langle l|
\end{equation} 
and 
\begin{widetext}
	\begin{gather}
		\nonumber\hat{I}^{(1)} = \sum\limits_{n,l}e^{i \omega_{nl}^{(0)}t}
		\bigg[ \delta \dot\varepsilon_\text{rf}(\partial_{\varepsilon }^{2}\hat{\mathcal{H}})_{nl}^{(0)}+
		i \delta \varepsilon_\text{rf}(t) \dfrac{\partial}{\partial \varepsilon}\Big(\omega_{nl}(\partial_{\varepsilon }\hat{\mathcal{H}})_{nl} \Big)^{(0)}
		\bigg] \left\vert n\right\rangle \left\langle l\right\vert-\\
		-\dfrac{1}{\hbar}\bigg[ \sum\limits_{n,m,l} i \omega_{nm}^{(0)}(\partial_\varepsilon \hat{\mathcal{H}})_{nm}^{(0)}  \Big[F_{ml}\dfrac{e^{i(\omega_{nl}^{(0)}-\omega_\text{rf}) t}-e^{i \omega_{nm}^{(0)} t}}{\omega_{ml}^{(0)}-\omega_\text{rf}} +F_{lm}^* \dfrac{e^{i(\omega_{nl}^{(0)}+\omega_\text{rf}) t}-e^{i\omega_{nm}^{(0)} t}}{\omega_{ml}^{(0)}+\omega_\text{rf}}\Big]|n\rangle \langle l| +\text{h.c.}\bigg]. \label{I1gen}
	\end{gather}
	
\end{widetext}
We are interested only in the part of $\hat{I}^{(1)}$ proportional to $e^{\pm i \omega_\text{rf} t}$. 
If we define $\hat{I}^{(1)}_{nm}$ such that
\begin{equation}
	\hat{I}^{(1)} = \sum\limits_{n,m} \hat{I}^{(1)}_{nm} e^{i \omega_{nm}^{(0)}t},
\end{equation}
the relevant part for the quantum current reads
\begin{gather}
	\hat{I}^{(1)}_\text{rf} = \sum\limits_{n} \hat{I}^{(1)}_{nn}.
\end{gather}
Such terms arise only for $n = l$ in Eq.~(\ref{I1gen}). 
Furthermore, the second term of Eq.~(\ref{I1gen}) vanishes at $n = m$ since $\omega_{nn}^{(0)} = 0$, so that only the off-diagonal elements of $\hat{F}$ contribute to the effective capacitance. 
From Eq.~(\ref{F_definition}) for $m \neq l$, we obtain
\begin{equation}
	F_{ml} =  \dfrac{\delta \varepsilon_\text{rf} \omega_\text{rf}}{2}\dfrac{(\partial_\varepsilon \hat{\mathcal{H}})_{ml}^{(0)}}{\omega_{ml}^{(0)}}. \label{F_matr_el}
\end{equation}

Using Eq.~(\ref{F_matr_el}), the first term of Eq.~(\ref{I1gen}) at $n=l$ is proportional to $(\partial_\varepsilon^2 \hat{\mathcal{H}})_{nn}^{(0)}$, while the second term is proportional to $|(\partial_\varepsilon \hat{\mathcal{H}})_{nm}^{(0)}|^2 / \omega_{nm}^{(0)}$, exactly as needed to obtain the second derivatives of the energy levels $\partial^2 E_n / \partial \varepsilon^2$ [see Eq.~(\ref{d2Edeps2})]. 
Then it is straightforward to obtain the final result
\begin{gather}
	\hat{I}^{(1)}_\text{rf} = \delta \dot \varepsilon_\text{rf}\sum\limits_n \left[ \dfrac{\partial^2 E_n}{\partial \varepsilon^2}+\dfrac{2}{\hbar}\sum\limits_{m \neq n} \dfrac{ \omega_\text{rf}^2|(\partial_\varepsilon \hat{\mathcal{H}})_{nm}^{(0)}|^2}{\omega_{nm}^{(0)}(\omega_{nm}^{(0)2}-\omega_\text{rf}^2)}\right]|n \rangle \langle n|. \label{CQ_isolated_qudit_no_simp}
\end{gather}
The second term vanishes for $\omega_\text{rf} \ll \omega_{nm}^{(0)}$. 
With this condition satisfied, for an arbitrary initial state written in the energy representation
\begin{equation}
	\ket{\psi_0} = \sum\limits_n c_n \ket{n} 
\end{equation}
we finally obtain 
\begin{equation}
	\dfrac{d  \langle \partial_\varepsilon \hat{\mathcal{H}} \rangle}{dt} =  \delta \dot \varepsilon_\text{rf}\sum\limits_n p_n \dfrac{\partial^2 E_n}{\partial \varepsilon^2}+\sum_{n \neq m}e^{i \omega_{nm}^{(0)}t}\langle \hat{I} \rangle_{nm}, \label{quantum_current_ans}
\end{equation}
where $p_n = |c_n|^2$ is the occupation probability of the $n$-th energy level, and the second term represents the high-frequency contributions to the quantum current associated with the internal oscillations of the quantum system at qu$d$it frequencies $\omega_{nm}^{(0)}$, which do not contribute to the rf response extracted on the timescale of the classical resonator.

\subsection{Effective capacitance}
\label{Appendix:IdQd:CQ}
From Eqs.~(\ref{quantum_current_exp}) and (\ref{quantum_current_ans}), we obtain
\begin{equation}
	C_\text{eff} \, \ddot{\varphi}_\text{cl} = - \frac{2e}{\hbar} \frac{\partial \varepsilon}{\partial V} 
	\overline{\frac{\partial^2 E}{\partial \varepsilon^2}}_\text{S} \, \delta \dot{\varepsilon}_\text{rf}.
\end{equation}
Since $\ddot{\varphi}_\text{cl} = 2 e \dot{V} / \hbar$ and $\delta \dot{\varepsilon}_\text{rf} = \dot{V} \, (\partial \varepsilon / \partial V)$, 
the effective capacitance of an \textit{arbitrary isolated qu$d$it} is
\begin{equation}
	C_\text{eff} = - \left( \frac{\partial \varepsilon}{\partial V} \right)^2 
	\overline{\frac{\partial^2 E}{\partial \varepsilon^2}}_\text{S} 
	= - \overline{\frac{\partial^2 E}{\partial V^2}}_\text{S}.
\end{equation}

This can be split into the geometric capacitance
\begin{equation}
	C_\text{geom} = - \dfrac{\partial^2 E_\text{geom}}{\partial V^2}
\end{equation}
and the quantum capacitance
\begin{equation}
	C_\text{Q} = - \overline{\frac{\partial^2 E'}{\partial V^2}}_\text{S}.
\end{equation}


\section{The effective capacitance and resistance for an open $d$-level system}
\label{Appendix:GKSL}
In this Appendix, we derive general analytical expressions for the effective capacitance and resistance for the dynamics governed by the GKSL master equation. Analogously to how Appendix~\ref{Appendix:IdQd} generalizes Appendix~\ref{Appendix:IdQb}, the present Appendix extends Appendices~\ref{Appendix_bloch_qubit} and \ref{Appendix:IdQd} to open quantum multi-level systems. 

First, in Appendix~\ref{Appendix:GKSL:inst}, we discuss the structure of the GKSL equation in the instantaneous energy representation. Then, in Appendix~\ref{Appendix:GKSL:Liouville}, we formulate the problem in the Liouville space. Next, in Appendix~\ref{Appendix:GKSL:properties}, we point out properties of the GKSL equation in Liouville space that will be used throughout Appendices~\ref{Appendix:GKSL} and \ref{Appendix:GKSL_GOOD_BAD}. 

In Appendix~\ref{Appendix:GKSL:td}, we derive expressions for the time derivatives of operators in Liouville space, which are used to calculate the expectation value of the quantum current operator. Finally, in Appendices~\ref{Appendix:GKSL:pert} and \ref{Appendix:GKSL:qc}, we develop a perturbation theory to analyze the system near thermal equilibrium and calculate the first-order corrections to the quantum current. The final expressions for the effective capacitance and conductance are presented in Appendix~\ref{Appendix:GKSL:Ceff_Reff}. In Appendix~\ref{Appendix:GKSL:Ceff} we briefly discuss how to introduce geometric, quantum and tunneling capacitances.

\subsection{GKSL master equation in the instantaneous energy representation}
\label{Appendix:GKSL:inst}
Let us write the quantum master equation~(\ref{Lindblad_eq}) in the instantaneous energy representation, where the instantaneous Hamiltonian $\hat{\mathcal{H}}_\text{inst}$ is defined in Eq.~(\ref{H_inst}).

In the absence of microwave driving and classical probing, the instantaneous Hamiltonian $\hat{\mathcal{H}}_\text{inst}$ is diagonal, and we expect the system to relax to a thermal equilibrium state $\hat{\rho}^{(0)}$ as $t \to \infty$. In this limit, $\dot{\rho}_{lm}=0$ and $[\hat{\rho}^{(0)},\hat{\mathcal{H}}_\text{inst}]=0$. This requirement constrains the relaxation rates for a given choice of the Lindblad jump operators.

For example, by choosing $\hat{L}_\alpha = \hat{L}_{ij}=|i\rangle \langle j|$ with $\gamma_\alpha = \gamma_{ij}$, we obtain
\begin{equation}
	\sum\limits_{j} \gamma_{lj}\delta_{lm}p_j^\text{th}=\dfrac{1}{2}\sum\limits_{j}(\gamma_{jl}+\gamma_{jm})\delta_{lm}p_l^\text{th}, \label{detailed_balance} 
\end{equation}
where we have used $\hat{\rho}^{(0)}_{lm} = \delta_{lm}p_l^\text{th}$, and $p_l^\text{th}$ are the thermal equilibrium populations. The detailed balance condition~(\ref{detailed_balance}) is satisfied if we impose the stronger condition [for $m=l$ and for all $j$]
\begin{equation}
	\gamma_{lj}p_j^\text{th}=\gamma_{jl}p_l^\text{th} \,\,\, \rightarrow \,\,\, \gamma_{lj} = \gamma_{jl}\exp \dfrac{E_j-E_l}{k_\text{B}T}. \label{detailed_balance1}
\end{equation}

For the case of a time-dependent bias, $\dot{\varepsilon} \neq 0$, we generalize Eq.~(\ref{detailed_balance1}) to an instantaneous detailed-balance condition:
\begin{equation}
	\gamma_{lj} = \gamma_{jl}\exp \left[ \dfrac{E_j(\varepsilon(t))-E_l(\varepsilon(t))}{k_\text{B}T}\right],
\end{equation}
analogously to Eq.~(\ref{two_level_detailed_balance}) in Sec.~\ref{Theory:GKSL}.

\subsection{Liouville space}
\label{Appendix:GKSL:Liouville}
It is convenient to work with the GKSL equation in the Liouville space \cite{Gyamfi2020}. The motivation for working in Liouville space is to construct a perturbative expansion and obtain the first-order correction to the density matrix. The ansatz~(\ref{Lindblad_anzatz}) leads to two coupled operator equations for $\hat{A}$ and $\hat{B}$, which in Liouville space reduce to a linear system for the superkets $\lket{A}$ and $\lket{B}$. 

In the Liouville space, the master equation becomes
\begin{equation}
	\lket{\dot \rho} = \mathscr{L}\!\! \lket{\rho},
\end{equation}
where $\mathscr{L} = \mathscr{H}_\text{inst}+\mathscr{D}$ is the Liouvillian superoperator. The Hamiltonian and dissipative superoperators are defined as
\begin{gather}
	\mathscr{H}_\text{inst}=-\frac{i}{\hbar} \llbracket \hat{\mathcal{H}}_\text{inst}, \hat{\mathbb{I}}_d \rrbracket, \label{exact_liouvillian}\\
	\nonumber\mathscr{D}= \sum_{\alpha} \gamma_{\alpha} \left[ \hat{L}_\alpha \otimes ( \hat{L}_\alpha ^\dagger)^\text{T} - \frac{1}{2}  \llbracket  \hat{L}_\alpha ^\dagger  \hat{L}_\alpha , \hat{\mathbb{I}}_d \rrbracket_+ \right].
\end{gather}
Here, the super-commutator and super-anti-commutator are defined as
\begin{gather}
	\llbracket \hat{O}_1,\hat{O}_2 \rrbracket = \hat{O}_1  \otimes \hat{O}_2^\text{T} - \hat{O}_2  \otimes \hat{O}_1^\text{T},\\
	\llbracket \hat{O}_1,\hat{O}_2 \rrbracket_+ = \hat{O}_1  \otimes \hat{O}_2^\text{T} + \hat{O}_2  \otimes \hat{O}_1^\text{T},
\end{gather}
where $\hat{O}_1^\text{T}$ denotes the transpose of the operator $\hat{O}_1$. In what follows, the following identities will be used:
\begin{gather}
	|O_1O_2O_3 \rangle\!\rangle = (\hat{O}_1\otimes \hat{O}_3^\text{T})|O_2 \rangle\!\rangle,\\
	\langle\!\langle O_1|O_2\rangle\!\rangle = \Tr(\hat{O}_1^\dagger \hat{O}_2), \label{trace_identity_AB}
\end{gather} 
and  the combination of the two previous identities gives
\begin{equation}
	\langle\!\langle O_1 | \hat{O}_2\otimes \hat{O}_4^\text{T}|O_3 \rangle\!\rangle = \Tr (\hat{O}_1^\dagger \hat{O}_2 \hat{O}_3 \hat{O}_4). \label{trace_identity_ABCD}
\end{equation}
\subsection{Important properties of the GKSL equation in the Liouville space}
\label{Appendix:GKSL:properties}

In this subsection, we summarize several important properties of the GKSL Liouvillian superoperator $\mathscr{L}$~\cite{Gyamfi2020} that will be used throughout the rest of the Appendices.

The Liouvillian superoperator $\mathscr{L}_0$ is generally non-Hermitian. As a result, the right and left eigenvectors are not related by Hermitian conjugation and instead form a biorthogonal basis. The eigenvalue problem is defined as
\begin{align}
	\mathscr{L}_0 \, |\ell_{\mathrm{R},n}\rangle\!\rangle &= \ell_n \, |\ell_{\mathrm{R},n}\rangle\!\rangle, \,\,\, \langle\!\langle \ell_{\mathrm{L},n}| \, \mathscr{L}_0 &= \ell_n \, \langle\!\langle \ell_{\mathrm{L},n}|.
\end{align}
The right and left eigenvectors satisfy the biorthonormality condition $\langle\!\langle \ell_{\mathrm{L},m} | \ell_{\mathrm{R},n} \rangle\!\rangle = \delta_{mn}$ and form a complete basis in the Liouville space,
\begin{equation}
	\sum_n |\ell_{\mathrm{R},n}\rangle\!\rangle \langle\!\langle \ell_{\mathrm{L},n}| = \mathbb{I}_{d^2}.
\end{equation}
Using this basis, the Liouvillian superoperator admits the spectral decomposition
\begin{equation}
	\mathscr{L}_0 = \sum_n \ell_n \, |\ell_{\mathrm{R},n}\rangle\!\rangle \langle\!\langle \ell_{\mathrm{L},n}|.
\end{equation}

The density matrix of the thermal-equillibrium state $\hat{\rho}^{(0)}$ satisfies
\begin{equation}
	\mathscr{L}_0 \, | \rho^{(0)}\rangle\!\rangle = 0.
	\label{right_kernel}
\end{equation}
This equation expresses the fact that $\hat{\rho}^{(0)}$ is a steady-state solution of the GKSL equation.

Trace preservation of the density matrix implies the existence of a left zero eigenvector of the unperturbed Liouvillian,
\begin{equation}
	\langle\!\langle 1| \mathscr{L}_0 = 0,
	\label{left_kernel}
\end{equation}
where $\langle\!\langle 1|$ corresponds to the identity operator in the Liouville space, $\langle\!\langle 1|\rho\rangle\!\rangle = \Tr \hat{\rho}$.

In what follows, we assume that the thermal equilibrium state is the unique stationary state of the unperturbed Liouvillian~\cite{Albert2014}, so that the dimensionality of its kernel is unity
\begin{equation}
	\dim \ker \mathscr{L}_0 = 1,
\end{equation}
which guarantees the ergodicity.

The dependence of the relaxation rates and jump operators on the bias parameter $\varepsilon$ is redundant and will not appear in the final results. To see this, we differentiate the stationary-state condition~(\ref{right_kernel}) with respect to the stationary bias $\varepsilon_0$, which yields
\begin{equation}
	\mathscr{L}_{\delta \varepsilon} | \rho^{(0)}\rangle\!\rangle 
	= - \mathscr{L}_0  |\rho^{(0)}_{\delta \varepsilon}\rangle\!\rangle.
	\label{de_rho_identity}
\end{equation}
where 
\begin{equation}
	| \rho^{(0)}_{\delta \varepsilon} \rangle\!\rangle 
	= \frac{\partial | \rho^{(0)} \rangle\!\rangle}{\partial \varepsilon_0},
\end{equation}
and $\mathscr{L}_{\delta \varepsilon} = \partial \mathscr{L}_0/\partial \varepsilon_0$. This quantity relates to the first-order correction term of $\mathscr{L}$ and will be subject of the next subsection.

\subsection{Time derivative of an operator}
\label{Appendix:GKSL:td}
For an arbitrary operator $\hat{A}_\text{inst}$ in the instantaneous energy representation, its time derivative can be obtained by differentiating the expectation value
$\langle \hat{A}_\text{inst}\rangle = \Tr(\hat{\rho}\, \hat{A}_\text{inst})$
and using the GKSL equation for $\hat{\dot{\rho}}$. The resulting time-derivative operator reads
\begin{gather}
	\nonumber 
	\hat{\dot A}_\text{inst} = \frac{\partial \hat{A}_\text{inst}}{\partial t} 
	+ \frac{i}{\hbar}[\hat{\mathcal{H}}_\text{inst}, \hat{A}_\text{inst}] \\
	+ \sum_{\alpha} \gamma_{\alpha} \left[ \hat{L}_\alpha^\dagger \hat{A}_\text{inst} \hat{L}_\alpha 
	- \frac{1}{2}\left\{\hat{L}_\alpha^\dagger \hat{L}_\alpha, \hat{A}_\text{inst}\right\} \right].
	\label{dAinstdt}
\end{gather}

Alternatively, in the Liouville space we have 
\begin{equation}
	d \langle \hat{A}_\text{inst}\rangle/dt = d \langle\!\langle A_\text{inst}| \rho \rangle\!\rangle/dt=\langle\!\langle \dot{A}_\text{inst}|\rho \rangle\!\rangle,
\end{equation}
with the definition of the super-bra time derivative
\begin{equation}
	\langle\!\langle \dot{A}_\text{inst} | = \lbra{\partial_t A_\text{inst}} + \lbra{A_\text{inst}} \mathscr{L}.
	\label{dAinstdtLiov}
\end{equation}

\subsection{Perturbation theory}
\label{Appendix:GKSL:pert}
Now we construct a perturbative solution for the density matrix. In principle, the evolution superoperator can be constructed following the procedure outlined in Appendix~\ref{Appendix:IdQd:pert}. Its detailed investigation will be reported elsewhere \cite{my_upcoming_work_relaxation}. However, the same results for the effective capacitance and resistance can be derived using an alternative approach, which we describe below.

We consider a small rf probing of the energy bias, $\varepsilon(t)=\varepsilon_0+\delta \varepsilon_\text{rf}(t) $ [see Eqs.~(\ref{QuditA_applicability_for_perturbation_theory}) and (\ref{pert_gamma})]. This allows us to expand the Liouvillian superoperator $\mathscr{L}$ up to the first order in $\delta \varepsilon_\text{rf}$, yielding Eq.~(\ref{Liov_upto1}). The corresponding superoperators are defined as
\begin{gather}
	\mathscr{L}_{0}=\mathscr{H}_{0}+\mathscr{D}_0, \,\,\,\,\,\, \mathscr{H}_0 = -\frac{i}{\hbar} \llbracket \hat{\mathcal{H}}_{\text{inst}}^{(0)}, \hat{\mathbb{I}}_d \rrbracket, \label{def1}\\
	\mathscr{L}_{\delta \varepsilon} = \mathscr{V}_{\delta  \varepsilon}+\mathscr{D}_{\delta \varepsilon}, \,\,\,\,\,\, \mathscr{L}_{\delta \dot \varepsilon} = \mathscr{V}_{\delta  \dot \varepsilon},\label{def2}\\ 
	\mathscr{V}_{\delta  \varepsilon}=-\dfrac{i}{\hbar} \llbracket \hat{V}_{\delta  \varepsilon}, \hat{\mathbb{I}}_d \rrbracket , \,\,\,\,\,\,
	\mathscr{V}_{\delta \dot \varepsilon}=-\dfrac{i}{\hbar} \llbracket \hat{V}_{\delta \dot \varepsilon}, \hat{\mathbb{I}}_d \rrbracket , \label{def3}\\
	\mathscr{D}_0 =  \mathscr{D}|_{\varepsilon = \varepsilon_0}, \,\,\,\,\,\, \mathscr{D}_{\delta \varepsilon} = \dfrac{\partial \mathscr{D}_0}{\partial \varepsilon_0}. \label{def4}
\end{gather}
We note that in the dissipation superoperator, both the relaxation rates $\gamma_\alpha$ and the jump operators $\hat{L}_\alpha$ may depend on the energy bias $\varepsilon$. The definitions of $\hat{\mathcal{H}}_{\text{inst}}^{(0)}$, $\hat{V}_{\delta \varepsilon}$, and the Berry connection $\hat{V}_{\delta \dot \varepsilon}$ are the same as in Eq.~(\ref{perturbative_operators}). Since the superoperator $\mathscr{L}_{\delta \dot \varepsilon}$ is derived from the Berry connection, it can be interpreted as a Berry superoperator.

As stated in Sec.~\ref{Sec:Qudit:GKSL}, the solution for the density matrix can be sought for in the form~(\ref{new_anzatz}). Substituting this ansatz into the first-order GKSL equation
\begin{equation}
	\lket{\dot \rho} = (\mathscr{L}_0+\mathscr{L}_{\delta \varepsilon} \delta \varepsilon_\text{rf}+\mathscr{L}_{\delta \dot \varepsilon} \delta \dot \varepsilon_\text{rf})\!\! \lket{\rho} 
\end{equation}
and retaining only the first-order terms, while using the fact that \(\delta \ddot \varepsilon_\text{rf}(t) = -\omega_\text{rf}^2 \, \delta \varepsilon_\text{rf}(t)\), one obtains a system of two linear equations for the superkets \(\lket{A}\) and \(\lket{B}\). The zeroth-order term vanishes exactly, since \text{\(\mathscr{L}_0 \lket{\rho^{(0)}} = 0\)}:
\begin{gather}
	\begin{cases}
		-\omega_\text{rf}^2 \lket{B} = \mathscr{L}_{0} \, \lket{A} + \mathscr{L}_{\delta \varepsilon} \, \lket{\rho^{(0)}},\\[2mm]
		\lket{A} = \mathscr{L}_{0} \, \lket{B} + \mathscr{L}_{\delta \dot\varepsilon} \, \lket{\rho^{(0)}},
	\end{cases}
	\label{system_AB}
\end{gather}
whose solution is
\begin{gather}
	\begin{cases}
		\lket{A} = {\left[\mathbb{I}_{d^2}+\left(\dfrac{\mathscr{L}_0}{\omega_\text{rf}}\right)^2 \right]}^{-1} \left[-\omega_\text{rf}^{-2}\mathscr{L}_0\mathscr{L}_{\delta \varepsilon}+\mathscr{L}_{\delta \dot \varepsilon} \right] \lvert  \rho^{(0)} \rangle\!\rangle,\\
		\lket{B}=-\omega_\text{rf}^{-2}{\left[\mathbb{I}_{d^2}+\left(\dfrac{\mathscr{L}_0}{\omega_\text{rf}}\right)^2 \right]}^{-1} \left[\mathscr{L}_0\mathscr{L}_{\delta \dot \varepsilon}+\mathscr{L}_{\delta \varepsilon} \right] \lvert  \rho^{(0)} \rangle\!\rangle.
	\end{cases} \label{sol_AB}
\end{gather}
The above result can be further simplified by noting that $\hat{V}_{\delta \varepsilon}$ is diagonal [see Eq.~(\ref{perturbative_operators})]. Therefore,
\begin{equation}
	\mathscr{V}_{\delta  \varepsilon} \lvert  \rho^{(0)} \rangle\!\rangle=-\dfrac{i}{\hbar}|[\hat{V}_{\delta \varepsilon}, \hat{\rho}^{(0)}]\rangle \! \rangle=0. \label{vde_identity}
\end{equation}
since $\hat{\rho}^{(0)}$ is also diagonal in the instantaneous energy basis. Consequently, in all expressions above, $\mathscr{L}_{\delta \varepsilon}$ can effectively be replaced by $\mathscr{D}_{\delta \varepsilon}$ alone, as the diagonal Hamiltonian contribution $\mathscr{V}_{\delta \varepsilon}$ vanishes.

\subsection{First-order corrections to the quantum current}
\label{Appendix:GKSL:qc}

As shown in Eq.~(\ref{quantum_current}), the quantum current $ \langle \hat{\mathcal{I}}\rangle$ contributes to both the effective reactance $C_\text{eff}$ and the effective resistance $R_\text{eff}$ for open quantum systems. Here, the calculations are performed in the Liouville space. The expression for the quantum current operator, Eq.~(\ref{Qudit_isolated_dhdeps_inst}), must be modified accordingly. The quantum current operator should therefore be defined using Eq.~(\ref{dAinstdtLiov}) as
\begin{equation}
	\langle \! \langle  (\partial_\varepsilon \dot{\mathcal{H}})_\text{inst}|  = \dot \varepsilon \langle \! \langle\partial_\varepsilon(\partial_\varepsilon \mathcal{H})_\text{inst}|+\langle \! \langle(\partial_\varepsilon \mathcal{H})_\text{inst}|\mathscr{L}.
\end{equation}

Analogously to Eq.~(\ref{quantum_current_intro}), we introduce the zeroth- and first-order quantum current superbras $\langle\!\langle I^{(0,1)}|$ such that
\begin{equation}
	\langle \! \langle  (\partial_\varepsilon \dot{\mathcal{H}})_\text{inst}| = \langle \! \langle I^{(0)}| + \langle \! \langle I^{(1)}|.
\end{equation}
Here, the zeroth-order term is
\begin{equation}
	\langle \! \langle I^{(0)}| = \langle \! \langle(\partial_\varepsilon \mathcal{H})_\text{inst}^{(0)}|\mathscr{L}_0,
\end{equation}
and the first-order term reads
\begin{gather}
	\nonumber\langle \! \langle I^{(1)}| = \delta \varepsilon_\text{rf} \left[\langle \! \langle\partial_\varepsilon(\partial_\varepsilon \mathcal{H})_\text{inst}^{(0)}|\mathscr{L}_0+\langle \! \langle(\partial_\varepsilon \mathcal{H})_\text{inst}^{(0)}|\mathscr{L}_{\delta \varepsilon} \right]+\\
	+\delta \dot \varepsilon_\text{rf} \left[\langle \! \langle\partial_\varepsilon(\partial_\varepsilon \mathcal{H})_\text{inst}^{(0)}|+\langle \! \langle(\partial_\varepsilon \mathcal{H})_\text{inst}^{(0)}|\mathscr{L}_{\delta \dot \varepsilon} \right].
\end{gather}

Thus, the expectation value of the quantum current reads
\begin{equation}
	\langle \hat{I} \rangle = \langle \! \langle I^{(0)}+I^{(1)}|\rho^{(0)}+\delta \varepsilon_\text{rf}A+\delta \dot \varepsilon_\text{rf} B \rangle\!\rangle.
\end{equation}
Using Eq.~(\ref{right_kernel}), this simplifies to
\begin{gather}
	\nonumber\langle \hat{I} \rangle = \delta \varepsilon_\text{rf} \left[\langle \! \langle(\partial_\varepsilon \mathcal{H})_\text{inst}^{(0)}|\mathscr{L}_{\delta \varepsilon}| \rho^{(0)}\rangle\!\rangle+ \langle \! \langle(\partial_\varepsilon \mathcal{H})_\text{inst}^{(0)}|\mathscr{L}_0|A \rangle\!\rangle \right]+\\
	\nonumber+\delta \dot\varepsilon_\text{rf} \Big[\langle \! \langle\partial_\varepsilon(\partial_\varepsilon \mathcal{H})_\text{inst}^{(0)}|\rho^{(0)}\rangle\!\rangle+\langle \! \langle(\partial_\varepsilon \mathcal{H})_\text{inst}^{(0)}|\mathscr{L}_{\delta \dot \varepsilon}|\rho^{(0)}\rangle\!\rangle+\\
	+\langle \! \langle(\partial_\varepsilon \mathcal{H})_\text{inst}^{(0)}|\mathscr{L}_0|B\rangle\!\rangle \Big].\label{quantum_current_ans_GKSL}
\end{gather}

\subsection{Effective capacitance and resistance}
\label{Appendix:GKSL:Ceff_Reff}
From Eqs.~(\ref{quantum_current_exp}), and (\ref{quantum_current_ans_GKSL}), for an \emph{arbitrary dissipative qu$d$it} we obtain the final expressions for the effective capacitance 
\begin{equation}
	\begin{aligned}
		&C_\text{eff} = -\left(\frac{\partial \varepsilon}{\partial V}\right)^2 
		\Big[  \langle\!\langle \partial_\varepsilon (\partial_\varepsilon \mathcal{H})_\text{inst}^{(0)}
		\,|\, \rho^{(0)} \rangle\!\rangle + \\
		& + \langle\!\langle (\partial_\varepsilon \mathcal{H})_\text{inst}^{(0)}
		\,|\, \mathscr{L}_{\delta \dot\varepsilon} | \rho^{(0)} \rangle\!\rangle + \langle\!\langle (\partial_\varepsilon \mathcal{H})_\text{inst}^{(0)}
		\,|\, \mathscr{L}_0 | B \rangle\!\rangle 
		\Big]
	\end{aligned} 
\end{equation}
and the effective conductance
\begin{equation}
	\begin{aligned}
		&R_\text{eff}^{-1} = -\left(\frac{\partial \varepsilon}{\partial V}\right)^2 
		\Big[\langle\!\langle (\partial_\varepsilon \mathcal{H})_\text{inst}^{(0)} 
		\,|\, \mathscr{L}_{\delta \varepsilon} | \rho^{(0)} \rangle\!\rangle +
		\\
		&  			+ \langle\!\langle (\partial_\varepsilon \mathcal{H})_\text{inst}^{(0)}
		\,|\, \mathscr{L}_0 | A \rangle\!\rangle \Big].
	\end{aligned} 
\end{equation}

These expressions can be further simplified. We observe that the superkets $|A\rangle\!\rangle$ and $|B\rangle\!\rangle$ appear only in combination with $\mathscr{L}_0$ in the final expressions. Using Eq.~(\ref{de_rho_identity}), we obtain
\begin{gather}
	\begin{cases}
		\mathscr{L}_0\lket{A} =\omega_\text{rf}^{-2} \mathscr{L}_0^3{\left[\mathbb{I}_{d^2}+\left(\dfrac{\mathscr{L}_0}{\omega_\text{rf}}\right)^2 \right]}^{-1}   |\rho^{(0)}_{\delta \varepsilon} \rangle\!\rangle+\\+
		\mathscr{L}_0{\left[\mathbb{I}_{d^2}+\left(\dfrac{\mathscr{L}_0}{\omega_\text{rf}}\right)^2 \right]}^{-1}  \mathscr{L}_{\delta \dot \varepsilon}|  \rho^{(0)} \rangle\!\rangle  ,\\
		\mathscr{L}_0\lket{B}=-\omega_\text{rf}^{-2} \mathscr{L}_0^2 {\left[\mathbb{I}_{d^2}+\left(\dfrac{\mathscr{L}_0}{\omega_\text{rf}}\right)^2 \right]}^{-1} \mathscr{L}_{\delta \dot \varepsilon}|  \rho^{(0)} \rangle\!\rangle +\\
		+\omega_\text{rf}^{-2}\mathscr{L}_0^2{\left[\mathbb{I}_{d^2}+\left(\dfrac{\mathscr{L}_0}{\omega_\text{rf}}\right)^2 \right]}^{-1}   |  \rho^{(0)}_{\delta \varepsilon} \rangle\!\rangle.
	\end{cases} 
\end{gather}
Using the identity
\begin{gather}
	\nonumber \mathscr{L}_0^n {\left[\mathbb{I}_{d^2}+\left(\dfrac{\mathscr{L}_0}{\omega_\text{rf}}\right)^2 \right]}^{-1} = \omega_\text{rf}^2 \mathscr{L}_0^{n-2} -\\
	- \omega_\text{rf}^2 \mathscr{L}_0^{n-2}{\left[\mathbb{I}_{d^2}+\left(\dfrac{\mathscr{L}_0}{\omega_\text{rf}}\right)^2 \right]}^{-1}, \label{Ln_identity}
\end{gather}
we can simplify the effective capacitance as
\begin{gather}
	C_\text{eff} = -(\alpha e)^2
	\Big[  \langle\!\langle \partial_\varepsilon (\partial_\varepsilon \mathcal{H})_\text{inst}^{(0)}
	\,|\, \rho^{(0)} \rangle\!\rangle +\label{Ceff_qd_ap} \\
	\nonumber + \langle\!\langle (\partial_\varepsilon \mathcal{H})_\text{inst}^{(0)}
	| \rho^{(0)}_{\delta \varepsilon} \rangle\!\rangle + \\
	\nonumber+
	\langle\!\langle (\partial_\varepsilon \mathcal{H})_\text{inst}^{(0)}
	| \left[\mathbb{I}_{d^2}+(\mathscr{L}_0/\omega_\text{rf})^2 \right]^{-1}  [\mathscr{V}_{\delta \dot \varepsilon}| \rho^{(0)} \rangle\!\rangle-| \rho^{(0)}_{\delta \varepsilon} \rangle\!\rangle]
	\Big].
\end{gather}
Similarly, the effective conductance reads
\begin{gather}
	R_\text{eff}^{-1} = -(\alpha e)^2 \cdot \label{Reff_qd_ap}\\
	\nonumber\cdot \langle\!\langle (\partial_\varepsilon \mathcal{H})_\text{inst}^{(0)}  | \mathscr{L}_0 \left[\mathbb{I}_{d^2}+(\mathscr{L}_0/\omega_\text{rf})^2 \right]^{-1}  [\mathscr{V}_{\delta \dot \varepsilon}| \rho^{(0)} \rangle\!\rangle-| \rho^{(0)}_{\delta \varepsilon} \rangle\!\rangle].
\end{gather}

In the next subsection we briefly discuss how to correctly split the effective capacitance into corresponding components.

\subsection{Geometric, quantum and tunneling capacitances}
\label{Appendix:GKSL:Ceff}
The first two terms in Eq.~(\ref{Ceff_qd_ap}) can be readily computed. 
The first term is obtained by differentiating Eq.~(\ref{He_inst}) with respect to \(\varepsilon\), using Eqs.~(\ref{dketEdeps}--\ref{d2Edeps2}). Applying identity~(\ref{trace_identity_AB}) and noting that \(\hat{\rho}^{(0)}\) is diagonal, only the diagonal elements \(\partial_\varepsilon (\partial_\varepsilon \hat{\mathcal{H}})_{\text{inst},nn}\), which are just the second derivatives of the energies, contribute to the final result
\begin{equation}
	\langle\!\langle \partial_\varepsilon (\partial_\varepsilon \mathcal{H})_\text{inst}^{(0)}
	\,|\, \rho^{(0)} \rangle\!\rangle 
	=\sum\limits_n  \dfrac{\partial^2 E_n}{\partial \varepsilon^2}p_n^\text{th}. \label{first_term_ceff_evaluated}
\end{equation}
Similarly, the second term in Eq.~(\ref{Ceff_qd_ap}) evaluates to
\begin{equation}
	\langle\!\langle (\partial_\varepsilon \mathcal{H})_\text{inst}^{(0)}
	| \rho^{(0)}_{\delta \varepsilon} \rangle\!\rangle = \sum\limits_n  \dfrac{\partial E_n}{\partial \varepsilon} \dfrac{\partial p_n^\text{th}}{\partial \varepsilon}. \label{second_term_ceff_evaluated}
\end{equation}

From Eq.~(\ref{first_term_ceff_evaluated}) we see that the first term of Eq.~(\ref{Ceff_qd_ap}) results in both geometric and quantum capacitances
\begin{equation}
	C_\text{geom}+C_\text{Q} = -\overline{ \dfrac{\partial^2 E}{\partial V^2 }}_\text{th} = -\dfrac{\partial^2 E_\text{geom}}{\partial V^2}-\overline{ \dfrac{\partial^2 E'}{\partial V^2 }}_\text{th}.
\end{equation}
Thus, the second and third terms of Eq.~(\ref{Ceff_qd_ap}) contribute to the tunneling capacitance
\begin{gather}
	C_\text{T} = -(\alpha e)^2
	\Big[   \langle\!\langle (\partial_\varepsilon \mathcal{H})_\text{inst}^{(0)}
	| \rho^{(0)}_{\delta \varepsilon} \rangle\!\rangle + \\
	\nonumber+
	\langle\!\langle (\partial_\varepsilon \mathcal{H})_\text{inst}^{(0)}
	| \left[\mathbb{I}_{d^2}+(\mathscr{L}_0/\omega_\text{rf})^2 \right]^{-1}  [\mathscr{V}_{\delta \dot \varepsilon}| \rho^{(0)} \rangle\!\rangle-| \rho^{(0)}_{\delta \varepsilon} \rangle\!\rangle]
	\Big].
\end{gather}

\section{Effective capacitance and resistance of good and bad qu\textit{d}its}
\label{Appendix:GKSL_GOOD_BAD}
This Appendix is devoted to the derivation of the effective capacitance and resistance in the limiting cases of the good and bad qu$d$its [Eqs.~(\ref{C_qd_BAD}) and (\ref{C_qd_GOOD})], starting from Eqs.~(\ref{QD_GKSL_Ceff}) and (\ref{QD_GKSL_Reff}).

First, let us estimate different terms appearing in the expressions for the effective capacitance and resistance. In particular, the expressions~(\ref{Ceff_qd_ap}) and (\ref{Reff_qd_ap}) involve the following relevant quantities: \(\mathscr{L}_0 = \mathscr{H}_0 + \mathscr{D}_0\), \(\mathscr{L}_{\delta \dot \varepsilon}\), \(\langle\!\langle (\partial_\varepsilon \mathcal{H})_\text{inst}^{(0)}|\), \(\langle\!\langle \partial_\varepsilon (\partial_\varepsilon \mathcal{H})_\text{inst}^{(0)}|\), \(| \rho^{(0)} \rangle\!\rangle\), and $| \rho^{(0)}_{\delta \varepsilon} \rangle\!\rangle$.

For any qu$d$it, the Hamiltonian and dissipation superoperators scale as 
\begin{equation}
	\mathscr{H}_0\sim \omega_\text{qd}, \,\,\,\,\,\, \mathscr{D}_0 \sim \gamma_\text{qd},
\end{equation}
where, analogous to the definition of \(\omega_\text{qd}\) in Eq.~(\ref{QuditA_applicability_for_perturbation_theory}), \(\gamma_\text{qd}\) is the minimal relaxation rate of the system:
\begin{equation}
	\gamma_\text{qd}\coloneqq \min_{\alpha,\varepsilon} \gamma_\alpha(\varepsilon). \label{Qudit_gamma}
\end{equation}
Since \(\Tr \hat{\rho}^{(0)} = 1\), we have \(|\rho^{(0)} \rangle\!\rangle \sim 1\). The superket \(|\rho^{(0)}_{\delta \varepsilon} \rangle\!\rangle\) can be estimated as \(|\rho^{(0)}_{\delta \varepsilon} \rangle\!\rangle \sim 1/k_\text{B} T\). There is no need to estimate \(\mathscr{V}_{\delta \varepsilon}\), since it always appears in the trivial combination \(\mathscr{V}_{\delta \varepsilon} |\rho^{(0)} \rangle\!\rangle = 0\). From Eqs.~(\ref{def3}) and (\ref{perturbative_operators}), the remaining part of the Liouvillian superoperator scales as \(\mathscr{L}_{\delta \dot \varepsilon} = \mathscr{V}_{\delta \dot \varepsilon} \sim 1 / \omega_\text{qd}\), because \(\partial \hat{\mathcal{H}} / \partial \varepsilon \sim \partial E_n / \partial \varepsilon \sim 1\). Similarly, we have \(\langle\!\langle (\partial_\varepsilon \mathcal{H})_\text{inst}^{(0)}| \sim 1\) and \(\langle\!\langle \partial_\varepsilon (\partial_\varepsilon \mathcal{H})_\text{inst}^{(0)}| \sim (\hbar \omega_\text{qd})^{-1}\). 

Altogether, we obtain
\begin{gather}
	\nonumber \mathscr{H}_0\sim \omega_\text{qd}, \,\,\,\,\,\,  \mathscr{D}_0 \sim \gamma_\text{qd}, \\
	\nonumber\mathscr{L}_{\delta \dot \varepsilon}\sim  \omega_\text{qd}^{-1}, \,\,\,\,\,\,
	\partial \hat{\mathcal{H}}/\partial \varepsilon \sim 1, \,\,\,\,\,\,  \langle\!\langle \partial_\varepsilon (\partial_\varepsilon \mathcal{H})_\text{inst}^{(0)}\,| \sim (\hbar \omega_\text{qd})^{-1}, \\
	|\, \rho^{(0)} \rangle\!\rangle \sim 1, \,\,\,\,\,\,  | \rho^{(0)}_{\delta \varepsilon} \rangle\!\rangle \sim 1/k_\text{B} T. \label{estimations}
\end{gather}

\subsection{Good qu\textit{d}it}

For a good qu$d$it, we consider the regime of very small relaxation rates, $ \gamma_\text{qd} \ll \omega_\text{rf} \ll \omega_\text{qd} $. To compute the inverse of the superoperator in Eqs.~(\ref{Ceff_qd_ap}) and (\ref{Reff_qd_ap}), we note that $\mathscr{H}_0$ can be decomposed in the basis \(\hat{P}_k \otimes \hat{P}_l\), where $\hat{P}_k = |k \rangle \langle k|$. Indeed, we can write the unperturbed instantaneous Hamiltonian as
\begin{equation}
	\hat{\mathcal{H}}_{\text{inst}}^{(0)}=\sum\limits_k E_k^{(0)} \hat{P}_k,
\end{equation}
and the identity operator as \(1 = \sum_k \hat{P}_k\). Substituting these into Eq.~(\ref{def1}) gives
\begin{gather}
	\mathscr{H}_0 =-i \sum\limits_{k,l}\omega_{kl}^{(0)} \hat{P}_k \otimes \hat{P}_l.
\end{gather}
Consequently, any function of \(\mathscr{H}_0\) can be evaluated as
\begin{equation}
	f(\mathscr{H}_0) = \sum_{k,l} f\big(-i \omega_{kl}^{(0)}\big) \, \hat{P}_k \otimes \hat{P}_l. \label{fh0}
\end{equation}
This decomposition will be needed when inverting \(\left[\mathbb{I}_{d^2} + (\mathscr{H}_0 / \omega_\text{rf})^2\right]\) in the expressions for the effective capacitance and resistance.

\subsubsection{The effective capacitance}

Using Eqs.~(\ref{def3}) and (\ref{fh0}), we can evaluate the last term in Eq.~(\ref{Ceff_qd_ap}):

\begin{gather}
	\nonumber\langle\!\langle (\partial_\varepsilon \mathcal{H})_\text{inst}^{(0)}
	| \left[\mathbb{I}_{d^2}+(\mathscr{H}_0/\omega_\text{rf})^2 \right]^{-1}  [\mathscr{V}_{\delta \dot \varepsilon}| \rho^{(0)} \rangle\!\rangle-| \rho^{(0)}_{\delta \varepsilon} \rangle\!\rangle]=\\
	= \Tr \sum_{k,l} \dfrac{(\partial_\varepsilon \hat{\mathcal{H}})_{\text{inst}}^{(0)}\hat{P}_k[-i/\hbar [\hat{V}_{\delta \dot \varepsilon},\hat{\rho}^{(0)}]-\hat{\rho}^{(0)}_{\delta \varepsilon}]\hat{P}_l}{1-\left(\omega_{kl}^{(0)}/\omega_\text{rf}\right)^2}=\label{third_term_simplified_ceff_good}\\
	\nonumber=-\sum\limits_k \dfrac{\partial E_n}{\partial \varepsilon} \dfrac{\partial p_n^\text{th}}{\partial \varepsilon}+\dfrac{2}{\hbar}\sum\limits_{k \neq l}\dfrac{|(\partial_\varepsilon \mathcal{H}_{kl}^{(0)})|^2}{\omega_{kl}^{(0)}\left[1-\left(\omega_{kl}^{(0)}/\omega_\text{rf}\right)^2\right]}p_l^{T}. 
\end{gather}
Here, the first term in Eq.~(\ref{third_term_simplified_ceff_good}) cancels the second term in Eq.~(\ref{Ceff_qd_ap}). The second term is proportional to \(\omega_\text{rf}^2 / \omega_\text{qd}^2\), which is much smaller than the first term in Eq.~(\ref{Ceff_qd_ap}) that scales as \(\omega_\text{qd}^{-1}\). 

Hence, for a good qu$d$it, the effective capacitance is dominated solely by the geometric and quantum components: $C_\text{eff}^{(\text{good})} = C_\text{geom}+C_\text{Q}$:
\begin{equation}
	C_\text{eff}^{(\text{good})} = -\left(\frac{\partial \varepsilon}{\partial V}\right)^2 
	\sum\limits_n  \dfrac{\partial^2 E_n}{\partial \varepsilon^2}p_n^\text{th} = -\overline{ \dfrac{\partial^2 E}{\partial V^2 }}_\text{th}. \label{C_eff_good_ap}
\end{equation}

\subsubsection{The effective resistance}

Since in Fig.~\ref{Fig_other}(a) both the effective capacitance and resistance are coupled in parallel, the total impedance of the quantum system can be written as in Eq.~(\ref{total_impedance}):
\begin{equation}
	Z_\text{eff} = [R_\text{eff}^{-1}+i \omega_\text{rf} C_\text{eff}]^{-1}. \label{imp_app}
\end{equation}
From Eq.~(\ref{C_eff_good_ap}), we have \(\omega_\text{rf} C_\text{eff}^{(\text{good})} \propto \omega_\text{rf}/\omega_\text{qd}\). Meanwhile, using Eq.~(\ref{Reff_qd_ap}) and a derivation analogous to Eq.~(\ref{third_term_simplified_ceff_good}), the effective conductance is zero in this order of approximation. 

The next-order correction can be estimated as
\begin{equation}
	R_\text{eff}^{-1} \propto \gamma_\text{qd}\left(\dfrac{\omega_\text{rf}}{\omega_\text{qd}}\right)^2 \max\left[\dfrac{1}{ \omega_\text{qd}},\dfrac{\hbar}{k_\text{B}T}\right]\ll \dfrac{\omega_\text{rf}}{\omega_\text{qd}}.
\end{equation}

Therefore, in the limiting case of a good qu$d$it, the response of the quantum system is dominated by the effective capacitance, and in the first order in $\omega_\text{rf}/\omega_\text{qd}$ we can write
\begin{equation}
	G_\text{eff}^{(\text{good})}=0.
\end{equation}

\subsection{Bad qu\textit{d}it}
For a bad qu$d$it, we consider the regime \text{$\omega_\text{rf} \ll  \omega_\text{qd} \ll \gamma_\text{qd}$}.  
To compute the effective capacitance and resistance, we need the eigen-decomposition of $\mathscr{D}_0$. As discussed in Appendix~\ref{Appendix:GKSL:properties}, there is only one zero eigenvalue $d_0 = 0$ [see Eqs.~(\ref{right_kernel}) and (\ref{left_kernel})].  
Thus, for any function of $\mathscr{D}_0$ we can write
\begin{gather}
	\nonumber f(\mathscr{D}_0) = \sum\limits_n f(d_n) |\mathscr{D}_{\text{R},n} \rangle\!\rangle \langle\!\langle \mathscr{D}_{\text{L},n}|=\\
	=f(0) |\rho^{(0)} \rangle\!\rangle \langle\!\langle 1|+\sum\limits_{n \neq 0} f(d_n) |\mathscr{D}_{\text{R},n} \rangle\!\rangle \langle\!\langle \mathscr{D}_{\text{L},n}|. \label{D0_decomposition}
\end{gather}

Applying Eq.~(\ref{Ln_identity}) to $\mathscr{D}_0$ and using the decomposition~(\ref{D0_decomposition}), we obtain
\begin{equation}
	{\left[\mathbb{I}_{d^2}+\left(\dfrac{\mathscr{D}_0}{\omega_\text{rf}}\right)^2 \right]}^{-1}\!\!\!\!\!\! =  |\rho^{(0)} \rangle\!\rangle \langle\!\langle 1|+O\left(\dfrac{\omega_\text{rf}^2}{\gamma_\text{qd}^2}\right) \label{d_for_ceff_bad}
\end{equation}
and 
\begin{equation}
	\mathscr{D}_0{\left[\mathbb{I}_{d^2}+\left(\dfrac{\mathscr{D}_0}{\omega_\text{rf}}\right)^2 \right]}^{-1} = O\left(\dfrac{\omega_\text{rf}^2}{\gamma_\text{qd}}\right), \label{d_for_reff_bad}
\end{equation}
since only the nonzero eigenvalues $d_n  \propto \gamma_\text{qd}$ contribute to the corrections.

\subsubsection{The effective capacitance}

Let us evaluate the last term of Eq.~(\ref{Ceff_qd_ap}) for a bad qu$d$it. Using Eq.~(\ref{d_for_ceff_bad}), we obtain
\begin{gather}
	\nonumber\langle\!\langle (\partial_\varepsilon \mathcal{H})_\text{inst}^{(0)}
	| \left[\mathbb{I}_{d^2}+(\mathscr{D}_0/\omega_\text{rf})^2 \right]^{-1}  [\mathscr{V}_{\delta \dot \varepsilon}| \rho^{(0)} \rangle\!\rangle-| \rho^{(0)}_{\delta \varepsilon} \rangle\!\rangle]=\\
	=\langle\!\langle (\partial_\varepsilon \mathcal{H})_\text{inst}^{(0)}
	| \rho^{(0)} \rangle\!\rangle \langle\!\langle 1|  [\mathscr{V}_{\delta \dot \varepsilon}| \rho^{(0)} \rangle\!\rangle-| \rho^{(0)}_{\delta \varepsilon} \rangle\!\rangle]=0, \label{third_term_cbad}
\end{gather}
because
\begin{equation}
	\langle\!\langle 1|\mathscr{V}_{\delta \dot \varepsilon}\lvert  \rho^{(0)}\rangle\!\rangle = -\dfrac{i}{\hbar}\Tr [\hat{V}_{\delta \dot \varepsilon},\rho^{(0)}]=0
\end{equation}
and 
\begin{equation}
	\langle\!\langle 1|  \rho^{(0)}_{\delta \varepsilon} \rangle\!\rangle =  \Tr \hat{\rho}^{(0)}_{\delta \varepsilon} = \partial_\varepsilon 1 = 0.
\end{equation}

Thus, the final expression for the effective capacitance~(\ref{Ceff_qd_ap}) of a bad qu$d$it contains only the first~(\ref{first_term_ceff_evaluated}) and second~(\ref{second_term_ceff_evaluated}) terms, giving
\begin{equation}
	C_\text{eff}^{(\text{bad})} = -\left(\frac{\partial \varepsilon}{\partial V}\right)^2 
	\partial_\varepsilon \sum\limits_n\dfrac{\partial E_n}{\partial \varepsilon}p_n^\text{th} = -\dfrac{\partial }{\partial V }\overline{ \dfrac{\partial E}{\partial V }}_\text{th}. \label{C_eff_bad_ap}
\end{equation}

\subsubsection{The effective resistance}

Using Eq.~(\ref{d_for_reff_bad}), we can estimate the effective conductance as
\begin{equation}
	R_\text{eff}^{-1} \,\,\, \propto  \,\,\, \dfrac{\omega_\text{rf}^2}{\gamma_\text{qd}} \max\left[\dfrac{1}{ \omega_\text{qd}},\dfrac{\hbar}{k_\text{B}T}\right] \,\,\, \ll \,\,\, \dfrac{\omega_\text{rf}}{\omega_\text{qd}}
\end{equation}
since $\gamma_\text{qd}$ is assumed to be much larger than all other relevant parameters. 

Therefore, in the limiting case of a bad qu$d$it, the effective conductance is negligible:
\begin{equation}
	G_\text{eff}^{(\text{bad})}=0.
\end{equation}

\section{From the GKSL equation to the rate equation}
\label{Appendix:GKSL_to_rate}

In this Appendix, we examine how to derive the quantum balance equation for a multi-level system from the GKSL equation and discuss the approximations involved in this procedure.

We start from the GKSL equation~(\ref{Lindblad_eq}) with the instantaneous Hamiltonian~(\ref{H_inst}). We choose the Lindblad jump operators as $\hat{L}_{ij} = |i\rangle \langle j|$. By substituting them, with relaxation rates $\gamma_{ij}$, into Eq.~(\ref{Lindblad_eq}), we obtain equations for every element of the density matrix:
\begin{equation}
	\dot\rho_{lm}=-\dfrac{i}{\hbar}([\hat{\mathcal{H}}_\text{inst},\hat{\rho}])_{lm}+\sum\limits_{j}\gamma_{lj}\delta_{lm}\rho_{jj}-\Gamma_{lm}\rho_{lm}, \label{GKSL_to_rate1}
\end{equation}
where $\Gamma_{lm} = \sum_j(\gamma_{jl}+\gamma_{jm})/2$. 

For the balance equation, only the diagonal entries of the density matrix are considered to be nontrivial, while all off-diagonal density-matrix elements are neglected. Furthermore, the balance equations do not include the unitary evolution term $[\hat{\mathcal{H}}_\text{inst},\hat{\rho}]$. If we choose 
\begin{equation}
	\omega_{km}, \,\,\,\, \dot{\varepsilon} \dfrac{(\partial_\varepsilon \hat{\mathcal{H}})_{mk}}{(\hbar \omega_{mk})} \,\, \ll \,\,  \gamma_{lm} ,
\end{equation}
the first term in the r.h.s. of Eq.~(\ref{GKSL_to_rate1}) can be neglected, so that for the diagonal elements the GKSL equation reduces to the Pauli master equation
\begin{equation}
	\dot\rho_{nn}=\sum\limits_{j}\gamma_{nj}\rho_{jj}-\rho_{nn}\sum\limits_j \gamma_{jn},
\end{equation}
while for the off-diagonal elements with $l \neq m$
\begin{equation}
	\dot\rho_{lm}=-\Gamma_{lm}\rho_{lm}
\end{equation}
the dynamics is purely dissipative. However, in order to neglect the off-diagonal matrix elements, one additionally needs to consider the solution on the timescale $t \gtrsim \Gamma_{lm}^{-1}$, and the contribution of the off-diagonal elements to any observable must be small compared to that of the diagonal elements.

\section{From qu\textit{d}its to qubits}
\label{Appendix:d=2}
\subsection{Super-bras, -kets, and -operators for a two-level system}
In this Appendix, we show how to obtain expressions for the effective capacitance and resistance of a two-level system from Eqs.~(\ref{Ceff_qd_ap}) and (\ref{Reff_qd_ap}) with the known structure of the Hamiltonian, relaxation rates, and jump operators.

For a two-level system, the Hamiltonian is given by Eq.~(\ref{Theory:2levelHam}). Relaxation rates and jump operators are the same as in Sec.~\ref{Theory:GKSL}.

To obtain the effective capacitance and resistance, we need to know the superbras $ \langle\!\langle (\partial_\varepsilon \mathcal{H})_\text{inst}^{(0)}
|$, $\langle\!\langle \partial_\varepsilon (\partial_\varepsilon \mathcal{H})_\text{inst}^{(0)}
|$, the superkets $| \rho^{(0)} \rangle\!\rangle$ , $|\rho^{(0)}_{\delta \varepsilon} \rangle\!\rangle$, and the superoperators $\mathscr{H}_0$, $\mathscr{D}_0$, $\mathscr{V}_{\delta \dot \varepsilon}$.

The superbras are
\begin{gather}
	\nonumber\langle\!\langle (\partial_\varepsilon \mathcal{H})_\text{inst}^{(0)} 
	\,| =\\ \begin{pmatrix}
		-\frac{\varepsilon_0}{2 \Delta E_0}+E_\text{geom}', &
		\frac{\Delta}{2 \Delta E_0}, &
		\frac{\Delta}{2 \Delta E_0}, &
		\frac{\varepsilon_0}{2 \Delta E_0}+E_\text{geom}' 
	\end{pmatrix}\label{deHinst_superket}
\end{gather}

and 

\begin{gather}
	\nonumber\langle\!\langle \partial_\varepsilon (\partial_\varepsilon \mathcal{H})_\text{inst}^{(0)}
	| =\\ \begin{pmatrix}
		-\frac{\Delta^2}{2 \Delta E_0^3}+E_\text{geom}'', &
		-\frac{\Delta \varepsilon_0}{2 \Delta E_0^3}, &
		-\frac{\Delta \varepsilon_0}{2 \Delta E_0^3}, &
		-\frac{\Delta^2}{2 \Delta E_0^3}+E_\text{geom}''
	\end{pmatrix}.
\end{gather}

The superkets are
\begin{gather}
	| \rho^{(0)}\rangle\!\rangle = \begin{pmatrix}
		\dfrac{1+\chi_\text{th}}{2}, &
		0, &
		0, &
		\dfrac{1-\chi_\text{th}}{2}
	\end{pmatrix}^\text{T}
\end{gather}
and
\begin{gather}
	| \rho^{(0)}_{\delta \varepsilon}\rangle\!\rangle = \begin{pmatrix}
		\dfrac{\varepsilon_0 (1-\chi_\text{th}^2)}{4 k_\text{B}T \Delta E_0}, &
		0, &
		0, &
		-\dfrac{\varepsilon_0 (1-\chi_\text{th}^2)}{4 k_\text{B}T \Delta E_0}
	\end{pmatrix}^\text{T},
\end{gather}
where $\chi_\text{th} = \tanh  (\Delta E_0/2k_\text{B}T)$. 

The unperturbed Liouvillian $\mathscr{L}_0$ and Berry superoperator $\mathscr{V}_{\delta \dot \varepsilon}$ are
\begin{equation}
	\mathscr{L}_0 =\begin{pmatrix}
		-\gamma_- & 0 & 0 & \gamma_+ \\
		0 & -T_2^{-1}+i \omega_\text{q0} & 0 & 0 \\
		0 & 0 & -T_2^{-1}-i \omega_\text{q0} & 0 \\
		\gamma_- & 0 & 0 & -\gamma_+
	\end{pmatrix}
\end{equation}
and
\begin{equation}
	\mathscr{V}_{\delta \dot \varepsilon} = \frac{\Delta}{2\Delta E_0^2}
	\begin{pmatrix}
		0 & -1 & -1 & 0 \\
		1 & 0 & 0 & -1 \\
		1 & 0 & 0 & -1 \\
		0 & 1 & 1 & 0
	\end{pmatrix}
\end{equation}
respectively.

To evaluate the effective capacitance and resistance in Eqs.~(\ref{QD_GKSL_Ceff}) and (\ref{QD_GKSL_Reff}), we need to compute the matrix inverse  $[\mathbb{I}_{4}+(\mathscr{L}_0/\omega_\text{rf})^2]^{-1}$:
\begin{gather}
	{\left[\mathbb{I}_{4}+\left(\dfrac{\mathscr{L}_0}{\omega_\text{rf}}\right)^2 \right]}^{-1}
	=\begin{pmatrix}
		1-\tilde{\gamma}_-& 0 & 0 &\tilde{\gamma}_+		\\
		0 & \dfrac{1}{1 +\Omega^2} & 0 & 0\\
		0 & 0 & \dfrac{1}{1 + \bar{\Omega}^2} & 0\\
		\tilde{\gamma}_- & 0 & 0 &1 - \tilde{\gamma}_+
	\end{pmatrix},\label{inverse_superoperator}
\end{gather}
where 
\begin{equation}
	\tilde{\gamma}_{\pm} = \dfrac{\gamma \gamma_\pm}{\gamma^2+\omega_\text{rf}^2},
\end{equation}
and 
\begin{equation}
	\Omega = \dfrac{-T_2^{-1}+i\omega_\text{q0}}{\omega_\text{rf}}, \,\,\, \bar{\Omega} = \dfrac{-T_2^{-1}-i\omega_\text{q0}}{\omega_\text{rf}}.
\end{equation}

\subsection{Effective capacitance}
Let us now calculate the effective capacitance. Since the eigenenergies of the Hamiltonian~(\ref{Theory:2levelHam}) are $E_i = E_\text{geom}\pm \Delta E_0/2$, the sum of the geometric and quantum capacitances from Eq.~(\ref{Cgeom+CQ}) is
\begin{equation}
	C_\text{geom}+C_\text{Q}
	= -\dfrac{\partial^2 E_\text{geom}}{\partial V^2}+ \left(\dfrac{\partial \varepsilon}{\partial V}\right)^2 \dfrac{\Delta^2}{2 \Delta E_0^3}\tanh \left(\dfrac{\Delta E_0}{2 k_\text{B}T}\right), \label{cgeom+cq}
\end{equation}
which with the previous definition of the $C_{\text{Q}0}$ (\ref{CQ0}) perfectly coincides with Eqs.~(\ref{C_geom}) and (\ref{qubit_CQ}).
The tunneling capacitance, from Eq.~(\ref{CT_qd_main}), is the sum of three terms
\begin{gather}
	C_\text{T} = -\left(\dfrac{\partial \varepsilon}{\partial V}\right)^2 \Big[ 
	-\dfrac{\varepsilon_0^2}{4 k_\text{B} T \Delta E_0^2}\cosh^{-2}\left(\dfrac{\Delta E_0}{2 k_\text{B}T}\right)+ \label{ct1}\\
	\nonumber+ \dfrac{\Delta^2}{2 \Delta E_0^3}\dfrac{T_2^2 \omega_\text{rf}^2\left[1-T_2^2(\omega_{\text{q}0}^2-\omega_\text{rf}^2)\right]}{1+T_2^4(\omega_{\text{q}0}^2-\omega_\text{rf}^2)^2+2 T_2^2 (\omega_{\text{q}0}^2+\omega_\text{rf}^2)} \tanh \left(\dfrac{\Delta E_0}{2 k_\text{B}T}\right)
	+\\
	\nonumber+\dfrac{\varepsilon_0^2 }{4 k_\text{B}T \Delta E_0^2}\dfrac{\omega_\text{rf}^2}{\omega_\text{rf}^2+\gamma^2}\cosh^{-2}\left(\dfrac{\Delta E_0}{2 k_\text{B}T}\right)\Big]. 
\end{gather}
The sum of geometric, quantum and tunneling parts~(\ref{cgeom+cq},~\ref{ct1}) coincides with the effective capacitance~(\ref{Appendix_bloch_qubit_Ceff}) obtained previously.

\subsection{Effective resistance}
Now we calculate the effective resistance from Eq.~(\ref{Reff_qd_ap}). The Hermes conductance (\ref{Rhrm_main}) is
\begin{gather}
	R_\text{Hrm}^{-1} =\left(\dfrac{\partial \varepsilon}{\partial V}\right)^2\dfrac{\Delta ^2 T_2\omega_\text{rf}^2}{2 \Delta E_0^3}\tanh \left(\dfrac{\Delta E_0}{2 k_\text{B}T}\right) \cdot \label{Rhrm_app}\\
	\nonumber\cdot \dfrac{1+T_2^2(\omega_\text{q0}^2+\omega_\text{rf}^2)}{1+T_2^4 (\omega_\text{q0}^2-\omega_\text{rf}^2)^2+2T_2^2(\omega_\text{q0}^2+\omega_\text{rf}^2)},
\end{gather}
while the Sisyphus conductance (\ref{Rsis_main}) is
\begin{gather}
	R_\text{Sis}^{-1}= \left(\dfrac{\partial \varepsilon}{\partial V}\right)^2\dfrac{\varepsilon_0^2  \gamma }{4 k_\text{B}T \Delta E_0^2}\dfrac{\omega_\text{rf}^2}{\omega_\text{rf}^2+\gamma^2}\cosh^{-2}\left(\dfrac{\Delta E_0}{2 k_\text{B}T}\right).\label{Rsis_app}
\end{gather}
The sum of Hermes and Sisyphus conductances~(\ref{Rhrm_app},~\ref{Rsis_app}) also agrees with the previously obtained expression~(\ref{Appendix_bloch_qubit_Reff}).
\section{Single-Cooper-pair transistor}
\label{Appendix:SCPT}
We now start from the classical Lagrangian of a single-Cooper-pair transistor \cite{Sillanpaeae2005} by choosing the phase differences across junctions to be $\varphi_1$ and $\varphi_2$, so that the phase on the superconducting island is defined symmetrically as $\varphi = (\varphi_1+\varphi_2)/2$, while the phase difference is constrained with the flux quantization condition \text{$\varphi_1-\varphi_2 = 2\pi \Phi/\Phi_0 = \phi$}, where $\Phi$ is the total time-independent magnetic flux and $\Phi_0=h/(2e)$ is the quantum of magnetic flux~[see Fig.~\ref{Fig_other}(b)]. Then the Lagrangian reads
\begin{gather}
	\left( \frac{2e}{\hbar }\right) ^{2}\mathcal{L}=\frac{C_\text{r}}{2}\dot{\varphi}
	_\text{rf}^{2}-\frac{1}{2L_\text{r}}\left( \varphi_\text{rf}-\varphi _{R
	}\right) ^{2}+\dfrac{C_1+C_2}{2}\dot \varphi^2+ \nonumber\\
	+\dfrac{C_\text{g}}{2}(\dot \varphi-\dot \varphi_\text{g})^2+\dfrac{C_{\text{g}0}}{2}(\dot \varphi-\dot \varphi_\text{g0})^2+ \nonumber \\
	+E_\text{J}(1+\delta)\cos\left(\varphi+\dfrac{\phi}{2}\right)+E_\text{J}(1-\delta)\cos\left(\varphi-\dfrac{\phi}{2}\right). \label{SCPT_Lagrangian}
\end{gather}

Here, the classical part of the Lagrangian $\mathcal{L}_\text{cl}$ consists of the first two terms of Eq.~(\ref{SCPT_Lagrangian}) and the quantum part $\mathcal{L}_\text{q}$ contains all the other terms. 
The Routhian is given by 
\begin{gather}
	\mathcal{R}=4E_{\text{C}}(n-n_{\text{g}})^2-\\
	-2 E_\text{J} \cos \dfrac{\phi}{2} \cos \varphi+2\delta E_\text{J}\sin \dfrac{\phi}{2}\sin \varphi-
	\notag\\
	-\left( \frac{\hbar }{2e}\right) ^{2}\left[ \frac{C_\text{r}+C_{\text{g}}}{2}\dot{
		\varphi}_\text{rf}^{2}+\frac{C_{\text{g}0}}{2}\dot{\varphi}_{\text{g0}}^{2}-
	\frac{1}{2L}\left( \varphi_\text{rf}-\varphi _{R}\right) ^{2}
	\right],  \notag
\end{gather}
where $n \hbar = \partial \mathcal{L}/\partial \dot \varphi $, $E_\text{C}=e^2/2C_\Sigma$ is the total charging energy of the island with the total capacitance $C_\Sigma=C_\text{g}+C_{\text{g}0}+C_1+C_2$ and the dimensionless charge is defined as
\begin{equation}
	n_\text{g}=-\frac{\hbar }{4e^{2}}\left( C_{\text{g}0}\dot{\varphi}_{\text{g0}
	}+C_{\text{g}}\dot{\varphi}_\text{rf}\right). 
\end{equation}
The Hamiltonian after the quantization $\hat{\tilde{\mathcal{H}}}$ becomes
\begin{gather}
	\hat{\tilde{\mathcal{H}}}=4E_{\text{C}}(\hat{n}-n_{\text{g}})^2
	-2 E_\text{J} \cos \dfrac{\phi}{2} \cos \hat{\varphi}+ \notag\\
	+2\delta E_\text{J}\sin \dfrac{\phi}{2}\sin \hat{\varphi}
	-\left( \frac{\hbar }{2e}\right) ^{2}\dfrac{C_{\text{g}}}{2}\dot{
		\varphi}_\text{rf}^{2}.
\end{gather}

Using $e^{\pm i \hat{\varphi}}\ket{n} = \ket{n \pm 1}$, in the two-level approximation we obtain $\cos \hat{\varphi} = \hat{\sigma}_x/2$, $\sin \hat{\varphi} = -\hat{\sigma}_y/2$ and the two-level Hamiltonian
\begin{equation}
	\hat{\tilde{\mathcal{H}}}=E_\text{geom} \hat{\mathbf{1}} -\dfrac{\Delta_x}{2}\hat{\sigma}_x -\dfrac{\Delta_y}{2}\hat{\sigma}_y -\dfrac{\varepsilon(t)}{2}\hat{\sigma}_z, 
\end{equation}
where the tunneling amplitudes along the $x$ and $y$ directions are  
\begin{equation}
	\Delta_x=2 E_\text{J} \cos( \phi/2), \,\,\,\,\,\, \Delta_y=2 \delta E_\text{J} \sin( \phi/2).
\end{equation}
The energy bias is $\varepsilon(t)=4E_\text{C}(1-2n_\text{g})$. The geometric energy is defined as
\begin{equation}
	E_\text{geom}=2E_\text{C}(1-2 n_\text{g}+2 n_\text{g}^2)-\dfrac{C_\text{g} }{2} \left(\dfrac{\hbar \dot \varphi_\text{rf}}{2e}\right)^2.\label{Egeom_SCPT}
\end{equation}
Thus the geometric capacitance is 
\begin{equation}
	C_\text{geom}=- \dfrac{\partial^2 E_\text{geom}}{\partial V_\text{rf}^2}=C_\text{g}\left(1-\dfrac{C_\text{g}}{C_\Sigma} \right). \label{Cgeom_SCPT}
\end{equation} 

With the help of the unitary rotation around $z$-axis 
\begin{equation}
	\hat{S}_0=\exp\left(\dfrac{i \zeta_z\hat{\sigma}_z}{2} \right)
\end{equation}
with 
\begin{equation}
	\tan(\zeta_z) = -\dfrac{\Delta_y}{\Delta_x}
\end{equation}
the Hamiltonian  $\hat{\mathcal{H}}=\hat{S}_0^\dagger  \hat{\tilde{\mathcal{H}}} \hat{S}_0$ can be turned into its standard pseudo-spin form, Eq.~(\ref{Theory:2levelHam}), with minimal energy distance $\Delta = \sqrt{\Delta_x^2+\Delta_y^2}$ given by Eq.~(\ref{delta_SCPT}).
The energy bias $\varepsilon$ is connected to the probing voltage $V_\text{rf}$ via the following relation
\begin{equation}
	\alpha_\text{SCPT}=\dfrac{1}{e}\dfrac{\partial \varepsilon}{\partial V_\text{rf}}=2 \dfrac{C_\text{g}}{C_\Sigma}.
\end{equation}

\section{Double quantum dot}
\label{Appendix:DQD}

We start from the classical Lagrangian of a double quantum dot [see Fig.~(\ref{Fig_other})]
\begin{gather}
	\left( \frac{2e}{\hbar }\right) ^{2}\mathcal{L}=\dfrac{C_\text{r}}{2}\dot{\varphi}
	_\text{rf}^{2}-\frac{1}{2L_\text{r}}\left( \varphi_\text{rf}-\varphi_{R
	}\right) ^{2}+\dfrac{C_\text{m}}{2}\left(\dot \varphi_1-\dot \varphi_2\right)^2+ \notag\\
	\frac{C_{\text{G}1}}{2}\left( \dot{\varphi}_1-\dot{\varphi}_\text{rf}\right) ^{2}+\frac{C_{\text{G}2}}{2}\left( \dot{\varphi}_2-\dot{\varphi}_\text{rf}\right) ^{2}+\dfrac{C_{\text{SD}1}}{2}\dot \varphi_1^2  +\dfrac{C_{\text{SD}2}}{2}\dot \varphi_2^2, \label{Lagrangian_DQD}
\end{gather}
where $C_{\text{SD}i}=C_{\text{S}i}+C_{\text{D}i}$ and the dissipation function $\mathcal{F}$ is the same as in Eq.~(\ref{A:dissipation}). In Eq.~(\ref{Lagrangian_DQD}), the classical part of the Lagrangian $\mathcal{L}_\text{cl}$ consists of the first two terms of the Lagrangian and  $\mathcal{L}_\text{q}$ contains all other terms. 

The Routhian is given by 
\begin{gather}
	\mathcal{R}=E_{\text{C}1}(n_1-n_{\text{g}1})^2+E_{\text{C}2}(n_2-n_{\text{g}2})^2+
	\\
	+2E_{\text{C}12}n_1 n_2-\left( \frac{\hbar }{2e}\right) ^{2}\left[ \frac{C_\text{r}+C_{\text{c}}}{2}\dot{
		\varphi}_\text{rf}^{2}-
	\frac{1}{2L_\text{r}}\left( \varphi_\text{rf}-\varphi_{R}\right) ^{2}
	\right] ,  \notag
\end{gather}
where $2 n_i \hbar = \partial \mathcal{L}/\partial \dot \varphi_i $, and the factor of 2 is due to the fact that only electrons rather than Cooper pairs tunnel onto the dots. We have also introduced charging energies of individual dots $E_{\text{C}i}$ and coupling energy $E_{\text{C}12}$:
\begin{gather}
	E_{\text{C}i}=\xi \dfrac{e^2}{2C_{\Sigma i}}, \,\,\, E_{\text{C}12}= \dfrac{\xi-1}{\xi} \dfrac{e^2}{2 C_\text{m}},
\end{gather}
where $\xi^{-1}=1-C_\text{m}^2/C_{\Sigma1}C_{\Sigma2}$ and the total capacitances of the dots are $C_{\Sigma i}=C_{\text{G}i}+C_{\text{SD}i}+C_{\text{m}}$. The dimensionless charges $n_{\text{g}i}$ are
\begin{gather}
	n_{\text{g}1}=-\dfrac{\hbar \dot \varphi_\text{rf}}{2e^2}\left(C_{\text{G}1}+C_\text{m}\alpha_2 \right), \\
	\notag n_{\text{g}2}=-\dfrac{\hbar \dot \varphi_\text{rf}}{2e^2}\left(C_{\text{G}2}+C_\text{m}\alpha_1 \right),
\end{gather}
where $\alpha_i = C_{\text{G}i}/C_{\Sigma i}$ and the additional capacitance $C_\text{c}$ is 
\begin{gather}
	C_\text{c}=2 \xi C_\text{m}\alpha_1 \alpha_2+\sum\limits_{i=1}^2 C_{\text{G}i}\left[1+(\xi-1)\alpha_i \right].
\end{gather}

The Hamiltonian after the quantization $\hat{\mathcal{H}}$ becomes
\begin{gather}
	\hat{\mathcal{H}}=E_{\text{C}1}(\hat{n}_1-n_{\text{g}1})^2+E_{\text{C}2}(\hat{n}_2-n_{\text{g}2})^2+
	\\
	\notag+2E_{\text{C}12}\hat{n}_1 \hat{n}_2-\hat{\mathbf{1}}\left( \frac{\hbar }{2e}\right) ^{2}\frac{C_{\text{c}}}{2}\dot{\varphi}_\text{rf}^{2} .
\end{gather}
As a next step we consider the two-level approximation formed by the nearest charge states $n_1=0$ and $n_2=1$ associated with $\ket{0}\coloneqq\ket{01}$ and $n_1=1$ and $n_2=0$ associated with $\ket{1}\coloneqq\ket{10}$, respectively. In the two-level approximation, the Hamiltonian is given by Eq.~(\ref{Theory:2levelHam}), where
\begin{gather}
	2E_\text{geom}=\sum\limits_{i=1}^2 E_{\text{C}i}(1-2n_{\text{g}i}+2n_{\text{g}i}^2)-C_\text{c}\left( \dfrac{\hbar \dot \varphi_\text{rf}}{2e}\right)^2 \label{Egeom_DQD}\!\!\!,\\
	\varepsilon(t)=E_{\text{C}1}(1-2n_{\text{g}1})-E_{\text{C}2}(1-2n_{\text{g}2}),
\end{gather}
and $\Delta$ is introduced as the tunnel-coupling energy between the nearest charge states. The dependence of the energy bias $\varepsilon$ on the probing voltage $V_\text{rf}$ is given by $\alpha_\text{DQD}=\xi \alpha'$, which leads to the geometric capacitance
\begin{gather}
	C_\text{geom}=\dfrac{\xi}{C_{\Sigma1}C_{\Sigma2}}\Big[C_{\text{G}1}C_{\text{SD}1}C_{\Sigma 2}+\nonumber\\+C_{\text{G}2}C_{\text{SD}2}C_{\Sigma 1}+C_\text{m}(C_{\text{G}1}+C_{\text{G}2})(C_{\text{SD}1}+C_{\text{SD}2}) \Big]. \label{C_geom_DQD}
\end{gather}
\bibliography{refs}

\begin{thebibliography}{86}%
\makeatletter
\providecommand \@ifxundefined [1]{%
 \@ifx{#1\undefined}
}%
\providecommand \@ifnum [1]{%
 \ifnum #1\expandafter \@firstoftwo
 \else \expandafter \@secondoftwo
 \fi
}%
\providecommand \@ifx [1]{%
 \ifx #1\expandafter \@firstoftwo
 \else \expandafter \@secondoftwo
 \fi
}%
\providecommand \natexlab [1]{#1}%
\providecommand \enquote  [1]{``#1''}%
\providecommand \bibnamefont  [1]{#1}%
\providecommand \bibfnamefont [1]{#1}%
\providecommand \citenamefont [1]{#1}%
\providecommand \href@noop [0]{\@secondoftwo}%
\providecommand \href [0]{\begingroup \@sanitize@url \@href}%
\providecommand \@href[1]{\@@startlink{#1}\@@href}%
\providecommand \@@href[1]{\endgroup#1\@@endlink}%
\providecommand \@sanitize@url [0]{\catcode `\\12\catcode `\$12\catcode
  `\&12\catcode `\#12\catcode `\^12\catcode `\_12\catcode `\%12\relax}%
\providecommand \@@startlink[1]{}%
\providecommand \@@endlink[0]{}%
\providecommand \url  [0]{\begingroup\@sanitize@url \@url }%
\providecommand \@url [1]{\endgroup\@href {#1}{\urlprefix }}%
\providecommand \urlprefix  [0]{URL }%
\providecommand \Eprint [0]{\href }%
\providecommand \doibase [0]{https://doi.org/}%
\providecommand \selectlanguage [0]{\@gobble}%
\providecommand \bibinfo  [0]{\@secondoftwo}%
\providecommand \bibfield  [0]{\@secondoftwo}%
\providecommand \translation [1]{[#1]}%
\providecommand \BibitemOpen [0]{}%
\providecommand \bibitemStop [0]{}%
\providecommand \bibitemNoStop [0]{.\EOS\space}%
\providecommand \EOS [0]{\spacefactor3000\relax}%
\providecommand \BibitemShut  [1]{\csname bibitem#1\endcsname}%
\let\auto@bib@innerbib\@empty
\bibitem [{\citenamefont {Haroche}\ and\ \citenamefont
  {Raimond}(2006)}]{Haroche2006}%
  \BibitemOpen
  \bibfield  {author} {\bibinfo {author} {\bibfnamefont {S.}~\bibnamefont
  {Haroche}}\ and\ \bibinfo {author} {\bibfnamefont {J.-M.}\ \bibnamefont
  {Raimond}},\ }\href
  {https://doi.org/10.1093/acprof:oso/9780198509141.001.0001} {\emph {\bibinfo
  {title} {Exploring the Quantum: Atoms, Cavities, and Photons}}}\ (\bibinfo
  {publisher} {Oxford University Press},\ \bibinfo {year} {2006})\BibitemShut
  {NoStop}%
\bibitem [{\citenamefont {Zagoskin}(2011)}]{Zagoskin2011}%
  \BibitemOpen
  \bibfield  {author} {\bibinfo {author} {\bibfnamefont {A.~M.}\ \bibnamefont
  {Zagoskin}},\ }\href {https://doi.org/10.1017/cbo9780511844157} {\emph
  {\bibinfo {title} {{Quantum Engineering}}}}\ (\bibinfo  {publisher}
  {Cambridge University Press},\ \bibinfo {year} {2011})\BibitemShut {NoStop}%
\bibitem [{\citenamefont {Huard}\ \emph {et~al.}(2014)\citenamefont {Huard},
  \citenamefont {Cugliandolo}, \citenamefont {Devoret},\ and\ \citenamefont
  {Schoelkopf}}]{Huard2014}%
  \BibitemOpen
  \bibinfo {editor} {\bibfnamefont {B.}~\bibnamefont {Huard}}, \bibinfo
  {editor} {\bibfnamefont {L.~F.}\ \bibnamefont {Cugliandolo}}, \bibinfo
  {editor} {\bibfnamefont {M.~H.}\ \bibnamefont {Devoret}},\ and\ \bibinfo
  {editor} {\bibfnamefont {R.}~\bibnamefont {Schoelkopf}},\ eds.,\ \href@noop
  {} {\emph {\bibinfo {title} {Quantum machines}}}\ (\bibinfo  {publisher}
  {Oxford University Press},\ \bibinfo {address} {Oxford},\ \bibinfo {year}
  {2014})\ \bibinfo {note} {lecture notes}\BibitemShut {NoStop}%
\bibitem [{\citenamefont {Shevchenko}(2019)}]{Shevchenko2019}%
  \BibitemOpen
  \bibfield  {author} {\bibinfo {author} {\bibfnamefont {S.~N.}\ \bibnamefont
  {Shevchenko}},\ }\href
  {https://www.worldscientific.com/worldscibooks/10.1142/11310} {\emph
  {\bibinfo {title} {{Mesoscopic Physics meets Quantum Engineering}}}}\
  (\bibinfo  {publisher} {World Scientific Pub Co Inc},\ \bibinfo {year}
  {2019})\BibitemShut {NoStop}%
\bibitem [{\citenamefont {Boutin}\ \emph {et~al.}(2025)\citenamefont {Boutin},
  \citenamefont {Karzig}, \citenamefont {Dandachi}, \citenamefont {Mishmash},
  \citenamefont {Gukelberger}, \citenamefont {Lutchyn},\ and\ \citenamefont
  {Bauer}}]{Boutin2025}%
  \BibitemOpen
  \bibfield  {author} {\bibinfo {author} {\bibfnamefont {S.}~\bibnamefont
  {Boutin}}, \bibinfo {author} {\bibfnamefont {T.}~\bibnamefont {Karzig}},
  \bibinfo {author} {\bibfnamefont {T.~E.}\ \bibnamefont {Dandachi}}, \bibinfo
  {author} {\bibfnamefont {R.~V.}\ \bibnamefont {Mishmash}}, \bibinfo {author}
  {\bibfnamefont {J.}~\bibnamefont {Gukelberger}}, \bibinfo {author}
  {\bibfnamefont {R.~M.}\ \bibnamefont {Lutchyn}},\ and\ \bibinfo {author}
  {\bibfnamefont {B.}~\bibnamefont {Bauer}},\ }\bibfield  {title} {\bibinfo
  {title} {Predictive simulations of the dynamical response of mesoscopic
  devices},\ }\href@noop {} {\  (\bibinfo {year} {2025})},\ \Eprint
  {https://arxiv.org/abs/2502.12960} {arXiv:2502.12960 [cond-mat.mes-hall]}
  \BibitemShut {NoStop}%
\bibitem [{\citenamefont {Johansson}\ \emph {et~al.}(2006)\citenamefont
  {Johansson}, \citenamefont {Tornberg}, \citenamefont {Shumeiko},\ and\
  \citenamefont {Wendin}}]{Johansson2006}%
  \BibitemOpen
  \bibfield  {author} {\bibinfo {author} {\bibfnamefont {G.}~\bibnamefont
  {Johansson}}, \bibinfo {author} {\bibfnamefont {L.}~\bibnamefont {Tornberg}},
  \bibinfo {author} {\bibfnamefont {V.~S.}\ \bibnamefont {Shumeiko}},\ and\
  \bibinfo {author} {\bibfnamefont {G.}~\bibnamefont {Wendin}},\ }\bibfield
  {title} {\bibinfo {title} {Readout methods and devices for
  {J}osephson-junction-based solid-state qubits},\ }\href
  {https://doi.org/10.1088/0953-8984/18/21/S14} {\bibfield  {journal} {\bibinfo
   {journal} {J. Phys. Cond. Matt.}\ }\textbf {\bibinfo {volume} {18}},\
  \bibinfo {pages} {S901} (\bibinfo {year} {2006})}\BibitemShut {NoStop}%
\bibitem [{\citenamefont {Sillanpää}\ \emph {et~al.}(2006)\citenamefont
  {Sillanpää}, \citenamefont {Lehtinen}, \citenamefont {Paila}, \citenamefont
  {Makhlin},\ and\ \citenamefont {Hakonen}}]{Sillanpaeae2006}%
  \BibitemOpen
  \bibfield  {author} {\bibinfo {author} {\bibfnamefont {M.}~\bibnamefont
  {Sillanpää}}, \bibinfo {author} {\bibfnamefont {T.}~\bibnamefont
  {Lehtinen}}, \bibinfo {author} {\bibfnamefont {A.}~\bibnamefont {Paila}},
  \bibinfo {author} {\bibfnamefont {Y.}~\bibnamefont {Makhlin}},\ and\ \bibinfo
  {author} {\bibfnamefont {P.}~\bibnamefont {Hakonen}},\ }\bibfield  {title}
  {\bibinfo {title} {Continuous-time monitoring of {Landau-Zener} interference
  in a {Cooper}-pair box},\ }\href
  {https://doi.org/10.1103/physrevlett.96.187002} {\bibfield  {journal}
  {\bibinfo  {journal} {Phys. Rev. Lett.}\ }\textbf {\bibinfo {volume} {96}},\
  \bibinfo {pages} {187002} (\bibinfo {year} {2006})}\BibitemShut {NoStop}%
\bibitem [{\citenamefont {Vigneau}\ \emph {et~al.}(2023)\citenamefont
  {Vigneau}, \citenamefont {Fedele}, \citenamefont {Chatterjee}, \citenamefont
  {Reilly}, \citenamefont {Kuemmeth}, \citenamefont {Gonzalez-Zalba},
  \citenamefont {Laird},\ and\ \citenamefont {Ares}}]{Vigneau2023}%
  \BibitemOpen
  \bibfield  {author} {\bibinfo {author} {\bibfnamefont {F.}~\bibnamefont
  {Vigneau}}, \bibinfo {author} {\bibfnamefont {F.}~\bibnamefont {Fedele}},
  \bibinfo {author} {\bibfnamefont {A.}~\bibnamefont {Chatterjee}}, \bibinfo
  {author} {\bibfnamefont {D.}~\bibnamefont {Reilly}}, \bibinfo {author}
  {\bibfnamefont {F.}~\bibnamefont {Kuemmeth}}, \bibinfo {author}
  {\bibfnamefont {M.~F.}\ \bibnamefont {Gonzalez-Zalba}}, \bibinfo {author}
  {\bibfnamefont {E.}~\bibnamefont {Laird}},\ and\ \bibinfo {author}
  {\bibfnamefont {N.}~\bibnamefont {Ares}},\ }\bibfield  {title} {\bibinfo
  {title} {{Probing quantum devices with radio-frequency reflectometry}},\
  }\href {https://doi.org/10.1063/5.0088229} {\bibfield  {journal} {\bibinfo
  {journal} {Appl. Phys. Rev.}\ }\textbf {\bibinfo {volume} {10}},\ \bibinfo
  {pages} {021305} (\bibinfo {year} {2023})}\BibitemShut {NoStop}%
\bibitem [{\citenamefont {Averin}\ and\ \citenamefont
  {Likharev}(1986)}]{Averin1986}%
  \BibitemOpen
  \bibfield  {author} {\bibinfo {author} {\bibfnamefont {D.~V.}\ \bibnamefont
  {Averin}}\ and\ \bibinfo {author} {\bibfnamefont {K.~K.}\ \bibnamefont
  {Likharev}},\ }\bibfield  {title} {\bibinfo {title} {Coulomb blockade of
  single-electron tunneling, and coherent oscillations in small tunnel
  junctions},\ }\href {https://doi.org/10.1007/bf00683469} {\bibfield
  {journal} {\bibinfo  {journal} {J. Low Temp. Phys.}\ }\textbf {\bibinfo
  {volume} {62}},\ \bibinfo {pages} {345} (\bibinfo {year} {1986})}\BibitemShut
  {NoStop}%
\bibitem [{\citenamefont {Likharev}(1996)}]{Likharev1996}%
  \BibitemOpen
  \bibfield  {author} {\bibinfo {author} {\bibfnamefont {K.~K.}\ \bibnamefont
  {Likharev}},\ }\href@noop {} {\emph {\bibinfo {title} {Dynamics of Josephson
  junctions and circuits}}},\ \bibinfo {edition} {3rd}\ ed.\ (\bibinfo
  {publisher} {Gordon and Breach},\ \bibinfo {address} {New York, N.Y.
  [u.a.]},\ \bibinfo {year} {1996})\BibitemShut {NoStop}%
\bibitem [{\citenamefont {Averin}\ \emph {et~al.}(1991)\citenamefont {Averin},
  \citenamefont {Korotkov},\ and\ \citenamefont {Likharev}}]{Averin1991}%
  \BibitemOpen
  \bibfield  {author} {\bibinfo {author} {\bibfnamefont {D.~V.}\ \bibnamefont
  {Averin}}, \bibinfo {author} {\bibfnamefont {A.~N.}\ \bibnamefont
  {Korotkov}},\ and\ \bibinfo {author} {\bibfnamefont {K.~K.}\ \bibnamefont
  {Likharev}},\ }\bibfield  {title} {\bibinfo {title} {Theory of
  single-electron charging of quantum wells and dots},\ }\href
  {https://doi.org/10.1103/physrevb.44.6199} {\bibfield  {journal} {\bibinfo
  {journal} {Phys. Rev. B}\ }\textbf {\bibinfo {volume} {44}},\ \bibinfo
  {pages} {6199} (\bibinfo {year} {1991})}\BibitemShut {NoStop}%
\bibitem [{\citenamefont {Likharev}(1999)}]{Likharev1999}%
  \BibitemOpen
  \bibfield  {author} {\bibinfo {author} {\bibfnamefont {K.}~\bibnamefont
  {Likharev}},\ }\bibfield  {title} {\bibinfo {title} {Single-electron devices
  and their applications},\ }\href {https://doi.org/10.1109/5.752518}
  {\bibfield  {journal} {\bibinfo  {journal} {Proc. IEEE}\ }\textbf {\bibinfo
  {volume} {87}},\ \bibinfo {pages} {606} (\bibinfo {year} {1999})}\BibitemShut
  {NoStop}%
\bibitem [{\citenamefont {Schön}\ and\ \citenamefont
  {Zaikin}(1990)}]{Schoen1990}%
  \BibitemOpen
  \bibfield  {author} {\bibinfo {author} {\bibfnamefont {G.}~\bibnamefont
  {Schön}}\ and\ \bibinfo {author} {\bibfnamefont {A.}~\bibnamefont
  {Zaikin}},\ }\bibfield  {title} {\bibinfo {title} {Quantum coherent effects,
  phase transitions, and the dissipative dynamics of ultra small tunnel
  junctions},\ }\href {https://doi.org/10.1016/0370-1573(90)90156-v} {\bibfield
   {journal} {\bibinfo  {journal} {Phys. Rep.}\ }\textbf {\bibinfo {volume}
  {198}},\ \bibinfo {pages} {237} (\bibinfo {year} {1990})}\BibitemShut
  {NoStop}%
\bibitem [{\citenamefont {Schoeller}\ and\ \citenamefont
  {Schön}(1994)}]{Schoeller1994}%
  \BibitemOpen
  \bibfield  {author} {\bibinfo {author} {\bibfnamefont {H.}~\bibnamefont
  {Schoeller}}\ and\ \bibinfo {author} {\bibfnamefont {G.}~\bibnamefont
  {Schön}},\ }\bibfield  {title} {\bibinfo {title} {Mesoscopic quantum
  transport: Resonant tunneling in the presence of a strong {C}oulomb
  interaction},\ }\href {https://doi.org/10.1103/physrevb.50.18436} {\bibfield
  {journal} {\bibinfo  {journal} {Phys. Rev. B}\ }\textbf {\bibinfo {volume}
  {50}},\ \bibinfo {pages} {18436} (\bibinfo {year} {1994})}\BibitemShut
  {NoStop}%
\bibitem [{\citenamefont {Makhlin}\ \emph {et~al.}(2001)\citenamefont
  {Makhlin}, \citenamefont {Schön},\ and\ \citenamefont
  {Shnirman}}]{Makhlin2001}%
  \BibitemOpen
  \bibfield  {author} {\bibinfo {author} {\bibfnamefont {Y.}~\bibnamefont
  {Makhlin}}, \bibinfo {author} {\bibfnamefont {G.}~\bibnamefont {Schön}},\
  and\ \bibinfo {author} {\bibfnamefont {A.}~\bibnamefont {Shnirman}},\
  }\bibfield  {title} {\bibinfo {title} {Quantum-state engineering with
  {J}osephson-junction devices},\ }\href
  {https://doi.org/10.1103/revmodphys.73.357} {\bibfield  {journal} {\bibinfo
  {journal} {Rev. Mod. Phys.}\ }\textbf {\bibinfo {volume} {73}},\ \bibinfo
  {pages} {357} (\bibinfo {year} {2001})}\BibitemShut {NoStop}%
\bibitem [{\citenamefont {Devoret}\ and\ \citenamefont
  {Schoelkopf}(2000)}]{Devoret2000}%
  \BibitemOpen
  \bibfield  {author} {\bibinfo {author} {\bibfnamefont {M.~H.}\ \bibnamefont
  {Devoret}}\ and\ \bibinfo {author} {\bibfnamefont {R.~J.}\ \bibnamefont
  {Schoelkopf}},\ }\bibfield  {title} {\bibinfo {title} {Amplifying quantum
  signals with the single-electron transistor},\ }\href
  {https://doi.org/10.1038/35023253} {\bibfield  {journal} {\bibinfo  {journal}
  {Nature}\ }\textbf {\bibinfo {volume} {406}},\ \bibinfo {pages} {1039}
  (\bibinfo {year} {2000})}\BibitemShut {NoStop}%
\bibitem [{\citenamefont {You}\ and\ \citenamefont {Nori}(2005)}]{You2005}%
  \BibitemOpen
  \bibfield  {author} {\bibinfo {author} {\bibfnamefont {J.~Q.}\ \bibnamefont
  {You}}\ and\ \bibinfo {author} {\bibfnamefont {F.}~\bibnamefont {Nori}},\
  }\bibfield  {title} {\bibinfo {title} {Superconducting circuits and quantum
  information},\ }\href {https://doi.org/10.1063/1.2155757} {\bibfield
  {journal} {\bibinfo  {journal} {Phys. Today}\ }\textbf {\bibinfo {volume}
  {58}},\ \bibinfo {pages} {42} (\bibinfo {year} {2005})}\BibitemShut {NoStop}%
\bibitem [{\citenamefont {You}\ and\ \citenamefont {Nori}(2011)}]{You2011}%
  \BibitemOpen
  \bibfield  {author} {\bibinfo {author} {\bibfnamefont {J.~Q.}\ \bibnamefont
  {You}}\ and\ \bibinfo {author} {\bibfnamefont {F.}~\bibnamefont {Nori}},\
  }\bibfield  {title} {\bibinfo {title} {Atomic physics and quantum optics
  using superconducting circuits},\ }\href
  {https://doi.org/10.1038/nature10122} {\bibfield  {journal} {\bibinfo
  {journal} {Nature}\ }\textbf {\bibinfo {volume} {474}},\ \bibinfo {pages}
  {589} (\bibinfo {year} {2011})}\BibitemShut {NoStop}%
\bibitem [{\citenamefont {Nigg}\ \emph {et~al.}(2012)\citenamefont {Nigg},
  \citenamefont {Paik}, \citenamefont {Vlastakis}, \citenamefont {Kirchmair},
  \citenamefont {Shankar}, \citenamefont {Frunzio}, \citenamefont {Devoret},
  \citenamefont {Schoelkopf},\ and\ \citenamefont {Girvin}}]{Nigg2012}%
  \BibitemOpen
  \bibfield  {author} {\bibinfo {author} {\bibfnamefont {S.~E.}\ \bibnamefont
  {Nigg}}, \bibinfo {author} {\bibfnamefont {H.}~\bibnamefont {Paik}}, \bibinfo
  {author} {\bibfnamefont {B.}~\bibnamefont {Vlastakis}}, \bibinfo {author}
  {\bibfnamefont {G.}~\bibnamefont {Kirchmair}}, \bibinfo {author}
  {\bibfnamefont {S.}~\bibnamefont {Shankar}}, \bibinfo {author} {\bibfnamefont
  {L.}~\bibnamefont {Frunzio}}, \bibinfo {author} {\bibfnamefont {M.~H.}\
  \bibnamefont {Devoret}}, \bibinfo {author} {\bibfnamefont {R.~J.}\
  \bibnamefont {Schoelkopf}},\ and\ \bibinfo {author} {\bibfnamefont {S.~M.}\
  \bibnamefont {Girvin}},\ }\bibfield  {title} {\bibinfo {title} {Black-box
  superconducting circuit quantization},\ }\href
  {https://doi.org/10.1103/physrevlett.108.240502} {\bibfield  {journal}
  {\bibinfo  {journal} {Phys. Rev. Lett.}\ }\textbf {\bibinfo {volume} {108}},\
  \bibinfo {pages} {240502} (\bibinfo {year} {2012})}\BibitemShut {NoStop}%
\bibitem [{\citenamefont {Gu}\ \emph {et~al.}(2017)\citenamefont {Gu},
  \citenamefont {Kockum}, \citenamefont {Miranowicz}, \citenamefont {Liu},\
  and\ \citenamefont {Nori}}]{Gu2017}%
  \BibitemOpen
  \bibfield  {author} {\bibinfo {author} {\bibfnamefont {X.}~\bibnamefont
  {Gu}}, \bibinfo {author} {\bibfnamefont {A.~F.}\ \bibnamefont {Kockum}},
  \bibinfo {author} {\bibfnamefont {A.}~\bibnamefont {Miranowicz}}, \bibinfo
  {author} {\bibfnamefont {Y.-X.}\ \bibnamefont {Liu}},\ and\ \bibinfo {author}
  {\bibfnamefont {F.}~\bibnamefont {Nori}},\ }\bibfield  {title} {\bibinfo
  {title} {Microwave photonics with superconducting quantum circuits},\ }\href
  {https://doi.org/10.1016/j.physrep.2017.10.002} {\bibfield  {journal}
  {\bibinfo  {journal} {Phys. Rep.}\ }\textbf {\bibinfo {volume} {718-719}},\
  \bibinfo {pages} {1} (\bibinfo {year} {2017})}\BibitemShut {NoStop}%
\bibitem [{\citenamefont {Krantz}\ \emph {et~al.}(2019)\citenamefont {Krantz},
  \citenamefont {Kjaergaard}, \citenamefont {Yan}, \citenamefont {Orlando},
  \citenamefont {Gustavsson},\ and\ \citenamefont {Oliver}}]{Krantz2019}%
  \BibitemOpen
  \bibfield  {author} {\bibinfo {author} {\bibfnamefont {P.}~\bibnamefont
  {Krantz}}, \bibinfo {author} {\bibfnamefont {M.}~\bibnamefont {Kjaergaard}},
  \bibinfo {author} {\bibfnamefont {F.}~\bibnamefont {Yan}}, \bibinfo {author}
  {\bibfnamefont {T.~P.}\ \bibnamefont {Orlando}}, \bibinfo {author}
  {\bibfnamefont {S.}~\bibnamefont {Gustavsson}},\ and\ \bibinfo {author}
  {\bibfnamefont {W.~D.}\ \bibnamefont {Oliver}},\ }\bibfield  {title}
  {\bibinfo {title} {A quantum engineer{\textquotesingle}s guide to
  superconducting qubits},\ }\href {https://doi.org/10.1063/1.5089550}
  {\bibfield  {journal} {\bibinfo  {journal} {Appl. Phys. Rev.}\ }\textbf
  {\bibinfo {volume} {6}},\ \bibinfo {pages} {021318} (\bibinfo {year}
  {2019})}\BibitemShut {NoStop}%
\bibitem [{\citenamefont {Kockum}\ and\ \citenamefont
  {Nori}(2019)}]{Kockum2019a}%
  \BibitemOpen
  \bibfield  {author} {\bibinfo {author} {\bibfnamefont {A.~F.}\ \bibnamefont
  {Kockum}}\ and\ \bibinfo {author} {\bibfnamefont {F.}~\bibnamefont {Nori}},\
  }\bibinfo {title} {Quantum bits with {J}osephson junctions},\ in\ \href
  {https://doi.org/10.1007/978-3-030-20726-7_17} {\emph {\bibinfo {booktitle}
  {Fundamentals and Frontiers of the Josephson Effect}}}\ (\bibinfo
  {publisher} {Springer International Publishing},\ \bibinfo {year} {2019})\
  pp.\ \bibinfo {pages} {703--741}\BibitemShut {NoStop}%
\bibitem [{\citenamefont {Kjaergaard}\ \emph {et~al.}(2020)\citenamefont
  {Kjaergaard}, \citenamefont {Schwartz}, \citenamefont {Braumüller},
  \citenamefont {Krantz}, \citenamefont {Wang}, \citenamefont {Gustavsson},\
  and\ \citenamefont {Oliver}}]{Kjaergaard2020}%
  \BibitemOpen
  \bibfield  {author} {\bibinfo {author} {\bibfnamefont {M.}~\bibnamefont
  {Kjaergaard}}, \bibinfo {author} {\bibfnamefont {M.~E.}\ \bibnamefont
  {Schwartz}}, \bibinfo {author} {\bibfnamefont {J.}~\bibnamefont
  {Braumüller}}, \bibinfo {author} {\bibfnamefont {P.}~\bibnamefont {Krantz}},
  \bibinfo {author} {\bibfnamefont {J.~I.-J.}\ \bibnamefont {Wang}}, \bibinfo
  {author} {\bibfnamefont {S.}~\bibnamefont {Gustavsson}},\ and\ \bibinfo
  {author} {\bibfnamefont {W.~D.}\ \bibnamefont {Oliver}},\ }\bibfield  {title}
  {\bibinfo {title} {Superconducting qubits: Current state of play},\ }\href
  {https://doi.org/10.1146/annurev-conmatphys-031119-050605} {\bibfield
  {journal} {\bibinfo  {journal} {Ann. Rev. Cond. Matt. Phys.}\ }\textbf
  {\bibinfo {volume} {11}},\ \bibinfo {pages} {369} (\bibinfo {year}
  {2020})}\BibitemShut {NoStop}%
\bibitem [{\citenamefont {Blais}\ \emph {et~al.}(2021)\citenamefont {Blais},
  \citenamefont {Grimsmo}, \citenamefont {Girvin},\ and\ \citenamefont
  {Wallraff}}]{Blais2021}%
  \BibitemOpen
  \bibfield  {author} {\bibinfo {author} {\bibfnamefont {A.}~\bibnamefont
  {Blais}}, \bibinfo {author} {\bibfnamefont {A.~L.}\ \bibnamefont {Grimsmo}},
  \bibinfo {author} {\bibfnamefont {S.~M.}\ \bibnamefont {Girvin}},\ and\
  \bibinfo {author} {\bibfnamefont {A.}~\bibnamefont {Wallraff}},\ }\bibfield
  {title} {\bibinfo {title} {Circuit quantum electrodynamics},\ }\href
  {https://doi.org/10.1103/RevModPhys.93.025005} {\bibfield  {journal}
  {\bibinfo  {journal} {Rev. Mod. Phys.}\ }\textbf {\bibinfo {volume} {93}},\
  \bibinfo {pages} {025005} (\bibinfo {year} {2021})}\BibitemShut {NoStop}%
\bibitem [{\citenamefont {Zwanenburg}\ \emph {et~al.}(2013)\citenamefont
  {Zwanenburg}, \citenamefont {Dzurak}, \citenamefont {Morello}, \citenamefont
  {Simmons}, \citenamefont {Hollenberg}, \citenamefont {Klimeck}, \citenamefont
  {Rogge}, \citenamefont {Coppersmith},\ and\ \citenamefont
  {Eriksson}}]{Zwanenburg2013}%
  \BibitemOpen
  \bibfield  {author} {\bibinfo {author} {\bibfnamefont {F.~A.}\ \bibnamefont
  {Zwanenburg}}, \bibinfo {author} {\bibfnamefont {A.~S.}\ \bibnamefont
  {Dzurak}}, \bibinfo {author} {\bibfnamefont {A.}~\bibnamefont {Morello}},
  \bibinfo {author} {\bibfnamefont {M.~Y.}\ \bibnamefont {Simmons}}, \bibinfo
  {author} {\bibfnamefont {L.~C.~L.}\ \bibnamefont {Hollenberg}}, \bibinfo
  {author} {\bibfnamefont {G.}~\bibnamefont {Klimeck}}, \bibinfo {author}
  {\bibfnamefont {S.}~\bibnamefont {Rogge}}, \bibinfo {author} {\bibfnamefont
  {S.~N.}\ \bibnamefont {Coppersmith}},\ and\ \bibinfo {author} {\bibfnamefont
  {M.~A.}\ \bibnamefont {Eriksson}},\ }\bibfield  {title} {\bibinfo {title}
  {Silicon quantum electronics},\ }\href
  {https://doi.org/10.1103/revmodphys.85.961} {\bibfield  {journal} {\bibinfo
  {journal} {Rev. Mod. Phys.}\ }\textbf {\bibinfo {volume} {85}},\ \bibinfo
  {pages} {961} (\bibinfo {year} {2013})}\BibitemShut {NoStop}%
\bibitem [{\citenamefont {Burkard}\ \emph {et~al.}(2023)\citenamefont
  {Burkard}, \citenamefont {Ladd}, \citenamefont {Pan}, \citenamefont
  {Nichol},\ and\ \citenamefont {Petta}}]{Burkard2023}%
  \BibitemOpen
  \bibfield  {author} {\bibinfo {author} {\bibfnamefont {G.}~\bibnamefont
  {Burkard}}, \bibinfo {author} {\bibfnamefont {T.~D.}\ \bibnamefont {Ladd}},
  \bibinfo {author} {\bibfnamefont {A.}~\bibnamefont {Pan}}, \bibinfo {author}
  {\bibfnamefont {J.~M.}\ \bibnamefont {Nichol}},\ and\ \bibinfo {author}
  {\bibfnamefont {J.~R.}\ \bibnamefont {Petta}},\ }\bibfield  {title} {\bibinfo
  {title} {Semiconductor spin qubits},\ }\href
  {https://doi.org/10.1103/revmodphys.95.025003} {\bibfield  {journal}
  {\bibinfo  {journal} {Rev. Mod. Phys.}\ }\textbf {\bibinfo {volume} {95}},\
  \bibinfo {pages} {025003} (\bibinfo {year} {2023})}\BibitemShut {NoStop}%
\bibitem [{\citenamefont {Schoelkopf}\ \emph {et~al.}(1998)\citenamefont
  {Schoelkopf}, \citenamefont {Wahlgren}, \citenamefont {Kozhevnikov},
  \citenamefont {Delsing},\ and\ \citenamefont {Prober}}]{Schoelkopf1998}%
  \BibitemOpen
  \bibfield  {author} {\bibinfo {author} {\bibfnamefont {R.~J.}\ \bibnamefont
  {Schoelkopf}}, \bibinfo {author} {\bibfnamefont {P.}~\bibnamefont
  {Wahlgren}}, \bibinfo {author} {\bibfnamefont {A.~A.}\ \bibnamefont
  {Kozhevnikov}}, \bibinfo {author} {\bibfnamefont {P.}~\bibnamefont
  {Delsing}},\ and\ \bibinfo {author} {\bibfnamefont {D.~E.}\ \bibnamefont
  {Prober}},\ }\bibfield  {title} {\bibinfo {title} {The radio-frequency
  single-electron transistor (rf-set): A fast and ultrasensitive
  electrometer},\ }\href {https://doi.org/10.1126/science.280.5367.1238}
  {\bibfield  {journal} {\bibinfo  {journal} {Science}\ }\textbf {\bibinfo
  {volume} {280}},\ \bibinfo {pages} {1238} (\bibinfo {year}
  {1998})}\BibitemShut {NoStop}%
\bibitem [{\citenamefont {House}\ \emph {et~al.}(2016)\citenamefont {House},
  \citenamefont {Bartlett}, \citenamefont {Pakkiam}, \citenamefont {Koch},
  \citenamefont {Peretz}, \citenamefont {van~der Heijden}, \citenamefont
  {Kobayashi}, \citenamefont {Rogge},\ and\ \citenamefont
  {Simmons}}]{House2016}%
  \BibitemOpen
  \bibfield  {author} {\bibinfo {author} {\bibfnamefont {M.}~\bibnamefont
  {House}}, \bibinfo {author} {\bibfnamefont {I.}~\bibnamefont {Bartlett}},
  \bibinfo {author} {\bibfnamefont {P.}~\bibnamefont {Pakkiam}}, \bibinfo
  {author} {\bibfnamefont {M.}~\bibnamefont {Koch}}, \bibinfo {author}
  {\bibfnamefont {E.}~\bibnamefont {Peretz}}, \bibinfo {author} {\bibfnamefont
  {J.}~\bibnamefont {van~der Heijden}}, \bibinfo {author} {\bibfnamefont
  {T.}~\bibnamefont {Kobayashi}}, \bibinfo {author} {\bibfnamefont
  {S.}~\bibnamefont {Rogge}},\ and\ \bibinfo {author} {\bibfnamefont
  {M.}~\bibnamefont {Simmons}},\ }\bibfield  {title} {\bibinfo {title}
  {High-sensitivity charge detection with a single-lead quantum dot for
  scalable quantum computation},\ }\href
  {https://doi.org/10.1103/physrevapplied.6.044016} {\bibfield  {journal}
  {\bibinfo  {journal} {Phys. Rev. Appl.}\ }\textbf {\bibinfo {volume} {6}},\
  \bibinfo {pages} {044016} (\bibinfo {year} {2016})}\BibitemShut {NoStop}%
\bibitem [{\citenamefont {Zirkle}\ \emph {et~al.}(2020)\citenamefont {Zirkle},
  \citenamefont {Filmer}, \citenamefont {Chisum}, \citenamefont {Orlov},
  \citenamefont {Dupont-Ferrier}, \citenamefont {Rivard}, \citenamefont
  {Huebner}, \citenamefont {Sanquer}, \citenamefont {Jehl},\ and\ \citenamefont
  {Snider}}]{Zirkle2020}%
  \BibitemOpen
  \bibfield  {author} {\bibinfo {author} {\bibfnamefont {T.~A.}\ \bibnamefont
  {Zirkle}}, \bibinfo {author} {\bibfnamefont {M.~J.}\ \bibnamefont {Filmer}},
  \bibinfo {author} {\bibfnamefont {J.}~\bibnamefont {Chisum}}, \bibinfo
  {author} {\bibfnamefont {A.~O.}\ \bibnamefont {Orlov}}, \bibinfo {author}
  {\bibfnamefont {E.}~\bibnamefont {Dupont-Ferrier}}, \bibinfo {author}
  {\bibfnamefont {J.}~\bibnamefont {Rivard}}, \bibinfo {author} {\bibfnamefont
  {M.}~\bibnamefont {Huebner}}, \bibinfo {author} {\bibfnamefont
  {M.}~\bibnamefont {Sanquer}}, \bibinfo {author} {\bibfnamefont
  {X.}~\bibnamefont {Jehl}},\ and\ \bibinfo {author} {\bibfnamefont {G.~L.}\
  \bibnamefont {Snider}},\ }\bibfield  {title} {\bibinfo {title} {Radio
  frequency reflectometry of single-electron box arrays for nanoscale voltage
  sensing applications},\ }\href {https://doi.org/10.3390/app10248797}
  {\bibfield  {journal} {\bibinfo  {journal} {Appl. Sci.}\ }\textbf {\bibinfo
  {volume} {10}},\ \bibinfo {pages} {8797} (\bibinfo {year}
  {2020})}\BibitemShut {NoStop}%
\bibitem [{\citenamefont {Filmer}\ \emph {et~al.}(2022)\citenamefont {Filmer},
  \citenamefont {Huebner}, \citenamefont {Zirkle}, \citenamefont {Jehl},
  \citenamefont {Sanquer}, \citenamefont {Chisum}, \citenamefont {Orlov},\ and\
  \citenamefont {Snider}}]{Filmer2022}%
  \BibitemOpen
  \bibfield  {author} {\bibinfo {author} {\bibfnamefont {M.~J.}\ \bibnamefont
  {Filmer}}, \bibinfo {author} {\bibfnamefont {M.}~\bibnamefont {Huebner}},
  \bibinfo {author} {\bibfnamefont {T.~A.}\ \bibnamefont {Zirkle}}, \bibinfo
  {author} {\bibfnamefont {X.}~\bibnamefont {Jehl}}, \bibinfo {author}
  {\bibfnamefont {M.}~\bibnamefont {Sanquer}}, \bibinfo {author} {\bibfnamefont
  {J.~D.}\ \bibnamefont {Chisum}}, \bibinfo {author} {\bibfnamefont {A.~O.}\
  \bibnamefont {Orlov}},\ and\ \bibinfo {author} {\bibfnamefont {G.~L.}\
  \bibnamefont {Snider}},\ }\bibfield  {title} {\bibinfo {title} {Gate
  reflectometry of single-electron box arrays using calibrated low temperature
  matching networks},\ }\href {https://doi.org/10.1038/s41598-022-06727-1}
  {\bibfield  {journal} {\bibinfo  {journal} {Sci. Rep.}\ }\textbf {\bibinfo
  {volume} {12}},\ \bibinfo {pages} {3098} (\bibinfo {year}
  {2022})}\BibitemShut {NoStop}%
\bibitem [{\citenamefont {Frake}\ \emph {et~al.}(2015)\citenamefont {Frake},
  \citenamefont {Kano}, \citenamefont {Ciccarelli}, \citenamefont {Griffiths},
  \citenamefont {Sakamoto}, \citenamefont {Teranishi}, \citenamefont {Majima},
  \citenamefont {Smith},\ and\ \citenamefont {Buitelaar}}]{Frake2015}%
  \BibitemOpen
  \bibfield  {author} {\bibinfo {author} {\bibfnamefont {J.~C.}\ \bibnamefont
  {Frake}}, \bibinfo {author} {\bibfnamefont {S.}~\bibnamefont {Kano}},
  \bibinfo {author} {\bibfnamefont {C.}~\bibnamefont {Ciccarelli}}, \bibinfo
  {author} {\bibfnamefont {J.}~\bibnamefont {Griffiths}}, \bibinfo {author}
  {\bibfnamefont {M.}~\bibnamefont {Sakamoto}}, \bibinfo {author}
  {\bibfnamefont {T.}~\bibnamefont {Teranishi}}, \bibinfo {author}
  {\bibfnamefont {Y.}~\bibnamefont {Majima}}, \bibinfo {author} {\bibfnamefont
  {C.~G.}\ \bibnamefont {Smith}},\ and\ \bibinfo {author} {\bibfnamefont
  {M.~R.}\ \bibnamefont {Buitelaar}},\ }\bibfield  {title} {\bibinfo {title}
  {Radio-frequency capacitance spectroscopy of metallic nanoparticles},\ }\href
  {https://doi.org/10.1038/srep10858} {\bibfield  {journal} {\bibinfo
  {journal} {Sci. Rep.}\ }\textbf {\bibinfo {volume} {5}},\ \bibinfo {pages}
  {10858} (\bibinfo {year} {2015})}\BibitemShut {NoStop}%
\bibitem [{\citenamefont {Zhang}\ \emph {et~al.}(2024)\citenamefont {Zhang},
  \citenamefont {Guan}, \citenamefont {Sheng}, \citenamefont {Chen},
  \citenamefont {Rogge}, \citenamefont {Du},\ and\ \citenamefont
  {Yin}}]{Zhang2024}%
  \BibitemOpen
  \bibfield  {author} {\bibinfo {author} {\bibfnamefont {Y.}~\bibnamefont
  {Zhang}}, \bibinfo {author} {\bibfnamefont {H.}~\bibnamefont {Guan}},
  \bibinfo {author} {\bibfnamefont {T.}~\bibnamefont {Sheng}}, \bibinfo
  {author} {\bibfnamefont {R.}~\bibnamefont {Chen}}, \bibinfo {author}
  {\bibfnamefont {S.}~\bibnamefont {Rogge}}, \bibinfo {author} {\bibfnamefont
  {J.}~\bibnamefont {Du}},\ and\ \bibinfo {author} {\bibfnamefont
  {C.}~\bibnamefont {Yin}},\ }\bibfield  {title} {\bibinfo {title} {Fast
  thermodynamic study on a silicon nanotransistor at cryogenic temperatures},\
  }\href {https://doi.org/10.1021/acs.nanolett.4c01424} {\bibfield  {journal}
  {\bibinfo  {journal} {Nano Letters}\ }\textbf {\bibinfo {volume} {24}},\
  \bibinfo {pages} {8859} (\bibinfo {year} {2024})}\BibitemShut {NoStop}%
\bibitem [{\citenamefont {van Loo}\ \emph {et~al.}(2026)\citenamefont {van
  Loo}, \citenamefont {Zatelli}, \citenamefont {Steffensen}, \citenamefont
  {Roovers}, \citenamefont {Wang}, \citenamefont {Van~Caekenberghe},
  \citenamefont {Bordin}, \citenamefont {van Driel}, \citenamefont {Zhang},
  \citenamefont {Huisman}, \citenamefont {Badawy}, \citenamefont {Bakkers},
  \citenamefont {Mazur}, \citenamefont {Aguado},\ and\ \citenamefont
  {Kouwenhoven}}]{Loo2026}%
  \BibitemOpen
  \bibfield  {author} {\bibinfo {author} {\bibfnamefont {N.}~\bibnamefont {van
  Loo}}, \bibinfo {author} {\bibfnamefont {F.}~\bibnamefont {Zatelli}},
  \bibinfo {author} {\bibfnamefont {G.~O.}\ \bibnamefont {Steffensen}},
  \bibinfo {author} {\bibfnamefont {B.}~\bibnamefont {Roovers}}, \bibinfo
  {author} {\bibfnamefont {G.}~\bibnamefont {Wang}}, \bibinfo {author}
  {\bibfnamefont {T.}~\bibnamefont {Van~Caekenberghe}}, \bibinfo {author}
  {\bibfnamefont {A.}~\bibnamefont {Bordin}}, \bibinfo {author} {\bibfnamefont
  {D.}~\bibnamefont {van Driel}}, \bibinfo {author} {\bibfnamefont
  {Y.}~\bibnamefont {Zhang}}, \bibinfo {author} {\bibfnamefont {W.~D.}\
  \bibnamefont {Huisman}}, \bibinfo {author} {\bibfnamefont {G.}~\bibnamefont
  {Badawy}}, \bibinfo {author} {\bibfnamefont {E.~P. A.~M.}\ \bibnamefont
  {Bakkers}}, \bibinfo {author} {\bibfnamefont {G.~P.}\ \bibnamefont {Mazur}},
  \bibinfo {author} {\bibfnamefont {R.}~\bibnamefont {Aguado}},\ and\ \bibinfo
  {author} {\bibfnamefont {L.~P.}\ \bibnamefont {Kouwenhoven}},\ }\bibfield
  {title} {\bibinfo {title} {Single-shot parity readout of a minimal {K}itaev
  chain},\ }\href {https://doi.org/10.1038/s41586-025-09927-7} {\bibfield
  {journal} {\bibinfo  {journal} {Nature}\ }\textbf {\bibinfo {volume} {650}},\
  \bibinfo {pages} {334} (\bibinfo {year} {2026})}\BibitemShut {NoStop}%
\bibitem [{\citenamefont {van Driel}\ \emph {et~al.}(2024)\citenamefont {van
  Driel}, \citenamefont {Roovers}, \citenamefont {Zatelli}, \citenamefont
  {Bordin}, \citenamefont {Wang}, \citenamefont {van Loo}, \citenamefont
  {Wolff}, \citenamefont {Mazur}, \citenamefont {Gazibegovic}, \citenamefont
  {Badawy}, \citenamefont {Bakkers}, \citenamefont {Kouwenhoven},\ and\
  \citenamefont {Dvir}}]{Driel2024}%
  \BibitemOpen
  \bibfield  {author} {\bibinfo {author} {\bibfnamefont {D.}~\bibnamefont {van
  Driel}}, \bibinfo {author} {\bibfnamefont {B.}~\bibnamefont {Roovers}},
  \bibinfo {author} {\bibfnamefont {F.}~\bibnamefont {Zatelli}}, \bibinfo
  {author} {\bibfnamefont {A.}~\bibnamefont {Bordin}}, \bibinfo {author}
  {\bibfnamefont {G.}~\bibnamefont {Wang}}, \bibinfo {author} {\bibfnamefont
  {N.}~\bibnamefont {van Loo}}, \bibinfo {author} {\bibfnamefont {J.~C.}\
  \bibnamefont {Wolff}}, \bibinfo {author} {\bibfnamefont {G.~P.}\ \bibnamefont
  {Mazur}}, \bibinfo {author} {\bibfnamefont {S.}~\bibnamefont {Gazibegovic}},
  \bibinfo {author} {\bibfnamefont {G.}~\bibnamefont {Badawy}}, \bibinfo
  {author} {\bibfnamefont {E.~P.}\ \bibnamefont {Bakkers}}, \bibinfo {author}
  {\bibfnamefont {L.~P.}\ \bibnamefont {Kouwenhoven}},\ and\ \bibinfo {author}
  {\bibfnamefont {T.}~\bibnamefont {Dvir}},\ }\bibfield  {title} {\bibinfo
  {title} {Charge sensing the parity of an {A}ndreev molecule},\ }\href
  {https://doi.org/10.1103/prxquantum.5.020301} {\bibfield  {journal} {\bibinfo
   {journal} {PRX Quantum}\ }\textbf {\bibinfo {volume} {5}},\ \bibinfo {pages}
  {020301} (\bibinfo {year} {2024})}\BibitemShut {NoStop}%
\bibitem [{\citenamefont {Shnyrkov}\ \emph {et~al.}(2006)\citenamefont
  {Shnyrkov}, \citenamefont {Wagner}, \citenamefont {Born}, \citenamefont
  {Shevchenko}, \citenamefont {Krech}, \citenamefont {Omelyanchouk},
  \citenamefont {Il'ichev},\ and\ \citenamefont {Meyer}}]{Shnyrkov2006}%
  \BibitemOpen
  \bibfield  {author} {\bibinfo {author} {\bibfnamefont {V.~I.}\ \bibnamefont
  {Shnyrkov}}, \bibinfo {author} {\bibfnamefont {T.}~\bibnamefont {Wagner}},
  \bibinfo {author} {\bibfnamefont {D.}~\bibnamefont {Born}}, \bibinfo {author}
  {\bibfnamefont {S.~N.}\ \bibnamefont {Shevchenko}}, \bibinfo {author}
  {\bibfnamefont {W.}~\bibnamefont {Krech}}, \bibinfo {author} {\bibfnamefont
  {A.~N.}\ \bibnamefont {Omelyanchouk}}, \bibinfo {author} {\bibfnamefont
  {E.}~\bibnamefont {Il'ichev}},\ and\ \bibinfo {author} {\bibfnamefont
  {H.~G.}\ \bibnamefont {Meyer}},\ }\bibfield  {title} {\bibinfo {title}
  {{Multiphoton transitions between energy levels in a phase-biased Cooper-pair
  box}},\ }\href {https://doi.org/10.1103/physrevb.73.024506} {\bibfield
  {journal} {\bibinfo  {journal} {Phys. Rev. B}\ }\textbf {\bibinfo {volume}
  {73}},\ \bibinfo {pages} {024506} (\bibinfo {year} {2006})}\BibitemShut
  {NoStop}%
\bibitem [{\citenamefont {Shevchenko}\ \emph {et~al.}(2008)\citenamefont
  {Shevchenko}, \citenamefont {van~der Ploeg}, \citenamefont {Grajcar},
  \citenamefont {Il'ichev}, \citenamefont {Omelyanchouk},\ and\ \citenamefont
  {Meyer}}]{Shevchenko2008}%
  \BibitemOpen
  \bibfield  {author} {\bibinfo {author} {\bibfnamefont {S.~N.}\ \bibnamefont
  {Shevchenko}}, \bibinfo {author} {\bibfnamefont {S.~H.~W.}\ \bibnamefont
  {van~der Ploeg}}, \bibinfo {author} {\bibfnamefont {M.}~\bibnamefont
  {Grajcar}}, \bibinfo {author} {\bibfnamefont {E.}~\bibnamefont {Il'ichev}},
  \bibinfo {author} {\bibfnamefont {A.~N.}\ \bibnamefont {Omelyanchouk}},\ and\
  \bibinfo {author} {\bibfnamefont {H.-G.}\ \bibnamefont {Meyer}},\ }\bibfield
  {title} {\bibinfo {title} {Resonant excitations of single and two-qubit
  systems coupled to a tank circuit},\ }\href
  {https://doi.org/10.1103/physrevb.78.174527} {\bibfield  {journal} {\bibinfo
  {journal} {Phys. Rev. B}\ }\textbf {\bibinfo {volume} {78}},\ \bibinfo
  {pages} {174527} (\bibinfo {year} {2008})}\BibitemShut {NoStop}%
\bibitem [{\citenamefont {Frey}\ \emph {et~al.}(2012)\citenamefont {Frey},
  \citenamefont {Leek}, \citenamefont {Beck}, \citenamefont {Faist},
  \citenamefont {Wallraff}, \citenamefont {Ensslin}, \citenamefont {Ihn},\ and\
  \citenamefont {B\"uttiker}}]{Frey2012}%
  \BibitemOpen
  \bibfield  {author} {\bibinfo {author} {\bibfnamefont {T.}~\bibnamefont
  {Frey}}, \bibinfo {author} {\bibfnamefont {P.~J.}\ \bibnamefont {Leek}},
  \bibinfo {author} {\bibfnamefont {M.}~\bibnamefont {Beck}}, \bibinfo {author}
  {\bibfnamefont {J.}~\bibnamefont {Faist}}, \bibinfo {author} {\bibfnamefont
  {A.}~\bibnamefont {Wallraff}}, \bibinfo {author} {\bibfnamefont
  {K.}~\bibnamefont {Ensslin}}, \bibinfo {author} {\bibfnamefont
  {T.}~\bibnamefont {Ihn}},\ and\ \bibinfo {author} {\bibfnamefont
  {M.}~\bibnamefont {B\"uttiker}},\ }\bibfield  {title} {\bibinfo {title}
  {Quantum dot admittance probed at microwave frequencies with an on-chip
  resonator},\ }\href {https://doi.org/10.1103/PhysRevB.86.115303} {\bibfield
  {journal} {\bibinfo  {journal} {Phys. Rev. B}\ }\textbf {\bibinfo {volume}
  {86}},\ \bibinfo {pages} {115303} (\bibinfo {year} {2012})}\BibitemShut
  {NoStop}%
\bibitem [{\citenamefont {Shnyrkov}\ \emph {et~al.}(2009)\citenamefont
  {Shnyrkov}, \citenamefont {Soroka},\ and\ \citenamefont
  {Krech}}]{Shnyrkov2009a}%
  \BibitemOpen
  \bibfield  {author} {\bibinfo {author} {\bibfnamefont {V.~I.}\ \bibnamefont
  {Shnyrkov}}, \bibinfo {author} {\bibfnamefont {A.~A.}\ \bibnamefont
  {Soroka}},\ and\ \bibinfo {author} {\bibfnamefont {W.}~\bibnamefont
  {Krech}},\ }\bibfield  {title} {\bibinfo {title} {Signal characteristics of
  charge-phase qubit detector with parametric energy conversion},\ }\href
  {https://doi.org/10.1063/1.3224724} {\bibfield  {journal} {\bibinfo
  {journal} {Low Temp. Phys.}\ }\textbf {\bibinfo {volume} {35}},\ \bibinfo
  {pages} {652} (\bibinfo {year} {2009})}\BibitemShut {NoStop}%
\bibitem [{\citenamefont {Shnyrkov}\ \emph {et~al.}(2012)\citenamefont
  {Shnyrkov}, \citenamefont {Soroka},\ and\ \citenamefont
  {Turutanov}}]{Shnyrkov2012}%
  \BibitemOpen
  \bibfield  {author} {\bibinfo {author} {\bibfnamefont {V.~I.}\ \bibnamefont
  {Shnyrkov}}, \bibinfo {author} {\bibfnamefont {A.~A.}\ \bibnamefont
  {Soroka}},\ and\ \bibinfo {author} {\bibfnamefont {O.~G.}\ \bibnamefont
  {Turutanov}},\ }\bibfield  {title} {\bibinfo {title} {Quantum superposition
  of three macroscopic states and superconducting qutrit detector},\ }\href
  {https://doi.org/10.1103/physrevb.85.224512} {\bibfield  {journal} {\bibinfo
  {journal} {Phys. Rev. B}\ }\textbf {\bibinfo {volume} {85}},\ \bibinfo
  {pages} {224512} (\bibinfo {year} {2012})}\BibitemShut {NoStop}%
\bibitem [{\citenamefont {Persson}\ \emph
  {et~al.}(2010{\natexlab{a}})\citenamefont {Persson}, \citenamefont {Wilson},
  \citenamefont {Sandberg}, \citenamefont {Johansson},\ and\ \citenamefont
  {Delsing}}]{Persson2010}%
  \BibitemOpen
  \bibfield  {author} {\bibinfo {author} {\bibfnamefont {F.}~\bibnamefont
  {Persson}}, \bibinfo {author} {\bibfnamefont {C.~M.}\ \bibnamefont {Wilson}},
  \bibinfo {author} {\bibfnamefont {M.}~\bibnamefont {Sandberg}}, \bibinfo
  {author} {\bibfnamefont {G.}~\bibnamefont {Johansson}},\ and\ \bibinfo
  {author} {\bibfnamefont {P.}~\bibnamefont {Delsing}},\ }\bibfield  {title}
  {\bibinfo {title} {Excess dissipation in a single-electron box: The
  {Sisyphus} resistance},\ }\href {https://doi.org/10.1021/nl903887x}
  {\bibfield  {journal} {\bibinfo  {journal} {Nano Lett.}\ }\textbf {\bibinfo
  {volume} {10}},\ \bibinfo {pages} {953} (\bibinfo {year}
  {2010}{\natexlab{a}})}\BibitemShut {NoStop}%
\bibitem [{\citenamefont {Gonzalez-Zalba}\ \emph {et~al.}(2015)\citenamefont
  {Gonzalez-Zalba}, \citenamefont {Barraud}, \citenamefont {Ferguson},\ and\
  \citenamefont {Betz}}]{GZ2015}%
  \BibitemOpen
  \bibfield  {author} {\bibinfo {author} {\bibfnamefont {M.~F.}\ \bibnamefont
  {Gonzalez-Zalba}}, \bibinfo {author} {\bibfnamefont {S.}~\bibnamefont
  {Barraud}}, \bibinfo {author} {\bibfnamefont {A.}~\bibnamefont {Ferguson}},\
  and\ \bibinfo {author} {\bibfnamefont {A.}~\bibnamefont {Betz}},\ }\bibfield
  {title} {\bibinfo {title} {Probing the limits of gate-based charge sensing},\
  }\href {https://doi.org/10.1038/ncomms7084} {\bibfield  {journal} {\bibinfo
  {journal} {Nature Comm.}\ }\textbf {\bibinfo {volume} {6}},\ \bibinfo {pages}
  {6084} (\bibinfo {year} {2015})}\BibitemShut {NoStop}%
\bibitem [{\citenamefont {Shevchenko}\ \emph {et~al.}(2015)\citenamefont
  {Shevchenko}, \citenamefont {Rubanov},\ and\ \citenamefont
  {Nori}}]{Shevchenko2015}%
  \BibitemOpen
  \bibfield  {author} {\bibinfo {author} {\bibfnamefont {S.~N.}\ \bibnamefont
  {Shevchenko}}, \bibinfo {author} {\bibfnamefont {D.~G.}\ \bibnamefont
  {Rubanov}},\ and\ \bibinfo {author} {\bibfnamefont {F.}~\bibnamefont
  {Nori}},\ }\bibfield  {title} {\bibinfo {title} {Delayed-response quantum
  back action in nanoelectromechanical systems},\ }\href
  {https://doi.org/10.1103/physrevb.91.165422} {\bibfield  {journal} {\bibinfo
  {journal} {Phys. Rev. B}\ }\textbf {\bibinfo {volume} {91}},\ \bibinfo
  {pages} {165422} (\bibinfo {year} {2015})}\BibitemShut {NoStop}%
\bibitem [{\citenamefont {Peri}\ \emph
  {et~al.}(2024{\natexlab{a}})\citenamefont {Peri}, \citenamefont {Benito},
  \citenamefont {Ford},\ and\ \citenamefont {Gonzalez-Zalba}}]{Peri2023b}%
  \BibitemOpen
  \bibfield  {author} {\bibinfo {author} {\bibfnamefont {L.}~\bibnamefont
  {Peri}}, \bibinfo {author} {\bibfnamefont {M.}~\bibnamefont {Benito}},
  \bibinfo {author} {\bibfnamefont {C.~J.~B.}\ \bibnamefont {Ford}},\ and\
  \bibinfo {author} {\bibfnamefont {M.~F.}\ \bibnamefont {Gonzalez-Zalba}},\
  }\bibfield  {title} {\bibinfo {title} {Unified linear response theory of
  quantum dot circuits},\ }\href
  {https://doi.org/https://doi.org/10.1038/s41534-024-00907-9} {\bibfield
  {journal} {\bibinfo  {journal} {npj Quantum Inf.}\ }\textbf {\bibinfo
  {volume} {10}},\ \bibinfo {pages} {114} (\bibinfo {year}
  {2024}{\natexlab{a}})}\BibitemShut {NoStop}%
\bibitem [{\citenamefont {Grajcar}\ \emph {et~al.}(2008)\citenamefont
  {Grajcar}, \citenamefont {van~der Ploeg}, \citenamefont {Izmalkov},
  \citenamefont {Il’ichev}, \citenamefont {Meyer}, \citenamefont {Fedorov},
  \citenamefont {Shnirman},\ and\ \citenamefont {Schön}}]{Grajcar2008}%
  \BibitemOpen
  \bibfield  {author} {\bibinfo {author} {\bibfnamefont {M.}~\bibnamefont
  {Grajcar}}, \bibinfo {author} {\bibfnamefont {S.~H.~W.}\ \bibnamefont
  {van~der Ploeg}}, \bibinfo {author} {\bibfnamefont {A.}~\bibnamefont
  {Izmalkov}}, \bibinfo {author} {\bibfnamefont {E.}~\bibnamefont
  {Il’ichev}}, \bibinfo {author} {\bibfnamefont {H.-G.}\ \bibnamefont
  {Meyer}}, \bibinfo {author} {\bibfnamefont {A.}~\bibnamefont {Fedorov}},
  \bibinfo {author} {\bibfnamefont {A.}~\bibnamefont {Shnirman}},\ and\
  \bibinfo {author} {\bibfnamefont {G.}~\bibnamefont {Schön}},\ }\bibfield
  {title} {\bibinfo {title} {Sisyphus cooling and amplification by a
  superconducting qubit},\ }\href {https://doi.org/10.1038/nphys1019}
  {\bibfield  {journal} {\bibinfo  {journal} {Nature Phys.}\ }\textbf {\bibinfo
  {volume} {4}},\ \bibinfo {pages} {612} (\bibinfo {year} {2008})}\BibitemShut
  {NoStop}%
\bibitem [{\citenamefont {Malinowski}\ \emph {et~al.}(2022)\citenamefont
  {Malinowski}, \citenamefont {Rupesh}, \citenamefont {Pavešić},
  \citenamefont {Guba}, \citenamefont {de~Jong}, \citenamefont {Han},
  \citenamefont {Prosko}, \citenamefont {Chan}, \citenamefont {Liu},
  \citenamefont {Krogstrup}, \citenamefont {Pályi}, \citenamefont {Zitko},\
  and\ \citenamefont {Koski}}]{Malinowski2022}%
  \BibitemOpen
  \bibfield  {author} {\bibinfo {author} {\bibfnamefont {F.~K.}\ \bibnamefont
  {Malinowski}}, \bibinfo {author} {\bibfnamefont {R.~K.}\ \bibnamefont
  {Rupesh}}, \bibinfo {author} {\bibfnamefont {L.}~\bibnamefont {Pavešić}},
  \bibinfo {author} {\bibfnamefont {Z.}~\bibnamefont {Guba}}, \bibinfo {author}
  {\bibfnamefont {D.}~\bibnamefont {de~Jong}}, \bibinfo {author} {\bibfnamefont
  {L.}~\bibnamefont {Han}}, \bibinfo {author} {\bibfnamefont {C.~G.}\
  \bibnamefont {Prosko}}, \bibinfo {author} {\bibfnamefont {M.}~\bibnamefont
  {Chan}}, \bibinfo {author} {\bibfnamefont {Y.}~\bibnamefont {Liu}}, \bibinfo
  {author} {\bibfnamefont {P.}~\bibnamefont {Krogstrup}}, \bibinfo {author}
  {\bibfnamefont {A.}~\bibnamefont {Pályi}}, \bibinfo {author} {\bibfnamefont
  {R.}~\bibnamefont {Zitko}},\ and\ \bibinfo {author} {\bibfnamefont {J.~V.}\
  \bibnamefont {Koski}},\ }\bibfield  {title} {\bibinfo {title} {Quantum
  capacitance of a superconducting subgap state in an electrostatically
  floating dot-island},\ }\href {https://arxiv.org/abs/2210.01519} {\
  (\bibinfo {year} {2022})},\ \Eprint {https://arxiv.org/abs/2210.01519}
  {arXiv:2210.01519 [cond-mat.mes-hall]} \BibitemShut {NoStop}%
\bibitem [{\citenamefont {Cohen-Tannoudji}\ and\ \citenamefont
  {Phillips}(1990)}]{CohenTannoudji1990}%
  \BibitemOpen
  \bibfield  {author} {\bibinfo {author} {\bibfnamefont {C.~N.}\ \bibnamefont
  {Cohen-Tannoudji}}\ and\ \bibinfo {author} {\bibfnamefont {W.~D.}\
  \bibnamefont {Phillips}},\ }\bibfield  {title} {\bibinfo {title} {New
  mechanisms for laser cooling},\ }\href {https://doi.org/10.1063/1.881239}
  {\bibfield  {journal} {\bibinfo  {journal} {Physics Today}\ }\textbf
  {\bibinfo {volume} {43}},\ \bibinfo {pages} {33} (\bibinfo {year}
  {1990})}\BibitemShut {NoStop}%
\bibitem [{\citenamefont {Bogolyubov}\ and\ \citenamefont
  {Mitropol'skii}(1962)}]{Bogolyubov-Mitropolskii}%
  \BibitemOpen
  \bibfield  {author} {\bibinfo {author} {\bibfnamefont {N.~N.}\ \bibnamefont
  {Bogolyubov}}\ and\ \bibinfo {author} {\bibfnamefont {Y.~A.}\ \bibnamefont
  {Mitropol'skii}},\ }\href@noop {} {\emph {\bibinfo {title} {Asymptotic
  Methods in the Theory of Nonlinear Oscillations}}}\ (\bibinfo  {publisher}
  {Gordon and Breach},\ \bibinfo {year} {1962})\BibitemShut {NoStop}%
\bibitem [{\citenamefont {Shevchenko}\ \emph
  {et~al.}(2012{\natexlab{a}})\citenamefont {Shevchenko}, \citenamefont
  {Omelyanchouk},\ and\ \citenamefont {Il'ichev}}]{Shevchenko2012}%
  \BibitemOpen
  \bibfield  {author} {\bibinfo {author} {\bibfnamefont {S.~N.}\ \bibnamefont
  {Shevchenko}}, \bibinfo {author} {\bibfnamefont {A.~N.}\ \bibnamefont
  {Omelyanchouk}},\ and\ \bibinfo {author} {\bibfnamefont {E.}~\bibnamefont
  {Il'ichev}},\ }\bibfield  {title} {\bibinfo {title} {{Multiphoton transitions
  in Josephson-junction qubits (Review Article)}},\ }\href
  {https://doi.org/10.1063/1.3701717} {\bibfield  {journal} {\bibinfo
  {journal} {Low Temp. Phys.}\ }\textbf {\bibinfo {volume} {38}},\ \bibinfo
  {pages} {283} (\bibinfo {year} {2012}{\natexlab{a}})}\BibitemShut {NoStop}%
\bibitem [{\citenamefont {Gonzalez-Zalba}\ \emph {et~al.}(2016)\citenamefont
  {Gonzalez-Zalba}, \citenamefont {Shevchenko}, \citenamefont {Barraud},
  \citenamefont {Johansson}, \citenamefont {Ferguson}, \citenamefont {Nori},\
  and\ \citenamefont {Betz}}]{Gonzalez-Zalba2016}%
  \BibitemOpen
  \bibfield  {author} {\bibinfo {author} {\bibfnamefont {M.~F.}\ \bibnamefont
  {Gonzalez-Zalba}}, \bibinfo {author} {\bibfnamefont {S.~N.}\ \bibnamefont
  {Shevchenko}}, \bibinfo {author} {\bibfnamefont {S.}~\bibnamefont {Barraud}},
  \bibinfo {author} {\bibfnamefont {J.~R.}\ \bibnamefont {Johansson}}, \bibinfo
  {author} {\bibfnamefont {A.~J.}\ \bibnamefont {Ferguson}}, \bibinfo {author}
  {\bibfnamefont {F.}~\bibnamefont {Nori}},\ and\ \bibinfo {author}
  {\bibfnamefont {A.~C.}\ \bibnamefont {Betz}},\ }\bibfield  {title} {\bibinfo
  {title} {{Gate-sensing coherent charge oscillations in a silicon field-effect
  transistor}},\ }\href {https://doi.org/10.1021/acs.nanolett.5b04356}
  {\bibfield  {journal} {\bibinfo  {journal} {Nano Lett.}\ }\textbf {\bibinfo
  {volume} {16}},\ \bibinfo {pages} {1614} (\bibinfo {year}
  {2016})}\BibitemShut {NoStop}%
\bibitem [{\citenamefont {Mizuta}\ \emph {et~al.}(2017)\citenamefont {Mizuta},
  \citenamefont {Otxoa}, \citenamefont {Betz},\ and\ \citenamefont
  {Gonzalez-Zalba}}]{Mizuta2017}%
  \BibitemOpen
  \bibfield  {author} {\bibinfo {author} {\bibfnamefont {R.}~\bibnamefont
  {Mizuta}}, \bibinfo {author} {\bibfnamefont {R.~M.}\ \bibnamefont {Otxoa}},
  \bibinfo {author} {\bibfnamefont {A.~C.}\ \bibnamefont {Betz}},\ and\
  \bibinfo {author} {\bibfnamefont {M.~F.}\ \bibnamefont {Gonzalez-Zalba}},\
  }\bibfield  {title} {\bibinfo {title} {Quantum and tunneling capacitance in
  charge and spin qubits},\ }\href {https://doi.org/10.1103/PhysRevB.95.045414}
  {\bibfield  {journal} {\bibinfo  {journal} {Phys. Rev. B}\ }\textbf {\bibinfo
  {volume} {95}},\ \bibinfo {pages} {045414} (\bibinfo {year}
  {2017})}\BibitemShut {NoStop}%
\bibitem [{\citenamefont {Esterli}\ \emph {et~al.}(2019)\citenamefont
  {Esterli}, \citenamefont {Otxoa},\ and\ \citenamefont
  {Gonzalez-Zalba}}]{Esterli2019}%
  \BibitemOpen
  \bibfield  {author} {\bibinfo {author} {\bibfnamefont {M.}~\bibnamefont
  {Esterli}}, \bibinfo {author} {\bibfnamefont {R.~M.}\ \bibnamefont {Otxoa}},\
  and\ \bibinfo {author} {\bibfnamefont {M.~F.}\ \bibnamefont
  {Gonzalez-Zalba}},\ }\bibfield  {title} {\bibinfo {title} {Small-signal
  equivalent circuit for double quantum dots at low-frequencies},\ }\href
  {https://doi.org/10.1063/1.5098889} {\bibfield  {journal} {\bibinfo
  {journal} {Appl. Phys. Lett.}\ }\textbf {\bibinfo {volume} {114}},\ \bibinfo
  {pages} {253505} (\bibinfo {year} {2019})}\BibitemShut {NoStop}%
\bibitem [{\citenamefont {Sillanpää}\ \emph {et~al.}(2005)\citenamefont
  {Sillanpää}, \citenamefont {Lehtinen}, \citenamefont {Paila}, \citenamefont
  {Makhlin}, \citenamefont {Roschier},\ and\ \citenamefont
  {Hakonen}}]{Sillanpaeae2005}%
  \BibitemOpen
  \bibfield  {author} {\bibinfo {author} {\bibfnamefont {M.~A.}\ \bibnamefont
  {Sillanpää}}, \bibinfo {author} {\bibfnamefont {T.}~\bibnamefont
  {Lehtinen}}, \bibinfo {author} {\bibfnamefont {A.}~\bibnamefont {Paila}},
  \bibinfo {author} {\bibfnamefont {Y.}~\bibnamefont {Makhlin}}, \bibinfo
  {author} {\bibfnamefont {L.}~\bibnamefont {Roschier}},\ and\ \bibinfo
  {author} {\bibfnamefont {P.~J.}\ \bibnamefont {Hakonen}},\ }\bibfield
  {title} {\bibinfo {title} {Direct observation of {J}osephson capacitance},\
  }\href {https://doi.org/10.1103/physrevlett.95.206806} {\bibfield  {journal}
  {\bibinfo  {journal} {Phys. Rev. Lett.}\ }\textbf {\bibinfo {volume} {95}},\
  \bibinfo {pages} {206806} (\bibinfo {year} {2005})}\BibitemShut {NoStop}%
\bibitem [{\citenamefont {Duty}\ \emph {et~al.}(2005)\citenamefont {Duty},
  \citenamefont {Johansson}, \citenamefont {Bladh}, \citenamefont {Gunnarsson},
  \citenamefont {Wilson},\ and\ \citenamefont {Delsing}}]{Duty2005}%
  \BibitemOpen
  \bibfield  {author} {\bibinfo {author} {\bibfnamefont {T.}~\bibnamefont
  {Duty}}, \bibinfo {author} {\bibfnamefont {G.}~\bibnamefont {Johansson}},
  \bibinfo {author} {\bibfnamefont {K.}~\bibnamefont {Bladh}}, \bibinfo
  {author} {\bibfnamefont {D.}~\bibnamefont {Gunnarsson}}, \bibinfo {author}
  {\bibfnamefont {C.}~\bibnamefont {Wilson}},\ and\ \bibinfo {author}
  {\bibfnamefont {P.}~\bibnamefont {Delsing}},\ }\bibfield  {title} {\bibinfo
  {title} {Observation of quantum capacitance in the {C}ooper-pair
  transistor},\ }\href {https://doi.org/10.1103/PhysRevLett.95.206807}
  {\bibfield  {journal} {\bibinfo  {journal} {Phys. Rev. Lett.}\ }\textbf
  {\bibinfo {volume} {95}},\ \bibinfo {pages} {206807} (\bibinfo {year}
  {2005})}\BibitemShut {NoStop}%
\bibitem [{\citenamefont {Ivakhnenko}\ \emph {et~al.}(2023)\citenamefont
  {Ivakhnenko}, \citenamefont {Shevchenko},\ and\ \citenamefont
  {Nori}}]{Ivakhnenko2023}%
  \BibitemOpen
  \bibfield  {author} {\bibinfo {author} {\bibfnamefont {O.~V.}\ \bibnamefont
  {Ivakhnenko}}, \bibinfo {author} {\bibfnamefont {S.~N.}\ \bibnamefont
  {Shevchenko}},\ and\ \bibinfo {author} {\bibfnamefont {F.}~\bibnamefont
  {Nori}},\ }\bibfield  {title} {\bibinfo {title} {{Nonadiabatic
  Landau-Zener-Stückelberg-Majorana transitions, dynamics, and
  interference}},\ }\href {https://doi.org/10.1016/j.physrep.2022.10.002}
  {\bibfield  {journal} {\bibinfo  {journal} {Phys. Rep.}\ }\textbf {\bibinfo
  {volume} {995}},\ \bibinfo {pages} {1} (\bibinfo {year} {2023})}\BibitemShut
  {NoStop}%
\bibitem [{\citenamefont {Xia}\ \emph {et~al.}(2009)\citenamefont {Xia},
  \citenamefont {Chen}, \citenamefont {Li},\ and\ \citenamefont
  {Tao}}]{Xia2009}%
  \BibitemOpen
  \bibfield  {author} {\bibinfo {author} {\bibfnamefont {J.}~\bibnamefont
  {Xia}}, \bibinfo {author} {\bibfnamefont {F.}~\bibnamefont {Chen}}, \bibinfo
  {author} {\bibfnamefont {J.}~\bibnamefont {Li}},\ and\ \bibinfo {author}
  {\bibfnamefont {N.}~\bibnamefont {Tao}},\ }\bibfield  {title} {\bibinfo
  {title} {Measurement of the quantum capacitance of graphene},\ }\href
  {https://doi.org/10.1038/nnano.2009.177} {\bibfield  {journal} {\bibinfo
  {journal} {Nature Nanotech.}\ }\textbf {\bibinfo {volume} {4}},\ \bibinfo
  {pages} {505} (\bibinfo {year} {2009})}\BibitemShut {NoStop}%
\bibitem [{\citenamefont {Ilani}\ \emph {et~al.}(2006)\citenamefont {Ilani},
  \citenamefont {Donev}, \citenamefont {Kindermann},\ and\ \citenamefont
  {McEuen}}]{Ilani2006}%
  \BibitemOpen
  \bibfield  {author} {\bibinfo {author} {\bibfnamefont {S.}~\bibnamefont
  {Ilani}}, \bibinfo {author} {\bibfnamefont {L.~A.~K.}\ \bibnamefont {Donev}},
  \bibinfo {author} {\bibfnamefont {M.}~\bibnamefont {Kindermann}},\ and\
  \bibinfo {author} {\bibfnamefont {P.~L.}\ \bibnamefont {McEuen}},\ }\bibfield
   {title} {\bibinfo {title} {Measurement of the quantum capacitance of
  interacting electrons in carbon nanotubes},\ }\href
  {https://doi.org/10.1038/nphys412} {\bibfield  {journal} {\bibinfo  {journal}
  {Nature Phys.}\ }\textbf {\bibinfo {volume} {2}},\ \bibinfo {pages} {687}
  (\bibinfo {year} {2006})}\BibitemShut {NoStop}%
\bibitem [{\citenamefont {{Microsoft Azure Quantum, Aghaee, M., Alcaraz
  Ramirez, A. et al.}}(2025)}]{Microsoft2025}%
  \BibitemOpen
  \bibfield  {author} {\bibinfo {author} {\bibnamefont {{Microsoft Azure
  Quantum, Aghaee, M., Alcaraz Ramirez, A. et al.}}},\ }\bibfield  {title}
  {\bibinfo {title} {Interferometric single-shot parity measurement in
  {InAs-Al} hybrid devices},\ }\href
  {https://doi.org/10.1038/s41586-024-08445-2} {\bibfield  {journal} {\bibinfo
  {journal} {Nature}\ }\textbf {\bibinfo {volume} {638}},\ \bibinfo {pages}
  {651} (\bibinfo {year} {2025})}\BibitemShut {NoStop}%
\bibitem [{\citenamefont {Jennings}\ \emph {et~al.}(2025)\citenamefont
  {Jennings}, \citenamefont {Grytsenko}, \citenamefont {Tian}, \citenamefont
  {Rybalko}, \citenamefont {Wang}, \citenamefont {Barabash},\ and\
  \citenamefont {Kawakami}}]{Jennings2025}%
  \BibitemOpen
  \bibfield  {author} {\bibinfo {author} {\bibfnamefont {A.}~\bibnamefont
  {Jennings}}, \bibinfo {author} {\bibfnamefont {I.}~\bibnamefont {Grytsenko}},
  \bibinfo {author} {\bibfnamefont {Y.}~\bibnamefont {Tian}}, \bibinfo {author}
  {\bibfnamefont {O.}~\bibnamefont {Rybalko}}, \bibinfo {author} {\bibfnamefont
  {J.}~\bibnamefont {Wang}}, \bibinfo {author} {\bibfnamefont {I.~J.}\
  \bibnamefont {Barabash}},\ and\ \bibinfo {author} {\bibfnamefont
  {E.}~\bibnamefont {Kawakami}},\ }\bibfield  {title} {\bibinfo {title}
  {Probing the quantum capacitance of {R}ydberg transitions of surface
  electrons on liquid helium via microwave frequency modulation},\ }\href
  {https://doi.org/10.1103/5y8p-qhb4} {\bibfield  {journal} {\bibinfo
  {journal} {Phys. Rev. Lett.}\ }\textbf {\bibinfo {volume} {135}},\ \bibinfo
  {pages} {087001} (\bibinfo {year} {2025})}\BibitemShut {NoStop}%
\bibitem [{\citenamefont {Prosko}\ \emph {et~al.}(2024)\citenamefont {Prosko},
  \citenamefont {Kulesh}, \citenamefont {Chan}, \citenamefont {Han},
  \citenamefont {Xiao}, \citenamefont {Thomas}, \citenamefont {Manfra},
  \citenamefont {Goswami},\ and\ \citenamefont {Malinowski}}]{Prosko2024}%
  \BibitemOpen
  \bibfield  {author} {\bibinfo {author} {\bibfnamefont {C.~G.}\ \bibnamefont
  {Prosko}}, \bibinfo {author} {\bibfnamefont {I.}~\bibnamefont {Kulesh}},
  \bibinfo {author} {\bibfnamefont {M.}~\bibnamefont {Chan}}, \bibinfo {author}
  {\bibfnamefont {L.}~\bibnamefont {Han}}, \bibinfo {author} {\bibfnamefont
  {D.}~\bibnamefont {Xiao}}, \bibinfo {author} {\bibfnamefont {C.}~\bibnamefont
  {Thomas}}, \bibinfo {author} {\bibfnamefont {M.~J.}\ \bibnamefont {Manfra}},
  \bibinfo {author} {\bibfnamefont {S.}~\bibnamefont {Goswami}},\ and\ \bibinfo
  {author} {\bibfnamefont {F.~K.}\ \bibnamefont {Malinowski}},\ }\bibfield
  {title} {\bibinfo {title} {{Flux-tunable hybridization in a double quantum
  dot interferometer}},\ }\href {https://doi.org/10.21468/SciPostPhys.17.3.074}
  {\bibfield  {journal} {\bibinfo  {journal} {SciPost Phys.}\ }\textbf
  {\bibinfo {volume} {17}},\ \bibinfo {pages} {074} (\bibinfo {year}
  {2024})}\BibitemShut {NoStop}%
\bibitem [{\citenamefont {Secchi}\ and\ \citenamefont
  {Troiani}(2022)}]{Secchi2023}%
  \BibitemOpen
  \bibfield  {author} {\bibinfo {author} {\bibfnamefont {A.}~\bibnamefont
  {Secchi}}\ and\ \bibinfo {author} {\bibfnamefont {F.}~\bibnamefont
  {Troiani}},\ }\bibfield  {title} {\bibinfo {title} {Multi-dimensional quantum
  capacitance of the two-site {H}ubbard model: The role of tunable interdot
  tunneling},\ }\href {https://doi.org/10.3390/e25010082} {\bibfield  {journal}
  {\bibinfo  {journal} {Entropy}\ }\textbf {\bibinfo {volume} {25}},\ \bibinfo
  {pages} {82} (\bibinfo {year} {2022})}\BibitemShut {NoStop}%
\bibitem [{\citenamefont {Secchi}\ and\ \citenamefont
  {Troiani}(2023)}]{Secchi2023_1}%
  \BibitemOpen
  \bibfield  {author} {\bibinfo {author} {\bibfnamefont {A.}~\bibnamefont
  {Secchi}}\ and\ \bibinfo {author} {\bibfnamefont {F.}~\bibnamefont
  {Troiani}},\ }\bibfield  {title} {\bibinfo {title} {Theory of
  multidimensional quantum capacitance and its application to spin and charge
  discrimination in quantum dot arrays},\ }\href
  {https://doi.org/10.1103/physrevb.107.155411} {\bibfield  {journal} {\bibinfo
   {journal} {Phys. Rev. B}\ }\textbf {\bibinfo {volume} {107}},\ \bibinfo
  {pages} {155411} (\bibinfo {year} {2023})}\BibitemShut {NoStop}%
\bibitem [{\citenamefont {Shevchenko}(2008)}]{Shevchenko2008b}%
  \BibitemOpen
  \bibfield  {author} {\bibinfo {author} {\bibfnamefont {S.~N.}\ \bibnamefont
  {Shevchenko}},\ }\bibfield  {title} {\bibinfo {title} {Impedance measurement
  technique for quantum systems},\ }\href
  {https://doi.org/10.1140/epjb/e2008-00061-9} {\bibfield  {journal} {\bibinfo
  {journal} {Eur. Phys. J. B}\ }\textbf {\bibinfo {volume} {61}},\ \bibinfo
  {pages} {187} (\bibinfo {year} {2008})}\BibitemShut {NoStop}%
\bibitem [{\citenamefont {Nori}(2008)}]{Nori2008}%
  \BibitemOpen
  \bibfield  {author} {\bibinfo {author} {\bibfnamefont {F.}~\bibnamefont
  {Nori}},\ }\bibfield  {title} {\bibinfo {title} {Atomic physics with a
  circuit},\ }\href {https://doi.org/10.1038/nphys1044} {\bibfield  {journal}
  {\bibinfo  {journal} {Nat. Phys.}\ }\textbf {\bibinfo {volume} {4}},\
  \bibinfo {pages} {589} (\bibinfo {year} {2008})}\BibitemShut {NoStop}%
\bibitem [{\citenamefont {Peri}\ \emph
  {et~al.}(2024{\natexlab{b}})\citenamefont {Peri}, \citenamefont {Oakes},
  \citenamefont {Cochrane}, \citenamefont {Ford},\ and\ \citenamefont
  {Gonzalez-Zalba}}]{Peri2023a}%
  \BibitemOpen
  \bibfield  {author} {\bibinfo {author} {\bibfnamefont {L.}~\bibnamefont
  {Peri}}, \bibinfo {author} {\bibfnamefont {G.~A.}\ \bibnamefont {Oakes}},
  \bibinfo {author} {\bibfnamefont {L.}~\bibnamefont {Cochrane}}, \bibinfo
  {author} {\bibfnamefont {C.~J.~B.}\ \bibnamefont {Ford}},\ and\ \bibinfo
  {author} {\bibfnamefont {M.~F.}\ \bibnamefont {Gonzalez-Zalba}},\ }\bibfield
  {title} {\bibinfo {title} {Beyond-adiabatic {Q}uantum {A}dmittance of a
  {S}emiconductor {Q}uantum {D}ot at {H}igh {F}requencies: {R}ethinking
  {R}eflectometry as {P}olaron {D}ynamics},\ }\href
  {https://doi.org/10.22331/q-2024-03-21-1294} {\bibfield  {journal} {\bibinfo
  {journal} {{Quantum}}\ }\textbf {\bibinfo {volume} {8}},\ \bibinfo {pages}
  {1294} (\bibinfo {year} {2024}{\natexlab{b}})}\BibitemShut {NoStop}%
\bibitem [{\citenamefont {Peri}\ \emph {et~al.}(2025)\citenamefont {Peri},
  \citenamefont {Gomez-Saiz}, \citenamefont {Ford},\ and\ \citenamefont
  {Gonzalez-Zalba}}]{Peri2025}%
  \BibitemOpen
  \bibfield  {author} {\bibinfo {author} {\bibfnamefont {L.}~\bibnamefont
  {Peri}}, \bibinfo {author} {\bibfnamefont {A.}~\bibnamefont {Gomez-Saiz}},
  \bibinfo {author} {\bibfnamefont {C.~J.~B.}\ \bibnamefont {Ford}},\ and\
  \bibinfo {author} {\bibfnamefont {M.~F.}\ \bibnamefont {Gonzalez-Zalba}},\
  }\bibfield  {title} {\bibinfo {title} {Compact quantum dot models for analog
  microwave co-simulation},\ }\href
  {https://doi.org/10.1038/s41534-025-01140-8} {\bibfield  {journal} {\bibinfo
  {journal} {npj Quantum Information}\ }\textbf {\bibinfo {volume} {11}},\
  \bibinfo {pages} {194} (\bibinfo {year} {2025})}\BibitemShut {NoStop}%
\bibitem [{\citenamefont {Breuer}\ and\ \citenamefont
  {Petruccione}(2010)}]{Breuer2010}%
  \BibitemOpen
  \bibfield  {author} {\bibinfo {author} {\bibfnamefont {H.-P.}\ \bibnamefont
  {Breuer}}\ and\ \bibinfo {author} {\bibfnamefont {F.}~\bibnamefont
  {Petruccione}},\ }\href@noop {} {\emph {\bibinfo {title} {The theory of open
  quantum systems}}},\ \bibinfo {edition} {repr.}\ ed.\ (\bibinfo  {publisher}
  {Clarendon Press},\ \bibinfo {address} {Oxford},\ \bibinfo {year}
  {2010})\BibitemShut {NoStop}%
\bibitem [{\citenamefont {Lidar}(2019)}]{Lidar2019}%
  \BibitemOpen
  \bibfield  {author} {\bibinfo {author} {\bibfnamefont {D.~A.}\ \bibnamefont
  {Lidar}},\ }\bibfield  {title} {\bibinfo {title} {Lecture notes on the theory
  of open quantum systems},\ }\href@noop {} {\  (\bibinfo {year} {2019})},\
  \Eprint {https://arxiv.org/abs/1902.00967} {arXiv:1902.00967 [quant-ph]}
  \BibitemShut {NoStop}%
\bibitem [{\citenamefont {Manzano}(2020)}]{Manzano2020}%
  \BibitemOpen
  \bibfield  {author} {\bibinfo {author} {\bibfnamefont {D.}~\bibnamefont
  {Manzano}},\ }\bibfield  {title} {\bibinfo {title} {A short introduction to
  the lindblad master equation},\ }\href {https://doi.org/10.1063/1.5115323}
  {\bibfield  {journal} {\bibinfo  {journal} {AIP Adv.}\ }\textbf {\bibinfo
  {volume} {10}},\ \bibinfo {pages} {025106} (\bibinfo {year}
  {2020})}\BibitemShut {NoStop}%
\bibitem [{\citenamefont {{Minganti, Fabrizio}}\ and\ \citenamefont {{Biella,
  Alberto}}(2026)}]{Minganti2024a}%
  \BibitemOpen
  \bibfield  {author} {\bibinfo {author} {\bibnamefont {{Minganti, Fabrizio}}}\
  and\ \bibinfo {author} {\bibnamefont {{Biella, Alberto}}},\ }\bibfield
  {title} {\bibinfo {title} {Open quantum systems - a brief introduction},\
  }\href {https://doi.org/10.1051/cipa/202602002} {\bibfield  {journal}
  {\bibinfo  {journal} {Cahiers de l\'{}IPa}\ }\textbf {\bibinfo {volume}
  {2}},\ \bibinfo {pages} {2} (\bibinfo {year} {2026})}\BibitemShut {NoStop}%
\bibitem [{\citenamefont {Shevchenko}\ \emph
  {et~al.}(2012{\natexlab{b}})\citenamefont {Shevchenko}, \citenamefont
  {Ashhab},\ and\ \citenamefont {Nori}}]{Shevchenko2012a}%
  \BibitemOpen
  \bibfield  {author} {\bibinfo {author} {\bibfnamefont {S.~N.}\ \bibnamefont
  {Shevchenko}}, \bibinfo {author} {\bibfnamefont {S.}~\bibnamefont {Ashhab}},\
  and\ \bibinfo {author} {\bibfnamefont {F.}~\bibnamefont {Nori}},\ }\bibfield
  {title} {\bibinfo {title} {{Inverse Landau-Zener-Stückelberg problem for
  qubit-resonator systems}},\ }\href
  {https://doi.org/10.1103/physrevb.85.094502} {\bibfield  {journal} {\bibinfo
  {journal} {Phys. Rev. B}\ }\textbf {\bibinfo {volume} {85}},\ \bibinfo
  {pages} {094502} (\bibinfo {year} {2012}{\natexlab{b}})}\BibitemShut
  {NoStop}%
\bibitem [{\citenamefont {Lundberg}\ \emph {et~al.}(2024)\citenamefont
  {Lundberg}, \citenamefont {Ibberson}, \citenamefont {Li}, \citenamefont
  {Hutin}, \citenamefont {Abadillo-Uriel}, \citenamefont {Filippone},
  \citenamefont {Bertrand}, \citenamefont {Nunnenkamp}, \citenamefont {Lee},
  \citenamefont {Stelmashenko}, \citenamefont {Robinson}, \citenamefont
  {Vinet}, \citenamefont {Ibberson}, \citenamefont {Niquet},\ and\
  \citenamefont {Gonzalez-Zalba}}]{Lundberg2024}%
  \BibitemOpen
  \bibfield  {author} {\bibinfo {author} {\bibfnamefont {T.}~\bibnamefont
  {Lundberg}}, \bibinfo {author} {\bibfnamefont {D.~J.}\ \bibnamefont
  {Ibberson}}, \bibinfo {author} {\bibfnamefont {J.}~\bibnamefont {Li}},
  \bibinfo {author} {\bibfnamefont {L.}~\bibnamefont {Hutin}}, \bibinfo
  {author} {\bibfnamefont {J.~C.}\ \bibnamefont {Abadillo-Uriel}}, \bibinfo
  {author} {\bibfnamefont {M.}~\bibnamefont {Filippone}}, \bibinfo {author}
  {\bibfnamefont {B.}~\bibnamefont {Bertrand}}, \bibinfo {author}
  {\bibfnamefont {A.}~\bibnamefont {Nunnenkamp}}, \bibinfo {author}
  {\bibfnamefont {C.-M.}\ \bibnamefont {Lee}}, \bibinfo {author} {\bibfnamefont
  {N.}~\bibnamefont {Stelmashenko}}, \bibinfo {author} {\bibfnamefont
  {J.~W.~A.}\ \bibnamefont {Robinson}}, \bibinfo {author} {\bibfnamefont
  {M.}~\bibnamefont {Vinet}}, \bibinfo {author} {\bibfnamefont
  {L.}~\bibnamefont {Ibberson}}, \bibinfo {author} {\bibfnamefont {Y.-M.}\
  \bibnamefont {Niquet}},\ and\ \bibinfo {author} {\bibfnamefont {M.~F.}\
  \bibnamefont {Gonzalez-Zalba}},\ }\bibfield  {title} {\bibinfo {title}
  {Non-symmetric {P}auli spin blockade in a silicon double quantum dot},\
  }\href {https://doi.org/10.1038/s41534-024-00820-1} {\bibfield  {journal}
  {\bibinfo  {journal} {npj Quantum Information}\ }\textbf {\bibinfo {volume}
  {10}},\ \bibinfo {pages} {28} (\bibinfo {year} {2024})}\BibitemShut {NoStop}%
\bibitem [{\citenamefont {Landau}\ and\ \citenamefont
  {Lifshitz}(1976)}]{Landau1976}%
  \BibitemOpen
  \bibfield  {author} {\bibinfo {author} {\bibfnamefont {L.~D.}\ \bibnamefont
  {Landau}}\ and\ \bibinfo {author} {\bibfnamefont {E.~M.}\ \bibnamefont
  {Lifshitz}},\ }\href@noop {} {\emph {\bibinfo {title} {{M}echanics}}},\
  \bibinfo {edition} {3rd}\ ed.,\ Vol.~\bibinfo {volume} {1}\ (\bibinfo
  {publisher} {Elsevier, Butterworth-Heinemann},\ \bibinfo {address}
  {Amsterdam},\ \bibinfo {year} {1976})\BibitemShut {NoStop}%
\bibitem [{\citenamefont {Persson}\ \emph
  {et~al.}(2010{\natexlab{b}})\citenamefont {Persson}, \citenamefont {Wilson},
  \citenamefont {Sandberg},\ and\ \citenamefont {Delsing}}]{Persson2010_2}%
  \BibitemOpen
  \bibfield  {author} {\bibinfo {author} {\bibfnamefont {F.}~\bibnamefont
  {Persson}}, \bibinfo {author} {\bibfnamefont {C.~M.}\ \bibnamefont {Wilson}},
  \bibinfo {author} {\bibfnamefont {M.}~\bibnamefont {Sandberg}},\ and\
  \bibinfo {author} {\bibfnamefont {P.}~\bibnamefont {Delsing}},\ }\bibfield
  {title} {\bibinfo {title} {{Fast readout of a single Cooper-pair box using
  its quantum capacitance}},\ }\href
  {https://doi.org/10.1103/physrevb.82.134533} {\bibfield  {journal} {\bibinfo
  {journal} {Phys. Rev. B}\ }\textbf {\bibinfo {volume} {82}},\ \bibinfo
  {pages} {134533} (\bibinfo {year} {2010}{\natexlab{b}})}\BibitemShut
  {NoStop}%
\bibitem [{\citenamefont {LaHaye}\ \emph {et~al.}(2009)\citenamefont {LaHaye},
  \citenamefont {Suh}, \citenamefont {Echternach}, \citenamefont {Schwab},\
  and\ \citenamefont {Roukes}}]{LaHaye2009}%
  \BibitemOpen
  \bibfield  {author} {\bibinfo {author} {\bibfnamefont {M.~D.}\ \bibnamefont
  {LaHaye}}, \bibinfo {author} {\bibfnamefont {J.}~\bibnamefont {Suh}},
  \bibinfo {author} {\bibfnamefont {P.~M.}\ \bibnamefont {Echternach}},
  \bibinfo {author} {\bibfnamefont {K.~C.}\ \bibnamefont {Schwab}},\ and\
  \bibinfo {author} {\bibfnamefont {M.~L.}\ \bibnamefont {Roukes}},\ }\bibfield
   {title} {\bibinfo {title} {Nanomechanical measurements of a superconducting
  qubit},\ }\href {https://doi.org/10.1038/nature08093} {\bibfield  {journal}
  {\bibinfo  {journal} {Nature}\ }\textbf {\bibinfo {volume} {459}},\ \bibinfo
  {pages} {960} (\bibinfo {year} {2009})}\BibitemShut {NoStop}%
\bibitem [{\citenamefont {Barajas}\ and\ \citenamefont
  {Campbell}(2026)}]{Barajas2026}%
  \BibitemOpen
  \bibfield  {author} {\bibinfo {author} {\bibfnamefont {K.~D.}\ \bibnamefont
  {Barajas}}\ and\ \bibinfo {author} {\bibfnamefont {W.~C.}\ \bibnamefont
  {Campbell}},\ }\bibfield  {title} {\bibinfo {title} {Quantum averaging theory
  for multitimescale driven quantum systems},\ }\href
  {https://doi.org/10.1103/jjvf-dw3l} {\bibfield  {journal} {\bibinfo
  {journal} {Phys. Rev. A}\ }\textbf {\bibinfo {volume} {113}},\ \bibinfo
  {pages} {032210} (\bibinfo {year} {2026})}\BibitemShut {NoStop}%
\bibitem [{\citenamefont {Barajas}\ and\ \citenamefont
  {Campbell}(2025)}]{Barajas2025a}%
  \BibitemOpen
  \bibfield  {author} {\bibinfo {author} {\bibfnamefont {K.~D.}\ \bibnamefont
  {Barajas}}\ and\ \bibinfo {author} {\bibfnamefont {W.~C.}\ \bibnamefont
  {Campbell}},\ }\bibfield  {title} {\bibinfo {title} {Quantum gate dynamics
  beyond the rotating-wave approximation using multi-timescale quantum
  averaging theory},\ }\href@noop {} {\  (\bibinfo {year} {2025})},\ \Eprint
  {https://arxiv.org/abs/2503.08886} {arXiv:2503.08886 [quant-ph]} \BibitemShut
  {NoStop}%
\bibitem [{\citenamefont {Landau}\ and\ \citenamefont
  {Lifshitz}(1965)}]{LandauLishitz_QM}%
  \BibitemOpen
  \bibfield  {author} {\bibinfo {author} {\bibfnamefont {L.~D.}\ \bibnamefont
  {Landau}}\ and\ \bibinfo {author} {\bibfnamefont {E.~M.}\ \bibnamefont
  {Lifshitz}},\ }\href
  {https://doi.org/https://doi.org/10.1016/0016-0032(77)90063-1} {\emph
  {\bibinfo {title} {Quantum {M}echanics, {N}on-{R}elativistic {T}heory}}},\
  \bibinfo {edition} {2nd}\ ed.\ (\bibinfo  {publisher} {Pergamon Press,
  Oxford},\ \bibinfo {year} {1965})\BibitemShut {NoStop}%
\bibitem [{\citenamefont {Yamaguchi}\ \emph {et~al.}(2017)\citenamefont
  {Yamaguchi}, \citenamefont {Yuge},\ and\ \citenamefont
  {Ogawa}}]{Yamaguchi2017}%
  \BibitemOpen
  \bibfield  {author} {\bibinfo {author} {\bibfnamefont {M.}~\bibnamefont
  {Yamaguchi}}, \bibinfo {author} {\bibfnamefont {T.}~\bibnamefont {Yuge}},\
  and\ \bibinfo {author} {\bibfnamefont {T.}~\bibnamefont {Ogawa}},\ }\bibfield
   {title} {\bibinfo {title} {Markovian quantum master equation beyond
  adiabatic regime},\ }\href {https://doi.org/10.1103/physreve.95.012136}
  {\bibfield  {journal} {\bibinfo  {journal} {Phys. Rev. E}\ }\textbf {\bibinfo
  {volume} {95}},\ \bibinfo {pages} {012136} (\bibinfo {year}
  {2017})}\BibitemShut {NoStop}%
\bibitem [{\citenamefont {Solgun}\ \emph {et~al.}(2014)\citenamefont {Solgun},
  \citenamefont {Abraham},\ and\ \citenamefont {DiVincenzo}}]{Solgun2014}%
  \BibitemOpen
  \bibfield  {author} {\bibinfo {author} {\bibfnamefont {F.}~\bibnamefont
  {Solgun}}, \bibinfo {author} {\bibfnamefont {D.~W.}\ \bibnamefont
  {Abraham}},\ and\ \bibinfo {author} {\bibfnamefont {D.~P.}\ \bibnamefont
  {DiVincenzo}},\ }\bibfield  {title} {\bibinfo {title} {Blackbox quantization
  of superconducting circuits using exact impedance synthesis},\ }\href
  {https://doi.org/10.1103/physrevb.90.134504} {\bibfield  {journal} {\bibinfo
  {journal} {Phys. Rev. B}\ }\textbf {\bibinfo {volume} {90}},\ \bibinfo
  {pages} {134504} (\bibinfo {year} {2014})}\BibitemShut {NoStop}%
\bibitem [{\citenamefont {{O. Yu. Kitsenko et
  al.}}(tion{\natexlab{a}})}]{my_upcoming_work_QAT}%
  \BibitemOpen
  \bibfield  {author} {\bibinfo {author} {\bibnamefont {{O. Yu. Kitsenko et
  al.}}},\ }\bibfield  {title} {\bibinfo {title} {Emergent quantum and
  tunneling capacitances from multi-timescale quantum dynamics},\ }\href@noop
  {} {\  (\bibinfo {year} {in preparation}{\natexlab{a}})}\BibitemShut
  {NoStop}%
\bibitem [{\citenamefont {Gyamfi}(2020)}]{Gyamfi2020}%
  \BibitemOpen
  \bibfield  {author} {\bibinfo {author} {\bibfnamefont {J.~A.}\ \bibnamefont
  {Gyamfi}},\ }\bibfield  {title} {\bibinfo {title} {Fundamentals of quantum
  mechanics in {L}iouville space},\ }\href
  {https://doi.org/10.1088/1361-6404/ab9fdd} {\bibfield  {journal} {\bibinfo
  {journal} {Eur. J. Phys.}\ }\textbf {\bibinfo {volume} {41}},\ \bibinfo
  {pages} {063002} (\bibinfo {year} {2020})}\BibitemShut {NoStop}%
\bibitem [{\citenamefont {{O. Yu. Kitsenko et
  al.}}(tion{\natexlab{b}})}]{my_upcoming_work_relaxation}%
  \BibitemOpen
  \bibfield  {author} {\bibinfo {author} {\bibnamefont {{O. Yu. Kitsenko et
  al.}}},\ }\bibfield  {title} {\bibinfo {title} {Dissipative dynamics impacts
  quantum reflectometry},\ }\href@noop {} {\  (\bibinfo {year} {in
  preparation}{\natexlab{b}})}\BibitemShut {NoStop}%
\bibitem [{\citenamefont {Paila}\ \emph {et~al.}(2009)\citenamefont {Paila},
  \citenamefont {Tuorila}, \citenamefont {Sillanpää}, \citenamefont
  {Gunnarsson}, \citenamefont {Sarkar}, \citenamefont {Makhlin}, \citenamefont
  {Thuneberg},\ and\ \citenamefont {Hakonen}}]{Paila2009}%
  \BibitemOpen
  \bibfield  {author} {\bibinfo {author} {\bibfnamefont {A.}~\bibnamefont
  {Paila}}, \bibinfo {author} {\bibfnamefont {J.}~\bibnamefont {Tuorila}},
  \bibinfo {author} {\bibfnamefont {M.}~\bibnamefont {Sillanpää}}, \bibinfo
  {author} {\bibfnamefont {D.}~\bibnamefont {Gunnarsson}}, \bibinfo {author}
  {\bibfnamefont {J.}~\bibnamefont {Sarkar}}, \bibinfo {author} {\bibfnamefont
  {Y.}~\bibnamefont {Makhlin}}, \bibinfo {author} {\bibfnamefont
  {E.}~\bibnamefont {Thuneberg}},\ and\ \bibinfo {author} {\bibfnamefont
  {P.}~\bibnamefont {Hakonen}},\ }\bibfield  {title} {\bibinfo {title}
  {Interband transitions and interference effects in superconducting qubits},\
  }\href {https://doi.org/10.1007/s11128-009-0102-4} {\bibfield  {journal}
  {\bibinfo  {journal} {Quantum Inf. Process.}\ }\textbf {\bibinfo {volume}
  {8}},\ \bibinfo {pages} {245} (\bibinfo {year} {2009})}\BibitemShut {NoStop}%
\bibitem [{\citenamefont {Cochrane}\ \emph {et~al.}(2024)\citenamefont
  {Cochrane}, \citenamefont {Seshia},\ and\ \citenamefont
  {Gonzalez-Zalba}}]{Cochrane2024}%
  \BibitemOpen
  \bibfield  {author} {\bibinfo {author} {\bibfnamefont {L.}~\bibnamefont
  {Cochrane}}, \bibinfo {author} {\bibfnamefont {A.~A.}\ \bibnamefont
  {Seshia}},\ and\ \bibinfo {author} {\bibfnamefont {M.~F.}\ \bibnamefont
  {Gonzalez-Zalba}},\ }\bibfield  {title} {\bibinfo {title} {Intrinsic noise of
  the single-electron box},\ }\href
  {https://doi.org/10.1103/physrevapplied.21.064066} {\bibfield  {journal}
  {\bibinfo  {journal} {Phys. Rev. Applied}\ }\textbf {\bibinfo {volume}
  {21}},\ \bibinfo {pages} {064066} (\bibinfo {year} {2024})}\BibitemShut
  {NoStop}%
\bibitem [{\citenamefont {Duncan}\ \emph {et~al.}(2025)\citenamefont {Duncan},
  \citenamefont {Poggi}, \citenamefont {Bukov}, \citenamefont {Zinner},\ and\
  \citenamefont {Campbell}}]{Duncan2025}%
  \BibitemOpen
  \bibfield  {author} {\bibinfo {author} {\bibfnamefont {C.~W.}\ \bibnamefont
  {Duncan}}, \bibinfo {author} {\bibfnamefont {P.~M.}\ \bibnamefont {Poggi}},
  \bibinfo {author} {\bibfnamefont {M.}~\bibnamefont {Bukov}}, \bibinfo
  {author} {\bibfnamefont {N.~T.}\ \bibnamefont {Zinner}},\ and\ \bibinfo
  {author} {\bibfnamefont {S.}~\bibnamefont {Campbell}},\ }\bibfield  {title}
  {\bibinfo {title} {Taming quantum systems: A tutorial for using
  shortcuts-to-adiabaticity, quantum optimal control, and reinforcement
  learning},\ }\href {https://doi.org/10.1103/j8c7-v2hd} {\bibfield  {journal}
  {\bibinfo  {journal} {PRX Quantum}\ }\textbf {\bibinfo {volume} {6}},\
  \bibinfo {pages} {040201} (\bibinfo {year} {2025})}\BibitemShut {NoStop}%
\bibitem [{\citenamefont {Albert}\ and\ \citenamefont
  {Jiang}(2014)}]{Albert2014}%
  \BibitemOpen
  \bibfield  {author} {\bibinfo {author} {\bibfnamefont {V.~V.}\ \bibnamefont
  {Albert}}\ and\ \bibinfo {author} {\bibfnamefont {L.}~\bibnamefont {Jiang}},\
  }\bibfield  {title} {\bibinfo {title} {Symmetries and conserved quantities in
  {L}indblad master equations},\ }\href
  {https://doi.org/10.1103/PhysRevA.89.022118} {\bibfield  {journal} {\bibinfo
  {journal} {Phys. Rev. A}\ }\textbf {\bibinfo {volume} {89}},\ \bibinfo
  {pages} {022118} (\bibinfo {year} {2014})}\BibitemShut {NoStop}%
\end{thebibliography}%
\end{document}